\documentclass[final,5p,times,twocolumn,number,round,longtitle]{elsarticle}

\usepackage{amsmath}
\usepackage{nccmath}
\usepackage{amssymb}
\usepackage[detect-all,mode=text]{siunitx}

\usepackage{graphicx}
\usepackage{subcaption}
\usepackage{epsfig}
\usepackage{multirow}

\usepackage[utf8]{inputenc}

\usepackage{booktabs}
\usepackage{xcolor}
\usepackage[dvipsnames]{xcolor}
\definecolor{codegreen}{rgb}{0,0.6,0}
\definecolor{codegray}{rgb}{0.5,0.5,0.5}
\definecolor{codepurple}{rgb}{0.58,0,0.82}
\definecolor{backcolour}{rgb}{0.95,0.95,0.92}

\usepackage{listings}
\lstdefinestyle{mystyle}{
    backgroundcolor=\color{backcolour},   
    commentstyle=\color{codegreen},
    keywordstyle=\color{magenta},
    numberstyle=\tiny\color{codegray},
    stringstyle=\color{codepurple},
    basicstyle=\ttfamily\footnotesize,
    breakatwhitespace=false,         
    breaklines=true,                 
    captionpos=b,                    
    keepspaces=true,                 
    numbers=left,                    
    numbersep=5pt,                  
    showspaces=false,                
    showstringspaces=false,
    showtabs=false,                  
    tabsize=2
}
\usepackage{hyperref}
\hypersetup{
    colorlinks=true,
    linkcolor=blue,
    filecolor=blue,      
    urlcolor=blue,
    }
\newtheorem{definition}{Definition}

\DeclareMathOperator*{\argmin}{\mathrm{arg\,min}}

\DeclareMathOperator*{\expectation}{\mathbb{E}}

\usepackage{lineno}
\usepackage{hyperref}
\usepackage[table]{xcolor}
\usepackage{stfloats}
\usepackage{cleveref}
\newcolumntype{s}{>{\columncolor{lightgray}} p{0.11\linewidth}}

\journal{Nuclear Physics B}

\begin{document}

\begin{frontmatter}

\title{On the Co-Design of Scientific Experiments and Industrial Systems}
\author[1,2,3,4]{Tommaso~Dorigo}
\author[21,31]{Pietro~Vischia}
\author[1,19]{Shahzaib~Abbas}
\author[2]{Tosin~Adewumi}
\author[2]{Lama~Alkhaled}
\author[38,39]{Lorenzo~Arsini}
\author[1,3,4,5]{Muhammad~Awais}
\author[11]{Maxim~Borisyak}
\author[2]{Andr\'as~B\'ota}
\author[1,36]{Florian~Bury}
\author[28]{Sascha~Caron}
\author[7]{James~Carzon}
\author[1,8]{Long~Chen}
\author[2]{Prakash~C.~Chhipa}
\author [2]{Paul~Christakopoulos}
\author[3,5]{Jacopo~De~Piccoli}
\author[5]{Andrea~De~Vita}
\author[25]{Zlatan~Dimitrov}
\author[5,3]{Michele~Doro}
\author[12]{Luigi~Favaro}
\author[1,2]{Francesco~Ferranti}
\author[21]{Santiago~Folgueras}
\author[2]{Rihab~Gargouri}
\author[2,8]{Nicolas~R.~Gauger}
\author[12]{Andrea~Giammanco}
\author[1,26,42,43]{Christian~Glaser}
\author[37]{Tobias~Golling}
\author[45,46]{João~A.~Gonçalves}
\author[2]{Hui~Han}
\author[1,15]{Hamza~Hanif}
\author[44]{Lukas~Heinrich}
\author [29]{Yan Chai Hum}
\author[2]{Florent~Imbert}
\author[18]{Andreas~Ipp}
\author[27]{Michael~Kagan}
\author[15]{Noor~Kainat~Syeda}
\author[34]{Rukshak~Kapoor}
\author[2]{Aparup~Khatua}
\author[23]{Eduard~J.~Kerkhoven}
\author[1,10]{Jan~Kieseler}
\author[1,8]{Tobias~Kortus}
\author[2]{Ashish~Kumar~Singh}
\author[40]{Marius~S.~Köppel}
\author[21]{Daniel~Lanchares}
\author[12]{Jeffrey~Lazar}
\author[2,7]{Ann~Lee}
\author[5]{Enrico Lupi}
\author[21]{Pelayo~Leguina}
\author[36]{Christos~Leonidopoulos}
\author[13,15]{Giuseppe~Levi}
\author[2]{Boying~Li}
\author[2]{Chang~Liu}
\author[1]{Marcus~Liwicki}
\author[1]{Karl~Lowenmark}
\author[38,39]{Carlo~Mancini-Terracciano}
\author[2]{Dominik~Maršík}
\author[2]{Leonidas~Matsakas}
\author[2]{Hamam~Mokayed}
\author[1,5,6]{Federico~Nardi}
\author[2]{Amirhossein~Nayebiastaneh}
\author[3,8]{Xuan~T.~Nguyen}
\author[35]{Aitor~Orio}
\author[1,10]{Jingjing~Pan}
\author[3]{Jigar~Patel}
\author[14]{Carmelo~Pellegrino}
\author[1,20]{María~Pereira~Martínez}
\author[48]{Karolos Potamianos}
\author[22]{Shah~Rukh~Qasim}
\author[42]{Martin~Ravn}
\author[47,5,3]{Luis~Recabarren~Vergara}
\author[17]{Humberto~Reyes-González}
\author[30]{Hipolito~A.~Riveros~Guevara}
\author[16]{Ippocratis~D.~Saltas}
\author[2]{Rajkumar~Saini}
\author[1]{Fredrik~Sandin}
\author[1,8]{Alexander~Schilling}
\author[10]{Kylian~Schmidt}
\author[22]{Nicola~Serra}
\author[24]{Saqib~Shahzad}
\author[2]{Foteini~Simistira~Liwicki}
\author[1]{Giles~C.~Strong}
\author[44]{Kristian~Tchiorniy}
\author[5]{Mia~Tosi}
\author[11]{Andrey~Ustyuzhanin}
\author[1,20]{Xabier~Cid~Vidal}
\author[37]{Kinga~A.~Wozniak}
\author[28]{Mengqing~Wu}
\author[12]{Zahraa~Zaher}

\affiliation[1]{organization={MODE Collaboration},
            }
\affiliation[2]{organization={Lule\aa \, University of Technology},
            city={Lule\aa},
            country={Sweden}}
\affiliation[3]{organization={Istituto Nazionale di Fisica Nucleare - Sezione di Padova},
            addressline={via Marzolo 8`}, 
            city={Padova},
            postcode={35131}, 
            country={Italy}
            }
\affiliation[4]{organization={Universal Scientific Education and Research Network (USERN)},
            country={Italy}
            }
\affiliation[21]{organization={Departamento de Física and ICTEA, Universidad de Oviedo},
            addressline={}, 
            city={Oviedo},
            postcode={33007}, 
            country={Spain}
            }
\affiliation[5]{organization={Dipartimento di Fisica e Astronomia G.Galilei, University of Padova},
            addressline={via Marzolo 8}, 
            city={Padova},
            postcode={35131}, 
            country={Italy}
            }
\affiliation[6]{organization={University Clermont Auvergne},
            country={France}
            }
\affiliation[7]{organization={Carnegie Mellon University},
            country={USA}
            }
\affiliation[8]{organization={RPTU University Kaiserslautern-Landau},
            country={Germany}
            }
\affiliation[9]{organization={Universidade Federal de Sao Carlos},
            country={Brazil}
            }
\affiliation[10]{organization={Karlsruhe Institute of Technology (KIT)},
            country={Germany}
            }
\affiliation[11]{organization={Constructor University Bremen},
            country={Germany}
            }
\affiliation[12]{organization={UCLouvain, Centre for Cosmology, Particle Physics and Phenomenology, CP3},
            addressline={Chemin du Cyclotron 2}, 
            city={Louvain-la-Neuve},
            postcode={1348}, 
            country={Belgium}
            }
\affiliation[13]{organization={Istituto Nazionale di Fisica Nucleare - Sezione di Bologna},
            addressline={viale Berti Pichat 6}, 
            city={Bologna},
            postcode={40127}, 
            country={Italy}
            }
\affiliation[14]{organization={Istituto Nazionale di Fisica Nucleare - CNAF},
            addressline={viale Berti Pichat 6}, 
            city={Bologna},
            postcode={40127}, 
            country={Italy}
            }
\affiliation[15]{organization={Dipartimento di Fisica - Università di Bologna},
            addressline={viale Berti Pichat 6}, 
            city={Bologna},
            postcode={40127}, 
            country={Italy}
            }
\affiliation[15]{organization={Simon Fraser University},
            addressline={}, 
            city={Burnaby},
            postcode={}, 
            country={Canada}
            }
\affiliation[16]{organization={Institute of Physics, Czech Academy of Sciences},
            addressline={Na Slovance 2, 182 21 Prague, Czechia}, 
            city={Burnaby},
            postcode={}, 
            country={Czechia}
            }

\affiliation[17]{organization={Institute for Theoretical Particle Physics and Cosmology, RWTH Aachen University},
            addressline={}, 
            city={Aachen},
            postcode={}, 
            country={Germany}
            }
\affiliation[18]{organization={Institute for Theoretical Physics, TU Wien},
            addressline={Wiedner Hauptstraße 8-10}, 
            city={Vienna},
            postcode={1040}, 
            country={Austria}
            }
\affiliation[19]{organization={University of Karachi},
            addressline={}, 
            city={Karachi},
            postcode={}, 
            country={Pakistan}
            }
\affiliation[20]{organization={Instituto Galego de Física de Altas Enerxías (IGFAE), Universidade de Santiago de Compostela},
            addressline={}, 
            city={Santiago de Compostela},
            postcode={}, 
            country={Spain}
            }
\affiliation[22]{organization={University of Zurich},
            addressline={}, 
            city={Zurich},
            postcode={}, 
            country={Switzerland}
            }
\affiliation[23]{organization={Chalmers University of Technology},
            addressline={}, 
            city={Göteborg},
            postcode={}, 
            country={Sweden}
            }
\affiliation[24]{organization={Florida Space Institute, University of Central Florida},
            city={Orlando},
            postcode={32826}, 
            country={USA}
            }
\affiliation[25]{organization={GATE Institute},
            city={Sofia},
            postcode={1164}, 
            country={Bulgaria}
            }
\affiliation[26]{organization={TU Dortmund University},
            city={Dortmund},
            postcode={44227}, 
            country={Germany}
            }
\affiliation[27]{organization={SLAC National Accelerator Laboratory},
            city={Menlo Park, CA},
            postcode={94025}, 
            country={USA}
            }
\affiliation[28]{organization={Institute for Mathematics, Astrophysics and Particle Physics, Radboud University/Nikhef},
            city={Nijmegen},
            postcode={6500 GL}, 
            country={The Netherlands}
            }
\affiliation[29]{organization={Universiti Tunku Abdul Rahman (UTAR)},
            city={Kuala Lumpur},
            postcode={31900}, 
            country={Malaysia}
            }
\affiliation[30]{organization={Universidad Nacional de Ingeniería},
            city={Lima},
            postcode={15333}, 
            country={Peru}
            }
\affiliation[31]{organization={The MODE Collaboration, https://mode-collaboration.github.io},
            city={},
            postcode={}, 
            country={Spain}
            }
\affiliation[32]{organization={Istituto Nazionale di Fisica Nucleare - Sezione di Cagliari},
            addressline={Complesso Universitario di Monserrato}, 
            city={Monserrato (Cagliari)},
            postcode={09042}, 
            country={Italy}
            }

\affiliation[33]{organization={University of Delaware},
            addressline={}, 
            city={Newark},
            postcode={}, 
            country={USA}
            }
\affiliation[34]{organization={Thapar Institute of Engineering \& Technology},
city={Patiala},
postcode = {147004},
country={India}
}
\affiliation[35]{organization={Department of Electronic Technology, Euskal Herriko Unibertsitatea},
            addressline={Ingeniero Torres Quevedo Square 1}, 
            city={Bilbao},
            postcode={48013}, 
            state={Basque Country},
            country={Spain}
            }
\affiliation[36]{organization={School of Physics \& Astronomy, University of Edinburgh},
            country={UK}
            }
\affiliation[37]{organization={Particle Physics Department, University of Geneva},
            country={Switzerland}
            }
\affiliation[38]{organization={Department of Physics, Sapienza University of Rome},
            country={Italy}
            }
\affiliation[39]{organization={Istituto Nazionale di Fisica Nucleare - Sezione di Roma},
            country={Italy}
            }
\affiliation[36]{organization={School of Physics, University of Bristol},
            country={UK}
}
\affiliation[40]{organization={Federal Institute of Technology Zurich (ETH)},
            city={Zurich},
            country={Switzerland}
            }
\affiliation[41]{organization={CERN - European Organization for Nuclear Research},
            country={Switzerland}
            }

\affiliation[42]{organization={Uppsala University},
            city={Uppsala},
            country={Sweden}
            }

\affiliation[43]{organization={Lamarr Institute for Machine Learning and Artificial Intelligence},
            city={Dortmund},
            country={Germany}
            }

\affiliation[44]{organization={Technical University of Munich, Department of Physics},
            city={Munich},
            country={Germany}
            }

\affiliation[45]{organization={Laboratório de Instrumentação e Física de Partículas, Pheno Group},
            city={Lisbon},
            country={Portugal}
            }

\affiliation[46]{organization={University of Lisbon, Instituto Superior Técnico, Department of Physics},
            city={Lisbon},
            country={Portugal}
            }
\affiliation[47]{organization={Centro di Ateneo di Studi e Attività Spaziali "Giuseppe Colombo"},
            addressline={Via Venezia 15}, 
            city={Padova},
            citysep={}, 
            postcode={ I-35131}, 
            country={Italy}}
\affiliation[48]{organization={University of Warwick},
            addressline={}, 
            city={Coventry},
            citysep={}, 
            postcode={CV4 7AL}, 
            country={United Kingdom}}

\begin{abstract} 
The optimization of large experiments in fundamental science, such as detectors for subnuclear physics at particle colliders, shares with the optimization of complex systems for industrial or societal applications the common issue of addressing the inter-relation between parameters describing the hardware used in data production and parameters used to analyse those data. While in many cases this coupling can be ignored -- when the problem can be successfully factored into simpler sub-tasks and the latter addressed serially -- there are situations in which that approach fails to converge to the absolute maximum of expected performance, as it results in a mis-alignment of the optimized hardware and software solutions. In this work we consider a few use cases of interest in fundamental science collected primarily from particle physics and related areas, and a pot-pourri of industrial and societal applications where the matter is similarly of relevance. We discuss the emergence of strong hardware-software coupling in some of those systems, as well as co-design procedures that may be deployed to identify the global maximum of their relevant utility functions.   
We observe how numerous opportunities exist to advance methods and tools for hardware-software co-design optimization, bridging fundamental science and industry through application- and challenge-driven projects, and shaping the future of scientific experiments and industrial systems.

\end{abstract}


\end{frontmatter}

\tableofcontents

\clearpage

\section{Introduction}\label{s:intro} 

The increasing complexity of modern scientific instrumentation in fundamental physics and related domains poses significant challenges for system design and optimization. 
Large-scale experiments, such as particle detectors and astrophysical observatories, involve high-dimensional hardware and software setups. 
As a result, the system-level performance---such as resolution, efficiency and cost---emerges from the coupled interplay of hardware components ({\it e.g.} sensors, electronics, cooling and trigger systems) and software elements, including data processing, calibration and reconstruction algorithms. In addition, the underlying stochastic physics processes make the resulting system behaviour highly non-linear and often require computationally expensive simulations.
The traditional approach to design and optimize such systems relies on sequential workflows, in which software development and tuning typically occur only after the hardware design has been finalized.

This limitation has motivated growing interest in \emph{co-design} approaches, where hardware and software parameters are treated as jointly optimizable variables. In co-design settings, design choices in one domain constrain and reshape attainable performance in the other through shared resource budgets, non-linear performance dependencies, algorithmic sensitivity to data characteristics, and system-level constraints. When such couplings are present, serial optimization strategies may become much more likely to converge to suboptimal solutions~\cite{946008}, and a holistic optimization treatment becomes necessary to approach system performance limits.

In parallel with this conceptual shift, recent years have seen a rapid expansion of artificial intelligence (AI) methods across a broad range of activities in fundamental physics~\citep{Karagiorgi2022,Carleo2019}, including particle and astroparticle physics, high-energy nuclear physics, neutrino experiments, gravitational-wave observatories, phenomenology, material science, and scientific computing. AI techniques have become central to data processing, event reconstruction, simulation acceleration, and inference, fostering new collaborations that bridge traditionally distinct communities with heterogeneous methodologies, data structures, and scientific objectives. Organized efforts such as the EUCAIF coalition\footnote{See \url{https://eucaif.org} .}, within which the present work was originally conceived, the MODE Collaboration\footnote{See \url{https://mode-collaboration.github.io} .}, and FastML\footnote{See \url {https://fastmachinelearning.org/} .}, among others, exemplify this broader trend toward shared AI-driven approaches across experimental and theoretical domains.

These developments have demonstrated the potential of AI to significantly enhance experimental sensitivity and to improve our understanding of physical phenomena at extreme scales, densities, and rarity. At the same time, they have highlighted the importance of viewing AI not merely as a downstream analysis tool but as a component interacting with upstream design choices. Design decisions that affect data quality, structure, and availability can only be meaningfully assessed in light of the full data-processing and inference pipeline, reinforcing the need for integrated hardware–software optimization strategies.

\paragraph{Relation to prior work and terminology}
Joint optimization of interacting system components has a long history in engineering, applied mathematics, and operations research, often developed under different names with different emphases. Closely related formulations appear, {\it e.g.}, in the literature on \emph{Engineering Design Optimization}~\citep{Martins2021} or \emph{Multidisciplinary Design Optimization (MDO)}~\citep{1994, Martins2013} that couple physical design variables with algorithmic or control parameters. The usage of AI methods for experiment design is also intensively studied in, {\it e.g.}, Ref.~\citep{Nolte2025}. In this work, we use the term \emph{co-design} to emphasize the explicit and systematic treatment of hardware and software parameters as jointly optimizable variables.
Specifically, in this work we formalize hardware–software co-design within a unified optimization framework, characterize when sequential strategies may fail to reach the Pareto front, and illustrate these concepts across representative scientific and industrial systems.
\subsection{Coupling, Decomposability and Separability}\label{s:sr} 

We formalize the co-design problem as an optimization task over two classes of design variables: hardware parameters $h \in \mathcal{H}$ and software parameters $s \in \mathcal{S}$. Let the feasible set be $\mathcal{F} \subseteq \mathcal{H} \times \mathcal{S}$, allowing for both domain-specific and coupled constraints. System performance is summarized by an objective (or loss) function
\[
\min_{(h,s) \in \mathcal{F}} f(h,s),
\]
where $f$ may be non-convex, non-smooth, noisy, and defined only implicitly through simulation or experimental evaluation. For example, the objective may be a monotonic function of a system-specific utility, and its global minimum corresponds with the optimal co-design. For clarity, we assume that the objective function has a unique global minimum, however, all conclusions are applicable for the multiple local minima settings.

To understand the necessity of joint optimization of hardware and software, we first consider optimization problems in which hardware can be optimized independently from the software or {\it vice versa}. Formally, for any sufficiently regular optimization problem, hardware and software optimizations can be decoupled if one considers marginalized objective functions:
\begin{align*}
    f_h(s) &= \min_h f(h, s); & f_s(h) &= \min_s f(h, s);\\
    s^* &= \argmin_s f_h(s); & h^* &= \argmin_h f_s(h).
\end{align*}
While such marginalized objectives formally exist, they are rarely practical: computing marginals via inner optimization effectively amounts to coupled optimization, and the bounds $f_h$ or $f_s$ are seldom known analytically. In such coupled settings, sequential strategies are known to tend to find suboptimal solutions~\cite{946008}. Nonetheless, in some scenarios, a monotone transformation of the bounds is known. As an example, consider a regression problem $x \to y$ where the target $y$ deterministically depends on the ground-truth event $\hat{x}$, $y = g_0(\hat{x})$, the measurement $x$ is noisy, with $x \sim \mathcal{N}\left(\hat{x}, \sigma^2(h)\right)$, where the noise variance depends on the design $h$. While the exact bounds are, in general, analytically intractable, it is clear that, for a sufficiently powerful estimator $g\left(x \mid h\right)$, the performance of the system strictly increases with decreasing $\sigma^2(h)$, thus, the optimal hardware design can be obtained without considering the estimator.
However, such examples are rare. Consider the following common extension of the example above --- a complex noise model with additive and multiplicative components: $x = \varepsilon_m \hat{x} + \varepsilon_a$, where $\varepsilon_m \sim \mathrm{LogNormal}\left(0, \sigma^2_m(h)\right)$ and $\varepsilon_a \sim \mathcal{N}(0, \sigma^2_a(h))$. In this case, the optimal hardware configuration can be determined independently only if $h^*$ can simultaneously minimize $\sigma_m^2(h)$ and $\sigma_a^2(h)$, otherwise the estimator must be considered to resolve the trade-off between different types of noises.

In practice, the marginalization bounds are frequently replaced by heuristic estimates derived from the expert knowledge. For example, A.~Baranov~et~al.~\citep{baranov2017optimising} optimize the muon shield design of the proposed CERN SHiP experiment~\cite{ahdida2022ship} (see \autoref{ss:pp} for details). The goal of the muon shield is to deflect a large background muon flux away from the detector, improving the signal significance. Evaluation of this metric is computationally costly and the authors use a loss function based on an analytical expression for muon ``importance weights'', effectively estimating impact on the signal significance by a simple, but well-motivated heuristic. As we show in this work, such heuristics often lead to improvements over the baseline designs, but often fall short of the principled co-design approach.

\begin{figure*}[h!t]
    \centering
    \begin{subfigure}{0.45\textwidth}
        \includegraphics[width=\linewidth]{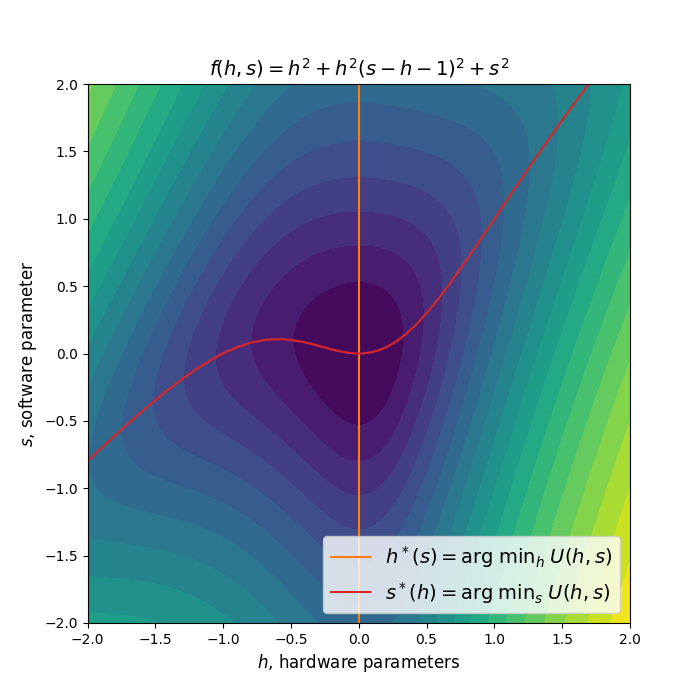}
        \caption{An example of sequential functions that admits only a single order of optimization steps.}
        \label{fig:not-fully-separable}
    \end{subfigure}
    ~
    \begin{subfigure}{0.45\textwidth}
        \includegraphics[width=\linewidth]{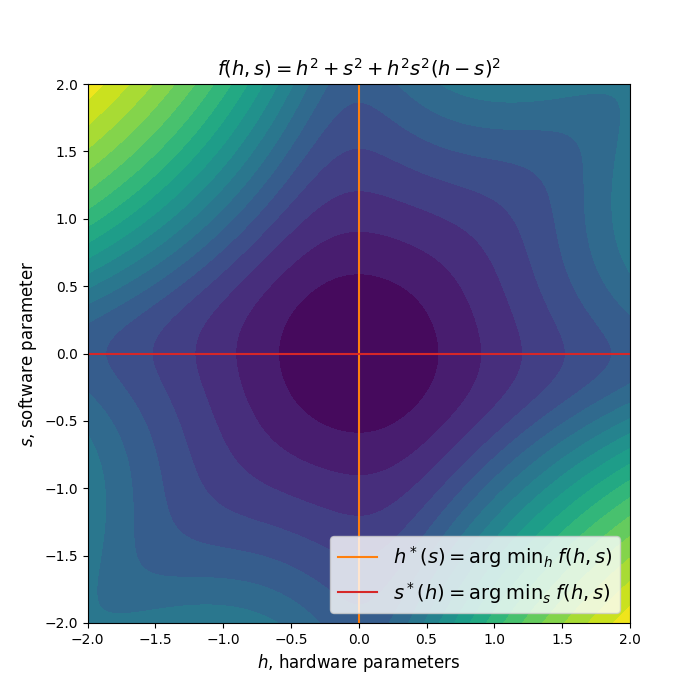}
        \caption{An example of sequential but not block-decomposable problem.}
        \label{fig:non-decomposable}
    \end{subfigure}
    \caption{Simple examples of different classes of optimization problems. For visual clarity, levels are scaled uniformly in \(\log\left(1 + f(h, s)\right)\).}
    \label{fig:counter-examples}
\end{figure*}

In general, marginalized objectives or their monotone transformations are not easily accessible, thus, we consider only \emph{practically} decoupled problems, to which we refer as sequential.
\begin{definition}\label{def:1}
An optimization problem $f(h, s) \to \min$ is called sequential if, for any initial $(h_0, s_0) \in \mathcal{H} \times \mathcal{S}$, either:
\begin{align*}
    h^* &= \argmin_{h} f(h, s_0);\\
    s^* &= \argmin_{s} f(h^*, s),
\end{align*}
or
\begin{align*}
    s^* &= \argmin_{s} f(h_0, s);\\
    h^* &= \argmin_{h} f(h, s^*).
\end{align*}
achieves the global minimum $(h^*, s^*) = \argmin f(h, s)$.
\end{definition}
It is worth noting that some problems admit only a single order of optimization steps. A simple analytical example of such a problem is shown in~\autoref{fig:not-fully-separable}.

A frequently appearing class of optimization problems is block-decomposable problems.
\begin{definition}
    An optimization problem is called block-decomposable if the objective function can be represented as:
    \begin{equation*}
        f(h, s) = \Phi\left(A(h), B(s)\right)\,;
    \end{equation*}
    where $\Phi(\cdot, \cdot)$ is monotonically increasing in each argument.
\end{definition}
A common subclass of block-decomposable problems is block-separable problems: $f(h, s) = \Psi\left(A(h) + B(s)\right)$, where $\Psi(\cdot)$ is an increasing function.

Due to the monotonicity of function \(\Phi\), a block-decomposable problem can be replaced by two independent problems:
\begin{align*}
    h^*, s^* = \argmin f(h, s) \;\Longleftrightarrow\;\begin{cases}
        h^* = \argmin_h A(h);\\
        s^* = \argmin_s B(s);
    \end{cases}
\end{align*}
therefore, block-decomposable problems are a subclass of sequential problems. It is worth noting that not all sequential problems are block-decomposable. An example of such a function is shown in ~\autoref{fig:non-decomposable}.

By contrast, in \emph{non-block-decomposable} problems, the objective or the feasible set contains genuine coupling between $h$ and $s$. A simple illustrative model is
\[
f(h,s) = A(h) + B(s) + h^\top M s,
\]
where $M$ is a coupling matrix, and the interaction term $h^\top M s$ encodes cross-dependence between hardware and software parameters. When such terms are present, improvements in one parameter block depend on the value of the other, and independent optimization is generally suboptimal. Non-zero mixed partial derivatives provide a local diagnostic of coupling in a given parameterization, although such tests are not invariant under monotone reparametrizations of the objective and therefore do not constitute a general separability criterion.

A useful quantitative measure of coupling impact is the suboptimality ratio
\begin{equation}
SR = \frac{f_{\mathrm{joint}}}{\mathbb{E}_{(h_0,s_0)\sim P}\!\left[f_{\mathrm{serial}}(h_0,s_0)\right]}\,,
\label{eq:sr}
\end{equation}
where $f_{\mathrm{joint}}$ is the optimal value obtained by joint optimization and the denominator denotes the expected objective value obtained by a prescribed serial procedure initialized from a reference distribution $P$ over feasible starting points. Values significantly above unity (for maximization problems, or below unity for minimization) indicate substantial performance loss from neglecting coupling.

\subsection{Optimization Strategies for Co-Design}

Different optimization strategies correspond to different assumptions about coupling strength and computational tractability.

\paragraph{Staged (one-pass) optimization}
In staged workflows, one block of variables is optimized once while the other is fixed, and the result is then frozen. For example, hardware parameters are first selected by minimizing $f(h,s_0)$ for a reference software configuration $s_0$, after which software parameters are optimized with hardware fixed. This approach reflects traditional engineering pipelines, and is appropriate only when coupling is weak or the problem is block-decomposable. In strongly coupled settings, staged optimization typically produces suboptimal designs because early decisions are not revisited.

\paragraph{Block-iterative optimization}
Block-iterative methods alternate between updates of hardware and software parameter blocks, enabling feedback between design layers. Given $(h^{(t)}, s^{(t)})$, a generic block coordinate scheme performs updates of the form
\begin{align*}
s^{(t+1)} & \approx \arg\min_{s:(h^{(t)},s)\in\mathcal{F}} f(h^{(t)}, s), \\    
h^{(t+1)} & \approx \arg\min_{h:(h,s^{(t+1)})\in\mathcal{F}} f(h, s^{(t+1)}),
\end{align*}
where block subproblems are solved either exactly or approximately. When each block is minimized to high accuracy, the method corresponds to block Gauss–Seidel alternating minimization; when only descent steps are taken ({\it e.g.}, gradient or surrogate-based steps), one obtains inexact block coordinate descent methods~\citep{Bertsekas2016, Beck2017}. These approaches are widely used when joint optimization is computationally prohibitive, but they may converge slowly or become trapped in poor stationary points in non-decomposable, non-convex problems.

\paragraph{Joint optimization}
Joint methods update hardware and software parameters simultaneously in the combined space. In smooth continuous settings, a prototype update is joint gradient descent,
\[
\begin{pmatrix} h^{(t+1)} \\ s^{(t+1)} \end{pmatrix}
=
\begin{pmatrix} h^{(t)} \\ s^{(t)} \end{pmatrix}
-
\alpha^{(t)}
\nabla f\!\big(h^{(t)}, s^{(t)}\big),
\]
while in black-box or mixed discrete–continuous settings, derivative-free, surrogate-based, or evolutionary strategies are used~\citep{Shahriari2016,Zhan2022}. Joint methods can follow coupled descent directions and are, in principle, best suited to strongly interacting design variables, but they are often the most computationally demanding.

\begin{figure*}[h!t]
\includegraphics[width=0.99\linewidth]{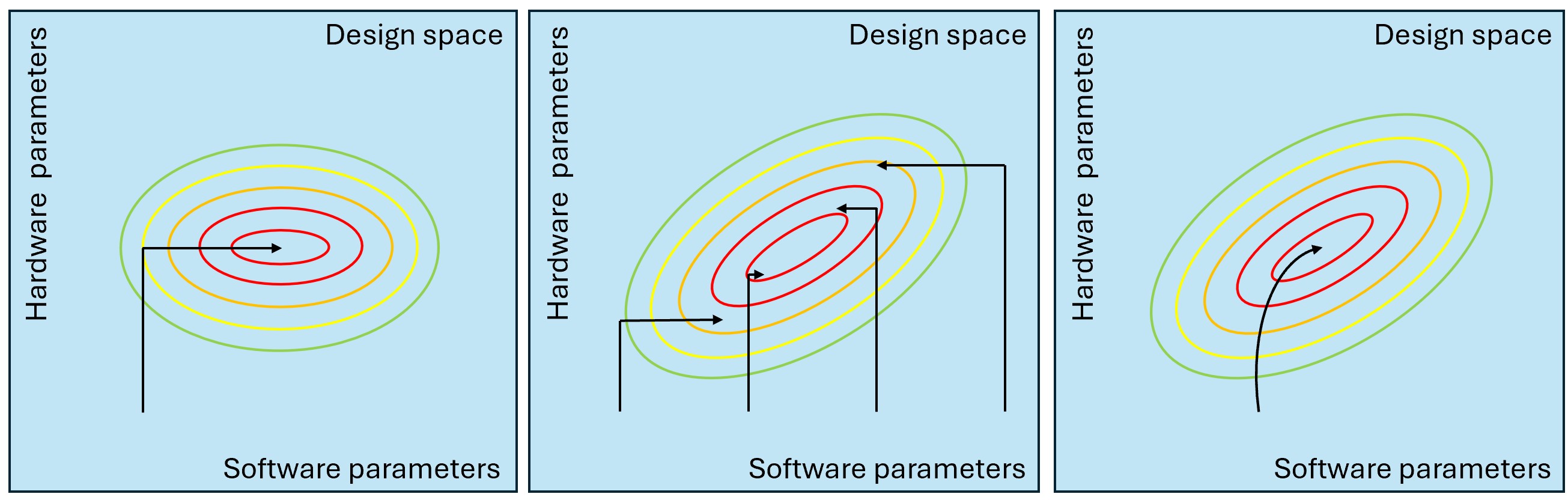}
\caption{Schematic of the result of optimization schemes on block-separable (left) and non-block-separable (center and right) problems. The elliptic contours identify points at same value of the utility. On the left, in a separable problem a hardware optimization (vertical arrow) followed by a software tuning (horizontal arrow) will always reach the highest value of the utility. At the center and on the right, in a non-separable problem the outcome of that {\em modus operandi} 
will depend on the initial software chosen for the hardware parameters scan -- here exemplified as four equispaced initial parameter values on the horizontal axis. On the right, gradient descent on the full space identifies the global extremum of the utility. }
\label{f:figure1}
\end{figure*}

An illustration of the results achievable by these optimization schemes in non-block-separable and block-separable problems is provided in \autoref{f:figure1}

\subsection{AI as an Enabler of Practical Co-Design}

Although the mathematical structure of coupled optimization problems is well understood, practical co-design faces challenges that go beyond formal optimization frameworks. Objectives are frequently simulation-based and noisy; evaluations are expensive; design spaces are high-dimensional and include discrete and constrained variables; and expert knowledge is often embedded in iterative human workflows rather than explicit models.

Recent advances in AI provide new tools to address these challenges. Machine Learning (ML) methods enable surrogate modeling of expensive simulations, learned performance predictors, adaptive sampling of large parameter spaces, and data-driven discovery of design trade-offs. These capabilities make joint and block-iterative optimization strategies computationally feasible in regimes where classical approaches are intractable. At the same time, ensuring that AI methods enforce physical constraints and are robust, interpretable, and trustworthy is critical in scientific and safety-relevant systems.

The purpose of this work is not to propose a single universal optimization algorithm, but to clarify the structure of hardware–software co-design problems, relate them to established optimization strategies, and assess where modern AI methods provide meaningful leverage. Through representative use cases, we identify when co-design is essential, when staged approaches suffice, and what open challenges remain at the interface of physics, engineering, and data-driven optimization. This will be the focus of \Cref{s:sota,s:cd_science,s:cd_industry}.

\subsection {Categories of non-decomposable problems} 

For problems that are not block-decomposable, non-block-diagonal elements exist in the corresponding objective functions. These elements often arise from common underlying causes that can be categorized into several principal cases, as outlined below.

\begin{itemize}
    \item presence of joint resource constraints, {\it e.g.} in a system where a complex algorithm performs an iterative optimization with final performance depending on the total number of iterations, and the latter is constrained by hardware clock and/or memory bandwidth; 
     or when software hyperparameters (such as batch size) interact with GPU memory size in the training of a ML model. In these situations, hardware and software parameters may still be separable in the utility, but the feasible set fulfilling the budget constraints is coupled;    
     \item factors in the utility that couple hardware and software parameters through a non-linear dependence, such as a ratio. {\it e.g.}, in a telescope where one wants the best possible angular resolution, aperture $D$ (a hardware parameter) defines the theoretical limit on the resolution, 
     but the end result depends on a noise-robust deblurring performed by software deconvolution. One may define \par
    
    \begin{equation}
    U(D,\xi) = \frac{1}{(\lambda/D)f({\xi})}\,,
    \end{equation}
    
    where $\lambda$ is light wavelength and $f(\xi)$ describes the deblurring factor. In reality, $f$ will depend on $D$, too, as deconvolution works better at higher signal-to-noise ratio. The interaction of the two terms does not allow decomposition of the objective; 
    \item multiplicative terms that couple hardware and software parameters also arise in many systems where energy consumption is relevant. The hardware power draw of an edge 
    device (such as a small sensor or a wearable device) may scale with clock cycles or voltage; software runtime $t$ will scale with the algorithm parameters $\xi$, but also on the hardware features $\theta$, {\it e.g.} in a ratio \par
    \begin{equation} 
    t(\theta,\xi) = O(\xi)/T(\theta)\,, 
    \end{equation}

    where $O(\cdot)$ defines the operations and $T(\cdot)$ the throughput, and when the latter depends on frequency, memory, and other hardware parameters. Energy consumption combines these factors in a non-decomposable product through the hidden cross-dependencies of  parameters. It should be noted that while a log-transform could make the utility additive, the presence of constraints will usually hamper serial optimization of these systems;
    \item systems where effective performance is the inverse of the maximum latency of a process which compounds a hardware-driven and a software-driven part: max/min operators couple the two domains and does not allow decomposition. A classical example in this category is the trigger system of a particle collider. The set of requirements posed by the data acquisition system on high-rate input data coming from particle collisions risk producing
    an impedance mismatch of the latency of a separately designed hardware-based low-level trigger system and the successively optimized software latency resulting from algorithms sorting out the pre-filtered data, resulting in suboptimal overall system performance;
    \item experiments where systematic uncertainties on a parameter of interest $x$ (the scientific target) depend on detector design parameters $\theta$ (such as segmentation, material budget, acceptance, resolutions, geometry, calibration hardware), as well as on software parameters $\xi$ (present in reconstruction algorithms, calibration constants, ML models). Utility in this case could be expressed as 
    \par
    \begin{equation}
    U(\theta, \xi)=\frac{1}{\sigma^2_{stat}(\theta,\xi)+\sigma^2_{syst}(\theta, \xi)}\,, 
    \end{equation}
    
    where statistical and systematic uncertainties both affect precision of inference on $x$. Coupling of hardware and software optimization exists even in cases when the statistical uncertainty is solely dependent on hardware parameters ({\it e.g.} efficiency, acceptance), as systematic uncertainties almost invariably produce a strong coupling between hardware and software. A classical example of this interplay in a particle collider setup is jet energy scale calibration, which is affected by hardware choices (both calorimeter design and tracking performance, as the latter influences inter-calibrations) and clustering and calibration algorithms. 
\end{itemize}

The above is an incomplete list, and should only be meant as an introductory glance at the diversity of situations where co-design is an unavoidable paradigm in the way to the search for a global optimum. Without the ambition of being exhaustive, in this document we will focus in more depth on a few key examples from scientific and industrial research, to unearth not only the existence of hardware-software coupling, but to try and assess how far from unity the ratio $SR$ of~\autoref{eq:sr} may result from a disjoint optimization of the different parts of the systems under consideration. While only a complete treatment of those optimization problems may allow precise estimates of the related value of the sub-optimality ratio $SR$, we believe that even a rougher back-of-the-envelope calculation is quite useful in directing researchers, given the resource-intensive nature of holistic optimization solutions.

\subsection{Decomposability and parametrization}

In some cases, a co-design optimization problem can be reduced to a decomposable problem even if the original problem is not decomposable. As a clear example, consider the straw tracker from the proposed Search for Hidden Particles (SHiP) experiment~\cite{ahdida2022ship}. The straw tracker is a gaseous drift detector based on straw tubes. Its design consists of two tracking telescopes located in the vacuum vessel, upstream and downstream of the magnet, and each composed of two tracking stations. Each station has several layers of five-meter-long tubes; within each station, the layers are arranged into four views, in a Y-U-V-Y pattern, where U and V are stereo views with straws rotated by a small angle $\pm \theta$ around the Z-axis with respect to the Y-measuring straws.
The tracker aims to accurately reconstruct the decay vertex, the mass, and the impact parameter of the hidden particle trajectory. Station positions and the stereo angle are the geometric variables that impact the detector performance most significantly.

Consider the task of optimizing the geometry of the tracker; the objective function can be written as following:
\begin{align*}
    f(h, s) = \expectation_{x, y \sim P(x, y \mid h)} l\left(g_s(x), y\right);
\end{align*}
where $y$ denotes the target properties ({\it e.g.}, decay vertex, mass), $x$ the response of the detector with geometry $h$, $g_s(\cdot)$ the reconstruction algorithm with parameters~$s$, and $l(\cdot, \cdot)$ the loss function ({\it e.g.}, mean squared error).

When using machine-learning algorithms for reconstruction, it is natural to represent the detector response as a vector $x = \{v_i\}^{N}_{i = 1}$ where each component $v_i$ represents the signal reported by the $i$-th straw, a binary indicator in the simplest case~\citep{gagnon2022machine}. In this parameterization the optimal reconstruction parameters $s$ change as the detector geometry changes; thus, the detector design $h$ is coupled with the reconstruction parameters $s$, making the problem, in general, non-decomposable~---~the reconstruction algorithm must be tuned to account for changes of the geometry.

There are general-purpose reconstruction algorithm, which typically fit particle trajectories to the observed points~\cite{rauch2021development, ai2022common}. Such algorithms often operate with a different parameterization, which for the straw tracker might take the form $x = \left\{(a_i, b_i, r_i, v_i)\right\}^N_{i = 1}$ where $a_i, b_i \in \mathbb{R}^3$ are the endpoints of the $i$-th tube, $r_i$ is its radius and $v_i$ is the recorded signal. In this representation, the geometry is included into the event description; and the algorithm is effectively ``pre-adjusted'' for any possible detector design. The parameters $s$ may depend on distributions of vertices, particle momenta or properties of the magnet, but these are independent of the detector geometry. As a result, such fit-based reconstruction algorithms are universal in this sense, and because the software does not need to be modified when the hardware changes, the problem becomes trivially decomposable. This example demonstrates that decomposability of a co-design optimization problem in some cases depends on the parametrization: in these scenarios, it may be possible to decouple hardware and software. However, as examples in this work demonstrate, such reparametrization is not always feasible or practically relevant.

\subsection{The situation outside fundamental physics} 

There is a clear convergence in the literature of the past decade on hardware-software co-design as a means to maximize performance~\cite{Ha2017}. In high-end scientific computing, for example, co-design ensures that the next generation of supercomputers and their software ecosystems evolve in tandem~\cite{ceed}. Another area where co-design has been identified as a crucial element for the design of cyber-physical and control systems, where a strong coupling between physical dynamics and digital infrastructure demands joint communication-control solutions for robustness~\cite{casestudy1,codesign_wireless}. Further, in emerging edge computing deployments, the need to align application logic and network and hardware capabilities is seen today as the logical pathway to meet the strict latency and reliability goals of these systems~\cite{codesign_edge}. Similarly, in the context of the internet of things, joint optimization of sensing hardware and algorithms pushes the boundary of efficiency and performance of these devices~\cite{sacod}. 

Overall, it is clear that the co-design paradigm has increasingly emerged over the years as a necessary ingredient in the conception and optimization of all systems complex enough to instantiate cross-talk between their hardware and software elements~\cite{Ha2017,Ahmad2017,sacod,grosjean}.
A growing body of related work, to which we will return in more detail in \autoref{s:sota} below, provides both methodological frameworks and case-study evidence to guide the effective design of future projects. 

\subsection {Structure of this document} 

This work is organized as follows. In \autoref{s:sota} we examine existing literature in co-design, and provide a brief survey of relevant optimization efforts for scientific experiments and for industrial use cases recently reported in the literature, with the purpose of taking a snapshot of the current level of complexity of the deployed models. The comparison highlights the existence of a significant gap, which is mainly due to the different amount of resources available in the two domains. We finally offer a concise description of the main methods that can be used for co-design optimization problems. 
In \autoref{s:cd_science} we discuss a number of use-cases in experimental physics, 
reviewing benefits and/or necessity of a holistic  hardware-software optimisation in detection and physics reconstruction with several examples: cases include particle physics tracking (a highly complex and CPU-demanding task), the usage of Field-Programmable Gate Arrays (FPGA) for real-time online selection, coding opportunities and risks of employing Large Language Models (LLM) in modern large experiments, the non-factorizability of the experimental setup in the calibration of Cherenkov telescope cameras, the joint optimisation of geometry and layout for an ultra-high-energy neutrino observatory, and the coherent optimisation of  hardware and software for muon tomography or for the detection of gravitational waves. 
In \autoref{s:cd_industry} we echo the structure of \autoref{s:cd_science} by dealing with a small set of industrial applications where co-design is important. These examples are not exhaustive, because the focus of this document is on fundamental science rather than industry; rather, they are presented for the purpose of illustrating the different issues arising in typical industrial settings, and the limited amount of synergy they offer with co-design problems in scientific experimentation. 
We provide some concluding remarks in \autoref{s:conclusions}.

\vskip 1cm

\section{Related Work \label{s:sota}}

The present work focuses on contributions to hardware-software co-design in fundamental physics and industrial applications. However, the concept of co-design has been applied to other fields as well, where the object of the co-design has been not only the interplay between hardware and software, but also the broader interplay between stakeholders, the one between experiment and simulation, and the one involved in AI-assisted automation. In this section, we provide therefore a overview of the existing works that we consider most relevant for both hardware-software co-design and co-design in the broader sense.

\subsection{Related work in hardware-software codesign}

We overview here relevant works in the field of hardware-software optimization in fundamental physics and industrial applications. These works are directly related to the activities we describe in~\autoref{s:cd_science} onwards.

\subsubsection{Satellites and other space vehicles}

Although not explicitly named as such, co-design was an essential component of the success of several {\em earth observation satellite programs} of the European Space Agency (ESA), such as the Phi-Sat and Sentinel programs. Satellites have strict limits on power, mass, thermal budget, communication bandwidth, processing power, and storage. Software, especially AI and ML applications, must be optimized to run efficiently under these constraints.

In Ref.~\cite{sat1}, the authors discuss onboard processing in satellite-borne synthetic aperture radar technology, highlighting the need to optimize the use of limited storage and bandwidth to reduce downlinked data volume. The authors review several techniques that can be used for this purpose, including deep learning. The authors conclude that onboard processing improves performance in the field of remote sensing, especially in real-time processing, impacting areas such as disaster monitoring and global security.

The study in Ref.~\cite{sat2} presents the data transformations and autocalibration of a Sentinel-2 (S-2) end-to-end model processor, an onboard cloud detection and atmospheric correction processor. The proposed approach employs two independent convolutional neural networks: the first one classifies clouds and cloud shadows, the second one emulates the operational software developed by the ESA. The proposed model is shown to provide up to a tenfold speed reduction, surpassing existing algorithms.

\subsubsection{Robotics}

Co-design can also be important in robotics, where good performance necessitates a tight coupling between physical mechanisms, sensing, and control. Hardware alone cannot guarantee agility, efficiency, or robustness, and software cannot compensate for physical limitations that could have been avoided through better design choices. Joint optimization of the various components involved in robots can achieve higher performance with lower energy use, reduced complexity, and greater reliability.

In Ref.~\cite{robo1} the authors employ co-design in the optimization of stochastic partial differential equations for the design of actuators. They propose an episodic reinforcement optimization strategy which combines policy network optimization with actuator design optimization. The applicability of the approach is demonstrated through several successful experiments on a complex non-linear 2D second-order stochastic partial differential equation model of a soft-robotic limb.

In a characteristic example of co-design, the study in Ref.~\cite{robo2} presents the joint optimization of hardware and control to improve the energy efficiency of complex legged robots. A stochastic programming approach was employed to co-optimize robot morphology, a nominal trajectory, and a feedback control policy, and can be employed across different locomotion scenarios. The optimization framework is demonstrated using the MIT mini cheetah robot, where it led to an up to 17.4\% energy reduction with the robot executing trotting gaits.

\subsubsection{Imaging}

Co-design is also emerging in imaging applications, where optical hardware and computational algorithms are jointly optimized for the overall system performance. Metasurfaces are ultrathin structures composed of subwavelength elements that have revolutionized light manipulation by enabling precise control over electromagnetic waves characteristics such as amplitude, phase, polarization, and spectral properties while maintaining a very compact form factor. Among the various planar devices based on metasurfaces, metalenses have been gaining prominence for the development of ultracompact imaging systems; they have enabled highly miniaturized optical systems compared to conventional optics, commonly known as metaoptics~\cite{Yu2014,Khorasaninejad2017}, which is essential for integration into portable devices for different applications  such as augmented/virtual reality, drones, and healthcare devices~\cite{Li2021,Chen22,Thomas2024}. Concurrently, computational imaging leverages algorithms to reconstruct images from optically processed signals, overcoming limitations of traditional imaging systems~\cite{Satat2017}. 

A synergistic co-design of metaoptics (hardware) and computational imaging (software) which combines the physical wavefront shaping ability of metalenses with advanced computational algorithms, has the potential to enhance imaging performance beyond conventional limits, enabling improved resolution, reduced system size, and enhanced functionality~\cite{Park2025,Roques-Carmes2025}. The design of metalenses requires the use of multiscale electromagnetic approaches as ray tracing and Maxwell's equations solvers~\cite{Thomas2024}. Metalenses can be represented at the ray tracing level as phase functions that need to be then implemented in actual hardware ({\it e.g.}, unit cell structures). Other optical technologies exist as refractive, diffractive and freeform optics which can be included into the hardware-software co-design framework. Working with mathematical phase functions, for example at the ray tracing level, enables a design step that decomposes the optical design into a set of phase functions that can be then implemented in hardware by exploring the set of existing technologies as metaoptics, refractive, diffractive and freeform optics or hybrid combinations~\cite{Nikolov2021, Thomas2024}. Such an approach can mitigate curse of dimensionality issues related to the number of design parameters, provide greater flexibility in selecting the appropriate hardware optical components and ensure that the optical system is optimized for performance, form factor, cost, and manufacturability.

\subsubsection{Particle Physics}\label{ss:pp}

In particle physics, detector optimization is an iterative process involving simulations of different detector geometries with various technology choices and reconstruction algorithms to evaluate trade-offs among detector performance, cost, and other system-level challenges such as cooling and readout, defining a multi-objective optimization problem.

\begin{figure}[!ht]
\includegraphics[width=0.99\linewidth]{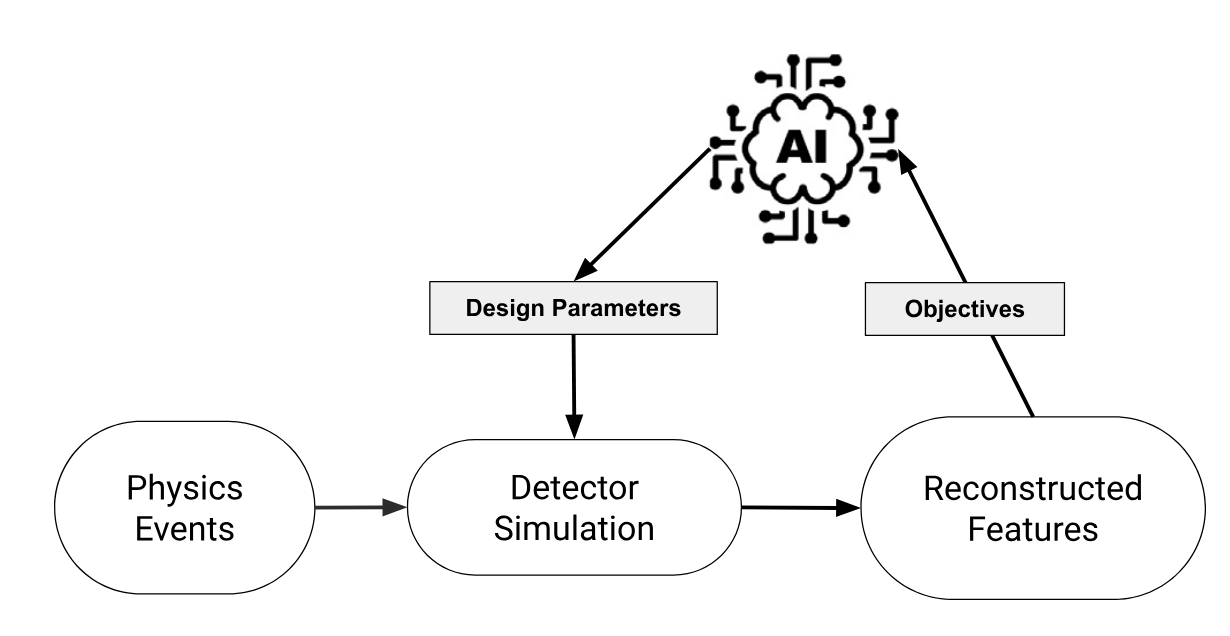}
\caption{Workflow of AI-assisted detector design~\cite{ai-tracking-eic}.}
\label{f:AI-PP-workflow}
\end{figure}

The first AI-based strategies for detector optimization were sequential, see~\autoref{f:AI-PP-workflow}: the AI elements of the pipeline are decoupled from the main simulation-reconstruction chain and act merely as an explorer of the phase space to suggest configurations that are then simulated and reconstructed by the main pipeline.
AID2E~\cite{ai-det-eic} is an AI-assisted detector design framework that has been established to realise the full potential of AI in detector design at Electron--Ion Collider (EIC). AID2E integrates a variety of AI techniques, including surrogate modeling, Bayesian optimization, and reinforcement learning, to optimize detector performance while minimizing the number of expensive simulations required. The framework has been successfully applied to optimize the design of the EIC tracking system~\cite{ai-tracking-eic}, demonstrating significant improvements in performance metrics such as momentum resolution and tracking efficiency compared to traditional design approaches, resulting, for example, in a 10\% improvement in the resolution of the $\pi^+ K^-$ invariant mass. The success of AID2E highlights the transformative potential of AI in accelerating the design process and achieving superior detector performance in particle physics experiments. 

One of the earliest concrete applications of machine learning to particle physics instrumentation was the optimization of the muon shield for the SHiP experiment~\cite{Akmete2017}. The goal of the muon shield is to suppress the muon flux emerging from a high-intensity proton beam dump by approximately six orders of magnitude using a sequence of tapered magnetic elements. The shield consists of multiple magnets, each described by several geometric parameters, resulting in a high-dimensional and strongly constrained optimization problem with highly non-linear dependencies. Initial design studies employed Bayesian optimization to efficiently explore this complex design space~\cite{baranov2017optimising}. Similar techniques were also applied at the EIC~\cite{Cisbani2020}, providing another early example of AI-driven detector design in particle physics.

Subsequent SHiP studies made a step forward by proposing a gradient-based methodology where a generative adversarial network served as a differentiable surrogate for stochastic simulations~\cite{shirobokov2020black} produced using the \textsc{Geant4} toolkit~\cite{Agostinelli2003, Geant4_1_1610988, Geant4_2_ALLISON2016186}. Gradients are propagated directly backwards through the reconstruction and simulation chain, using the surrogate. This methodology was demonstrated on many examples, including the aforementioned muon shield, and enabled full end-to-end optimization via backpropagation across the design workflow, substantially reducing iteration time. Building on this work, further efforts demonstrated making the simulation code itself differentiable~\cite{aehle2024geant4_diff} through operator overloading, as well as the continued use of surrogates~\cite{Schmidt2025}. Two recent reviews explored the use of differentiable programming for end-to-end optimization, following work produced by members of the MODE Collaboration~\cite{mode1,mode2}.
On the software design for detector optimization, gradient based adaptation of AI-based particle reconstruction across detector geometries has also been explored~\cite{PhysRevD.111.092015}. 

More recently, to complement differentiable methods and to address certain limitations, such as sensitivity to local optimization and challenges in handling discrete design choices, the use of reinforcement learning was explored in Ref.~\cite{Qasim_2025}. These studies considered the optimization of both calorimeter and spectrometer designs.

Subsequently, LLMs have also been investigated in this context~\cite{zoccheddu2026largelanguagemodelsphysics}, demonstrating their potential both as standalone design agents and in combination with a trust-region optimizer. This approach suggests a role for LLMs as high-level planners that guide specialized optimization methods such as reinforcement learning.

\subsubsection{Plasma Physics}

The vast majority of baryonic matter in the universe exists in a fourth state of matter known as plasma, spanning systems from the Sun and nearby stars to the interstellar and intergalactic media. Accurate modeling of these ubiquitous plasma states is therefore essential. Traditionally, plasma dynamics are modeled using the magnetohydrodynamics (MHD) framework~\cite{Kulsrud1983}, in which collisional processes dominate, allowing the plasma to be treated as a quasi-neutral fluid. This description is essentially an extension of neutral fluid dynamics, augmented by the Lorentz force and Maxwell’s equations. However, many plasma systems often exist in weakly collisional or collisionless regimes, where the assumptions underlying MHD break down. Examples include the solar wind, planetary magnetospheres, the interstellar medium, and the intergalactic medium. This necessitates the use of a kinetic description, in which the statistical dynamics of ions and electrons are tracked separately as particles, rather than being represented by a single fluid model. 

The particle-in-cell (PIC) method is a widely used numerical approach for simulating such kinetic plasmas in space and astrophysical systems. In PIC simulations, the dynamics of macro particles are self-consistently tracked in the presence of electric and magnetic fields. However, as system size grows, simulations of these kinetic systems become computationally expensive due to the large separation between the spatial and temporal scales of interest. A significant portion of the community effort is therefore devoted to accelerating these simulations through massive parallelization across multiple compute nodes, with recent efforts leveraging graphics processing units (GPUs) acceleration to further enhance performance. 
While the traditional PIC method maps well to the highly parallel structure of modern computing architectures, inter-accelerator communication remains a significant bottleneck. To address these issues, more recently, studies have explored co-design approaches that optimize algorithms jointly with hardware and underlying physics to accelerate the complex dynamics inherent in kinetic plasma simulations~\cite{Payne2014, Glinskiy2014}. These approaches are becoming increasingly relevant as modern computational systems grow more heterogeneous, combining CPUs with GPUs, tensor processing units (TPUs) and other domain‑specific processors. Consequently, there is a growing need for co-design approaches that can fully exploit the potential of these accelerators. 

A recent study~\cite{Payne2014} co-designed a PIC method while explicitly considering the architecture of the available accelerators such as many-integrated cores (MIC) coprocessors and GPUs. In their design, they employed a scale-bridging technique that couples a lower-order (LO) and higher-order (HO) moment-based scheme, isolating the maximum amount of computation on a single accelerator and minimizing inter-device communication. In this formulation, the LO scheme essentially solves the discretized fluid equation, while the HO scheme handles the Vlasov equation coupled with Maxwell's equations. This hierarchical approach aligns naturally with the architecture of modern computational systems and accurately predicts macroscopic plasma behavior arising from microscopic effects. While the framework is designed to support accelerators like GPUs and MIC co-processors, the initial results demonstrated the efficacy of this approach on multi-core CPUs, showing the ability to isolate large amounts of work with minimal communication. Overall, the study illustrates that through a co-design process, kinetic PIC simulations can achieve high performance on multi-core systems without sacrificing portability.

\subsection{Broader applications of co-design by field of application}

When we don't limit ourselves to hardware-software co-design, we can identify several fields of application where co-design is applied to the interplay between stakeholders, between experiment and simulation, and the one involved in AI-assisted automation. In this section, we highlight the most relevant examples of this broader definition of co-design, categorized by main field of application.

One of the main reasons of the widespread popularity of co-design across several fields of science is its value in interdisciplinary research, where combining the knowledge of scientists from different fields is essential for the success of the project. Involving stakeholders, such as end users and policy makers, is especially important in research with direct applications. As a result, co-design approaches are commonly used in large-scale projects, {\it e.g.}, in medical and health sciences~\cite{vargascodesign}, and environmental and climate research~\cite{climatereview}.

\subsubsection{Medical and health sciences}

Applied research in the {\em medical and health sciences} has several constraints that must be considered for results to be applicable in practice. Health infrastructure, including both equipment and qualified personnel, is always a limiting factor in the applicability of the research. Health solutions have to be accessible to all intended users, especially patients with disabilities and chronic health problems. Finally, policy makers influence health infrastructure at multiple levels. Successful health interventions require the involvement of all these stakeholders, making co-design approaches desirable.
In recent years, funding bodies and policy makers have became more aware of the importance of involving stakeholders to ensure the outcomes of research projects stay relevant to end users, resulting in several successful examples in Europe and beyond. An early, but widely used approach is the Experience-based co-design (ECBD) framework~\citep[{\it e.g.},][]{health2, health1, health3}, which has many successful applications.

One example aims at improving post-stroke rehabilitation in Denmark~\cite{health2}, involving stroke survivors, their relatives, medical professionals, app developers, and researchers, and employing the ECBD framework to design two mobile applications. Co-design is crucial in this application because it directly addresses a long-standing problem in stroke care: fragmented rehabilitation across hospitals, municipal services, and home settings. Participants reported that their views were heard and valued, contributing to outcomes better aligned with end-user needs. Another application focuses on cardiovascular diseases, one of the leading causes of morbidity and mortality globally. 

In Ref.~\cite{health1} the authors employ ECBD to tackle the challenges in cardiovascular and chronic disease nursing. Traditional approaches in this area often rely on professional perspectives or quantitative data alone. The introduction of co-design was essential to bridge the gap between clinical expertise and patient lived experiences. The findings of the study created greater empathy and engagement for staff, as well as insight into patient priorities allowing for more person-centered support.

Developing actionable research outputs in {\em environmental and climate research} requires collaboration between academia, industry and public policy. Employing co-design in this area makes it possible to integrate experiences from these stakeholders, building trust with them, resulting in better applicability of the results~\citep[{\it e.g.},][]{climate1, climate2, climate3}. Environmental and climate research affects multiple aspects of life, including public health, and safe and sustainable industry practices, and is becoming increasingly relevant with the ongoing climate crisis. 

A multi-year international European project has created recommendations for managing occupational heat stress~\cite{climate1}. Weather extremes pose a challenge for a  broad range of workers, but especially those who work outdoors and/or with protective equipment. Potential complications include cardiovascular conditions, respiratory problems, and kidney disease. The authors highlight the necessity of employing a co-design approach that involves
multi-disciplinary experts from academia, policy makers in the public sector, and stakeholders from industrial collaborators, in order to create recommendations that follow state-of-the-art scientific standards, but are also relevant for and applicable to actual, real-life working environments.

In Ref.~\cite{climate2}, the authors demonstrate their work on designing an app aiming at reducing energy consumption to mitigate the climate footprint of individual users in Switzerland. Taking a participatory approach, the authors involve the intended users of the app to co-design it, allowing the project to capture user motivations, constraints, insights, and address knowledge gaps. Co-design is combined with the Model of Action Phases framework~\cite{actionphases} from the behavioral sciences to create an app that supports the motivation of users, guides them towards concrete action, and helps sustain behavioral changes. The authors highlight that the co-design process enabled them to design better interventions that feel more personal, contextual, and achievable.

\subsubsection{Large language models} 

Large Language Models (LLMs) have recently emerged as a powerful cognitive tool to assist and optimize several aspects of human life and thinking. In the field of co-design, LLMs can be agents or tools to co-design a certain system, or can be themselves objects of co-design. 

When used for hardware-software co-design and system-level optimizaiton, LLMs enable reasoning over heterogeneous sources such as hardware specifications, software kernels, compiler intermediate representations, operational logs, and design constraints to support joint optimization strategies. Unlike traditional workflows that rely primarily on manual exploration of design spaces or heuristic optimization, LLMs can interpret and generate code in hardware description languages, assist in hardware‑aware software generation, and provide design suggestions that balance performance, power, latency, and resource utilization~\cite{GuoSurvey2025, Abdollahi2025, HeLLMEDA2025}. For example, LLM fine‑tuning on Verilog datasets has been shown to improve Verilog code generation quality compared to untuned models, demonstrating potential for automating parts of digital design tasks such as register-transfer level generation and functional templates~\cite{ThakurVeriGen2024, Abdollahi2025}. 

Recent work has also introduced LLM‑based high-level synthesis frameworks that combine LLM reasoning with design space exploration tools to translate high-level software code into synthesizable hardware accelerators, reducing manual effort in mapping sequential code to hybrid CPU–FPGA implementations~\cite{XiongHLSPilot2024}. Broader survey studies on LLM-centric hardware and software co-design highlight how LLMs can be integrated into design automation workflows, covering algorithm optimization, hardware generation, and system-level innovations tailored to the unique characteristics and constraints of large neural models and their deployment environments~\cite{GuoSurvey2025}. In addition, benchmarks and tooling efforts for evaluating LLM performance on hardware design tasks ({\it e.g.}, Verilog code generation and reasoning about post-synthesis metrics) are emerging, providing structured ways to assess how well models can support hardware-software co-design decisions~\cite{AbdelattyMetRex2025, KhanSAGEHLS2025, AbiKaramHLSEval2025}. These developments indicate that LLMs -- especially when adapted to hardware domains through fine-tuning, domain-specific datasets, or hybrid reasoning pipelines -- are becoming an important component of hardware-software co-design workflows, helping bridge high-level design intent and low-level implementation choices.

From a methodological perspective, LLMs can be objects of co-design themselves: in addition to deployment-time hardware–software optimization, we can therefore consider training-time co-design principles that jointly couple model scale, hardware compute, and data volume, as established by recent industrial LLM studies. These works demonstrate that compute-optimal training requires balanced scaling of parameters, data, and hardware resources, with direct implications for energy consumption and system efficiency~\cite{hoffmann2022training, wu2024towards, pearce2024scaling}. The detailed treatment of training-scale co-design, including scaling laws and energy–CO$_2$ considerations, is provided in~\autoref{llm_training}.

\vskip 1cm

\section{Co-Design in Fundamental Physics \label{s:cd_science}}

In this Section we examine a variety of typical design problems arising in experiments for fundamental physics studies, with the purpose of characterizing them according to presence of non-diagonal elements in their specifiable utility. For some of the less complex applications we also manage to present quantitative results on the effectiveness of different optimization approaches.

\subsection{Particle tracking} \label{sec:particle-tracking}

We define {\em tracking detector} a device designed to detect the trajectory of charged particles by means of ionization produced when they interact with the atoms of the detector material. These devices are essential to nearly all modern particle-physics experiments, but they are also of primary importance for detectors finding applications in space, nuclear physics, and medical domains. Here we wish to discuss what are the hardware-software co-design principles that govern their performance. We will deliberately restrict our consideration to silicon-based detectors to allow a semi-quantitative discussion, meanwhile trying to keep the insights as general as possible. 
The key question we are interested in discussing here is, in typical situations, whether sensor geometry, materials, electromagnetic fields, and readout schemes (which are elements required to define the detector hardware during the design phase) can be effectively optimized individually, while software aspects are left to be defined at a later stage. Depending on the specific application, the latter usually includes  track seeding procedures, pattern recognition, track fitting and trajectory cleaning, data-driven detector alignment procedures, background identification and removal. 
On the other hand, if the precise definition of detector hardware limits the space of software solutions, thereby limiting the final attainable performance of the instrument, then sequential, separate optimization of these two components is not optimal, and hence co-design procedures becomes crucial.

\subsubsection {Tracking as an optimization problem}

Particle tracking can be viewed as a fitting and inference task performed on discrete, noisy sensor data. The complexity of this task stems from confounding factors from physical effects. 
Incident particles interact with the sensing material to generate signals for tracking, however, they also interact with other detector components, left unwanted effects on the trajectory to be corrected. Moreover, all these interactions are inherently stochastic.
Detector inefficiencies, which must be accounted for in tracking, arise from imperfections in the sensing materials, radiation damage, or construction tolerances.
Hit ambiguities result from multiple particles traversing the same detection elements, as is common in high-density environments where those detection elements operate ({\it e.g.}, in particle colliders). 
Background noise, which affect occupancy, induce ambiguities, and can lead to loss of resolution, often come from nuclear interactions, as well as from leakage current or other readout-related effects.  

While hardware design sets the information content we wish to access, through the definition of nominal hit resolution, granularity, and number of expected hits per trajectory, the software tools operating on the produced data define how efficiently that information is extracted, and ultimately determine the statistical sufficiency of the dimensionality reduction they operate in transforming a very large,
raw input dataset (which includes position, timing, and/or deposited charge of each recorded hit) into high-level output features (particle trajectories, electric charge, momentum, and sometimes particle type). A coupling between hardware and software in a particle tracker is therefore expected to arise through the subtle yet tight interplay that connects the two parts through the effect of the confounding factors listed above. 

It is worth noting that, in an idealized situation---devoid of background noise, inefficiencies, misalignments, and all the other mentioned ancillary physical phenomena affecting the problem---the hardware specifications of a few layers of silicon pixel detectors ({\it e.g.} number of layers, vertical spacing and lateral dimensions) become effectively decoupled from the specific methods used to infer trajectories from the set of hits produced by a single charged particle passing through the layers. In this situation, the problem  becomes well-defined and falls back within the range of exact statistical inference; a likelihood function can be formulated, co-design becomes unnecessary, and the optical solution is, in principle, unambiguous. 

We offer the following dumbed-down example to clarify the situation, pushing the abstraction to its extreme boundary. 

\begin{figure}[!ht]
\includegraphics[width=0.99\linewidth]{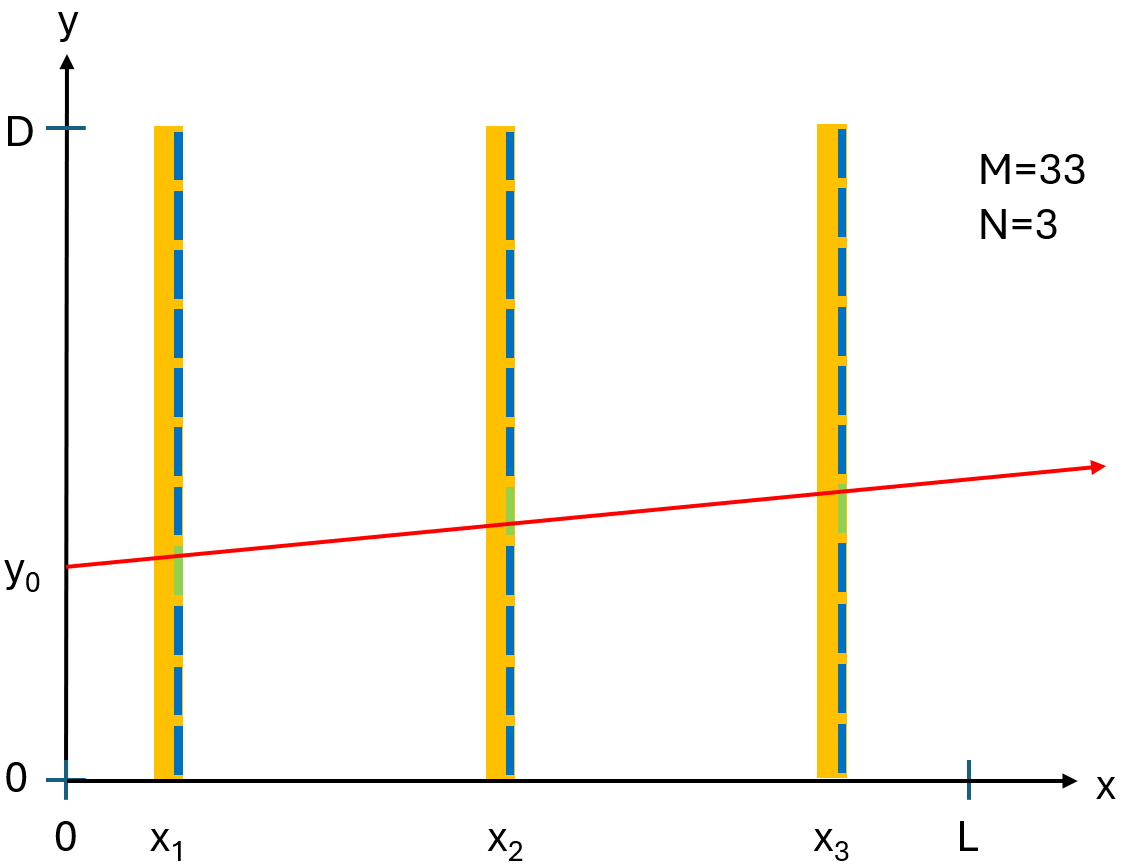}
\caption{ Scheme of a simplified tracker made of 33 silicon strips (blue or green segments along $y$) organized into three layers (yellow) . The passage of a straight-going particle (red) may be recorded by the three green-coloured strips.}
\label{f:simplifiedtracker}
\end{figure}

Suppose we want to measure with the highest possible precision the vertical position $y_0 \in [0,D]$ hit by a particle traveling approximately along $x$, for $x=0$; here $D$ is the width of a relevant region wherein we want to be sensitive. We may use for that task a detector comprising a fixed number $M$ of silicon strips, each read out by one of $M$ electronic channels. Strips  are aligned along the direction orthogonal to the $xy$ plane, and are arranged in a set of $N$ layers set vertically along $y$ (see~\autoref{f:simplifiedtracker}) and positioned at $N$ locations in $x \in [0,L]$. In this simplified setup we assume that hit positions are determined from single-strip hits (as would happen in the absence of magnetic fields, electric fields orthogonal to the strips, and nearly orthogonal crossing of particles on the layers), so that the vertical resolution of each hit corresponds to the RMS of a uniform distribution, and is simply \par

\begin{equation}
\sigma_{y}=P/\sqrt{12}\,,
\label{eq:sigmay}
\end{equation}

\noindent
where $P$ is the strip pitch. It is worth noting that, this simplified resolution estimate assumes that the distribution of measurement errors are Gaussian, while in reality we know they are better represented by a box distribution; we will come back to this detail {\em infra}. With a fixed number of strips, and the need to instrument the $[0,D]$ region in $y$, $P$ is determined as \par

\begin{equation}
    P = (D/M) \times N\,.
\end{equation}

\noindent 
To estimate the vertical coordinate $\hat{y}$ where the particle beam is passing, we now imagine performing a straight line fit to the measured vertical coordinates $y_i$ of the set of $N \geq 2$ points in $x \in [0,L]$, where horizontal uncertainties are assumed null and vertical uncertainties are all equal to $\sigma_{y}$ as above. The maximum likelihood estimate is uniformly minimum variance and unbiased for this linear Gaussian model. Ignoring constants, we write the negative log-likelihood as\par

\begin{equation}
    - \log {\cal{L}}(y_0,m) = \frac{1}{2 \sigma_y^2} \sum_{i=1}^{N} (y_i-y_0-m x_i)^2\,,
\end{equation}

\noindent
where $m$ is the slope of the trajectory in the $xy$ plane. The solution $\hat{y}_0$ follows by finding the extremum of $- \log {\cal{L}}$. The inverse of the Fisher information is \par

\begin{equation}
    \mathcal{I}^{-1}
= \frac{\sigma_y^{2}}{\,N\sum x_i^{2} - (\sum x_i)^{2}\,}
\begin{pmatrix}
\displaystyle \sum x_i^{2} & -\displaystyle\sum x_i \\[6pt]
-\displaystyle\sum x_i & N
\end{pmatrix}\,,
\end{equation}

\noindent from which the uncertainty on the $\hat{y}_0$ estimate can be obtained by using the Rao-Cramer-Frechet bound: \par

\begin{equation}
    \operatorname{Var}(\hat{y}_0)
= \sigma_y^{2}\!\left(
\frac{1}{N} + \frac{\bar{x}^{2}}{S_{xx}}
\right),
\quad
S_{xx} = \sum_{i=1}^{N}(x_i - \bar{x})^{2},
\quad
\bar{x} = \frac{1}{N}\sum_{i=1}^{N}x_i\,.
\end{equation}

\noindent
By recalling the scaling form of $\sigma_y$ (\autoref{eq:sigmay}), we thus see that in this setup the uncertainty on the $\hat{y}_0$ estimate scales with the square root of the number of layers: we see it if, {\it e.g.}, we place ``reasonably'' the silicon layers in the $[0,L]$ interval ({\it i.e.}, uniformly), such that $\bar{x}=1/2$ and $S_{xx}=N/12$: in that case \par

\begin{equation}
    \operatorname{Var}(\hat{y}_0)
= \frac{4}{N} \left( \frac{D}{\sqrt{12}} \frac{N}{M}\right)^2 = 
\frac{D^2}{3M^2} N\,.
\end{equation}

\noindent
We thus discover that the optimal number of layers of such a device is $N=2$: more layers would increase the position uncertainty in proportion to $\sqrt{N}$. Also, the optimal placement of these layers is no mystery - one has to be at $x=0$, the other at $x=L$, so that $S_{xx}$ is maximised.

In the simplified setup above, the optimal condition is workable analytically and is uniquely determined, hence the problem trivially factorizes, and hardware choices ({\it i.e.}, number of strips/electronic channels, connected to total cost) do not modify the software solution. 

Consider a slightly more complex situation where all simplifying elements of the above problem remain unchanged, except that the exact form of the vertical uncertainties is non analytical, and can only be obtained through expensive simulations. A precise estimate of $\hat{y}$ would now require a non-analytical method to be extracted; it could be a simple one based on template histograms of hit residual distributions, or a more complex machine-learning-based one. Further assume that the number of electronic channels is not fixed, but must obey to a constraint on the global cost of the experiment. The cost would not only include the sheer cost of hardware components, but also the CPU cost of simulations and inference procedures. In the same fixed-budget scenario  of the previous version of this example, an optimization based on co-design would become unavoidable, due to the introduced coupling: spending too much for the hardware would reduce the attainable precision of the simulated residual distributions, with a resulting impact on the quality of the inference on $\hat{y}$. 

From the above examples, we conclude that it is the simulation-based, likelihood-free inference nature of the problem of determining particle trajectories from noisy hits that necessitates the need for a simultaneous optimization of the hardware design and software methods used to extract the desired information. The underlying complexity, arising from the stochastic nature of the physical processes at the core of the problem, precludes the analytical inference. 
Such a situation is common to many of the problems in the rest of this section and in~\autoref{s:cd_industry}.
Whenever the data-generation process admits a closed-form likelihood and the associated inference problem is analytically solvable, the dependence of statistical performance on detector design can be expressed explicitly ({\it e.g.}, via Fisher information), enabling a factorized optimization of hardware and inference. In contrast, when inference relies on simulations or approximate models, this explicit mapping is lost, and hardware and software become intrinsically coupled through computational and modeling constraints.

\subsubsection {Persistent co-design dimensions in tracking}

The end-to-end optimization of a particle tracker is a highly demanding task, and a worked-out example where quantitative assessments of the sub-optimality ratio defined in~\autoref{s:intro} could be made lays beyond the scope of this work. Here we limit ourselves to consider six different situations where a co-design dimension arises in tracking detectors, and discuss their generalities.
\begin {enumerate}
\item {\bf MuOnE detector} This is one example where clear co-design elements were demonstrated quantitatively for a full system. The MuOnE detector is a silicon-strip particle tracker designed for the purpose of measuring with high accuracy the differential distribution of muon-electron scattering at high energy, with special focus in the high-$q^2$ kinematic region~\cite{muone}. The three particles to be tracked in scattering events (incoming and outgoing muon, and outgoing electron) travel nearly parallel to the beam axis $z$, but the all-important precision of the $q^2$ determination rests on a precise (micrometric) knowledge of the positioning of the silicon strip layers along $z$. A determination of the layers positions $z_i$ and tilt angles $\theta_i$ from survey measurements is trumped by a data-driven measurement, where real scattering events can be fit with in turn each $z_i$ or $\theta_i$ be left out as a free parameter, obtaining $1.5 \mu m$ precision on the layer $z$ position from 10 million reconstructed events (which can be collected in a few minutes of run time in the considered experimental conditions). An optimization of the geometry of the detector layout (relative positioning of sensors and targets) gets therefore coupled to the software extraction of the position uncertainties. For more detail see Ref.~\cite{dorigo2020}. 

\begin{figure}[!ht]
\includegraphics[width=0.99\linewidth]{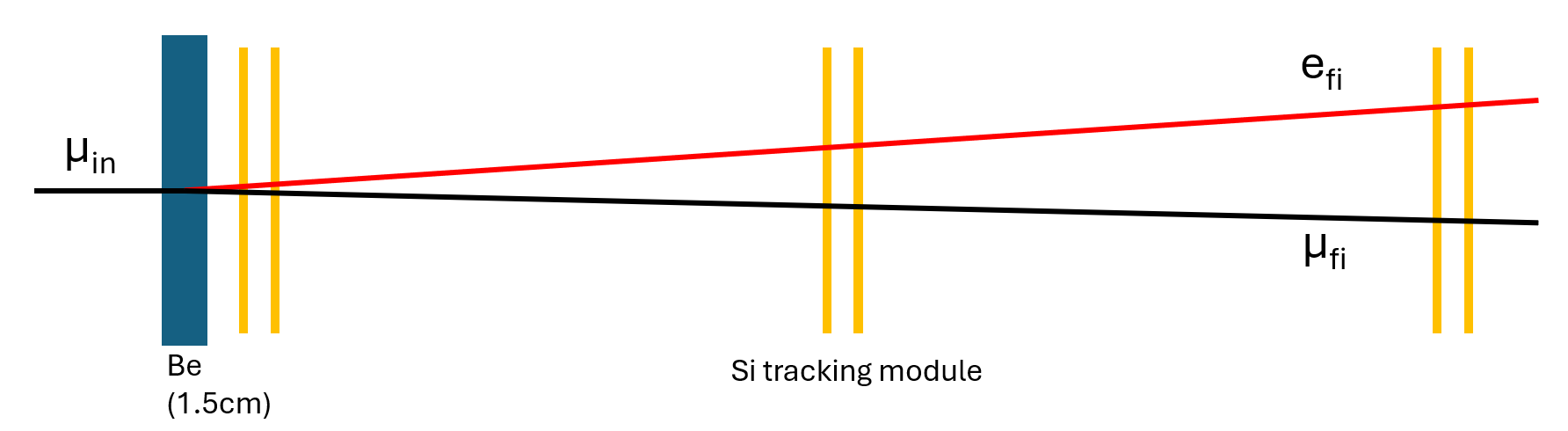}
\caption{Layout of one of 40 1m-long stations composing the MuOnE detector as envisioned by the MuOnE Collaboration~\cite{muone}. Muons (of 150 GeV momentum) enter from the left, impinge on a beryllium target of 1.5 cm thickness, and the resulting scattered particles are then tracked by a set of double-sided silicon strip tracking layers.}
\end{figure}
A further coupling of hardware and software in the MuOnE example comes from the reconstruction of the three particle trajectories, and it is instructive for its demonstration of how consideration of the whole problem leads to possible drastic improvements in the design. Since most of the scattering interactions occur within a thick beryllium target\footnote{Other small-$Z$ materials could be similarly envisioned for the target material.}, there is very little or no advantage in $q^2$ resolution from the simultaneous fitting of the three particles to a common three-dimensional vertex, as the driving uncertainty propagates to $q^2$ from the $z$ position of the vertex while the vertex-constrained fit can only improve the $x,y$ residuals (again, due to the trajectories being almost parallel to the beam axis). One might then be tempted to adopt a reconstruction scheme where each particle is treated separately; an advantage of this {\em modus operandi} would come from reducing potential smearing effects to the kinematics from multiple scattering in the target. This was indeed what the collaboration did in their early performance studies. However, the realization that a strong dependence of the precision of $q^2$ estimate on the precision of $z$ location of the vertex is only present if a vertex-constrained reconstruction is adopted may lead researchers to change the hardware design to exploit the fact: a $z$-segmented target, made of {\it e.g.} thin carbon-fiber sheets (50 $\mu m$ or below) spaced in air, or even better in vacuum, generates a strong reduction of the $z$-position uncertainty, and consequently on the $q^2$ resolution particularly in the high-$q^2$ region of specific focus of the experiment (see \autoref{f:muoneq2}). Here, the key aspect of co-design is the precise impedance matching of the spacing to the precision of the vertex-constrained fit as guaranteed by the silicon hits. This creates a strong coupling of chosen geometry and vertex segmentation detail. Again, the interested reader is referred to Ref.~\cite{dorigo2020} for more detail. 

\begin{figure}[h!]
\includegraphics[width=0.99\linewidth]{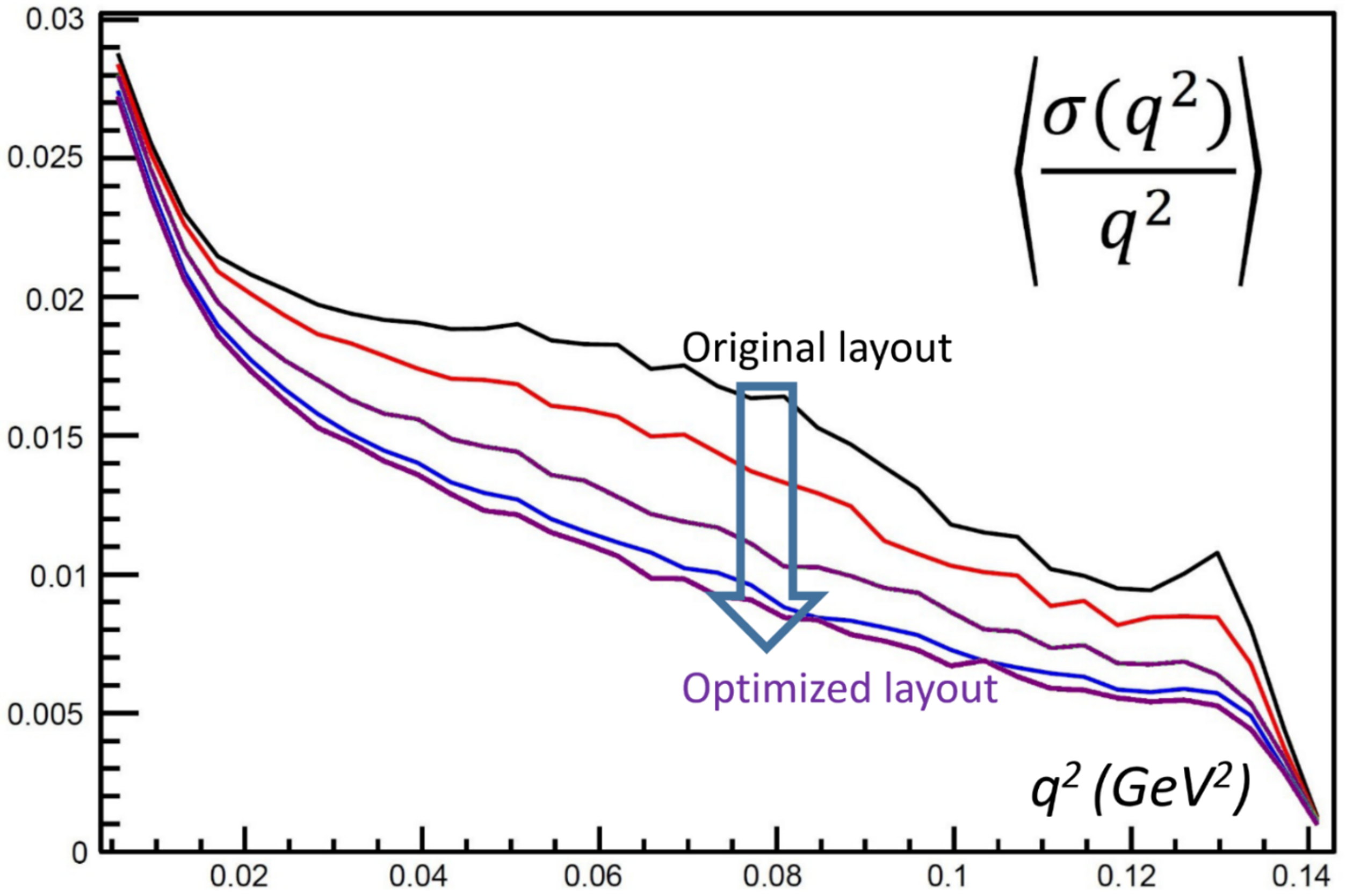}
\caption{Progressive improvement of the resolution in muon-electron scattering $q^2$ as a function of $q^2$, obtained by the staged optimization discussed in Ref.~\cite{dorigo2020}.}
\label{f:muoneq2}
\end{figure}

\item {\bf Sensor positions and mechanical tolerances} The themes considered above for the placement of silicon sensors in the MuOnE detector are common of all charged particle trackers, so it is useful to generalize their discussion here. Sensor resolution is one of the hardware factors driving the performance of a tracker, but its impact is modulated by the precision of positioning and alignment of the sensors. Depending on the specific goals of a detector, silicon strips or pixels may be chosen to detect the passage of particles; in general, the higher precision on the $z$ coordinate (usually the one along which strips are aligned) provided by the latter guarantees higher precision to the overall reconstruction task, but this may be not crucial when an axial magnetic field guarantees that transversal hit positions in the $xy$ plane dominate the precision on particle momenta. The much larger number of electronic channels required by pixel sensors induces the need of a compromise when global cost is a tight constraint. A further element materializing a non-trivial optimization frontier arises when mechanical stability couples with global fit strategies for sensors positioning determination: mechanical cost and calibration effort both play in the global resources budget. Again mentioning the MuOnE example, a determination of the position of silicon layers to within a few micrometers was demonstrated with the use of a complex (and expensive) holographic laser system; even in the absence of the data-driven determination discussed above, the resources required for that special system determine a Pareto boundary – although in this case between hardware-only parameters.
\\
\item {\bf Material budget and interaction length}
A further structural co-design dimension in tracking detectors is the amount and distribution of passive materials inside the tracking volume, usually parameterized in units of radiation length $X_0$ (governing electromagnetic interactions such as bremsstrahlung and photon conversion) and nuclear interaction length $\lambda_I$~\cite{PDG2022} (governing hadronic interactions). Modern silicon trackers for general-purpose collider experiments typically accumulate of order one radiation length and a sizeable fraction of an interaction length along the path from the interaction point to the electromagnetic calorimeter in at least part of their pseudorapidity~\footnote{Pseudorapidity $\eta$ is a function of the angle $\theta$ of final state particles with the initial beam direction, $\eta = -\log{\tan{\frac{\theta}{2}}}$.} coverage, with a large fraction of the material coming from services~\cite{ATLAS:Material2017}, cooling, supports and front-end electronics rather than the sensors themselves. Detailed R\&D for HL-LHC upgrades, for instance, has pushed inner trackers towards targets of $\mathcal{O}(0.1\,X_0)$ per layer~\footnote{This is based on expected tracking and related performance with the updated ATLAS Inner Tracker layout at the High-Luminosity LHC (2021).} in the central region~\cite{CMS:Phase2TrackerTDR}, 
but even in those designs the integrated thickness to the calorimeter remains a non-negligible fraction of $X_0$ and $\lambda_I$.

From a physics point of view, the material budget couples directly to several distinct performance drivers: multiple Coulomb scattering, longitudinal and transverse energy loss for electrons, photon conversion probability, and the rate of nuclear interactions of hadrons. For low- and intermediate-momentum tracks, multiple scattering rather than single-hit resolution often dominates the curvature and impact-parameter resolutions; the characteristic scattering angle scales as~\cite{Highland:1975}:

\begin{equation}
    \theta_0 \simeq \frac{13.6~\mathrm{MeV}}{\beta p}\sqrt{\frac{x}{X_0}}\left[1 + 0.038 \ln\!\left(\frac{x}{X_0}\right)\right]\,,
\end{equation}

so that each additional layer, cooling pipe, or support traversed by the track alters the balance between measurement noise and process noise in the track fit. Whether it is preferable, for a fixed material budget, to distribute that material in a larger number of thin layers or a smaller number of thicker layers depends on the track model employed by the reconstruction ({\it e.g.}, simple Kalman filter versus more sophisticated treatments~\cite{Fruhwirth:1987} of multiple scattering, adaptive outlier rejection, or trajectory smoothing) and on the seeding and pattern-recognition strategy. The same hardware configuration can thus yield markedly different effective resolutions and efficiencies depending on the software choices, and the hardware optimum in terms of $x/X_0$ and $\lambda_I$ cannot be determined in isolation from them.

Nuclear interactions in the tracker material represent another important dimension of coupling. When a charged hadron undergoes a hadronic interaction within the tracking volume, its original track may terminate or kink, and a spray of secondaries is produced. These secondaries can seed fake tracks, create secondary vertices that mimic heavy-flavour decays, and degrade jet and missing-transverse-momentum reconstruction. At the same time, exactly the same interactions can be exploited by dedicated reconstruction to map the distribution of tracker material \textit{in-situ}, and to validate or tune the detector simulation~\cite{ATLAS:Material2017}. The design decision to tolerate (or not) a certain integrated $\lambda_I$ in front of the calorimeters therefore depends on how powerful and robust the downstream algorithms for nuclear-interaction tagging, vertex reconstruction, and material mapping are expected to be. A detector with more material but a reconstruction chain that can efficiently tag and veto nuclear interactions might outperform, for some physics goals, a lighter detector operated with a simpler reconstruction.

This interplay becomes even sharper when the tracking system is required to provide additional functionality such as real-time track-based triggering. For example, modern outer trackers designed with ``transverse momentum modules ($p_T$-modules)"~\cite{CMS:Phase2TrackerTDR} performing on-detector stub finding and coarse transverse-momentum selection necessarily incorporate extra silicon, electronics and services within the tracking volume. That additional material degrades offline tracking, electron and photon reconstruction, and calorimeter performance, but it enables sophisticated Level-1 trigger algorithms that would be impossible without on-detector preprocessing. The net utility of such a design cannot be written as a monotonic function of the material budget alone: it depends in a strongly non-factorizable way on the firmware and software stack that exploits the real-time information. Hardware choices that look suboptimal in a purely offline tracking metric may be preferred once trigger performance and bandwidth constraints are folded in.

Thus, the material budget and interaction length are persistent co-design dimensions because the same hardware parameter $x/X_0$ (and $\lambda_I$) simultaneously controls multiple, qualitatively different reconstruction challenges, which are addressed with different classes of algorithms. The location and amount of material cannot be optimized independently from assumptions about pattern recognition, track fitting, nuclear-interaction and conversion reconstruction, and even triggering; conversely, algorithm design is constrained by the material profile inherited from hardware. The corresponding utility is therefore inherently non-decomposable in the sense discussed in~\autoref{s:intro}. The complex, bidirectional hardware-software dependencies are shown schematically in Fig. \ref{Diagram_budget}.

\begin{figure}[h]
    \centering
    \includegraphics[width=\linewidth]{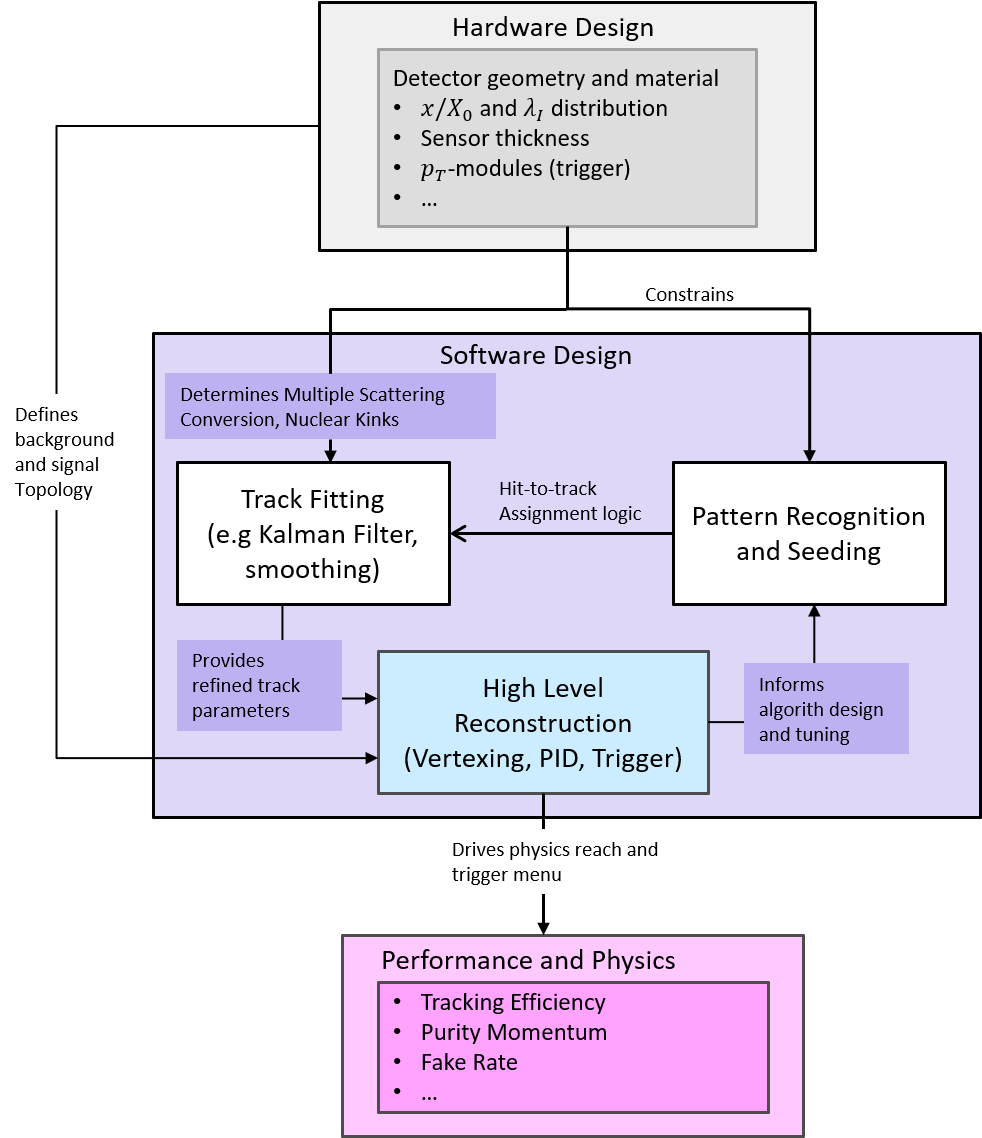}
    \caption{Schematic illustration of the material budget as a co-design axis in tracking detectors. The diagram shows the bidirectional, non-factorizable coupling between hardware and software. The detector geometry and material distribution ($x/X_0$, $\lambda_I$) constrain the design space of pattern recognition, track fitting, and high-level reconstruction algorithms. These algorithms, in turn, determine the final tracking performance metrics (efficiency, resolution, fake rate), which inform iterative hardware optimization. The cycle highlights that the same material configuration can yield different physics outcomes depending on the algorithmic choices, and conversely, that algorithm development is constrained by the inherited material profile.}
    \label{Diagram_budget}
    
\end{figure}

\item {\bf Electron reconstruction and tracker-calorimeter interplay}

Electrons exemplify in a particularly acute form the non-decomposable coupling between tracking and calorimetry. As they traverse the tracker material, electrons radiate bremsstrahlung photons with a probability that increases with $x/X_0$ and with their energy; in an LHC-like detector an electron of tens of GeV typically crosses a substantial fraction of a radiation length before reaching the electromagnetic calorimeter, and may therefore lose a large and highly fluctuating fraction of its energy in the tracker volume. The resulting momentum-loss distribution is strongly non-Gaussian~\cite{adam2005reconstruction}, with a long tail of hard emissions, and the associated photons can themselves convert in the tracker material. This induces abrupt changes in track curvature, large $E/p$ tails, and extended energy deposits in the calorimeter along the bending direction of the magnetic field.

Modern general-purpose detectors respond to this situation by reconstructing electrons through an explicitly joint use of tracker and calorimeter information. On the calorimeter side, algorithms group multiple neighbouring cells or crystals into ``superclusters'' whose shape is tuned to the expected pattern of bremsstrahlung in the magnetic field and to the segmentation of the calorimeter. In CMS, for instance, elongated superclusters in $\phi$ are built around a high-energy seed crystal~\cite{cms2020electron} to collect radiated photons that have been swept sideways in the 3.8~T solenoidal field, with the parametrization of the clustering window depending on pseudorapidity, energy, and local material distribution in front of the ECAL. The optimal granularity and longitudinal segmentation of the electromagnetic calorimeter, and even the choice of absorber and scintillator material (which fix $X_0$ and the Molière radius), are therefore co-determined with the clustering and energy-regression algorithms that will exploit the shower shapes (Fig.\ref{pipe2}). A calorimeter design that is ``best'' under a simplistic assumption of compact, un-radiated showers may cease to be optimal once realistic bremsstrahlung and conversion patterns, and the corresponding reconstruction, are taken into account.

On the tracking side, the same bremsstrahlung complicates the track model. Standard Kalman filtering, which assumes approximately Gaussian process noise, is suboptimal~\cite{fruhwirth1997track} for electrons because the occasional large energy losses and kinks are poorly described by a single Gaussian term. Gaussian-sum filter (GSF) algorithms address this~\cite{adam2005reconstruction} by representing the electron state as a weighted sum of hypotheses with different momenta, effectively modelling the bremsstrahlung loss distribution as a Gaussian mixture. This improves the momentum resolution and, crucially, the reliability of the estimated uncertainties, but at the price of substantially increased computational cost and algorithmic complexity. The benefit of deploying a GSF (and the amount of complexity that can profitably be invested in it) depends sensitively on the actual amount and localization of material in the tracker, as well as on the magnetic field strength and layer radii. A hardware design that concentrates most of the material in a few well-defined layers may make the bremsstrahlung pattern easier for a GSF to capture than a design with the same integrated $x/X_0$ but a complicated, service-dominated distribution. Conversely, the availability of a powerful electron-specific track model may relax otherwise stringent constraints on the detailed homogeneity of the tracker material.

The final electron momentum (or energy) estimate typically results from a non-trivial combination of tracking and calorimeter measurements. At low and intermediate momenta the track curvature can dominate the resolution, while at higher energies the calorimeter measurement is superior; in the presence of substantial bremsstrahlung, however, even at moderate energies the calorimeter often provides a more stable estimator of the electron energy, and the track serves primarily to determine the direction and charge. State-of-the-art reconstruction chains therefore employ multivariate regressions that take as input track parameters~\cite{cms2020electron}, the pattern of bremsstrahlung along the track, supercluster shapes, and local material indicators, and return an optimally weighted estimate of the electron kinematics. The performance of such regressions depends both on hardware (which fixes the feature space: segmentation, depth, noise, material profile) and on software (which fixes the class of functions that can be represented and learned), and the utility of changes in one space cannot be assessed without modelling the other.

Because electrons are key objects for precision measurements ({\it e.g.}, of the $W$ mass, electroweak mixing angle, and various differential cross sections)~\cite{atlas2017measurement} and for many searches, the electron energy and angle resolutions, tails, and identification efficiencies often set the sensitivity floor of an experiment. Small modifications of the tracker material in front of the calorimeter, or of the ECAL granularity or longitudinal segmentation, can therefore have disproportionate impact on physics reach once they are propagated through realistic electron reconstruction and calibration. In design and upgrade studies, it is precisely this strongly coupled electron channel that frequently drives the choice among alternative layouts for supports, services, and calorimeter modules. As a result, electron reconstruction and the tracker–calorimeter system that enables it, form a persistent co-design dimension: hardware and software choices influence each other through genuinely non-linear and non-factorizable dependencies, and meaningful optimization requires treating them together rather than in a hardware-first, software-second sequence.

\begin{figure}
    \centering
    \includegraphics[width=\linewidth]{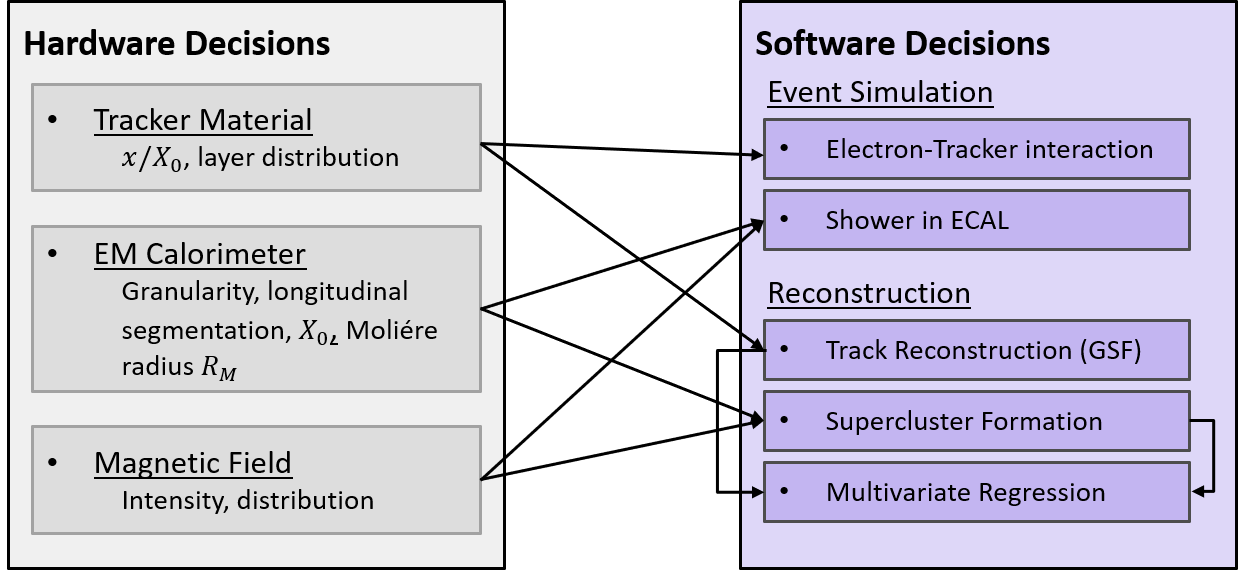}
    \caption{Hardware–software co‑design in electron reconstruction. Tracker material and ECAL granularity (hardware) determine bremsstrahlung patterns and shower shapes, which in turn drive algorithmic choices: GSF for tracking, superclusters for calorimetry, and multivariate regression for combined estimation. Performance feedback closes the loop, underscoring the non‑separable nature of the optimization problem.}
    \label{pipe2}
\end{figure}

\item {\bf 4D track seeding vs. pile-up occupancy}
In the HL-LHC environment, with up to $\mathcal{O}(200)$ proton--proton interactions per bunch crossing and a vertex time spread of order $\sigma_{\mathrm{PU}} \sim 180~\mathrm{ps}$~\cite{HLLHC:TDR2017}, track seeding in the innermost tracker layers is challenged by combinatorial background arising from hits produced by pile-ups, {\it i.e.}., simultaneous collisions overlaid on top of the interaction of interest. Precision timing layers such as the CMS MIP Timing Detector (MTD) and the ATLAS High-Granularity Timing Detector (HGTD), which target at track-time resolutions of $\sigma_t \simeq 30$--$50~\mathrm{ps}$~\cite{CMS:MTD2019, ATLAS:HGTD2020}, introduce an additional handle in the time dimension: hits can be filtered using a temporal gate of width $\Delta t$ around a predicted track or vertex time. This effectively reduces the instantaneous occupancy seen by the seeding and track-building algorithms, at the price of potentially discarding in-time hits when the gate is too narrow.

A simple scaling argument illustrates the trade-off. Consider a region of interest in $(\eta,\phi)$ on a given layer where, without timing, the average number of hits per bunch crossing is $\lambda_0$. If the hit times are approximately uniform over a relevant time span $T_{\mathrm{eff}}$ (set by the luminous region and readout response), restricting to a time window $\Delta t$ reduces the mean hit multiplicity to $\lambda(\Delta t) \simeq \lambda_0\,\Delta t/T_{\mathrm{eff}}$. For seeds built from $k$ layers with similar occupancies, the number of hit combinations to be tested scales roughly as
\[
N_{\mathrm{seed}}(\Delta t) \propto \bigl[\lambda(\Delta t)\bigr]^k \;\propto\; \Delta t^{\,k}.
\]
In a CKF-style combinatorial track finding, branchings at each layer also depend on the number of compatible hits~\cite{CMS:MTD2019}, so the total CPU time or latency for track building grows rapidly with $\Delta t$. Narrowing the timing gate can therefore reduce seeding and building time by orders of magnitudes in dense events.

On the other hand, the seeding efficiency for true tracks falls if the gate is too tight compared to the timing resolution. For a track originating at time $t_0$, the measured time of a hit on the timing layer can be modeled as
\[
t_{\mathrm{hit}} = t_0 + \delta t, \qquad \delta t \sim \mathcal{N}(0,\sigma_{\mathrm{eff}}^2),
\]
where $\sigma_{\mathrm{eff}}$ collects intrinsic sensor resolution $\sigma_t$ and additional contributions from time-of-flight modeling, calibration, and vertex-time uncertainty. If a hit is accepted only when $|t_{\mathrm{hit}} - t_0| < \Delta t/2$, the single-hit acceptance probability is
\[
p_{\mathrm{hit}}(\Delta t, \sigma_{\mathrm{eff}}) =
\operatorname{erf}\!\left(\frac{\Delta t}{2\sqrt{2}\,\sigma_{\mathrm{eff}}}\right).
\]

\begin{figure*}[h]
    \centering
    \includegraphics[width=\linewidth]{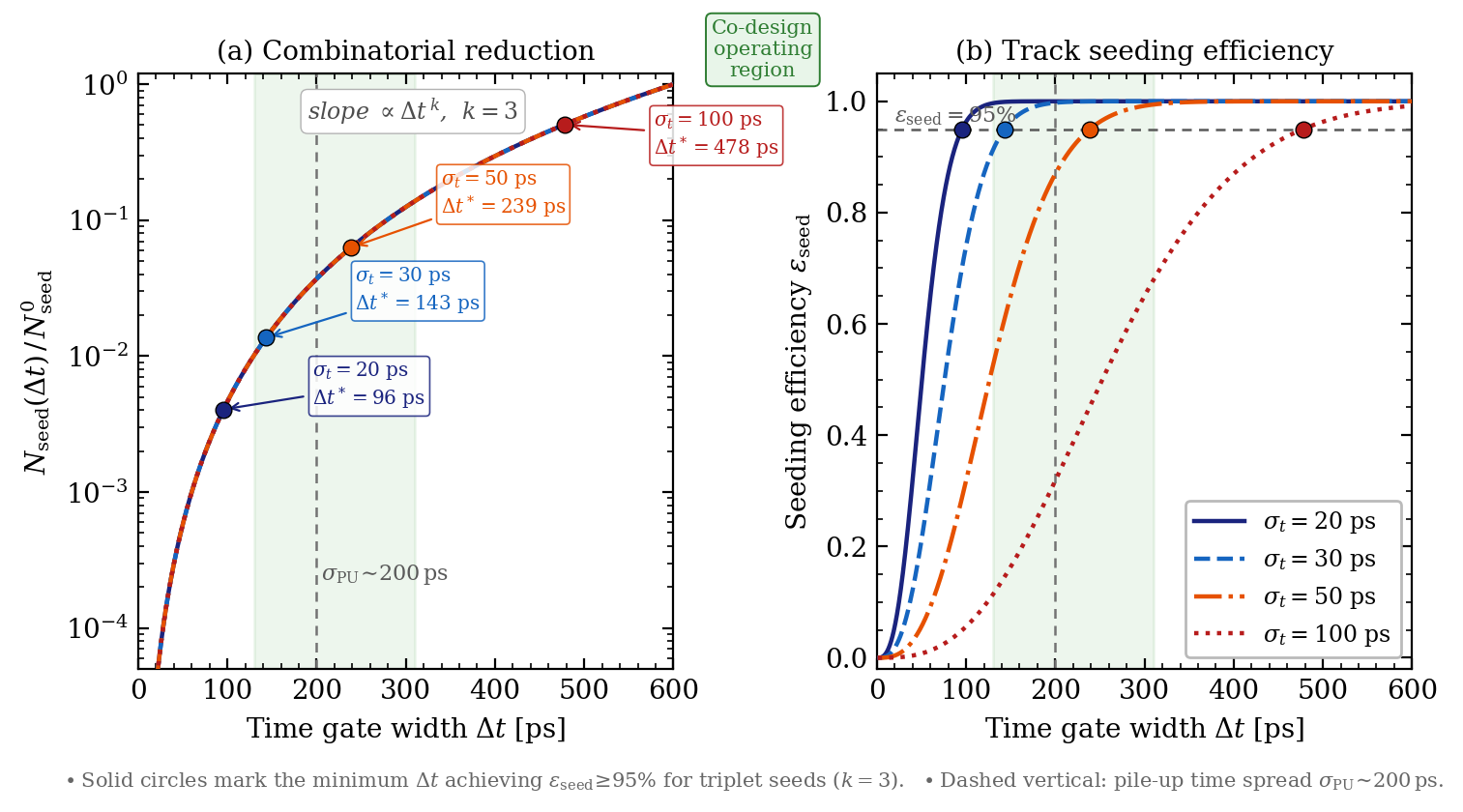}
    \caption{(a) Ratio of seed combinations as a function of the timing gate width $\Delta t$ for triplet seeds (k=3), relative to the untimed case. (b) Corresponding seeding efficiency for four values of the detector timing resolution $\sigma_t$. Solid circles indicate the minimum $\Delta t$ satisfying $\epsilon_{seed} \geq$ 95$\%$ in each case; the dashed vertical marks the pile-up time spread $\sigma_{PU}$ ~ 200 ps. The shaded band highlights the co-design operating region where improvements in hardware timing resolution translate into a superlinear reduction of combinatorics only when the algorithmic gate is adapted accordingly.}
    \label{fig:placeholder}
\end{figure*}

For a seed requiring $k$ such hits to survive the time gate, a crude approximation to the seeding efficiency is $ \epsilon_{\mathrm{seed}} \simeq p_{\mathrm{hit}}^k$. Achieving, say, $\epsilon_{\mathrm{seed}} \gtrsim 95\%$ for triplet seeds typically implies $\Delta t \sim (5$--$6)\,\sigma_{\mathrm{eff}}$ (see \autoref{fig:placeholder}): for a per-hit resolution of $30~\mathrm{ps}$ this corresponds to $\Delta t$ of order $150$--$200~\mathrm{ps}$, while for $50~\mathrm{ps}$ timing it requires $\Delta t \sim 250$--$300~\mathrm{ps}$. Thus, the optimal time gate is set not by the pile-up timescale ($\sim 200~\mathrm{ps}$) alone, but by the ratio $\Delta t/\sigma_{\mathrm{eff}}$; improving the hardware timing resolution allows a proportional reduction of $\Delta t$ at fixed efficiency~\cite{Sadrozinski:4DTracking2018}, which translates into a superlinear reduction of combinatorics because $N_{\mathrm{seed}} \propto \Delta t^{\,k}$.

This establishes 4D seeding as a genuine co-design problem. On the hardware side, the choice of timing technology, number and location of timing layers, achievable $\sigma_t$ under irradiation, and digitization granularity determine $\sigma_{\mathrm{eff}}$ and the stability of time measurements over the detector lifetime. On the software side, the tracking pipeline must decide how aggressively to exploit time in seeding and CKF gates: whether to apply a single global $\Delta t$, to use layer- or region-dependent windows, or to propagate a full 4D state (including time and its covariance) through a time-aware Kalman filter~\cite{ATLAS:HGTD2020} that predicts the expected hit time at each layer and adapts the gate accordingly. More sophisticated time modeling enables narrower, track-dependent gates and hence larger combinatorial gains, but at the cost of higher per-candidate CPU and more complex calibration; conversely, simplistic use of timing ({\it e.g.}\ a loose fixed window) leaves much of the potential combinatorial reduction on the table, making investments in better $\sigma_t$ less valuable.
In this scenario the utility depends on cross-terms between hardware and software parameters. For example, a tracker with $\sigma_t \simeq 30~\mathrm{ps}$ operated with conservative $\Delta t$ comparable to the pile-up spread may see little improvement in seeding time over a $50~\mathrm{ps}$ design, whereas an algorithm that dynamically sets $\Delta t \approx c\,\sigma_{\mathrm{eff}}$ for each seed (with $c$ tuned to the physics requirements) can translate a factor $\sim 1.5$ improvement in timing resolution into a factor $\sim 3$–$5$ reduction in the number of seed combinations at fixed efficiency. Similarly, the design of hardware triggers or streaming-reconstruction farms sets an upper bound on acceptable track-finding latency; whether that bound is met depends jointly on the timing performance of the detectors and on how aggressively 4D information is exploited in seeding and fitting. As a result, optimal choices of $\sigma_t$, timing-layer layout, and seeding strategy cannot be obtained by separating hardware-first and software-second optimizations: they must be treated as a coupled 4D design space~\cite{Sadrozinski:4DTracking2018}.

\item {\bf Converted photons in the Higgs search} 
In the search for the Higgs boson decay to photon pairs performed by CMS and ATLAS for their 2012 discovery, the mass resolution played a pivotal role, as higher mass resolution directly translates in a higher significance of the small collectable signal. Mass resolution is dominated by how well each photons energy and direction are measured. The CMS detector has superb energy resolution, so angular resolution becomes a key limiter for the di-photon mass. As photons do not leave a track in the particle tracker, the identification of their direction is a tricky issue in high-luminosity conditions where multiple proton-proton interactions take place within the same event; it must leverage information from the primary vertex activity. However, converted photons, producing a electron-positron pair detectable in the inner tracker, have a degraded ECAL energy containment, but a precisely reconstructed direction---because the conversion vertex and the electrons trajectories provide strong geometric constraints. CMS exploited this by combining ECAL clusters with tracker-based conversion fits, improving the di-photon mass resolution for events with conversions. 
As the tracker material budget increases, the photon conversion probability also rises, leading to a trade-off in design between material budget and the number of conversion events in Higgs study.
However, the same material causes increased multiple scattering, which affects track parameter resolution; it adds bremsstrahlung, with a degraded electron momentum measurement; and adds secondary interactions, occupancy, and fake rate. This highlights the existence of an optimization frontier between conversion statistics and tracking quality. 
The CMS tracker has about 0.4 to 1.0 radiation lengths of material in the barrel, and up to $1.8 X_0$ in the endcaps. This makes the conversion probability as high as 40 to 60 percent. Events containing the resulting conversions did constitute a significant portion of the Higgs candidate signal. However, the benefits from the special characteristics of these events did not outweigh the global disadvantages of tracker material. The CMS Collaboration notes that reducing tracker material would improve the overall photon energy and resolution more than the gain resulting from conversions~\cite{CMS-DP-2011-008}. Indeed, subsequent CMS upgrade efforts for the Phase-1 and Phase-2 of the experiment prioritized material reduction, even though that reduced the conversions yield. This example thus highlights how a relevant question that had to be considered at design stage, once a careful appraisal of the relative importance of the observation of a Higgs signal with the diphoton decay mode were made with respect to the other scientific goals of the experiment, was the balancing of negative and positive effects of a significant amount of tracker material. 
\end{enumerate}

\subsubsection{Emerging directions}
The above examples spotlight a few common strategies in the design of tracking detectors, which may inspire synergies with design tasks in {\em e.g.,} medical imaging or muon tomography applications (see also ~\autoref{sec:tomography}). One of these is the advantage arising from data-driven alignment loops which may exploit the collected data to monitor the position of the sensing elements of a system. If the improvement in position resolution that such systems may offer at run time is disregarded during the design of a detector, this may lead to an incorrect assessment of its overall performance, with a resulting impedance mismatch of hardware- and software-driven resolutions. Another strategy that may have common use is the potential extraction of crucial extra information from physical effects that increase complexity rather than simplify the original task: that is the case of nuclear interactions in the detector material, which could in principle be exploited to help in particle identification, and of bremsstrahlung radiation, which is known to significantly improve the precision of electron trajectories and their energy and momentum measurement if the tracker is coupled to a granular electromagnetic calorimeter, in the presence of a strong magnetic field. Further, the mentioned exploitation of photon conversions in $H \to \gamma \gamma$ searches by CMS highlights how reconstruction ingenuity can extract information from detector imperfections, but they do not always justify those imperfections. Co-design aims to minimize such trade-offs by aligning hardware transparency with software sophistication.

\subsection{Calorimeters}
\label{sec:calorimetry}
Designing calorimeters for energy measurement and particle identification requires balancing sufficient absorber material, {\it e.g.}, iron, lead, or tungsten, to fully contain a particle shower with adequate sensitive volumes, {\it e.g.}, scintillating plastics or crystals, to infer the relevant physical properties from the resulting signals. The most straightforward solution is a homogeneous calorimeter, constructed either from a compact block of expensive scintillating material with a short radiation length ({\it e.g.}, PbWO$_4$) or from a much larger volume of inexpensive scintillator with a longer radiation length ({\it e.g.}, polystyrene). In practice, both approaches quickly encounter geometric constraints that limit detector size or cost constraints that restrict the feasible detector volume.

Sampling calorimeters address these limitations by alternating absorber layers with sensitive material, allowing efficient shower containment while remaining compatible with geometric and cost constraints. However, this introduces additional design parameters, such as the choice of materials and layer thicknesses, significantly increasing the complexity of detector optimization. Exhaustively exploring this parameter space through full detector simulations and reconstruction studies is computationally prohibitive~\cite{Jones1998EfficientGO}, motivating the use of more sophisticated optimization strategies.

\subsubsection{Methods}
\label{sec:methods}

In the following sections, we describe and showcase the co-design of a sampling calorimeter utilizing local generative surrogates (\autoref{sec:aido}), Bayesian optimization (\autoref{sec:sampling_calorimeter_bo}), and mutual information surrogates (\autoref{sec:calorimeters:mutual_inf_surrogates}).

\paragraph{Geometry}

We parameterize a simplified sampling calorimeter geometry, similar to the one described in Ref.~\citep{Schmidt2025}, consisting of five layers, each with a passive absorber and an active scintillator component. The absorber materials function as a cost-effective material with short radiation length to ensure shower containment, while the scintillators enable an electronic readout of the high-energy particles induced by emittance of optical photons. Both the absorbers and scintillators are constructed as rectangular cuboids with \SI{50}{\centi\meter} side length and variable depth, stacked along the beam axis. Each scintillator layer is segmented into 25 cells in each direction, giving each cell an area of $2\times 2\,\si{\square\centi\meter}$, a typical value used in experiments. 

Parameters $t_{A,i}$ and $t_{S,i}$ describe absorber and scintillator thickness at layer $i$, respectively. Further, the material of each layer is parameterized as $m_{A,i}$ and $m_{S,i}$. The selection of absorber material is restricted to iron (Fe) and lead (Pb), while the scintillator can be constructed out of lead tungstate (PbWO$_4$) or polystyrene ((C$_8$H$_8$)$_n$). The list of all design variables and their respective possible choices are summarized in \autoref{tab:design_variables_calo}.

\begin{table}[!ht]
\caption{Design variables for the calorimeter co-design optimization problem, including layer thicknesses and material selections.}
\label{tab:design_variables_calo}
\centering
\begin{tabular}{lll}
\hline
Param & Value Range & Description \\
\hline
$[t_{A,i}]$ & $\mathbb{R}^{+}$ &  Absorber thickness       \\
$[t_{S,i}]$ & $\mathbb{R}^{+}$ &  Scintillator thickness    \\
$[m_{A,i}]$ & $\{\mathrm{Fe}, \mathrm{Pb}\}$     &  Absorber material    \\
$[m_{S,i}]$ & $\{\mathrm{PbWO_4}, \mathrm{(C_8H_8)_n}\}$ &  Scintillator material \\
\hline
\end{tabular}
\end{table}

We subsequently define the full mixed discrete-continuous hardware design space of the calorimeter as the Cartesian product of the individual design variables
\begin{equation}
\begin{split}
    h &\in \mathcal{H}\,, \quad \text{where} \quad 
    \mathcal{H} \coloneqq \prod_i \mathcal{H}_i, \;\text{and} \\
    \mathcal{H}_i &\coloneqq 
    \mathbb{R}^+ \times \mathbb{R}^+ \times \{\mathrm{Fe}, \mathrm{Pb}\} \times \{\mathrm{PbWO_4}, \mathrm{(C_8H_8)_n}\}\,.
\end{split}
\end{equation}

Here $\mathcal{H}_i$ and $\mathcal{H}$ denote the admissible design domain of layer $i$ and the full calorimeter, respectively.

\paragraph{Simulation}

The simulations are performed through \textsc{Geant4}. Each event consists of a single particle ($\gamma$ or $e^+$) with energy uniformly sampled between 5 and \SI{100}{\giga\electronvolt}. The incident position of the particle is sampled around the center of the calorimeter such that the entire shower is recorded. The cells deposited energy, their center position, and their size are used in the reconstruction.

\paragraph{Reconstruction} We target a calorimeter layout that excels both in terms of particle identification and energy resolution. To estimate the particle type and its energy, we utilize independent neural networks for energy prediction and particle identification, which constitute the software parameterization of our co-design problem. We therefore define the software design space as

\begin{equation}
    s \in \mathcal{S}, \quad \mathcal{S} \coloneqq \Theta_{reg} \times \Theta_{class}\,,
\end{equation}

where $\Theta_{reg}$ and $\Theta_{class}$ are the admissible regression and classification parameter spaces, respectively. For the purposes of this case study, we limit the complexity of the optimization task by considering fixed architectures, even though auto-ML and architectural search describe the possibility of incorporating the search for optimal architectures into the co-design process~\citep{He2021,Elsken2019}. Below we provide a comprehensive description of the architectures for both tasks.

\begin{enumerate}
    
    \item \textbf{Energy regression} To predict the initial particle energy, we utilize an energy regression network similar to the one described in Ref.~\citep{Schmidt2025}. The regression network is a  feed-forward network composed of three layers of 100 neurons $\phi_{reg}: \mathbb{R}^{\vert\theta\vert + \vert h\vert} \to \mathbb{R}$ that takes the concatenated vector of detector parameters and shower features. The shower features contain the log-scaled total energy deposited per layer, the z-position of the layer, and the cell dimensions in the three directions. Both detector parameters and shower features are standardized to stabilize the training, and the target is the true particle energy. Given a dataset of simulations for various detector geometries in a local neighborhood, the network is optimized by minimizing
    
    \begin{equation}
        \label{eq:obj_reg_calo}
        \mathcal{L}_{Reg} = \frac{1}{N}\sum_{i=1}^{N} \left(\frac{(E_{rec} - E_{true})^2}{E_{true} +1}\right)\,.
    \end{equation}

    The loss is a scaled mean squared error (MSE) loss, where the scaling factor $1/(E_{true} + 1)$ regulates the importance of events with higher initial energy~\cite{Schmidt2025}.
    
    \item \textbf{Particle classification} 
    To take advantage of the transverse shower shape, a convolutional neural network (CNN) is used to treat the deposited energy per cell as an image~\cite{Kasieczka:2017nvn}. The longitudinal shape of the shower is therefore the only direction that the optimization can affect. The optimization could be extended to include the transverse shape by allowing the granularity as an optimizable parameter; however, in that case the CNN is not suitable anymore, and graph neural networks should be used instead. We leave that to future work. 
    
    Each calorimeter layer image is processed by a dedicated CNN, whose output is then concatenated and passed to a small feed-forward network of two layers of 32 neurons. In addition, layer contextual information, such as total layer deposited energy, layer position and cell size, and detector parameters are preprocessed by another small feed-forward network with two layers of 64 and 32 neurons to be added to the CNN encodings.
    
    The CNN setup was chosen to allow different granularity in each layer, though for this study it is fixed to the same value. Each CNN consists of three blocks of two convolution operations preceded by \textit{GroupNorm}~\cite{wu2018groupnorm} 
    normalization layers, with residual connections between each block. The first two blocks have kernel sizes of (3,3) with the second one using a dilation parameter of 2, and the last one has a kernel size of (5,5), all using padding such that the image size does not change and each also expands the number of channels by a factor of 2. This architecture was chosen to allow the extraction of higher-level information as the image goes through the blocks. At the end of each CNN, both mean and max pooling are applied to the processed image in a $5\times5$ grid that is then flattened and concatenated to be fed to the feed-forward network. A dropout rate of 0.3 is applied to all layers to mitigate overfitting, and early stopping is used.

    The energy deposited per cell in each layer serves as the image fed to the different CNNs. Several preprocessing steps are performed to mitigate the wide range of values in each shower. First, the image is centered and normalized to avoid any bias from the position and total deposited energy. To mitigate the effect of high-energy tails, the fourth root is used on the image pixels, which are then standardized. A second channel is included as a binary mask to encode the information that some cells are empty or not.

    The classification network $\phi_{class}: \mathbb{R}^{\vert\theta\vert + l \times (\vert g \vert \times \vert g \vert + \vert c\vert)}\to \mathbb{R}$ objective is a binary classification over the particle type ($\gamma$ versus $e^+$), with classical cross-entropy loss 
    \begin{equation}
        \label{eq:obj_class_calo}
        \mathcal{L}_{Class} = -\frac{1}{n} \sum_{i=1}^{n}
\left[ y_i \log(\hat{y}_i) + (1 - y_i)\log(1 - \hat{y}_i) \right]\,,
    \end{equation}
    where we define the binary label $y\in \{0,1\}$ such that $y=1$ corresponds to the particle being an $e^+$ and $y=0$ corresponds to the particle being a $\gamma$.
  
\end{enumerate}

\paragraph{Optimization target} To obtain an optimized calorimeter layout, we perform a joint optimization of software and hardware parameters to maximize both energy resolution and $e^+,\gamma$ classification performance subject to additional geometry constraints.  

We write the joint optimization as a multi-objective optimization problem with additional constraints as

\begin{equation}
\label{eq:objective_sampling_calo}
\begin{split}
    \underset{\theta \in \mathcal{H}\times\mathcal{S}}{\arg\min} \; f(\theta) &= (\mathcal{L}_{Class}(\theta), \mathcal{L}_{Reg}(\theta)) \\
    \text{s.t.} \quad C(h) &\leq 100,000\,\text{EUR} \\
                      T(h) &\leq \SI{100}{\cm}\,,
\end{split}
\end{equation}

where $\mathcal{L}_{Class}$ and $\mathcal{L}_{Reg}$ are the validation losses of energy reconstruction and particle classification networks outlined in~\autoref{eq:obj_reg_calo} and~\autoref{eq:obj_class_calo}. Further, $C(h)$ and $T(h)$ are geometry constraints, limiting both the total thickness and cost of the calorimeter to ensure practical feasibility of the design. Both constraint functions $T(h)$ and $C(h)$ are defined as 

\begin{equation}
\begin{split}
    T(h) &= \sum_{i=1}^{5} t_{A,i} + t_{S,i} \quad \text{and} \\
    C(h) &= \sum_{i=1}^{5} c(m_{A,i})t_{A,i} + c(m_{S,i})t_{S,i}\,,
\end{split}
\end{equation}

respectively. For the cost function $c(m)$, we use EUR/cm costs for absorber and scintillator materials presented in Ref.~\cite{Schmidt2025}.

\begin{figure}[ht]
    \centering
    \begin{subfigure}{0.8\columnwidth}
        \centering
        \includegraphics[width=\columnwidth]{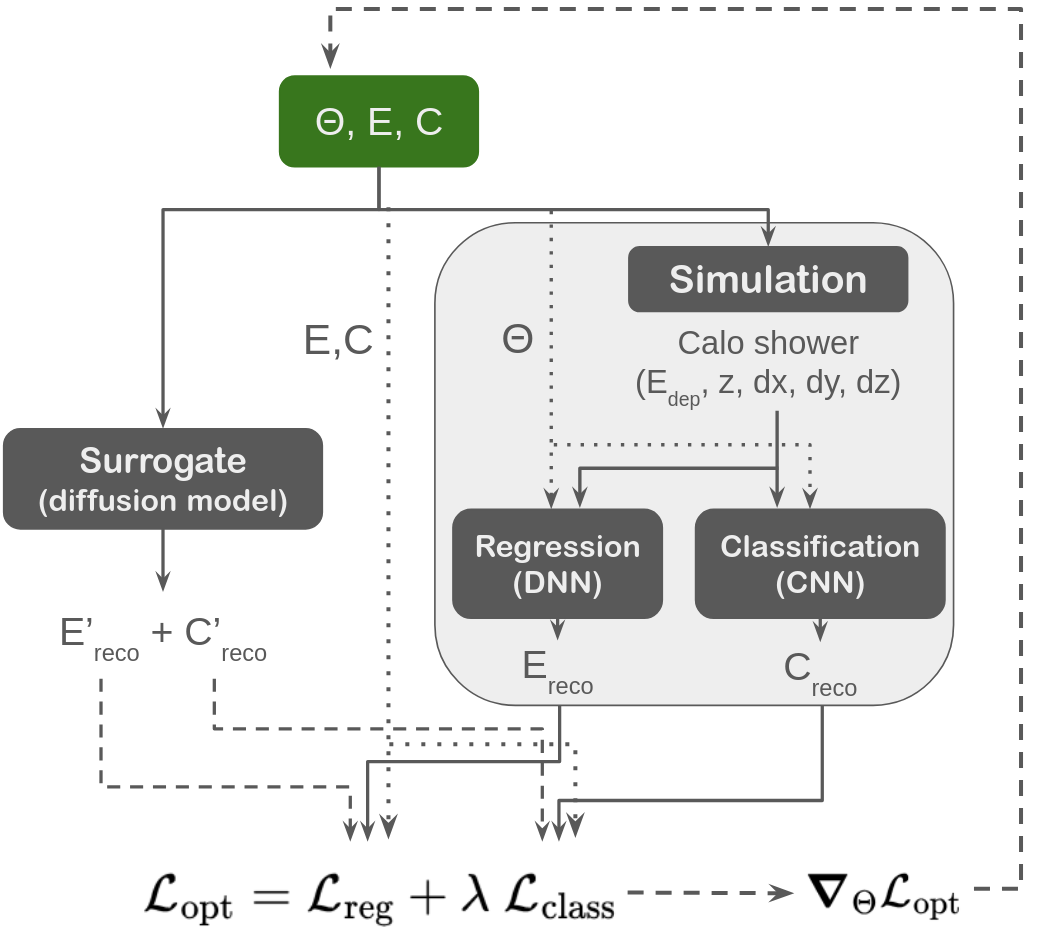}
        \caption{AIDO optimization scheme}
        \label{fig:calo_AIDO_scheme}
    \end{subfigure}
    \begin{subfigure}{0.8\columnwidth}
        \centering
        \includegraphics[width=\columnwidth]{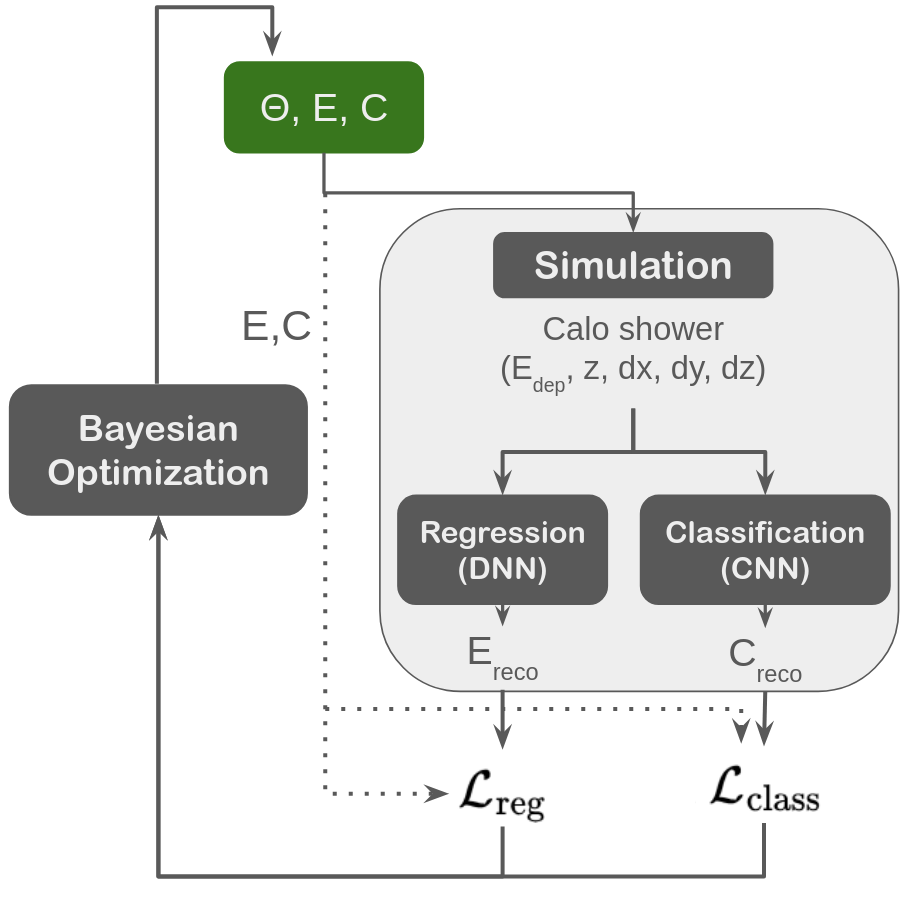}
        \caption{BO optimization scheme}
        \label{fig:calo_BO_scheme}
    \end{subfigure}
    \caption{Optimization schemes for both AIDO and BO methods. $\Theta$ represents the set of optimizable detector parameters, $E$ and $C$ are the true targets for particle energy and type, $E_{reco}$ and $C_{reco}$ their reconstructed quantities. While AIDO (top) uses a generative surrogate model---trained to emulate the reconstruction---from which a gradient can be obtained to optimize $\Theta$, BO (bottom) uses a Bayesian surrogate to map the loss space as a function of $\Theta$.}
    \label{fig:calo_schemes}
\end{figure}

\subsubsection{End-to-end Optimization with AIDO}
\label{sec:aido}

Given the parameterized sampling calorimeter geometry introduced in \autoref{sec:methods}, the detector design can be described by a set of parameters $\boldsymbol{\theta}$, including absorber and scintillator layer thicknesses as well as material choices. 
In principle, one could directly optimize the detector design by minimizing a physics performance metric
$
\mathcal{L}(\boldsymbol{\theta}),
$
but performing a full detector simulation for every parameter update quickly becomes computationally prohibitive when exploring the large design space of sampling calorimeters.

To address this challenge, the \textit{AI Detector Optimization} (AIDO) framework introduces a differentiable surrogate model acting as a digital twin of the simulation--reconstruction chain. Instead of directly optimizing the expensive function $\mathcal{L}(\boldsymbol{\theta})$, AIDO learns an approximation
$
\mathcal{S}(\boldsymbol{\theta}) \approx \mathcal{L}(\boldsymbol{\theta})
$
from a dataset of simulated detector configurations. Once trained, this surrogate predicts detector performance for new parameter values at negligible computational cost and provides gradients that can be used by the optimizer.

The optimization proceeds through an iterative \textit{Simulation--Reconstruction--Surrogate--Optimizer} (SRSO) loop, as shown in \autoref{fig:calo_AIDO_scheme}. 
Starting from an initial configuration $\boldsymbol{\theta}_0$, a set of nearby detector configurations is sampled from a multivariate normal distribution
$
\boldsymbol{\theta}_i \sim \mathcal{N}(\boldsymbol{\theta}, K_{\boldsymbol{\theta}}),
$
where $K_{\boldsymbol{\theta}}$ denotes the covariance matrix controlling the sampling scale in the detector parameter space. Each sampled configuration is evaluated using the full simulation and reconstruction pipeline to obtain the corresponding performance metric $\mathcal{L}_i$. The resulting dataset $\{(\boldsymbol{\theta}_i, \mathcal{L}_i)\}$ is then used to train the surrogate model.
After training, the surrogate is used to update the detector parameters using a gradient-based optimizer,
$$
\boldsymbol{\theta}_{n+1} = \boldsymbol{\theta}_n - \eta \nabla_{\boldsymbol{\theta}} \mathcal{S}(\boldsymbol{\theta}_n),
$$
where $\eta$ denotes the learning rate. Additional constraints, such as detector size, material cost, or engineering limits, can be incorporated into the optimization objective through penalty terms in the loss function.

An important feature of this framework is its ability to treat mixed continuous and discrete detector parameters. Continuous parameters such as layer thicknesses are optimized directly, while categorical parameters such as material choices are represented by probability distributions over possible categories. The optimizer updates the corresponding logits, which are mapped to probabilities through a softmax transformation, enabling gradient-based optimization even in the presence of discrete design choices.

Specifically, the optimizer loss is expressed as a trade-off between regression and classification losses $\mathcal{L}_{opt} = \mathcal{L}_{reg} + \lambda \ \mathcal{L}_{class}$, a technique called linear scalarization. Setting different values of $\lambda$ can be used to explore the Pareto front. In this section, however, lambda is initially set to zero, such that only the regression is optimized, then annealed to one so the optimizer gradually orients to another regression loss space minimum that corresponds to a classification loss space minimum. The sigmoid annealing is chosen to provide a smooth transition, without which the optimizer can lose the already established improvement on the regression task. The starting configuration for the calorimeter is chosen as a very suboptimal one, with 5 layers of absorber with 10~\si{\cm} depth, each followed by a scintillator layer of 3~\si{\cm} depth. Both absorbers and scintillators are initialized with 50\% probability in each material to provide a neutral starting point.

\begin{figure}[ht]
    \centering
    \includegraphics[width=\columnwidth]{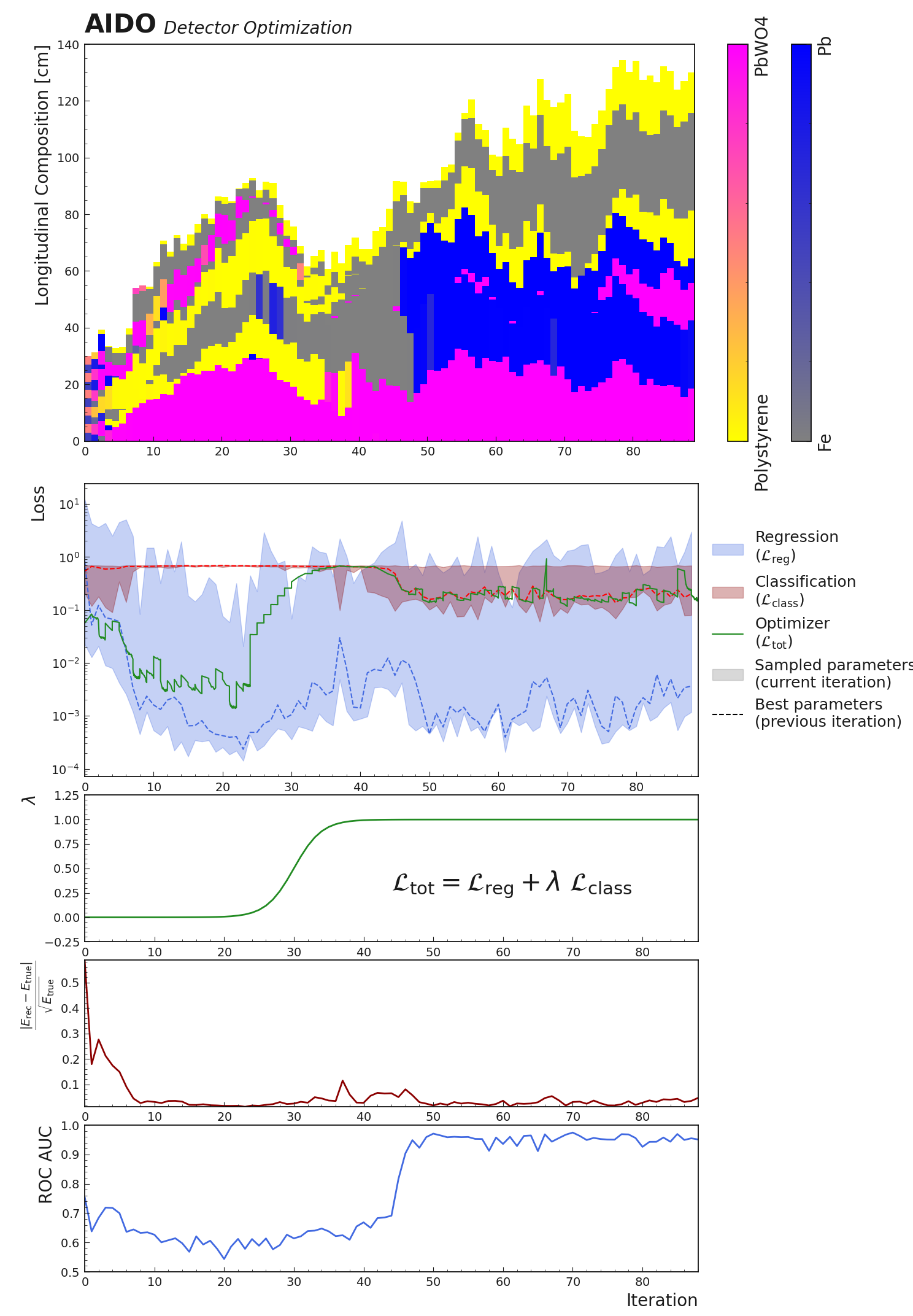}
    \caption{Evolution of the optimization using AIDO for both regression and classification as a function of the iteration number. First panel: longitudinal composition of the calorimeter. Second panel: evolution of the different losses with spread of the different sampled detector parameters in the shaded area, and among them the optimizer parameter set from the previous iteration in the dotted line. Third panel: annealing parameter dictating the trade-off between regression and classification performance. Fourth panel: regression performance as the energy resolution. Fifth panel: classification performance as the Receiver Output Characteristic Area Under the Curve (AUC) score.}
    \label{fig:aido_sampling_calorimeter}
\end{figure}

\begin{figure}[ht]
    \centering
    \begin{subfigure}{0.8\columnwidth}
        \centering
        \includegraphics[width=\columnwidth]{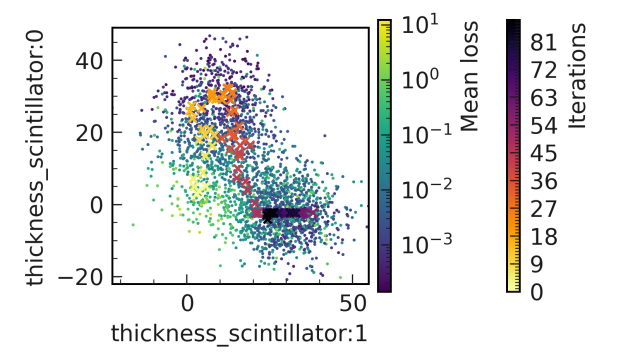}
        \caption{Regression loss space}
    \end{subfigure}
    \vspace{0.5cm}
    \begin{subfigure}{0.8\columnwidth}
        \centering
        \includegraphics[width=\columnwidth]{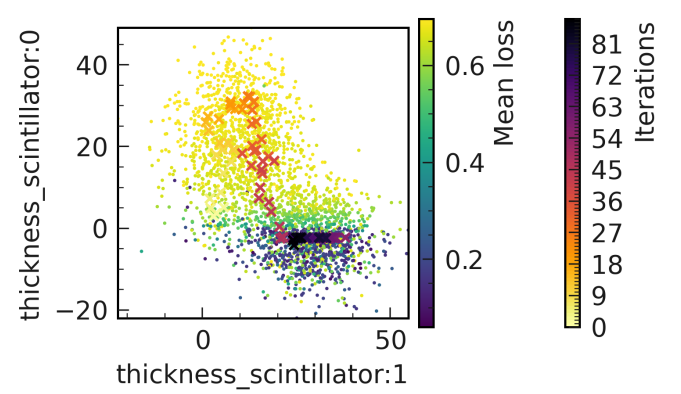}
        \caption{Classification loss space}
    \end{subfigure}
    \caption{Loss space for regression (top) and classification (bottom) as a function of two particular parameters: the first and second scintillator layer lengths. The dots represent the mean loss values for each set of detector parameters, and the crosses represent the state of the optimizer as a function of the algorithm iterations.}
    \label{fig:calo_AIDO_loss_space}
\end{figure}

This evolved experiment is illustrated in \autoref{fig:aido_sampling_calorimeter}. The first phase depicts that when $\lambda=0$, the regression loss decreases as expected, whereas the classification performance degrades and the spread of sampled parameters loss values collapses. It shows both regression and classification as two clear directions. In this phase, the classifier converges first to the dense scintillator layer to catch most of the shower, after collapsing other layers to zero length. In the second phase during the annealing of $\lambda$, the optimizer loss increases as the classification loss values get included but no improvement is observed as there is no proper direction in the classification loss space from the sampled parameters. After overcoming this, in the third phase when $\lambda=1$, the classification loss decreases while the regression loss remains at stable minimal value, which shows that the regression loss has reached a new local equivalent minimum. Due to the expansion of the previously collapsed layers, the calorimeter gets access to the longitudinal shape of the shower. Specifically, the first layer provides crucial data as $e^+$ particles produce an immediate shower, whereas $\gamma$ particles start their shower upon the first interaction through pair production later in the calorimeter.

These three phases can be observed in the parameter space of \autoref{fig:calo_AIDO_loss_space}. Initially, the optimizer favours a deep first scintillator layer to catch most of the shower and provide good energy resolution. This local regression loss minimum however coincides with a classification loss maximum, as this first deep layer prevents access to the information about the shower starting depth. After annealing, the optimizer moves to another point where both losses have a local minimum, and the first scintillator is reduced to a very small depth.

\begin{figure}[ht]
    \centering
    \begin{subfigure}{0.8\columnwidth}
        \centering
        \includegraphics[width=\columnwidth]{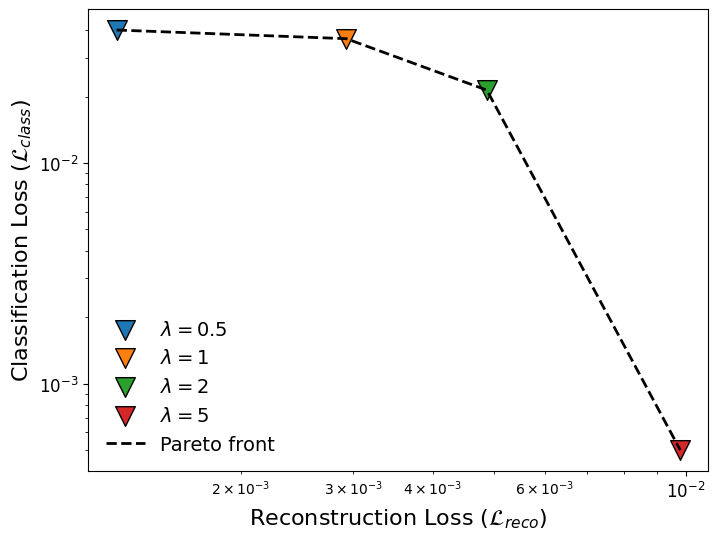}
        \caption{Pareto loss space}
    \end{subfigure}
    \vspace{0.5cm}
    \begin{subfigure}{0.8\columnwidth}
        \centering
        \includegraphics[width=\columnwidth]{ 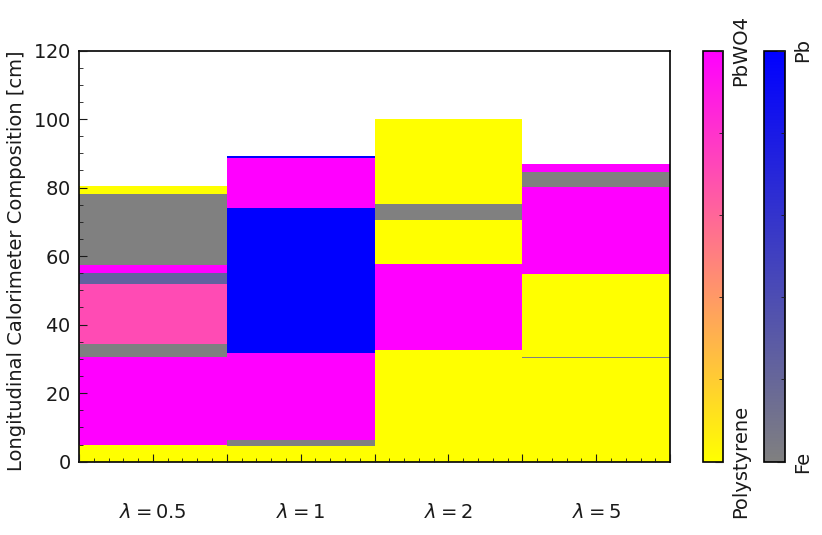}
        \caption{Optimal detector longitudinal view}
    \end{subfigure}
    \caption{Four optimization scenarios are run with constant values of the trade-off parameter $ \lambda$. Top: evolution in the loss space during optimization and the Pareto front. Bottom: resulting calorimeter longitudinal view.}
    \label{fig:calo_AIDO_pareto}
\end{figure}

The study of the effect of the trade-off parameter $\lambda$ between the regression and classification losses and the visualization of its Pareto front is in~\autoref{fig:calo_AIDO_pareto}. For small values of $\lambda$ where the regression objective is emphasized, AIDO will tend to place dense and thick scintillator layers early on to catch most of the shower, which are expensive, and therefore has to rely on larger absorber layers and thinner scintillator layers at the end of the shower. This situation is reversed for large values of $\lambda$ where the classification objective is emphasized and the calorimeter has cheaper and thicker scintillator layers to obtain meaningful longitudinal shower profiles, at the expense of potentially losing the tail of the shower, which impacts the energy estimation. In all cases, however, each detector retains a thin first scintillator layer that is beneficial to detect the early $e^+$ showers.

This surrogate-based approach enables efficient exploration of the calorimeter design space while maintaining compatibility with realistic detector simulations and reconstruction algorithms.

\begin{figure*}[ht!]
    \centering
    \includegraphics[width=\linewidth]{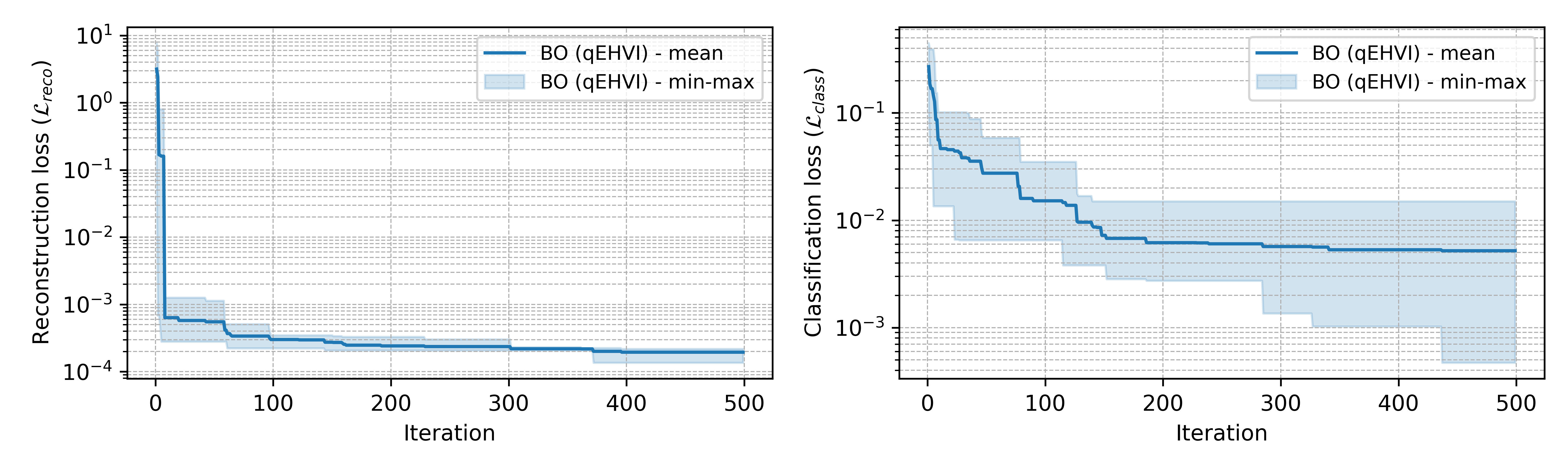}
    \caption{Best‑so‑far reconstruction loss $\mathcal{L}_{reco}$ and classification loss $\mathcal{L}_{class}$ over 500 iterations of Bayesian optimization. Solid curves show the mean best‑observed objective value across optimization runs, and shaded regions indicate the corresponding min–max envelope over five independent calorimeter design studies.}
    \label{fig:objective_envelope}
\end{figure*}

 \begin{figure}[ht!]
     \centering
     \includegraphics[width=\columnwidth]{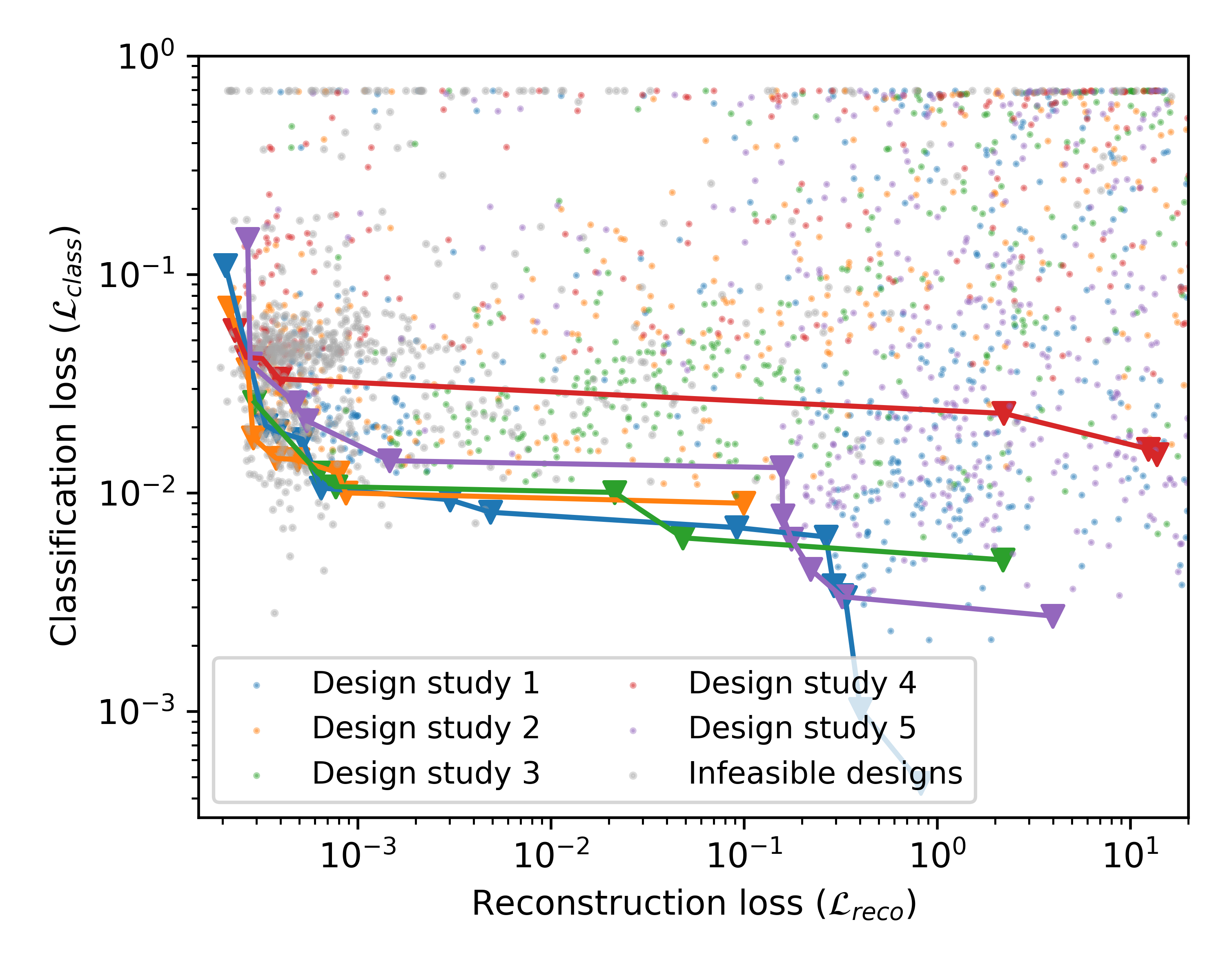}
     \caption{Estimated Pareto fronts for $\mathcal{L}_{reco}$ and $\mathcal{L}_{class}$ obtained from five independent calorimeter design studies. Each colored curve represents the non‑dominated frontier of feasible designs. Gray and colored markers denote feasible and infeasible configurations. }
     \label{fig:bo_sampling_calorimeter}
 \end{figure}

\subsubsection{End-to-end Bayesian Optimization}
\label{sec:sampling_calorimeter_bo}

In alternative to the gradient-based optimization scheme in \autoref{sec:aido}, which relies on local surrogate models, we study a global blackbox optimization approach using Bayesian optimization (kriging), as illustrated in \autoref{fig:calo_BO_scheme}. To perform this reference optimization, several adjustments to the previously described setup are required:

\begin{itemize}
    \item As this method does not require optimizing a local surrogate over multiple sampled detector layers, both classification and regression models are simplified and only trained on a single detector layout without requiring conditioning the model on the target detector layout. The previously introduced features are therefore removed from the network architectures. 
    \item Further, Bayesian optimization requires a well-defined, bounded search domain for continuous variables. For the detector layer thicknesses, the admissible design space is therefore restricted to box constraints

    \begin{equation}
        t_{\cdot,i} \in \{t \in \mathbb{R}^{+} \vert \SI{0}{\centi\meter} \leq t \leq \SI{50}{\centi\meter} \}\,,
    \end{equation}
    
    limiting the search space to layer thicknesses of up to \SI{50}{\centi\meter}.  We further define a log-uniform prior over the box constraint interval, defining a search space in which both small and large thicknesses are adequately explored, reflecting their disproportionate impact on detector performance and cost.
\end{itemize}

 With those changes, we can perform an optimization of the objective in~\autoref{eq:objective_sampling_calo}, where new detector layouts are sampled according to an acquisition function computed from a Gaussian process surrogate model.

We implement Bayesian optimization using BOTorch~\cite{Balandat2020} with a \emph{Quasi MC-based batch Expected Hypervolume Improvement} (qEHVI) acquisition function~\cite{Daulton2020} minimizing over the  multi-objective criterion in~\autoref{eq:objective_sampling_calo}. The geometric constraints outlined in~\autoref{eq:objective_sampling_calo} are handled by the qEHVI acquisition function as soft constraints, weighting the expected hypervolume improvement by the posterior probability of feasibility. To ensure a comparable optimization budget while accounting for the parallelization evaluation structure of the gradient-free approach, we increase the number of iterations to 500, with 100 of them allocated for the initial design of experiments (DoE). We further perform five independent design studies, measuring the uncertainty and variation  in both objectives across optimizations, quantifying the stability of the obtained design.

\paragraph{Results} Figure~\ref{fig:objective_envelope} presents the best-so-far utilities for classification and regression losses for the optimization, with the mean best-so-far results marked together with the min-max interval across design studies. We find for both objectives a significant improvement in both objective dimensions over time. Yet, we find for the reconstruction objective a stronger convergence compared to the classification objective, presented by a much narrower band of the min-max interval. Consequently, the classification task demonstrates being much more sensitive to initialization and the resulting design specifications sampled during the design of the experiment phase. Finding consistently well-performing detector layouts for high-quality particle identification likely requires a larger optimization budget or further tuning, enabling a more thorough exploration of the search space. This observation is consistent with the AIDO use case: although the detector can be initialized to a feasible starting point, reducing sensitivity of initialization, the higher complexity of the classification task amplifies surrogate modeling errors, leading to instabilities in the optimization convergence.

\begin{figure}
    \centering
    \includegraphics[width=\linewidth]{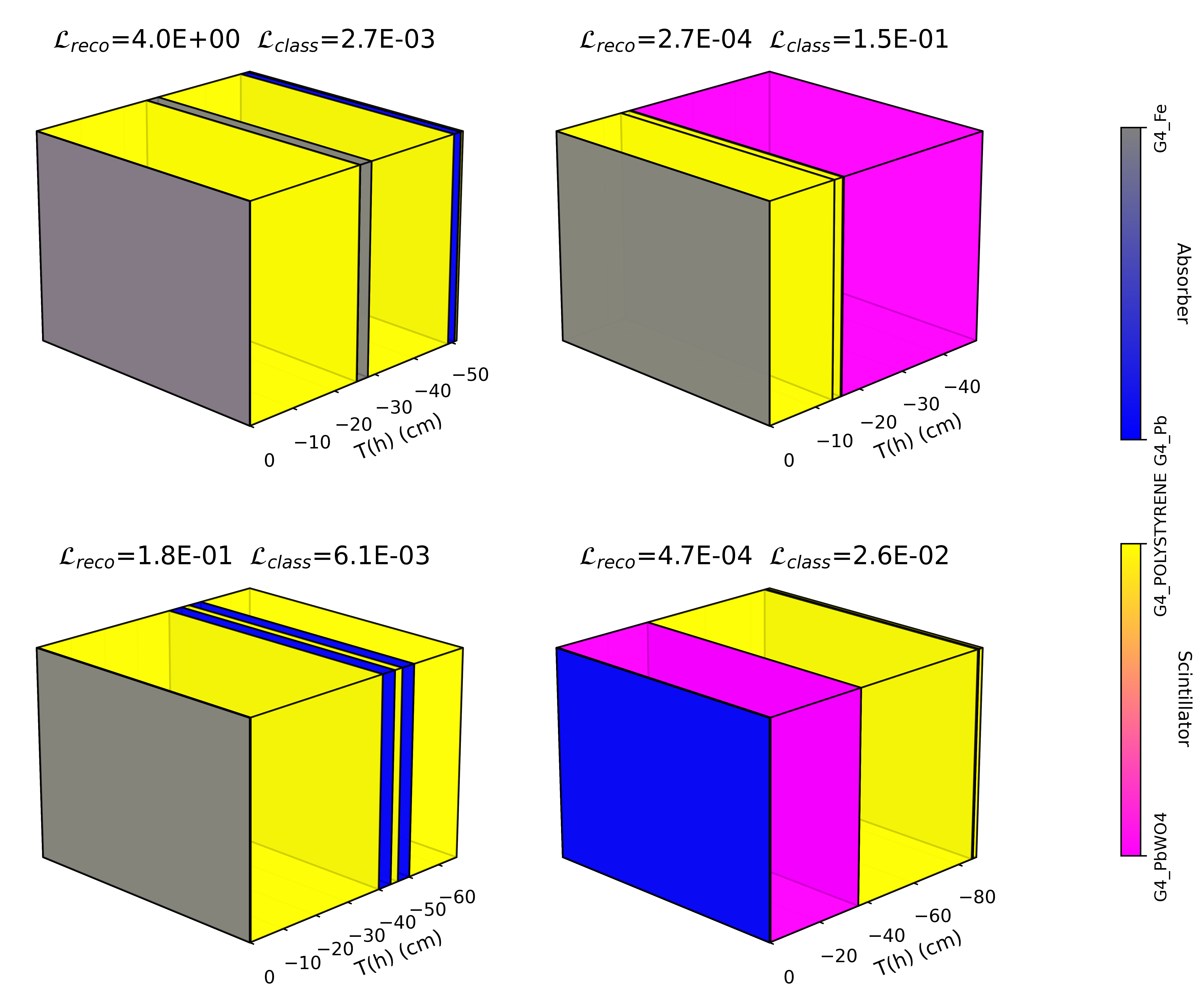}
    \caption{Selected number of detector layouts that are located along the Pareto frontier obtained during Bayesian optimization.}
    \label{fig:calo_geometries}
\end{figure}

Building upon those results, the joint performance of the optimization studies is presented and analysed in \autoref{fig:bo_sampling_calorimeter}. Each coloured curve represents an estimated Pareto frontier of non-dominated, feasible designs obtained by jointly optimizing detector and reconstruction-software parameters. All sampled feasible and infeasible designs for each study are marked in the design studies colour and gray, respectively.

Figure~\ref{fig:calo_geometries} visualizes a subset of Pareto-optimal designs, highlighting the variability of layouts that achieve competitive trade‑offs between reconstruction and classification performance. We find, similar to the results in~\autoref{fig:objective_envelope}, that regression‑dominated solutions remain highly consistent across optimization runs, whereas designs that emphasize classification performance show increasing variability.

\subsubsection{End-to-end calorimeter optimization with mutual information surrogates}\label{sec:calorimeters:mutual_inf_surrogates}

In addition to the studies described {\em supra}, we investigate the optimization of a sampling calorimeter detector using an information theoretic objective. The sampling calorimeter under investigation is composed of three layers, each consisting of a passive absorber and an active scintillator element. The materials are fixed in this study to lead (Pb) for the absorber and polystyrene ((C$_8$H$_8$)$_n$) followed by two lead tungstate (PbWO$_4$) segments for the scintillators. The detector parameters for optimization are the layer segment thicknesses $t_{A,i}$ and $t_{S,i}$ for $i \in \{1,2,3\}$.

\paragraph{Proposed solution: Mutual Information Surrogate}

Here we adopt the Mutual Information Surrogate approach presented in Ref.~\cite{Wozniak2025EndToEnd}, which is based on two central components. First, we train a deep learning–based \textit{surrogate} model that approximates the performance metric \textit{locally} in the detector design parameter space~\cite{shirobokov2020black}. Second, we employ mutual information~\cite{mackay2003information,cover2006elements} as the performance metric.
This information-theoretic objective measures the amount of information about the underlying physical quantities retained by the detector configuration $\{t_{A,i}$ and $t_{S,i}\}^i$, independent of any specific reconstruction algorithm or downstream analysis. The co-design aspect here is \textit{implicit}: By maximizing mutual information, the framework provides a task-agnostic and formally well-defined generic criterion for detector design.
The goal is to identify the design configuration of the detector $D_h$ that optimizes mutual information $MI$ for a set of physics processes $X$, yielding
\begin{equation}
\begin{split}
h^* = \argmin_{h \in \mathcal{H}} MI\!\left(X, D_{h}(X)\right)\,, \\
\text{where} \quad \mathcal{H} \coloneqq \prod_i \mathcal{H}_i \\
\text{and} \quad \mathcal{H}_i \coloneqq \mathbb{R}^+ \times \mathbb{R}^{+}\,.
\end{split}
\end{equation} 

As in the previous approaches, practical constraints, such as material budgets or cost limitations, can be incorporated directly into the optimization procedure. \\ 
By providing a differentiable approximation of the simulation-based objective, this approach enables efficient, gradient-based optimization of the detector design.

\paragraph{Data}
We generalize the experimental setup introduced in Ref.~\cite{Wozniak2025EndToEnd} to a \textit{multi-particle} scenario. As above, we generate a dataset by simulating single incident photons ($\gamma$) and positrons ($e^+$$)$ with energy uniformly sampled between 5 and \SI{20}{\giga\electronvolt} using \textsc{Geant4}. Only the energy deposited in the active scintillator layers is recorded.

Layer thickness values in centimeters define the design parameter vector $t_A, t_S \in \mathbb{R}^F$, where $F$ denotes the number of independent geometric parameters per layer type ({\it e.g.}\ $F=3$ for three layers with separately optimized absorber and scintillator thicknesses). In the following we will use a single symbol for the concatenated absorber and scintillator thicknesses denoted as $\theta := t_A | t_S $. Around a nominal configuration $\theta$, we sample $K$ perturbed designs
\[
\theta^k \sim \mathcal{N}(\theta, \sigma)
\]
with $\sigma=1.5$cm chosen on the order of one radiation length~\cite{pwo_alice_ALEKSANDROV2005169}. This localized sampling follows a trust-region strategy~\cite{shirobokov2020black, trust_region_doi:10.1137/0719026}. For each $\theta^k$, simulations yield true particle features $x^k$ and corresponding detector responses $y^k$.

\paragraph{Methods \& Models}
The objective $\delta$ for each candidate design is defined as the mutual information between true particle features and detector responses
\[
\delta^k = MI(x^k; y^k).
\]
Mutual information is estimated using the objective model $R$ - a neural mutual information estimator ~\cite{belghazi2018mutual} based on variational bounds.

A surrogate model $S$ is trained to approximate the mapping
\[
S : \theta^k \mapsto \delta^k.
\]
We use a multilayer perceptron to model $S$. After training, gradients $\nabla_{\theta} S(\theta)$ are used to update the design parameters via gradient descent within the trust region. A penalty is added if the total detector thickness exceeds \SI{35}{cm}. The updated parameter vector defines the next nominal configuration, and the procedure is repeated for a fixed number of iterations.

\begin{figure}
    \centering
    \includegraphics[width=0.35\textwidth]{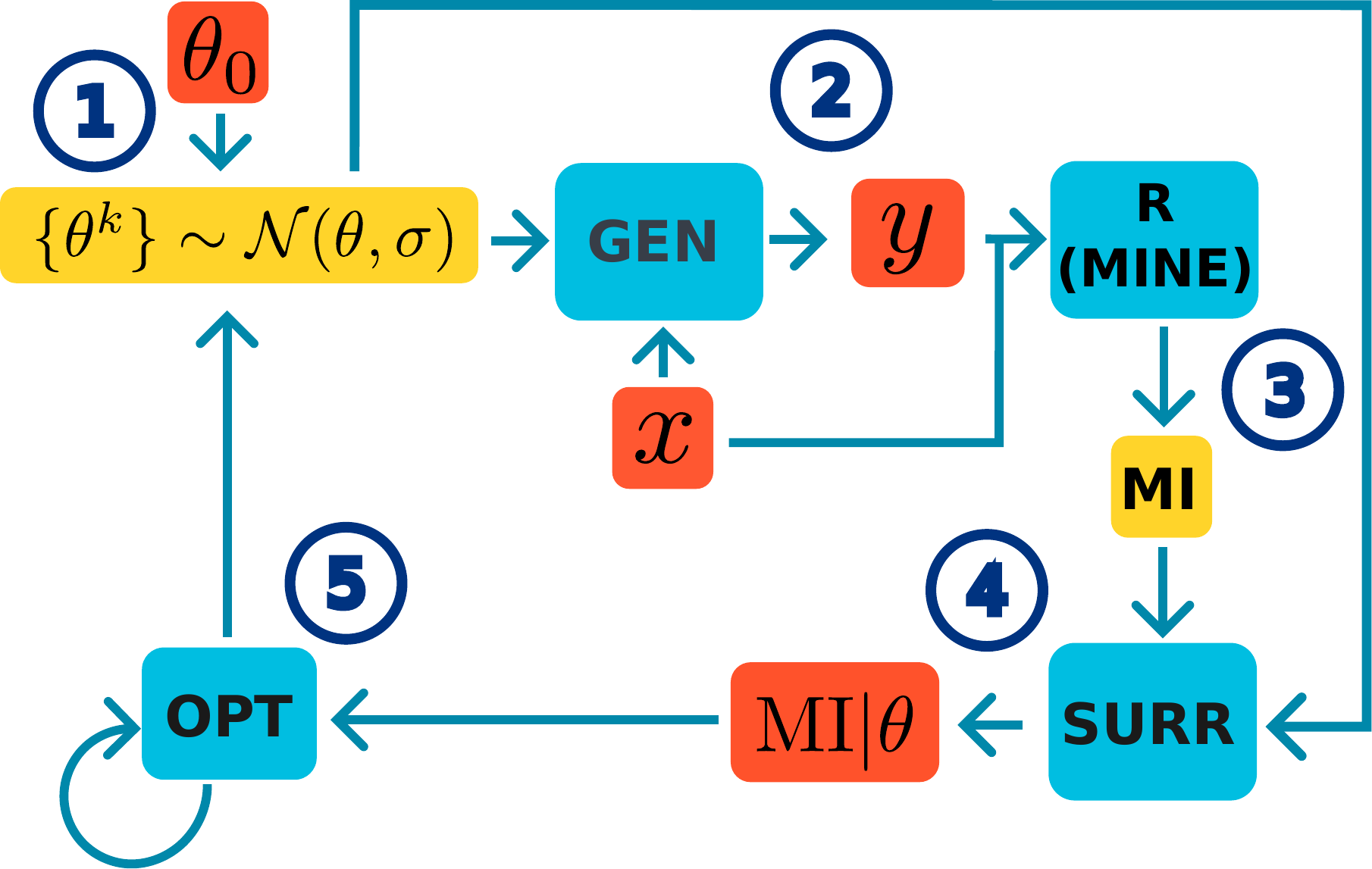}
    \caption{Optimization scheme for the mutual-information-based surrogate method.}
    \label{fig:mu-inf-calo-opt:opti-loop-pipe}
\end{figure}

The entire optimization loop is displayed in~\autoref{fig:mu-inf-calo-opt:opti-loop-pipe} and is composed of five steps: Generation of test points from the nominal parameter's local neighborhood (step~\raisebox{.5pt}{\textcircled{\raisebox{-.9pt} {1}}}); simulation of $M$ photons and positrons incident on the detector, generating $U$ true features per particle $x^k \in \mathbb{R}^{M \times U}$, as well as their corresponding $V$ measured calorimeter energy deposits $y^k  \in \mathbb{R}^{M \times V}$ (step~\raisebox{.5pt}{\textcircled{\raisebox{-.9pt} {2}}}); application of the objective model $R$ on the tuples $\{x^k, y^k\}_{k=1}^K$ to estimate the mutual information objective $\delta^k$ for each neighborhood design configuration $\theta^k$ (step~\raisebox{.5pt}{\textcircled{\raisebox{-.9pt} {3}}}); The MLP surrogate model $S$ is trained to predict this mutual information metric from the detector parameters: $\hat{\delta^k} = S(\theta^k)$ (step~\raisebox{.5pt}{\textcircled{\raisebox{-.9pt} {4}}}); Finally SGD is applied to identify the next optimal design parameters by following the gradient of the surrogate $-\frac{\partial S}{\partial \theta}$, (step~\raisebox{.5pt}{\textcircled{\raisebox{-.9pt} {5}}}). The identified new $\theta$ restarts the next iteration or is returned if the maximum number of iterations has been reached.

\paragraph{Experiments}
We evaluate our surrogate framework in the optimization of the described simplified calorimeter detector where the optimization parameters $\theta$ correspond to the thicknesses of each layer segment.\\
The set of studies investigates a three-layered calorimeter and 350 events of a photon and positron with an energy range of \qtyrange[range-units=single,range-phrase=-]{1}{20}{\GeV} per parameter set candidate. The nominal parameter neighborhood is covered by 70 candidates. \\
Initial segment thicknesses are \SI{1.0}{\cm} for both scintillators and absorbers. All results are presented as the mean evolution with standard deviation, computed from three experimental runs.

\begin{figure}
    \centering
    \includegraphics[width=0.35\textwidth]{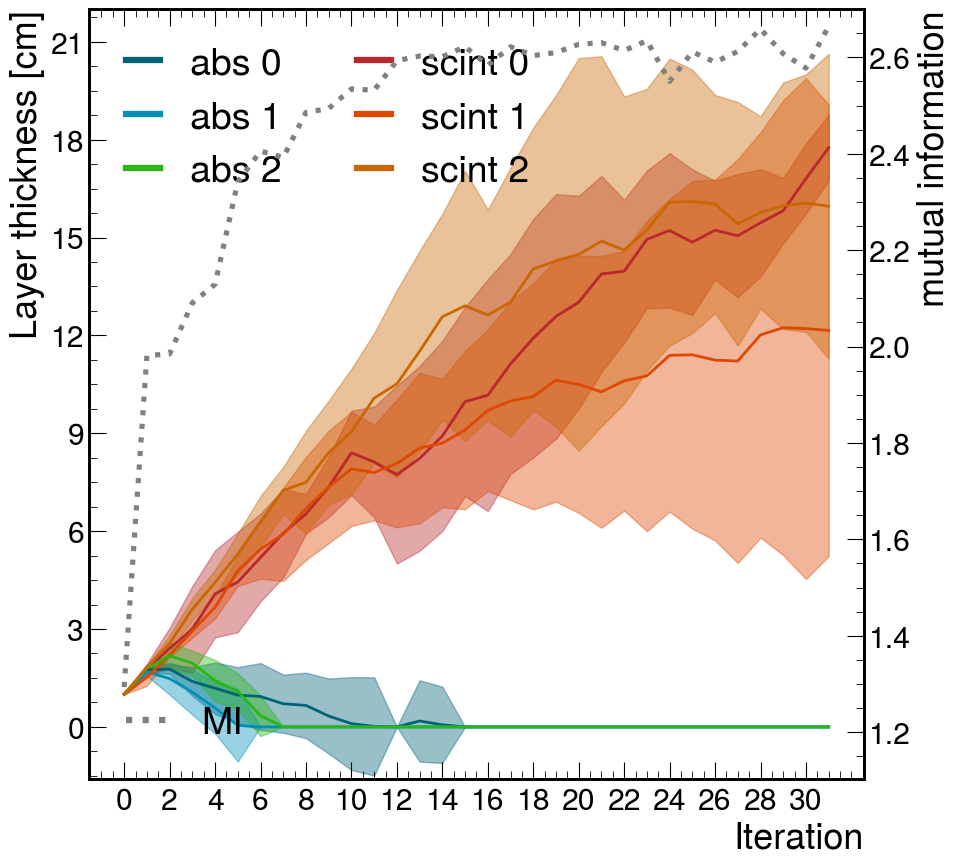}
    \caption{Layer segment thickness evolution (solid lines) and mutual information evolution (gray dashed) averaged over three runs for a three-layered calorimeter. Absorbers are shown in blue, scintillators in red.}
    \label{fig:mu-inf-calo-opt:evo_scint_abs}
\end{figure}

\begin{figure*}
    \centering
    \begin{subfigure}{0.35\textwidth}
        \includegraphics[width=\textwidth]{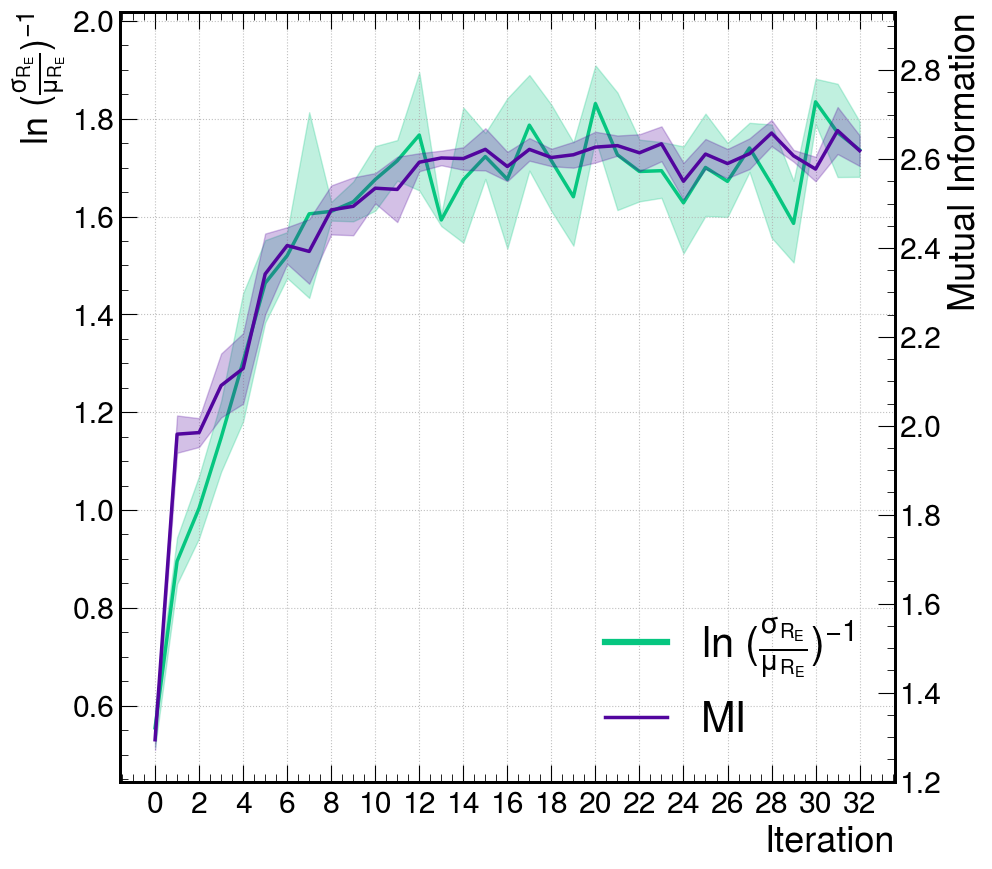}
        \caption{Mutual information vs. energy resolution, photon}
        \label{fig:energy_res_photon}
    \end{subfigure}%
    \hfil
    \begin{subfigure}{0.35\textwidth}
        \includegraphics[width=\textwidth]{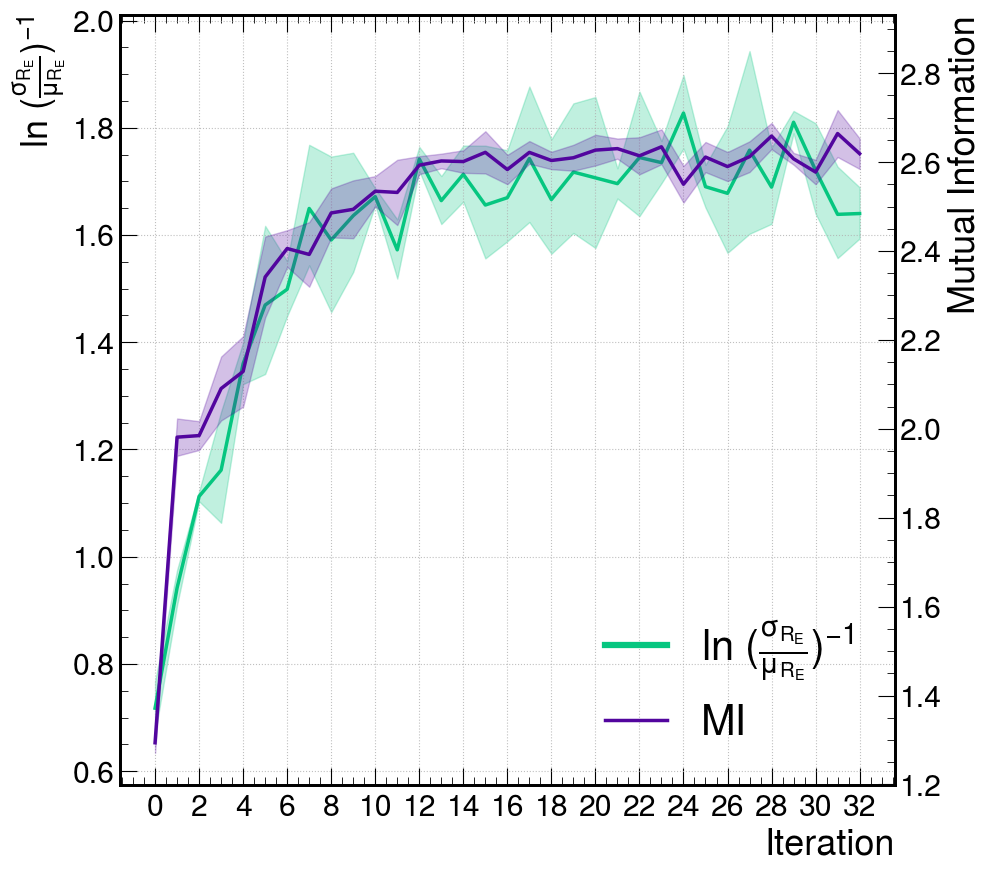}
        \caption{Mutual information vs. energy resolution, positron}
        \label{fig:energy_res_3_layers}
    \end{subfigure}
\caption{Mutual information and energy ratio resolution: Comparison of photon (left) and positron (right) events. Results shown are averages over three runs.}
\label{fig:mu-inf-calo-opt:energy_res_vs_mi}
\end{figure*}

The thickness evolution of scintillators and absorbers are displayed in~\autoref{fig:mu-inf-calo-opt:evo_scint_abs}.
We observe that the optimization maximizes scintillator thicknesses while minimizing absobers, essentially reducing the latter to zero. This strategy enhances sensitivity to the physics process and absorbers are only of use when scintillators are too thin to fully contain high-energy particles (first iterations).\\
Another interesting study looks at how mutual information relates to concrete physics‑informed metrics commonly used at the LHC: for each of the particle types - photons and positrons - we compute the energy resolution and compare it to the evolution in mutual information (see~\autoref{fig:mu-inf-calo-opt:energy_res_vs_mi}). 
 \\
We observe that mutual information scales proportionally with the logarithm of the inverse resolution. The two metrics show a largely monotonic relationship for both particle types of photons and positrons.

\subsubsection{Conclusion}

Further studies are required for a fully quantitative comparison; however, several qualitative conclusions can already be drawn. Both local-surrogate optimization via AIDO and global optimization with BO identify comparable regions of the Pareto front, indicating that similar solution spaces are explored. Further identified solutions on the estimated Pareto front show similar geometric patterns, demonstrating the robustness and suitability of both approaches for calorimeter codesign. Bayesian optimization provides a diverse set of non-dominated solutions, naturally enabling trade-off analysis between different designs. In contrast, AIDO achieves faster per-run convergence as only a single linear combination of objectives is optimized. By linear scalarization, weighting each contribution of objectives, the estimates of the Pareto frontier can also be obtained. Previous results in~\citep{shirobokov2020black} indicate an improved sampling efficiency of local-generative surrogates over BO, yet confirming this effect for multi-objective optimization is non-trivial, as the approaches fundamentally differ in terms of optimization strategy and handling of multiple objective functions. \\ 
Moreover, surrogate stability is a central limiting factor in AIDO, requiring further understanding of the internal mechanisms. We found that scheduling strategies during optimization are a central factor in stabilizing surrogate behavior.

Finally, also the approach based on mutual-information surrogates successfully optimizes detector parameters in a multiparticle setting with photons and positrons. The optimization consistently favors increased scintillator thickness and reduced absorber thickness, demonstrating the feasibility of end-to-end, black-box deep-learning–based detector design using a single scalar objective. Since this objective is given by mutual information, it does not require explicit task definition, rendering the method task-agnostic and user-independent. While the mutual-information estimator must be retrained under distribution shifts, depends on sufficiently sampled marginal distributions, and does not naturally accommodate multi-objective weighting, the strong correlation observed with physics-informed performance metrics supports its role as a robust proxy for detector performance in optimization.

\subsection{Ground arrays for gamma-ray astronomy 
\label{s:swgo}}

Here we consider the problem of optimizing the design of an array of  detectors tasked with measuring the flux of high-energy gamma rays, their energy, and direction, using the detected secondary particles produced by their extended air showers (EAS) or the Cherenkov light emitted with an EAS that propagates to the ground. 

\subsubsection{Optimization of layout of shower front detectors: the case of SWGO} 

One such project is the SWGO array, a proposed experiment which might consist of about 6,000 water tanks of 3.6~m diameter, deployed at 4,700~m altitude in the Chilean Andes~\cite{BarresdeAlmeida:2020hkh}. A detailed study of the optimization of the geometrical layout of the tanks on the ground~\cite{swgoopt}, performed with the SWGOLO software~\cite{swgolo}, demonstrated how significant gains could be obtained by employing a three-component utility function summarizing the flux precision ${\phi_\gamma}$ --resulting from effective discrimination of gamma rays from proton-induced showers and maximization of the flux of signal events-- along with the accuracy of energy and angular resolution ${\sigma_E},{\sigma_\theta}$, \par

\begin{equation}
    U = U_{\phi_\gamma} + U_{\sigma_E} + U_{\sigma_\theta}\,,
\label{eq:utilswgo}
\end{equation}

\noindent
in conjunction with a gradient-descent optimization approach relying on a fully differentiable model of EAS development, detection of secondary particles, and reconstruction of gamma-ray parameters.

In the above-cited source authors considered as optimization parameters the $(x_i,y_i)$ positions on the ground of the centers of the tanks, when the latter were aggregated for simplicity into multi-tanks consisting of hexagonal arrangements of 7 or 19 tanks to reduce the number of free parameters. While already a rather complex task, the maximization of the utility $U$ in ~\autoref{eq:utilswgo} does not fully solve the design problem, as a number of details --concerning, {\it e.g.}, the performance of the tanks in terms of time resolution and particle discrimination capabilities, or the triggering strategy for data collection-- were fixed in the model to suitable approximations. The reason for that choice was a focus of the cited work on the exploration of the possible geometrical configurations and their effect on shower reconstruction, rather than on the bigger problem of finding a complete solution of the experiment optimization task. Here we extend that model to include in it consideration of a few parameters that control the performance of single tanks as well as triggering schemes for data collection, setting to ourselves the task of appraising the gains in performance that can be brought by the inclusion of those additional parameters in the design space during an end-to-end optimization, with respect to a two-stage optimization that only deals with the added parameters after the optimization of the layout.

For the sake of the present study we also find it useful to upgrade the utility definition of ~\autoref{eq:utilswgo}, to overcome the requirement of some manual tuning of component weights ( $\eta_{GF}, \eta_{IR}, \eta_{PR}$) and the lack of a natural encoding of the coupled nature of flux sensitivity, angular resolution, and energy resolution for the scientific goals of the experiment~\cite{dorigo2025_util}. A more unified treatment that takes inspiration from typical search methods for astrophysical sources in the sky is suggested by the so-called  "on-off problem", where the signal rate $S$ from a point source in the sky is affected by an additional diffuse background rate $B$. By looking "off-source", a telescope or other detection apparatus may estimate $B$, allowing a subtraction of background contributions in the "on-source" region, provided that a proper normalization depending on the ratio of areas and integration time is performed~\cite{LiMa}. 
As the detection of high-energy gamma-ray sources is one of the primary goals of SWGO, we consider the statistical significance of the signal resulting from a monochromatic (fixed energy) point source in the sky, for a given integration time, to be a sound proxy to the overall performance of the detection apparatus. We define the point-source utility as \par

\begin{equation}
    U_{PS} = \frac{1}{t_{5\sigma}}\,,
\end{equation}

\noindent
where $t_{5\sigma}$ is the data-taking time required to integrate a signal sufficiently large to correspond to a significance of five standard deviations, once accounting for backgrounds; this requires to specify a signal rate (in terms of number of photons per square km per year), which can be set to a small fraction of the total generated flux; the exact value of the signal rate does not affect significantly the optimization task. We also use the following simplified formula for the pseudo-significance $Z$ resulting from an excess of $N_S$ signal events over a Poisson-distributed background contamination of mean $N_B$: \par

\begin{equation}
    Z = 2(\sqrt{N_B+N_S} - \sqrt{N_B}).
\end{equation}

\noindent 
From the above expression we can solve for $N_{S_{5\sigma}}$ by setting $Z=5$; expressing $N_S$ as a function of effective area and signal efficiency yields \par

\begin{equation}
    U_{PS} = \frac{4 A_{eff} \epsilon_{\gamma}}{25+20\sqrt{N_B}}\,,
    \label{eq:Ups}
\end{equation}

\noindent
where $A_{eff}$ is the effective area wherein the array is sensitive to showers, $\epsilon_{\gamma}$ is the reconstruction efficiency of gamma-ray showers falling within $A_{eff}$, and where $N_B$ is computed by adding together the estimated contribution of real gamma rays from diffuse sources and the noise from hadronic showers reconstructed as gamma rays. The former is extracted from counting observed showers in the angular sideband around the signal region, while the latter is estimated from hadrons reconstructed in the signal region, multiplied by the misclassification probability of an optimized selection cut on the likelihood ratio between the two hypotheses (gamma ray or hadron) on the primary particle: \par

\begin{equation}
    N_B = N_{had}^{SR} \times P_{fake}(h\to\gamma) + N_{\gamma}^{SB} \times \tau\,,
\end{equation}

\noindent
where $N_{had}^{SR}$ the number of protons in the signal region, $P_{fake}(h \to \gamma)$ is the probability that a hadronic shower is classified as a gamma shower, $N_{\gamma}$ is the number of gamma rays in the sidebands, and $\tau$ the extrapolation factor equal to the ratio betweeen the area of signal and sideband regions in the sky. 

Defined as in ~\autoref{eq:Ups}, the utility unifies flux sensitivity, pointing accuracy, and energy resolution in a single expression without the need to assign different arbitrary weights to each component as in ~\autoref{eq:utilswgo}. Energy resolution dependence comes from the inclusion as signal event candidates of showers in an energy range spanning the range  $[-1.4\sigma_E,+1.4\sigma_E]$ around the hypothetical energy of the source. For the results discussed in this work, we set $E=4~PeV$ as the energy of photons from the source in our optimization runs.

We consider the following software parameters in our extended model:\par

\begin{itemize} 
\item The minimum number of tanks detecting at least one secondary particle in order for the event to be recorded and reconstructed, $N_{trigger}$. Because of its direct effect in determining which showers are included in the calculation of the utility $U_{PS}$, a variation of $N_{trigger}$ will affect the optimality condition of any array. More in detail, a higher value of $N_{trigger}$ will generally improve the energy and angular resolution of detected showers (as showers with an insufficient number of involved tanks will be removed from the sample where an average of these resolutions is evaluated), but it will reduce the flux (as the detection efficiency will globally decrease). 
\item One further extension of the original SWGOLO modeling concerns the consideration of a distributed data acquisition scheme, whereby instead of sending their signal to a single central counting room for triggering and reconstruction, units are connected to a number of decentralized counting rooms, each of which has independently the capability to trigger data acquisition for showers that produce a signal in the nearby tanks. A decentralized scheme has the benefit of reducing the cost of cabling (signal and high voltage) from counting rooms to detection units, as well as improving triggering capabilities, by exploiting the topology of showers, thus becoming less dependent on random background counts in far-away detection units. The number of counting rooms can be jointly optimized with the other parameters, once their position is independently assessed by a clustering algorithm that minimizes the sum of linear distances from the units.
\item The time resolution of secondary particles detection in each tank, $\sigma_t$. A conservative and fixed value of $10~ns$ was assumed in the original version of the SWGOLO optimization code. This number affects the precision of shower parameters reconstruction, as the time of arrival of secondaries is directly involved in the estimation of the angle of incidence of the primary cosmic or gamma ray. While timing resolution is more properly defined as a hardware parameter rather than a software one, its value results from the quadrature sum of hardware as well as software effects, some of which are not accessible at design time; hence an exploration of this parameter is useful. 
\item The precision in the counting of secondary particles $N_s$ of each tank is also modeled by assuming a value for the relative uncertainty $R_c=\sigma_{N_s}/N_s$ delivered by a careful reconstruction of the photomultiplier readout signals.
\end{itemize}

In order to further enrich the co-design aspects of the optimization task, we also extend the capabilities of SWGOLO by allowing at each iteration for the appearance or disappearance of detection tanks. The cost of the $N$ deployed tanks $C_{dets} = N C_{tank}$ (or $C_{dets} = N M C_{tank}$, if $N$ denotes the number of macro-tanks comprising $M$ tanks each) depends on the tank performance parameters $R_c$ and $\sigma_{t}$. The cost of each tank is specified by a baseline cost (8000\$) and by the timing and counting resolutions as \par

\begin{equation}
C_{tank} = 8000 \$ + \frac{10 \$}{R_c + 0.001} + \frac{5000 \$}{\sigma_t/ns + 1}\,.
\end{equation}

\noindent The above is a very rough and arbitrary model, which however encodes the correct monotonicity of cost on the specified parameters: that is to first order sufficient to acquire an understanding of the effect of the two parameters on the optimality conditions, and to observe how they play in the end-to-end optimization task. Other costs are connected with the area of the array ($C_{land}$), the total length of cables connecting each tank to the closest counting room ($C_{cables}$), the cost of counting rooms $C_{CR}$, and the perimeter (whose corresponding cost $C_{fence}$ is due to the need of fencing the facility), as detailed in \autoref{t:costs_swgo}. 

\begin{table}[!ht]
\begin{tabular}{l|l|l}
\bf{Cost item} & \bf{Cost in USD} & \bf{Applies to} \\
\hline
Land cost     & 0.5 \$ $/m^2$ & whole array\\
\hline
Leveling cost & 1.4 \$ $/m^2$ & times unit area, \\
              &                & per detection unit\\
\hline
Fencing cost  & 250 \$ $/m$    & total perimeter\\
\hline
Cabling cost  & 5 \$ $/m$      & per unit and per meter \\
              &                & to closest count. room \\
\hline
Counting room cost & 30,000 \$ & per counting room \\
\hline
Baseline unit cost &  8,000 \$ & per unit \\
                   &           & ($\sigma_t=10ns, R_c=0.05)$ \\
\hline
\end{tabular}
\caption{Ingredients of the cost model for the SWGO array used in the optimization by the upgraded SWGOLO code. While only approximate, for the sake of the presented study these numbers are a reasonable guess of the actual cost of the various components.}
\label{t:costs_swgo}
\end{table}

The total experiment cost is defined as \par

\begin{equation}
    TC = C_{land} + C_{level} + C_{fence} + C_{cables} + C_{CR} + C_{dets}\,.
\end{equation}

\noindent
Rather than enforcing a hard budget constraint, we model cost as a differentiable soft constraint via a smooth penalty function $U_{TC}(C)$, enabling gradient-based optimization.  The penalty is a multiplier that takes the value of $1.0$ in a range exceeding a minimum value and below a maximum budget, and decreases quickly for values outside. The function has the following form: \par

\begin{equation}
U_{TC}(C) = \frac{1+Erf[(C-m)/\sigma]}{2}\,,\qquad C<\frac{m+M}{2}\,;
\end{equation}
\begin{equation}
U_{TC}(C) = \frac{1-Erf[(C-M)/\sigma]}{2}\,, \qquad C \geq \frac{m+M}{2}\,,
\label{eq:Utc}
\end{equation}

\noindent
where $m$ and $M$ are minimum and maximum cost, respectively, and $\sigma$ is a smoothing width. Taking stock, we are setting up an optimization task where synchronously with the gradient descent procedure searching for the optimal layout of all tanks, the software jointly looks for the optimal value of a triggering threshold $N_{trigger}$, of the number of counting rooms $N_{CR}$, and of the most effective number of deployed tanks $N_{unit}$, accounting for their cost in the global utility;  the model is also enriched by consideration of the effect of the resolution in time and number of secondary particles in each detection tank, which jointly affect cost from the software design side of the problem. The utility target is \par

\begin{equation}
    U = U_{PS} \times U_{TC}\,.
    \label{eq:U}
\end{equation}

The integration of a search for the best value of the considered co-design parameters $N_{det}, N_{trigger}, N_{CR}, \sigma_t, R_c$ in the upgraded version of SWGOLO we employ here is performed by adopting a reinforcement learning approach. This is cast as a batched structured multi-armed bandit formulation, where each arm corresponds to a discrete detector-configuration choice (number of units, trigger threshold, and number of trigger zones). Arm selection is guided by a Thompson-sampling–inspired score~\cite{thompsonsampling}, 
combined with $\epsilon$-greedy exploration~\cite{epsilongreedy} 
to ensure continued sampling of sub-optimal configurations. Multiple arms are evaluated per iteration, placing the method in a batched bandit setting. For each selected arm, a local continuous refinement of detector performance parameters is performed via stochastic perturbations by gradient descent. This results in a hybrid discrete–continuous optimization strategy inspired by structured and combinatorial bandit methods.

At selected gradient descent iterations (every iteration in the first third of the run; every third iteration until the second third of the run; and every six iterations until the end), besides computing the gradient of the total utility with respect to the movement of each detection unit, a set of up to 8 different evaluations of the utility are performed by varying the five additional parameters according to proposed distributions that are updated depending on their outcome; the parameter set resulting in the best utility value is compared to the baseline, and if it provides an improved utility it is chosen for the next two validation iterations, along with updated unit positions; if the two further evaluations of the utility confirm the improvement, the new parameter set becomes the new baseline. When $N_{det}$ is varied (always by multiples of 3 units, to preserve the symmetry of the layout --see below), new triplets of units sharing the same radius and arranged in an equilateral triangle arrangement are removed or added; when they are added, their square radius is chosen uniformly within $[0,(1.5r)^2]$, where $r$ is the current radius of the array being considered, and the azimuthal position of one of the three units is chosen uniformly in $[0,2\pi/3]$.

For our tests we consider an array initially set in a configuration of $N_{det}=684$ detectors arranged into 36 macro-units\footnote{A macro-unit is a tightly packed hexagonal arrangement of 19 units.}. The array is made up of concentric circles of 6+12+18 equispaced macro-units, with an initial radius of the array of $1200~m$; this rather large-radius array is already close to optimality in terms of its extension on the ground, considering the problem of effectively detecting a 4-PeV gamma-ray source. Units are defined to be in the baseline performance configuration ($R_c=0.05, \sigma_t=10ns$); the other parameters are set to $N_{CR}=3$ and $N_{trigger}=80$. These parameters define the initial cost $C = 10,307,700$ US dollars of the array, which in turn defines the maximum cost (\autoref{eq:Utc}) as $M=2.0C$, and the minimum cost as $m=0.2C$). In such a way, we give to the optimization scan the flexibility to increase the total cost in the way that maximizes the value of the combined utility $U$ of ~\autoref{eq:U}. 

As was done in the studies of the stem publication~\cite{swgoopt}, we exploit the azimuthal symmetry of the problem by enforcing that the same configuration of units and counting rooms is repeated in three 120-degree sections. 

\begin{figure*}
\includegraphics[width=0.99\linewidth]{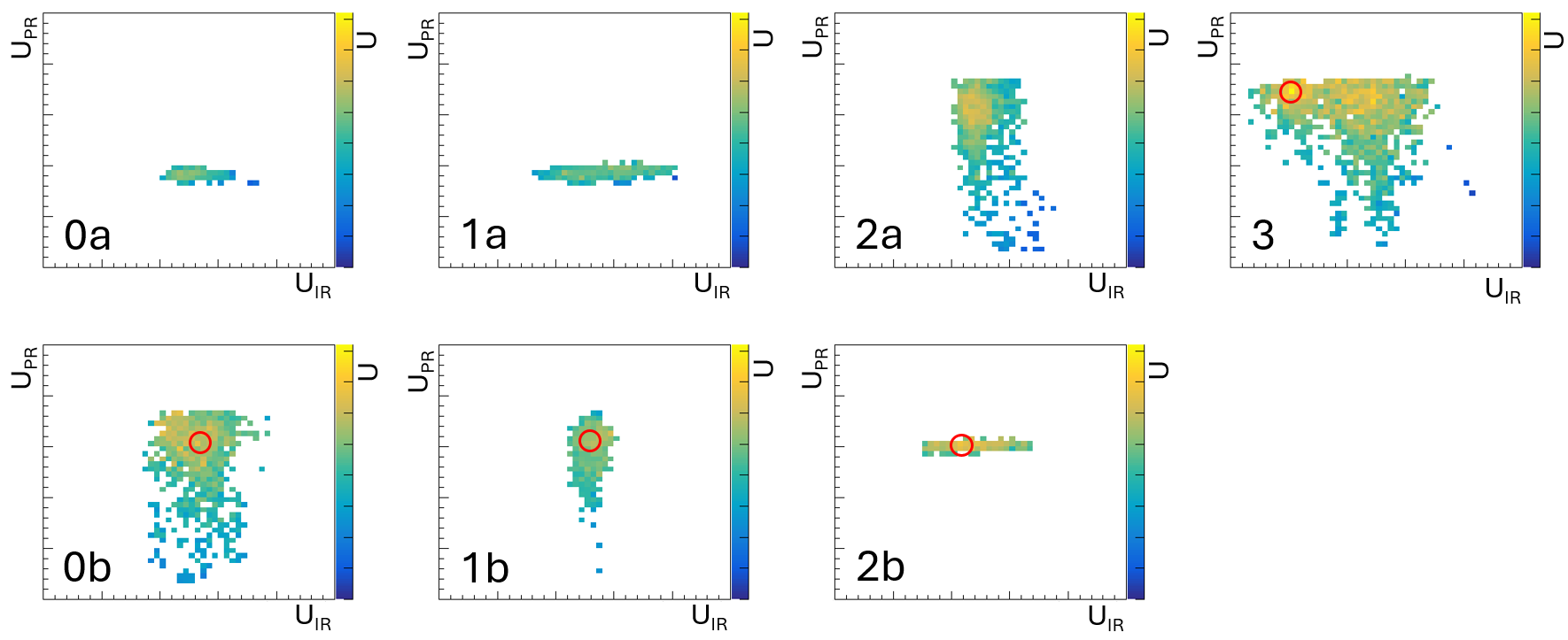}
\caption{The highest value of the total utility $U$ (\autoref{eq:U}) obtained by the layouts during the scans reported in ~\autoref{t:swgo_cdr} is shown as a function of the value of utility components $U_{IR}$ (horizontal axes) and $U_{PR}$ (vertical axes). For each of the investigated bins in the plane the graphs report the highest $U$ value found at those coordinates, such that the colour of the bins map the Pareto front. The different areas in the space of components $U_{IR}$ and $U_{PR}$ (which respectively provide an indirect measure of energy and angular resolution) sampled by the different serial scans (0a/0b, 1a/1b, 2a/2b) reflects their limited ability of working around cost constraints, due to fixed parameters. The full scan of the design parameters (3, top right) can instead reach a non-trivial solution (which accepts a lower energy resolution in exchange of higher flux sensitivity and angular resolution) and a higher total $U$ value. Red circles show the value of $U_{IR}$ and $U_{PR}$ of the highest-$U$ layout for each of the four optimization strategies.}
\label{f:paretoswgo}
\end{figure*}

We perform 500-epoch runs with different choices for the co-design parameters varied in the optimization; following these runs, we follow up with new ones where previously variable parameters are fixed to their optimal value, to search for the optimal value of the ones initially left fixed; in the new runs we still allow for position updates by gradient descent. A summary of the results is provided in \autoref{t:swgo_cdr}, while~\autoref{f:paretoswgo} shows Pareto fronts of the optimization scans.

\begin{table*}
\small
\begin{tabular}{l|rcccc|c|c|l}
\hline
Run ID & Epochs & $\delta xy$ & $\delta N_{det}$ & $\delta N_{trigger}, \delta N_{CR}$ & $\delta \sigma_t, \delta R_c$ & $U/U_0$ gain & $SR$ & Description\\
\hline
0a  & 1-500 & y & y & n & n & $1.220 \pm 0.037$ & & SGD scan varying only $N_{units}$ \\
0b & 501-1000 & y & n & y & y & $1.485 \pm 0.056$ & $1.220 \pm 0.059 $ & Subsequent scan of other pars\\
\hline
1a  & 1-500 & y & y & y & n   & $1.052 \pm 0.040$ &  & SGD scan varying all but detector pars  \\
1b & 501-1000 & y & n & n & y & $1.498 \pm 0.042$ & $1.210 \pm 0.049$ & Subsequent scan of detector pars \\
\hline
2a  & 1-500 & y & y & n & y &  $1.659 \pm 0.048$  &  & SGD scan varying all but trigger pars  \\
2b & 501-1000 & y & n & y & n & $1.707 \pm 0.052$ & $1.062 \pm 0.045$ & Subsequent scan of trigger pars \\
\hline
3  & 1-500 & y & y & y & y & $ 1.812 \pm 0.054$ & {\bf 1.0}& Full co-design scan \\

\hline
\end{tabular}
\normalsize
\caption{Results of optimization runs with different choices of optimizable parameters. The quoted $U/U_0$ ratios in all cases take as reference $U_0$ the same default parameters ($N_{trigger}=80$; $N_{CR}=3$; $\sigma_t=10~ns$; $R_c=0.05$) and layout conditions (a circular array of 36 initial macro-units).  $SR$ indicates the estimated sub-optimality ratio of ~\autoref{eq:sr}. See the text for detail.}
\label{t:swgo_cdr}
\end{table*}

As shown in ~\autoref{t:swgo_cdr}, the simultaneous scanning of all design parameters --"hardware" ones as well as "software" ones-- outperforms serial approaches that simplify the search by reducing it to the scanning of two subspaces one after the other. 
The sub-optimality ratio defined in \autoref{s:intro} is not computed in full by averaging over all possible initial conditions in this example, as the full optimization scans involve a large CPU load (one to two weeks of running on a 32-core machine per each of the four tests of ~\autoref{t:swgo_cdr}). Still, the single test we run suffices to exemplify how there is a significant decrease in performance of serial scans with respect to the co-design approach, with the single-instance SR evaluation taking variable values ranging between $1.06$ and $1.22$. In addition, ~\autoref{f:paretoswgo} shows distributions of the area scanned by the different optimization strategies, highlighting the wider reach of the full scan. In particular, from those distributions one notices that scans 2a/2b, which leave out of the initial search the trigger strategy parameters $N_{trigger}$ and $N_{CR}$, fail to discover the small-$N_{trigger}$ solution picked up by the full scan.
These results demonstrate how co-design enables the exploration of regions of the design space inaccessible to serial optimization, leading to consistent improvements in global utility. Serial scans fail to achieve full optimality because cost couples discrete and continuous parameters non-separably, so optimizing one subspace shifts the optimum in the other.

\subsubsection{{CTAO array layout optimization via Monte Carlo co-design}
\label{subsec:ctao-layout-mc}}
\begin{figure*}[h!t]
    \centering
    \includegraphics[height=7cm]{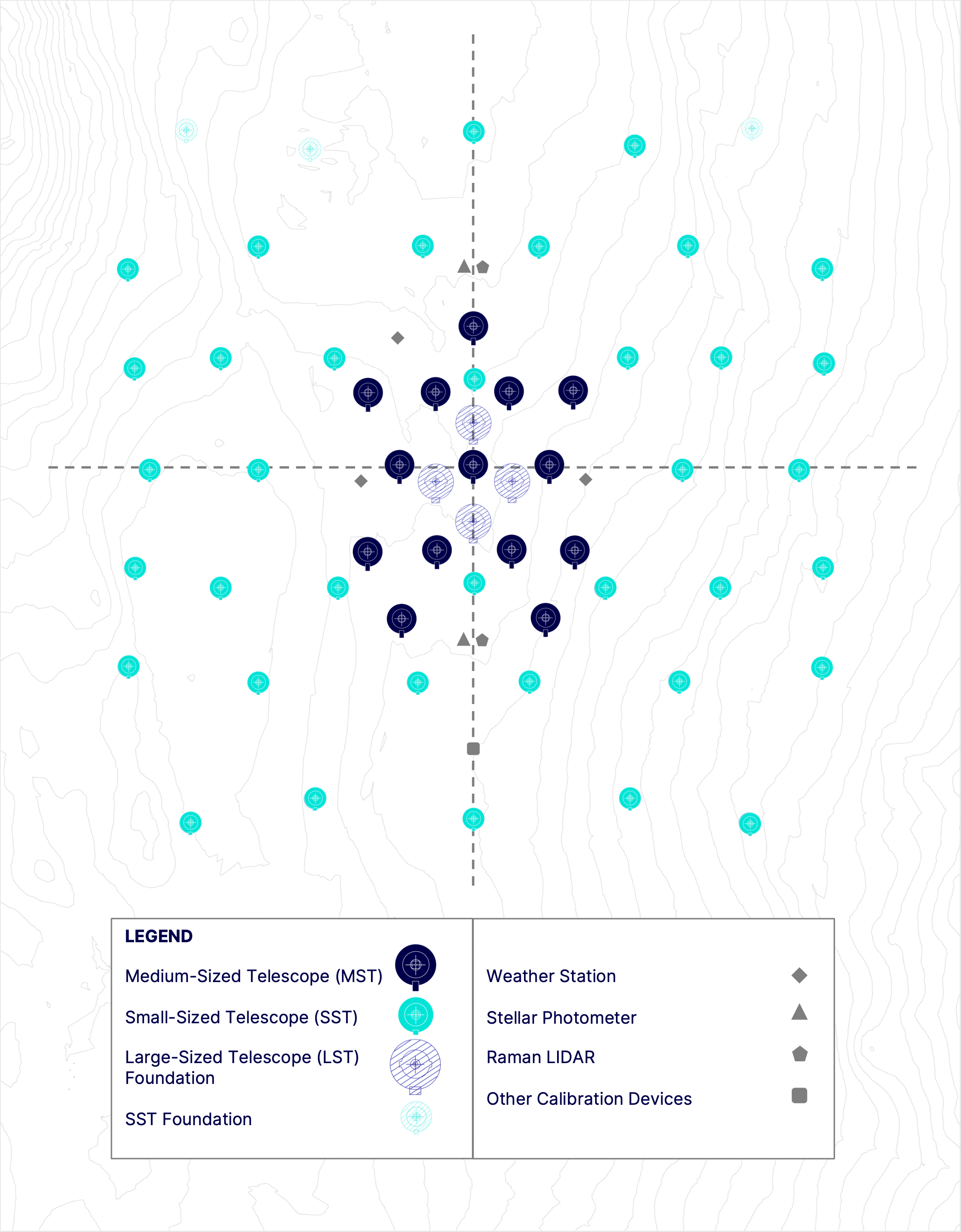}
    \includegraphics[height=7cm]{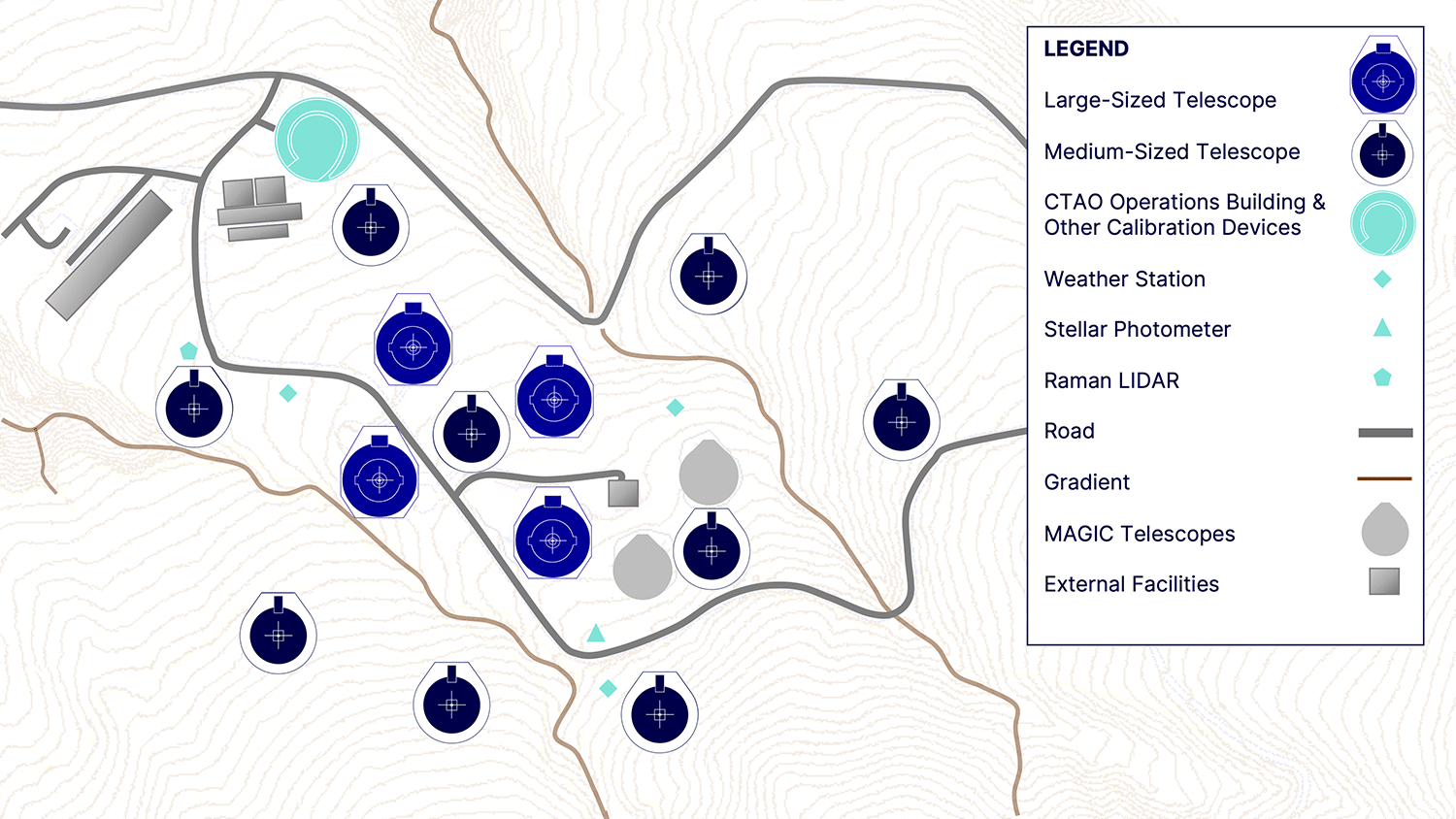}
    \caption{Illustrative CTAO baseline layouts for the Southern (left) and Northern (right) sites. Circles/squares indicate LST/MST/SST classes as in the Consortium documentation.}
    \label{fig:CTA-layouts}
\end{figure*}

An additional example of non-decomposable optimization in ground-based gamma-ray astronomy is provided by the design of the layout of the Cherenkov Telescope Array Observatory (CTAO)~\cite{ctaoweb}, constituted by Imaging Atmospheric Cherenkov Telescopes (IACTs) that detect the Cherenkov radiation (UV-optical light) emitted within an EAS at the passage of a cosmic gamma ray at about 10~km heigh and propagating to the ground. CTAO is a next-generation facility for very-high-energy gamma-ray astronomy, consisting of two arrays, one located in the southern hemisphere (Atacama Desert, Chile), and one in the northern hemisphere (La Palma, Spain). The southern array features a mix of large-sized telescopes (LSTs, 28~m diameter, optimized for low energies), medium-sized telescopes (MSTs, 12~m optmized for intermediate energies), and small-sized telescopes (SSTs, 6~m optimized for high energies), while the northern array consists primarily of LSTs and MSTs~\cite{ctaoweb}. Their preliminary layouts are shown in~\autoref{fig:CTA-layouts}.

The final layout of the CTAO arrays (North and South) is currently under definition by comparing a finite set of candidate configurations through full detector Monte Carlo (MC) simulations~\cite{Bernlohr:2008}. Each candidate layout $L$ specifies (i) the number of telescopes per class $\{N_{\rm LST},N_{\rm MST},N_{\rm SST}\}$, (ii) their positions $\{\mathbf{r}_t\}$ subject to site constraints (terrain, exclusion zones, infrastructure), and (iii) class-dependent hardware options (optics, camera, trigger) coupled to reconstruction settings. All has fixed budget. This is an inherently \emph{non-decomposable} problem: array geometry, camera/trigger design, and software reconstruction (image cleaning, direction/energy estimators, event selection) jointly determine performance.

For each $L$, the MC chain simulates atmospheric air showers, Cherenkov light propagation, optics and camera response, array trigger and readout, followed by reconstruction and event classification. From these, performance summaries are derived as functions of energy $E$ and zenith angle $\theta$:
\begin{equation}
A_{\rm eff}(E,\theta;L),\quad \sigma_\theta(E,\theta;L),\quad \sigma_{\log E}(E,\theta;L),\quad b_E(E,\theta;L)\,,
\end{equation}
where $A_{\rm eff}(E,\theta;L)$ is the effective area, $\sigma_\theta(E,\theta;L)$ the angular resolution, $\sigma_{\log E}(E,\theta;L)$ the energy resolution, and  $B(E,\theta;L)$ the residual background rate.

To compare layouts against a diversified science program $\{\mathcal{S}_k\}$ (Galactic/Extragalactic transients, extended sources, survey), one can adopt a weighted utility that aggregates key metrics over energy and sky:
\begin{align}
U(L) \;=\; 
\sum_k w_k \!\!\int\!\!\!\int 
\Bigg[
  & \frac{w_A}{A_{\rm eff}(E,\theta;L)} 
    + w_\Omega\,\sigma_\theta(E,\theta;L) \nonumber \\
  & + w_E\,\sigma_{\log E}(E,\theta;L)
    + w_b\,|b_E(E,\theta;L)| \nonumber \\
  & + w_B\,\frac{B(E,\theta;L)}{A_{\rm eff}(E,\theta;L)}
\Bigg] \, d\mu_k(E,\theta) 
+ w_C\,C(L),
\label{eq:ctao-utility}
\end{align}

In~\autoref{eq:ctao-utility}, the term $A_{\rm eff}(E,\theta;L)$ denotes the effective collection area, whose inverse is weighted by $w_A$ to penalize small apertures that reduce sensitivity. The factor $\sigma_\theta(E,\theta;L)$ represents the angular resolution, scaled by $w_\Omega$, since narrower point-spread functions enhance source localization. The energy resolution $\sigma_{\log E}(E,\theta;L)$, multiplied by $w_E$, accounts for the precision of spectral measurements, while the energy bias $b_E(E,\theta;L)$ enters as an absolute value weighted by $w_b$, reflecting the importance of unbiased energy reconstruction. The background rate $B(E,\theta;L)$ is normalized by the effective area and weighted by $w_B$, capturing the trade-off between sensitivity and contamination. The integration measure $d\mu_k(E,\theta)$ encodes the spectral and angular distributions of events relevant to science case $k$, with $w_k$ as the corresponding priority weight. Finally, $C(L)$ represents the cost associated with layout $L$, scaled by $w_C$, ensuring that resource constraints are included in the optimization.

The optimization proceeds iteratively:
\begin{enumerate}
  \item \emph{Design generation:} propose a set of layouts $L^{(m)}$ that satisfy site constraints and minimum separations (within-class and inter-class).
  \item \emph{MC and reconstruction:} for each $L^{(m)}$, run the full MC and \emph{retune} trigger thresholds, image cleaning, gamma–hadron separation, and energy/direction estimators (software is optimized \emph{per layout}).
  \item \emph{Scoring and selection:} compute $U(L^{(m)})$ via~\autoref{eq:ctao-utility}; retain Pareto-optimal layouts when multiple objectives are considered explicitly ({\it e.g.}, maximize survey speed while minimizing point-source sensitivity threshold).
  \item \emph{Refinement:} around high-performing $L^{(m)}$, generate local perturbations in telescope positions and counts, repeat MC, and update the Pareto front until improvements saturate.
\end{enumerate}

Because site conditions and science drivers differ, the optimal \emph{North} layout typically favors LST/MST configurations tuned for low-to-middle energies and extragalactic targets, whereas the \emph{South} layout leverages large MST/SST footprints for Galactic sources and the multi-TeV regime. The weights $\{w_k\}$ and exposure measures $d\mu_k$ are thus site-specific, ensuring that~\autoref{eq:ctao-utility} reflects the intended program of each hemisphere.

To reduce MC cost across large layout banks, surrogate models (Gaussian processes, gradient-boosted regressors, or small neural surrogates) can be trained to predict $(A_{\rm eff},\sigma_\theta,\sigma_{\log E},B)$ from geometric descriptors (radial density profiles, inter-telescope spacings, class ratios). These surrogates guide the search and are periodically recalibrated against full MC points to control modeling bias.
If reconstruction parameters were fixed across layouts, or if geometry were chosen without re-optimizing trigger/cleaning/classifiers, the resulting solutions would be systematically sub-optimal. The sub-optimality ratio $SR=U_{\rm serial}/U_{\rm codesign}$ (see~\autoref{eq:sr}) empirically departs from unity in such “serial” choices, highlighting that geometry and software must be co-optimized within the same loop.

\subsection{Calibration of Cherenkov Telescope Cameras} 
\label{sec:cherenkov-telescope}
\begin{figure}[h!t]
    \centering
    \includegraphics[width=0.9\linewidth]{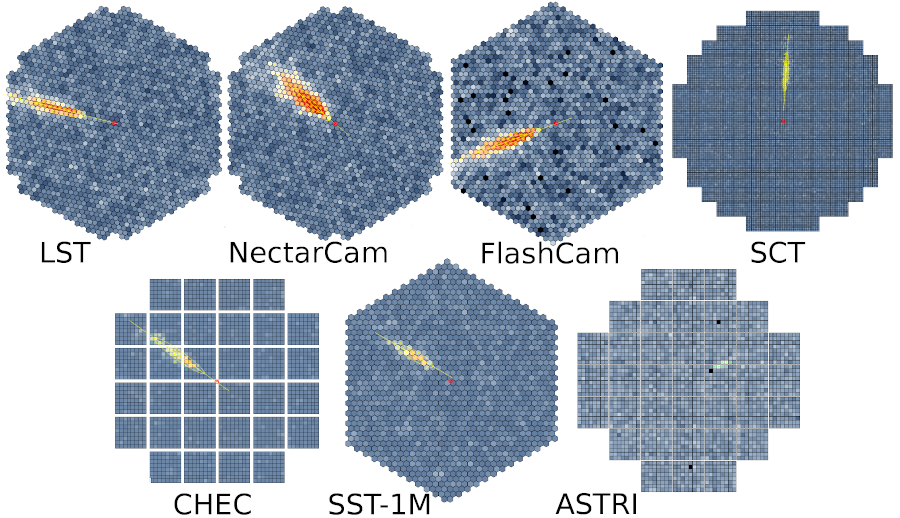}
    \caption{Examples of simulated Cherenkov shower images for different camera designs used in CTA and SWGO, including LST, NectarCam, FlashCam, SCT, CHEC, SST-1M, and ASTRI. Reproduced from Ref.~\cite{Bernlohr.web}.}
    \label{fig:ctao_cameras}
\end{figure}
The calibration of Cherenkov telescopes cameras (see ~\autoref{fig:ctao_cameras} for some camera proposed for the the mentioned CTAO) offers another paradigmatic example of a non-decomposable problem, where hardware (spectral quantum efficiency of PMTs, ageing, calibration hardware) and software (weighting of spectra, drift correction models) must be jointly optimized. Each camera hosts on the order of $\mathcal{O}(10^3)$ photomultiplier tubes (PMTs), or $\mathcal{O}(10^4)$ silicon photomultipliers (SiPMs). Focusing on the formers, each PMTs has a wavelength-dependent quantum efficiency $Q_i(\lambda)$ and a time-dependent degradation factor $A_i(\lambda,t)$ due to ageing effects. The effective photon detection efficiency of channel $i$ at time $t$ can thus be modeled as
\begin{equation}
\eta_i(\lambda,t) = Q_i(\lambda) \cdot A_i(\lambda,t)\,.
\end{equation}

The relative calibration of the PMTs is extremely important to reduce bias in energy and angular reconstruction. The calibration typically relies on reference light pulses produced by a dedicated HW, which includes LEDs at selected wavelengths~\cite{CTA-Calib}. How many wavelengths and which wavelengths, remains an optimization problem.

If calibration is performed at a single wavelength $\lambda_c$, one estimates an effective response $\eta_i(\lambda_c,t)$ and rescales accordingly. However, the true science signal is integrated over a broad Cherenkov spectrum $S(\lambda)$, so the reconstructed response depends on the spectral overlap:
\begin{equation}
R_i(t) = \frac{\int S(\lambda)\,\eta_i(\lambda,t)\, d\lambda}{\eta_i(\lambda_c,t)}\,.
\end{equation}

When only one calibration wavelength is used, the residual bias scales with the slope of $\eta_i(\lambda,t)$ around $\lambda_c$ and the mismatch between $\lambda_c$ and the effective mean $\langle\lambda\rangle$ of the Cherenkov spectrum~\cite{DoroBSc}. To first order, the error can be approximated as
\begin{equation}
\Delta R_i \propto \left. \frac{\partial \eta_i}{\partial \lambda}\right|_{\lambda_c} (\langle \lambda \rangle - \lambda_c)\,.
\end{equation}

Using two calibration wavelengths $\lambda_1, \lambda_2$ allows one to also constrain the first derivative of the spectral response, effectively correcting for slope-induced biases. Extending to three wavelengths $\lambda_1, \lambda_2, \lambda_3$ further enables estimation of curvature, mitigating residuals due to higher-order spectral distortions or ageing-induced changes in spectral shape. This strategy has been discussed both in CTA calibration design~\cite{CTA-PMT,CTA-Calib} and in earlier studies~\cite{DoroBSc}.

A practical wavelength choice protocol could be as follows: select a central pivot wavelength near the median of the Cherenkov spectrum; add two calibration wavelengths near the 20\% and 80\% quantiles of the detected spectrum; and optionally add a third wavelength close to the pivot to stabilize slope estimates. This selection balances sensitivity to spectral slope and curvature against hardware complexity. Calibration LEDs in CTA boxes currently span UV to near-IR ranges, allowing such strategies to be implemented in practice~\cite{CTA-Calib}.

To quantify the advantage of multi-wavelength calibration, we define a utility function $U$ balancing calibration accuracy and cost:
\begin{equation}
U(N_{\lambda}) = \frac{1}{w_{\text{calib}}\,\sigma^2_{\text{calib}}(N_{\lambda}) + w_{\text{drift}}\,\sigma^2_{\text{drift}}(N_{\lambda})} - C(N_{\lambda})\,,
\end{equation}
where $N_{\lambda}$ is the number of calibration wavelengths, $\sigma^2_{\text{calib}}$ the residual calibration error, $\sigma^2_{\text{drift}}$ the long-term stability error, $w_{\text{calib}},w_{\text{drift}}$ their relative weights, and $C$ the operational cost (increasing with $N_{\lambda}$). The sub-optimality ratio (SR) introduced in \autoref{s:sr} can then be estimated by comparing $U(N_{\lambda}=1)$ to $U(N_{\lambda}=2,3)$, showing that single-$\lambda$ calibration can be significantly sub-optimal (SR $<0.7$ in toy models~\cite{DoroBSc}).

Finally, a possible calibration protocol integrating hardware and software aspects is: periodic calibration at 2--3 wavelengths spanning the sensitive band; continuous monitoring of LED stability and PMT gain; software modeling of spectral drift using historical calibration data; and periodic re-optimization of wavelength choices as the system ages.

This case study highlights the co-design nature of Cherenkov telescope calibration: hardware QE, spectral ageing, and calibration sources interact with software reconstruction and drift models, making joint optimization unavoidable~\cite{PMT-Handbook,DoroBSc}.

\subsection{Neutrino Astronomy}
\label{sec:icecube}

Neutrino astronomy has developed into a major pillar of astroparticle physics since the discovery of a diffuse astrophysical neutrino flux by IceCube in 2013~\cite{IceCube:2013cdw,IceCube:2013low}, followed by the first evidence for individual high-energy neutrino sources~\cite{IceCube:2018cha,IceCube:2018dnn,IceCube:2022der,IceCube:2023ame} and the recent observation of an extremely energetic cosmic neutrino by KM3Net~\cite{KM3NeT:2025npi}. Because neutrinos interact only weakly and are not deflected by magnetic fields, they uniquely probe dense and distant astrophysical environments, motivating the construction of increasingly large detectors. Several optical neutrino observatories are currently under construction or in development, including KM3NeT~\cite{KM3NET2}, Baikal-GVD~\cite{Baikal-GVD}, P-ONE~\cite{P-ONE}, and IceCube-Gen2~\cite{Gen2-TDR}, aiming to extend sensitivity in the TeV–PeV range. Complementary to this, radio detection techniques exploiting the Askaryan effect enable cost-effective instrumentation of the vastly larger volumes required for EeV energies, as demonstrated by experiments such as ARA, ARIANNA, and RNO-G~\cite{Barwick:2022vqt,RNO-G:2020rmc}, and foreseen for IceCube-Gen2~\cite{Gen2-TDR}. 
The optimization of future neutrino telescopes bears huge potential to accelerate progress in astroparticle physics. However, traditional optimization is not tractable due to the high-dimensional design parameter space that cannot be explored efficiently with forward Monte Carlo simulations~\cite{cranmer2020frontier}.
Consequently, detector layouts have historically been guided by physical intuition and incremental scaling rather than systematic end-to-end optimization. Here, we outline how such an end-to-end optimization could be achieved and present a proof-of-concept using a toy detector setup. 

The parameters of interest in high-energy neutrino telescopes are the properties of the detected neutrino events. These are not directly accessible but must be inferred from the electronic signals measured by the detectors. Furthermore, neutrinos are only detectable by measuring secondary particles produced in neutrino-matter interactions. 
The electronic signal ("data") is related to the properties $\vec{\theta}_\nu$ of the original neutrino (such as direction, interaction position, neutrino type, energy) by the likelihood:
\begin{align}
\mathcal{L}(\vec\theta_\nu \mid \mathrm{data}, \, \vec\theta_\mathrm{det.}) = p(\mathrm{data} \mid \theta_\nu, \,\vec\theta_\mathrm{det.}),
\end{align}
where $\vec\theta_\mathrm{det.}$ is the parameters of the detector ({\it e.g.}, the positions of the optical modules/antennas) and $p(\mathrm{data} \mid \theta_\nu, \,\vec\theta_\mathrm{det.})$ is the probability density function (PDF) of the data given the neutrino and detector parameters.
If the likelihood can be formulated (or approximated sufficiently well), the reconstruction resolution can be estimated through the Fisher information and the Rao-Cramer-Fréchet bound. 

The Fisher information matrix (here for the neutrino parameters) is defined as
 \begin{align} \label{eq:fisher}
   \mathcal{I}_{ij}(\vec{\theta}_\nu \mid \,\vec\theta_\mathrm{det.}) = E \left [ \frac{\partial{\log \mathcal{L}} }{\partial{\theta_{\nu,i}}} \cdot \frac{\partial{\log\mathcal{L}}}{\partial{\theta_{\nu,j}}} \, \middle | \, \vec{\theta}_\mathrm{det.} \right].
 \end{align} 
The Rao-Cramér-Frechet bound then states that the covariance of any unbiased estimator $T(\vec{\theta}_\nu\mid \,\vec\theta_\mathrm{det.})$ of the parameters $\vec{\theta}_\nu$ is bounded by:
 \begin{align}
    \text{cov}(T(\vec{\theta}_\nu\mid \,\vec\theta_\mathrm{det.})) \geq \mathcal{I}^{-1}(\vec{\theta}_\nu\mid \,\vec\theta_\mathrm{det.}) .
 \end{align}
The Rao-Cramér-Frechet bound thus provides a lower bound on the achievable detector resolution for measuring the neutrino energy and direction, which are the parameters of primary interest.

Using the uncertainty estimate through the Fisher information substantially simplifies the problem. We no longer need to perform a reconstruction of simulated data, which is time-consuming and challenging to make differentiable. Instead, as shown in~\autoref{eq:fisher}, only a differentiable likelihood function is needed. In the case of an in-ice radio neutrino detector, such as RNO-G or IceCube-Gen2,~\autoref{eq:fisher} takes a particularly simple form. The likelihood can be modelled as a multivariate Gaussian~\cite{Ravn:2025puy}, which reduces the Fisher information calculation to
\begin{equation}
	\begin{aligned}
		\mathcal{I}_{ij}(\vec{\theta}_\nu \mid \,\vec\theta_\mathrm{det.}) &= \sum_{\text{antennas}} \frac{\partial \boldsymbol{\mu}(\vec{\theta}_\nu ,\,\vec\theta_\mathrm{det.})^\text{T}}{\partial \theta_{\nu, i}} \boldsymbol{\Sigma}^{-1}  \frac{\partial \boldsymbol{\mu}(\vec{\theta}_\nu ,\,\vec\theta_\mathrm{det.})}{\partial \theta_{\nu, j}}\,,
	\end{aligned}
\end{equation}
where $\boldsymbol{\Sigma}$ is the covariance matrix of the noise, $\boldsymbol{\mu}(\vec{\theta}_\nu ,\,\vec\theta_\mathrm{det.})$ is the signal model, {\it i.e.}., it predicts the measured signal vs. time in an antenna based on the neutrino parameters $\theta_{\nu,i}$, and it can easily be summed over the antennas measuring the signal. Thus, the problem reduces to constructing a differentiable signal model, which can be readily obtained through analytic approximations and/or neural network surrogate models. An initial proof-of-concept was presented in Ref.~\cite{Ravn:2025jzs}, using the surrogate Askaryan emission model presented in Ref.~\cite{Pilar:2025psc}.

A similar strategy can be formulated for optical Cherenkov neutrino telescopes, where the measured data consist of photon arrival times and amplitudes recorded by photomultiplier tubes distributed throughout a large detection volume. In this case, the likelihood is determined by the expected spatio-temporal light distribution produced by charged secondary particles propagating through the medium, which depends on the neutrino interaction parameters as well as on detector properties such as sensor positions, angular acceptance, and optical properties of the medium. As discussed in Ref.~\cite{Haack:2023uwd}, the photon propagation and detection process can be described by differentiable light-yield and transport models that predict the expected photoelectron rates at each sensor. Under commonly used assumptions of Poissonian photon counting statistics with approximately independent sensor responses, the Fisher information can again be expressed in terms of derivatives of the predicted sensor responses with respect to both event and detector parameters. This enables a computationally efficient estimation of achievable reconstruction resolutions without explicit event reconstruction and provides a natural framework for gradient-based optimization of detector geometries, such as string spacing and module density.

The optimization procedure closely follows the paradigm used in training neural networks. In practice, simulated neutrino interactions are processed in batches, for which the gradients of a chosen performance objective, derived from the Fisher information, with respect to detector parameters (such as sensor positions or deployment depths) are computed through automatic differentiation. These gradients are then used to iteratively update the detector configuration via stochastic gradient–descent–type algorithms, enabling an efficient exploration of the high-dimensional design space. To ensure realistic solutions, the objective function is augmented by additional cost terms that encode practical constraints, including the increased financial and logistical cost associated with deeper drillings or denser instrumentation, as well as engineering and deployment limitations. This results in a constrained optimization framework that naturally balances physics performance against feasibility considerations.

Comprehensive end-to-end optimization studies for next-generation neutrino observatories are currently in progress. In particular, a full co-design framework is being developed for the optical P-ONE detector, targeting the joint optimization of array geometry and sensor configuration for best reconstruction performance, as well as for the planned radio array component of IceCube-Gen2, where the station and array layout is optimized with respect to physics-driven performance metrics in the NuRadioOpt project~\cite{Glaser:2023udy}.

\begin{figure}[t!]
    \centering
    \includegraphics[width=0.7\linewidth]{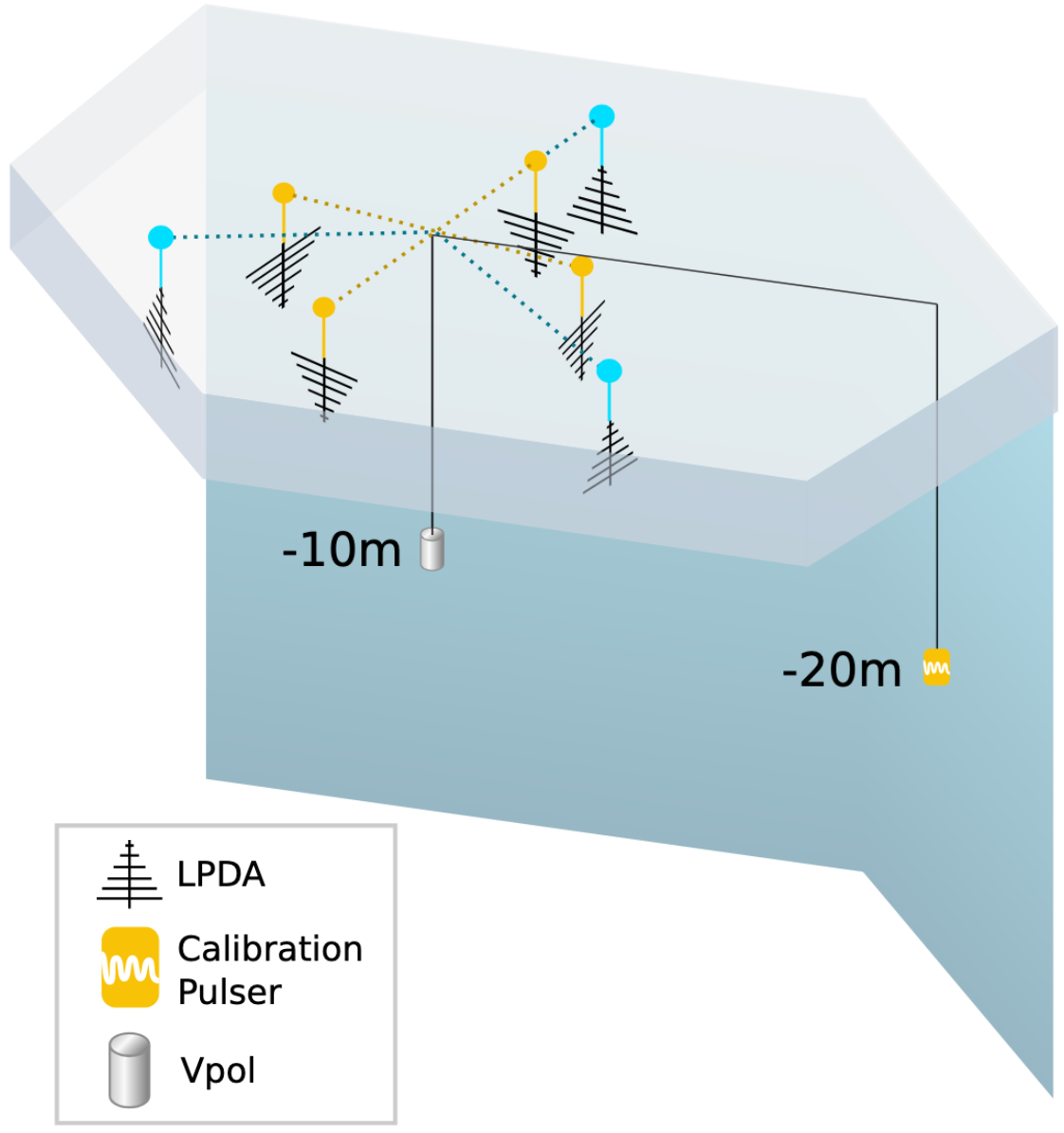}
    \caption{Schematic of a shallow IceCube-Gen2 radio detector station. Only the downward-facing LPDAs and Vpol antennas are considered in this study. Figure adapted from Ref.~\cite{Gen2-TDR}.}
    \label{fig:neutrino_Gen2_shallow}
\end{figure}

\begin{figure}[t!]
    \centering
    \includegraphics[width=0.9\linewidth]{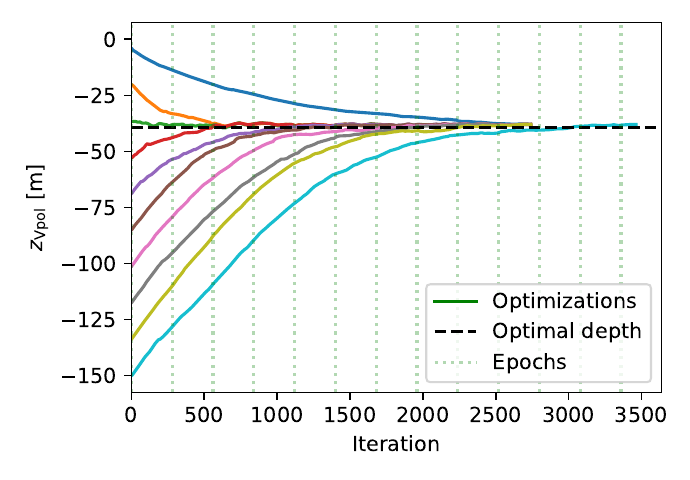}
    \caption{Repeated optimizations of the depth of the Vpol antenna ($z_\mathrm{Vpol}$) for a shallow IceCube-Gen2 radio detector station with different initial depths. The optimal depth is determined from the scan shown in \autoref{fig:neutrino_Gen2_shallow_scan}.}
    \label{fig:neutrino_Gen2_shallow_optimizations}
\end{figure}

\begin{figure}[t!]
    \centering
    \includegraphics[width=0.9\linewidth]{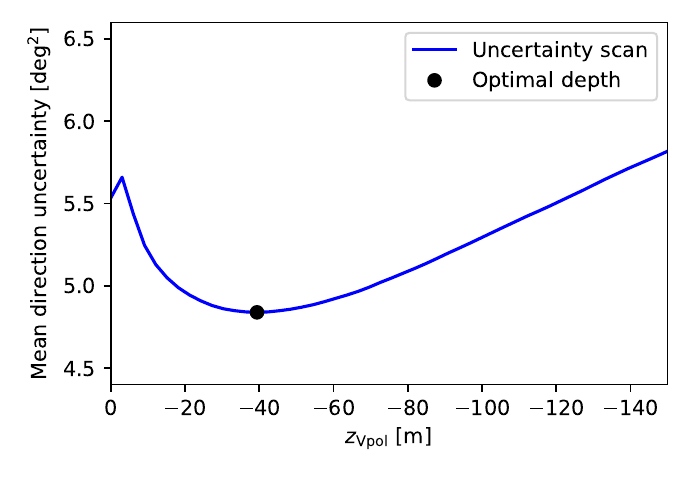}
    \caption{Rao-Cramér-Fréchet lower bound estimate of the neutrino direction reconstruction uncertainty as a function of depth of the Vpol antenna ($z_\mathrm{Vpol}$) calculated with the differentiable simulation and reconstruction pipeline.}
    \label{fig:neutrino_Gen2_shallow_scan}
\end{figure}

To illustrate the technique discussed in this section, we present a simple toy optimization experiment that can easily be validated. The study focuses on the one-dimensional optimization of a single antenna depth for an IceCube-Gen2 radio detector station. The nominal design of one of the station types is shown in~\autoref{fig:neutrino_Gen2_shallow}, which has four downward-facing log-periodic dipole antennas (LPDAs) and one vertically polarized dipole antenna (Vpol). The optimal depth of the Vpol is expected to be a trade-off between two competing effects: at greater depths, the triangulation of the neutrino interaction vertex improves, but the signal strength decreases. In this toy study, the depth of the Vpol antenna is optimized 10 times with different initial positions to validate that the same optimum is found for any initialization. We employ the prototype fully differentiable simulation and reconstruction pipeline presented in Ref.~\cite{Ravn:2025jzs}, which was later expanded to include most physical effects of state-of-the-art radio neutrino simulations, see, {\it e.g.}, NuRadioMC~\cite{Glaser:2019cws}. The dataset used for the optimization contains 100,000 neutrino neutral-current interactions with deposited energy in the $10^{16.5}$~eV to $10^{17.5}$~eV range that fulfil the trigger condition. The dataset is then split and processed in batches of 356 events. 
For simplicity, we treat the trigger optimization independently from the antenna position optimization. In this example, the trigger is only calculated from the four LPDA antennas, {\it i.e.}., the depth of the Vpol antenna does not impact whether the neutrino interaction is detected or not. In future studies the trigger antennas can be incorporated in the optimization by assigning every event a weight corresponding to its trigger efficiency. 
The loss function is then the $\log_{10}$ of the Rao-Cramér-Fréchet lower bound estimate of the neutrino direction reconstruction uncertainty in square degrees. For simplicity, we refrain from adding a cost function that would penalize increased costs due to the drilling of deeper holes. The optimizations are run with an initial learning rate of 0.2, an exponential learning rate decay, and a convergence criterion of no improvement for two epochs. The resulting optimizations are shown in~\autoref{fig:neutrino_Gen2_shallow_optimizations}, which demonstrates stable convergence to the same solution for any starting depth. In order to validate that the solution is optimal, a one-dimensional scan of the neutrino direction uncertainty is shown in~\autoref{fig:neutrino_Gen2_shallow_scan}, which was calculated with the differentiable pipeline. The minimum of the scan is consistent with the solution found by the optimizations. Hence, the method discussed in this section reliably finds the optimal detector layout for this simple toy example.

Additionally, we test optimised geometry generation using a simple toy experiment inspired by the P-ONE optical neutrino detector concept in deep sea water. The aim is to generate multiple geometries that satisfy differentiable engineering and cost constraints. The ability to generate multiple feasible geometries is essential for exploring Pareto fronts, which are increasingly relevant in multi-objective optimisation problems arising in large-scale neutrino detector design.

An optical neutrino detector is composed of long cables that hold optical modules (OMs), referred to as a string. The constraints considered here are (i) the maximum number of deployable strings, determined by cost considerations, and (ii) the total instrumented area available for detector construction. We assume a maximum of 70 strings and constrain the detector footprint to a circular area with a radius of 1~km. For P-ONE, a single string holds in chain up to 20 OMs, each housing multiple PMTs. For simplicity we assume only a single PMT per OM and do not consider the its directionality.

\begin{figure}[!ht]
    \centering
    \includegraphics[width=0.7\linewidth]{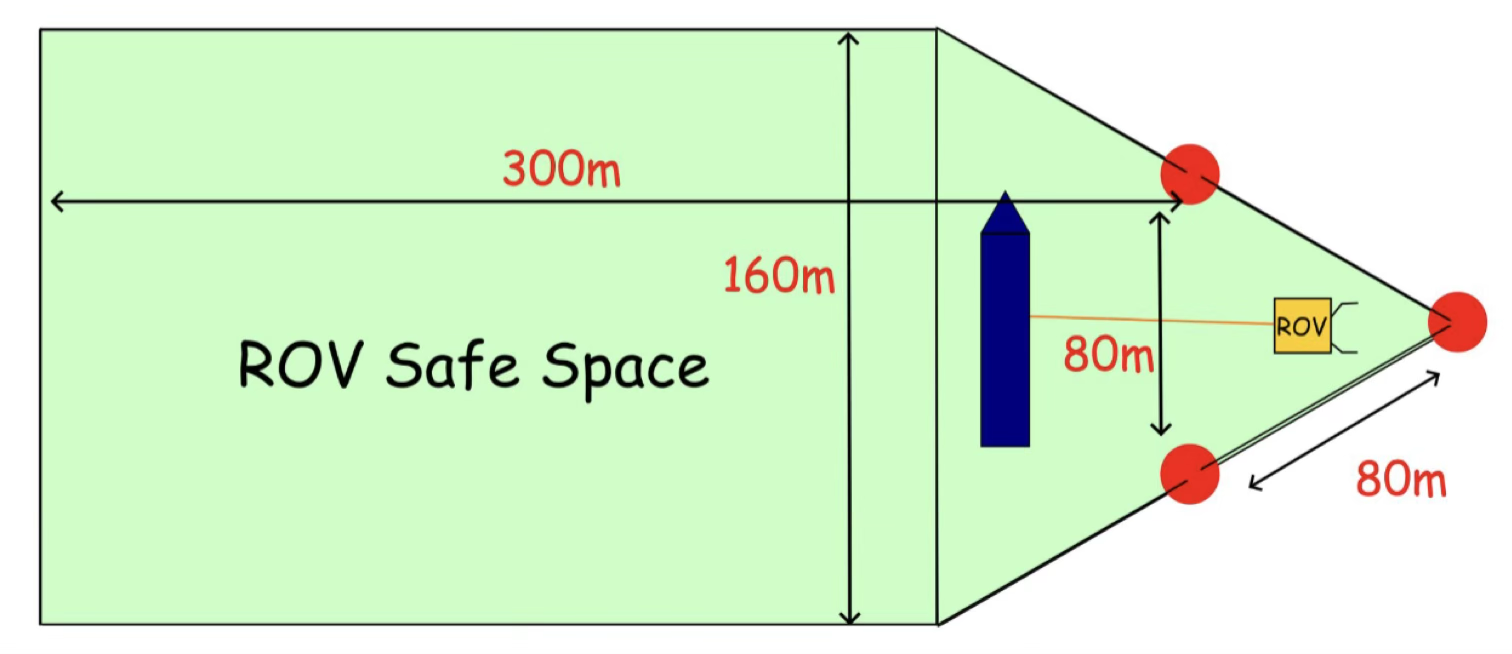}
    \caption{Top-down view of the ROV safe-space required for the deployment and maintenance of a P-ONE detector array/string (right-most red dot), in a constellation of 3 strings. Image provided by the P-ONE Collaboration.}
    \label{fig:rov_safespace}
\end{figure}

A further engineering constraint arises from deployment and maintenance operations involving remotely operated vehicles (ROVs) on the Pacific Ocean seabed. A minimum safe-space must be maintained to allow for contingency procedures. In the event of loss of ROV control (“dead sub”), a predefined safe heading combined with winch control enables extraction from within a cluster of strings. Similarly, in the case of dynamic positioning (DP) failure of the support vessel, manual control can be used to guide the ROV into a safe evacuation region. These operational requirements impose a geometric constraint on string placement: for any constellation of three strings forming an 80~m equilateral triangle, a rectangular clearance region of 300~m × 160~m must be available in at least one direction around a given string. Figure~\ref{fig:rov_safespace} illustrates the ROV safe-space configuration.

In this toy study, we assume a spatially uniform signal distribution and do not employ neutrino event surrogates, as the primary focus is the generation of geometries under multiple differentiable constraints. The uniform signal acts as a driver favouring a larger number of active strings, counterbalanced by string production and deployment costs.

Optimisation is performed on a dense base configuration consisting of 1000 candidate string positions arranged in a hexagonal packing. Each string is assigned a continuous weight $w_i$, which parametrizes the geometry. Strings with weights above a chosen threshold are considered active. This relaxed parameterization enables gradient-based optimisation in a continuous design space and is inspired by similar approaches used in the optimisation of gravitational-wave detector networks~\cite{differometor2025github}.

For the ROV constraint applied to the $i$-th string, we define a differentiable penalty function $C_{\mathrm{ROV},\,i}$ that evaluates the minimum weight-sum of sets of neighbouring strings $S$ within candidate safe-space regions scanned around the string. This minimum value is multiplied by $w_i$ to penalise configurations that violate the clearance requirement:

\begin{align}
    C_{\mathrm{ROV},\,i} = w_i \cdot \sum_{k \in \min(S)}w_k
\end{align}

\begin{figure}[!ht]
    \centering
    \includegraphics[width=0.8\linewidth]{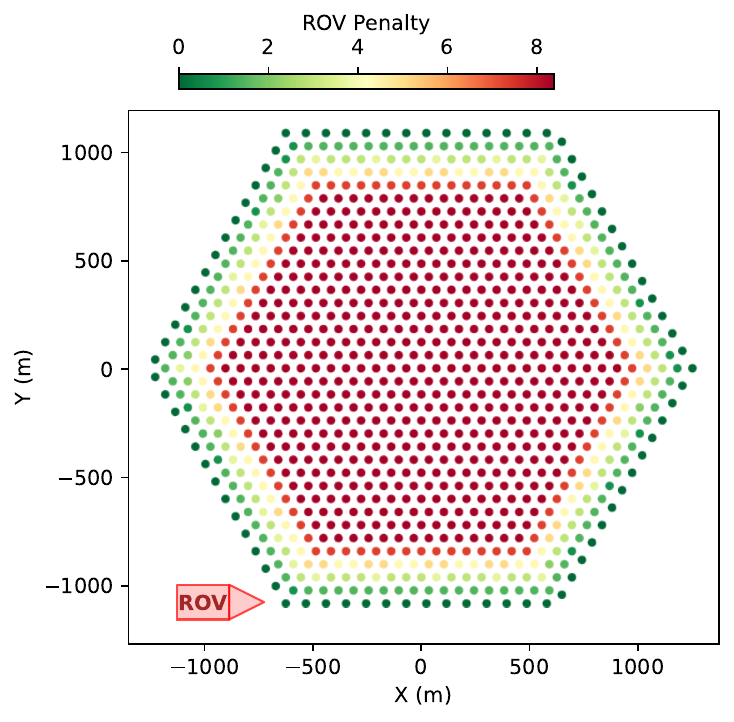}
    \caption{Top-down view of the 1000-string base of candidate P-ONE arrays/strings. The strings are colours of string show how much they violate the ROV safe-space constraint for deployment (sketch for scale in the bottom left). Clearly, strings that lie at the perimeter of the base are already not in violation of the constraint.}
    \label{fig:whole_geom_rov}
\end{figure}

In each optimisation iteration, 36 equally spaced angular checks are performed around each string, resulting in constraint evaluations across all 1000 candidate positions. Figure~\ref{fig:whole_geom_rov} shows the base 1000-string geometry and the corresponding ROV penalty per string for uniformly weighted strings. A penalty value of zero corresponds to ROV-compliant placement.

\begin{figure*}
    \centering
    \includegraphics[width=0.8\linewidth]{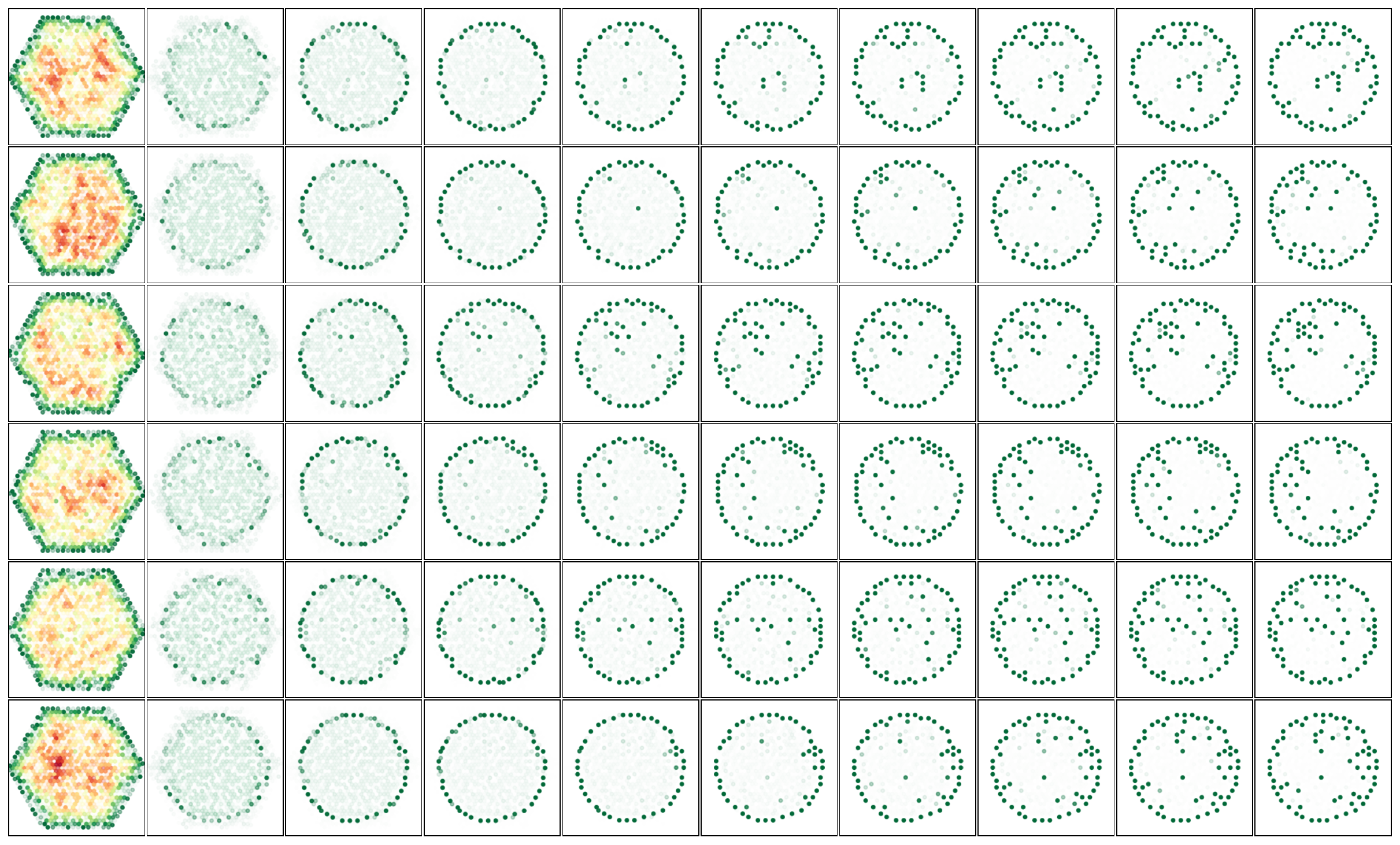}
    \caption{Top-down views of snapshots of iterations of geometries during optimization, similar to \autoref{fig:whole_geom_rov} and also coloured by the same scale of ROV safe-space penalty. Faded dots represent strings with low weights (inactive). Each row is a separate geometry optimization for 6 separate random weight-initializations. Starting from the 0th iteration, each consecutive column is 100 iterations later until the 1000th iteration.}
    \label{fig:iter_geoms_rov}
\end{figure*}

We train six independent geometries from random weight initialisations using the ADAM optimiser with a learning rate of 0.1 over 1000 epochs. Figure~\ref{fig:iter_geoms_rov} illustrates intermediate optimisation stages for the resulting geometries, which converge to approximately 70 active strings in each case. Since the ROV penalty is naturally smallest near the boundary of the base geometry, optimal solutions tend to form a perimeter of active strings near the 1~km radial limit. Within this perimeter, distinct ROV-compliant string clusters emerge, shaped by asymmetries introduced through random initial weights.

The emergence of multiple, distinct geometries that equally satisfy engineering constraints demonstrates the viability of differentiable co-design approaches for generating diverse feasible solutions. This capability is particularly relevant for constructing Pareto fronts in multi-objective neutrino detector optimisation. Future work could explore adaptive constraint-weighting strategies such as the one in Ref.~\cite{berzins2025ginn}, as well as fully generative models for constrained geometry synthesis.

\subsection{A ground array for high-energy tau neutrino detection}
\label{sec:tambo} 

\begin{figure}
    \centering
    \includegraphics[width=\linewidth]{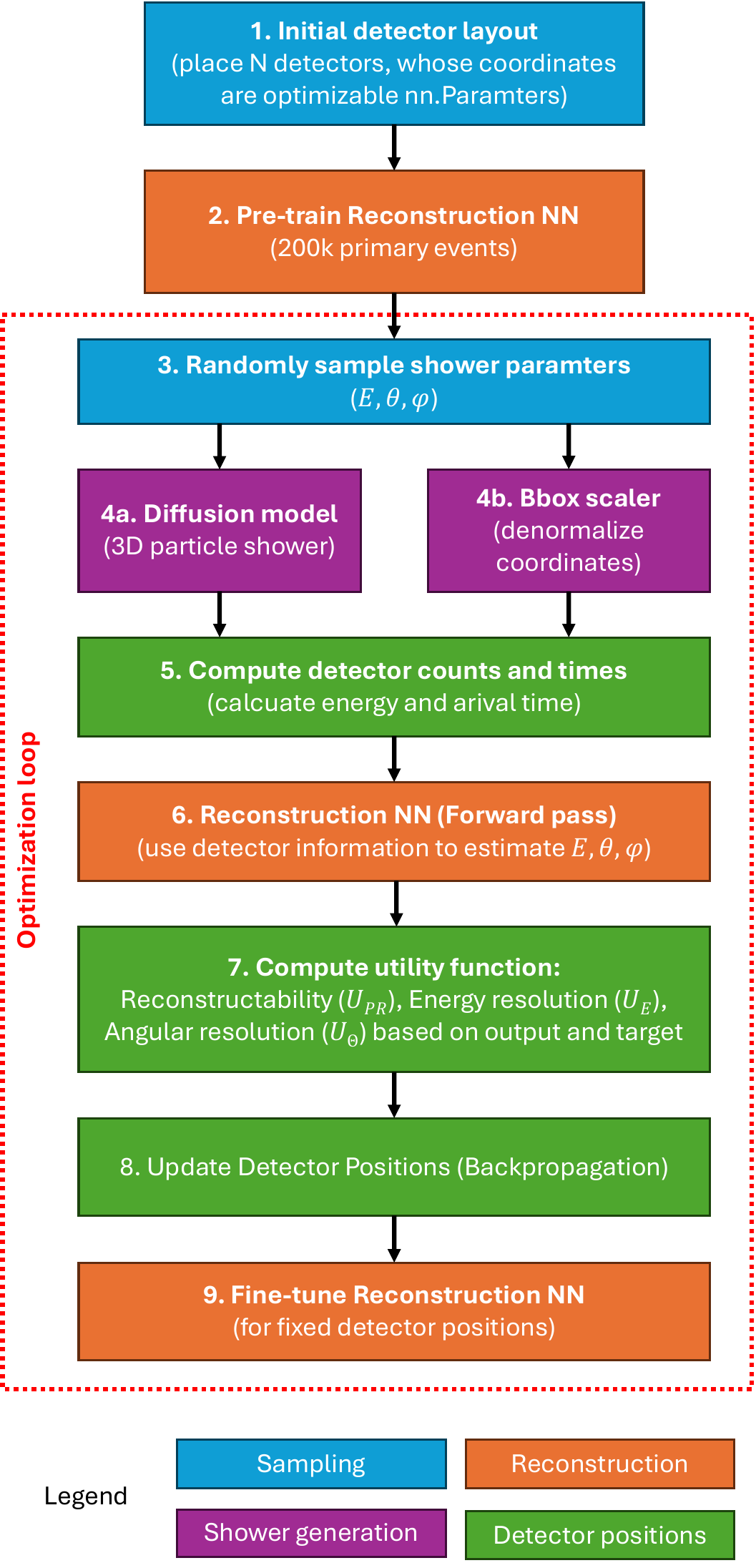}
    \caption{Flow chart outlining the main stages of the optimisation of the array for high-energy tau neutrino detection.}
    \label{fig:tambo_optimization_flow_chart}
\end{figure}

An astrophysics use case that shares a number of similarities with the SWGO optimization problem discussed in~\autoref{s:swgo} above is the one of the TAMBO (Tau Air-Shower Mountain-Based Observatory) experiment.
The idea of this experiment is to exploit the opaqueness of matter to ultra-high-energy neutrinos, combined with the peculiar topography of a wide U-shaped valley in the Peruvian Andes.
TAMBO will search for a signal produced when an Earth-skimming tau neutrino interacts with the rock of the mountain on one side of the valley by means of the charged-current interaction. The resulting tau lepton decays, producing a shower of secondary particles that can be detected by instrumenting the other side of the valley with a large number of scintillation planes. 
Due to the screening effect of the target mountain on cosmic-ray backgrounds, the layout guarantees low backgrounds and thus high discovery potential.
The framework supports both an idealized straight-walled valley geometry and a realistic terrain mesh of the Colca Valley, the nominal deployment site, derived from geographic data; validation studies show good agreement with the effective areas computed in the original TAMBO proposal, with peak sensitivity in the design-target range of 1--100~PeV.

One point of contact with the SWGO optimization problem is the general question we need to answer: where do we place each of a large number of detection units, so that together they enable the most precise reconstruction of intercepted showers?
Given the large number of detection units, gradient-based methods are appropriate.
As in the SWGO case, the simulation of atmospheric showers---for this case, generated by any one of the several possible decay final states of the tau lepton, rather than single gamma rays---is highly stochastic, so a gradient-based optimization scheme requires a suitable differentiable surrogate.
This can be produced with a diffusion model trained to reproduce the three-dimensional fluxes and arrival times of secondary particles on the detectors, as simulated with \texttt{CORSIKA}~\cite{corsika} for different values of incident neutrino energy and angle.
Triggering schemes can also be devised in a manner analogous to those discussed for SWGO, and a similar cost model could be developed.

Two features make the TAMBO optimization problem richer than the SWGO one. The first is the directionality of backgrounds: the cosmic-ray flux reaching the detector depends non-trivially on the variable thickness of the target mountain along different lines of sight, so the background rate is a function of both the detector position and the viewing direction. The second is that the detection surface is intrinsically three-dimensional---detector units must be positioned on the flank of a mountain rather than on a flat plane---and the complexity of deployment varies significantly with the local steepness of the terrain, directly affecting construction and maintenance costs. Importantly, these topographic effects couple to one another: the same mountain geometry that screens cosmic-ray backgrounds also constrains the set of feasible detector positions.

Several diffusion models are trained on between 100k and 600k particle showers to generate the 3D particle distribution from the parameters of a primary event. The selection of the best performing model is in progress. The optimization pipeline leverages on this diffusion model to generate a set of particles showers for randomly sampled primary event parameters (direction, energy and type of primary particle). The distribution is then scaled from the 32x32x24 output coordinates of the diffusion model to global coordinates using a Feedforward Neural Network (FNN). This two-part architecture is selected such that it retains a high resolution for the shower, while being able to generate showers within a wide range of global coordinates. The output of the diffusion model is then used to train a Reconstruction NN, which reconstructs the primary particle parameters from the 3D shower distribution. A utility is computed on the ability of the Reconstruction NN to recreate the primary particle parameters. The main parts of the optimization process are outlined in~\autoref{fig:tambo_optimization_flow_chart}. 

A global utility function for TAMBO has not yet been fully defined, as the experiment plans to both search for evidence of tau neutrinos independently and provide candidate source positions to IceCube and other experiments, which may use this information to cut down backgrounds to a manageable level. The utility will depend strongly on a triggering and reconstruction strategy that includes the effective treatment of backgrounds---which depend on direction as well as on layout---with associated tunable software parameters. Cost components affect both online data acquisition and local access to energy, and computing power, and have to be modulated effectively in conjunction with the cost of the detector hardware. The current plan is that the utility function will be implemented as a two-stage model: the first stage will be a NN trained to convert secondary particle observations to primary particle features; the reconstruction accuracy of those features will be combined with the detector position in the calculation of the final utility score. The separate stages of the optimization pipeline can be seen in~\autoref{fig:tambo_optimization_flow_chart}. The current form of the utility function -- considering reconstructability of the tau particle, which triggered the particle shower -- is:
\begin{align}
U_{\mathrm{PR}}(r) &= \sqrt{\sum_{i} r_i} \\[6pt]
U_{E}(\hat{E}, E, r) &= \sum_{i} \frac{r_i}{(\hat{E}_i - E_i)^2 + \epsilon_E} \  \quad \epsilon_E = 0.01 \\[6pt]
U_{\Theta}(\hat{\theta}, \theta, r) &= \sum_{i} \frac{r_i}{(\hat{\theta}_i - \theta_i)^2 + \epsilon_\Theta} \qquad \epsilon_\Theta = 10^{-5} \\[10pt]
U &= 10^{-2}\, U_{\Theta}(\hat{\theta}, \theta, r) + U_{E}(\hat{E}, E, r) + \frac{U_{\mathrm{PR}}(r)}{\sqrt{\rho}},
\end{align}
where $r_i$ is the reconstructability score for event $i$, $\rho$ is the shower core density, and $\hat{\cdot}$ are the outputs of the Reconstruction NN. The utility function is based on prior work~\cite{Frat2025}.

An alternative formulation is based on current work by the TAMBO collaboration, where the shower direction and core are reconstructed based on the distribution and arrival time of secondary particles. 
This is done through likelihood maximization using a set of probabilistic distributions for the time of arrival of each particle and number of hits per detector. The utility then depends on the angle between the original and reconstructed shower. This analytical approach enables better explainability of the the optimization gradients. 

In summary, the TAMBO experiment optimization constitutes a rich and complex problem, which involves both continuous (three-dimensional detector positions, voids in instrumented regions) and discrete parameters (number of detectors, number of triggering stations, reconstruction model hyperparameters), suggesting a hybrid evolutionary/gradient-based exploration strategy similar to the one described in \autoref{s:swgo}. The availability of a full-chain Monte Carlo framework provides the foundation on which the differentiable surrogates required by such an optimization scheme can be built. A co-design approach in any case appears necessary to guarantee convergence to truly optimal solutions, provided that the experiment is capable of defining in a precise way its utility function.

\subsection{Co-design of a Dynamic Collimation System for Spot Scanning Proton Therapy via Differentiable Optimization}

External Beam Radiotherapy (EBRT) is one of the most common cancer treatment modalities. Within this field, proton therapy exploits the Bragg peak depth-dose characteristic to improve the sparing of Organs At Risk (OARs). To further refine dose delivery, Dynamic Collimation Systems (DCS), such as the discrete spot scanning system proposed by Hyer et al.~\cite{Hyer2014-gw}, have been introduced to sharpen the lateral beam penumbra using trimmer bars.

Standard Treatment Planning Optimization (TPO) software determines beam parameters, such as spot fluences and collimator positions, to deliver a dose distribution compliant with the medical prescription. In this proof of concept, we propose a hardware-software co-design framework based on differentiable programming. Adapted from a pipeline originally developed for TPO~\cite{Arsini2025-dm}, this method allows for the concurrent optimization of treatment parameters and hardware geometry, specifically the thickness of collimators and energy degraders.

We employ a Deep Learning (DL) model as a differentiable surrogate for Monte Carlo simulations. By leveraging the differentiability of the surrogate, we optimize both hardware and software variables using gradient-based algorithms. Finally, we demonstrate the ability to incorporate hardware-related radioprotection constraints into the optimization loop, such as the minimization of secondary neutron production.
Reducing this secondary dose contribution can be particularly relevant for specific patient cohorts, such as pregnant women, due to the high radiosensitivity of the fetus~\cite{Xu2007-sl,Wang2016-do}.

\subsubsection{System Geometry and Monte Carlo Simulations}
\begin{figure}[t!]
    \centering
    \includegraphics[width=0.8\linewidth]{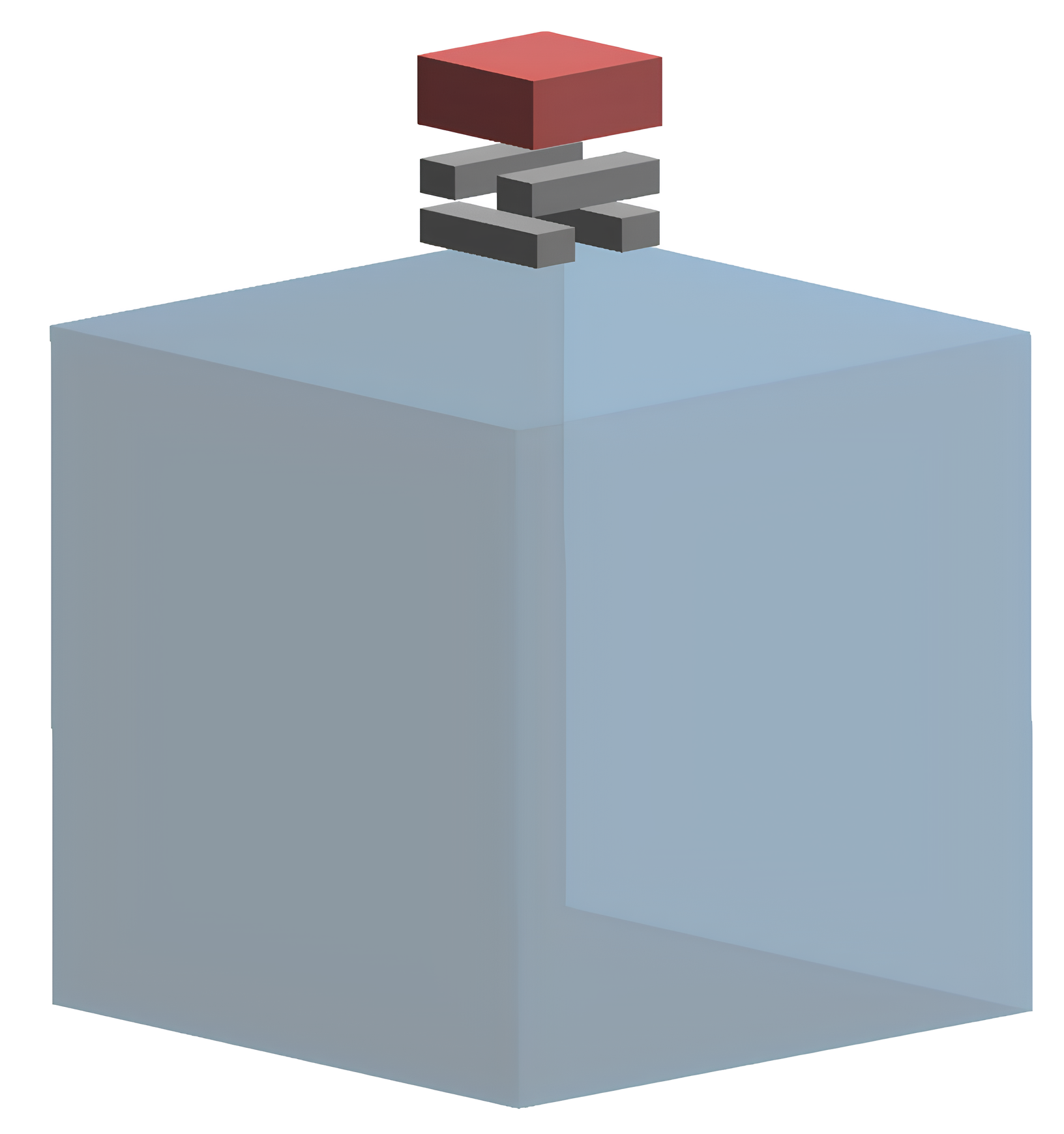}
    \caption{Schematic of the DCS geometry including the Beryllium degrader, Nickel collimators, and the water phantom.}
    \label{fig:geo}
\end{figure}

The system geometry, schematically illustrated in \autoref{fig:geo}, is based on the design proposed by Hyer et al.~\cite{Hyer2014-gw}. The beam source is modeled as a $3\times3$ matrix of proton pencil beams with a mean energy of 125 MeV, an energy spread $\sigma_E = 1$ MeV, and a spatial spread $\sigma_{pos} = 0.5$ cm.
The beam line comprises a Beryllium (Be) energy degrader with a $10\,\text{cm} \times 10\,\text{cm}$ cross-section and variable thickness. Downstream of the degrader, two pairs of Nickel (Ni) trimmer bars ($10\,\text{cm} \times 3\,\text{cm}$, variable thickness) act as X and Y collimators. The beam enters a $40\,\text{cm}$ cubic water phantom where the energy deposition occurs.

Monte Carlo (MC) simulations were performed using the aforementioned \textsc{Geant4} toolkit, which is regularly benchmarked, including for medical applications~\cite{G4med1,G4med2}. A training dataset of $1.5 \cdot 10^4$ examples was generated. Each example consists of a simulation of a single central pencil beam with $3 \cdot 10^4$ primary protons. For each simulation, four input parameters were sampled from uniform distributions:
\begin{itemize}
    \item Collimator thickness: $t_{coll} \in [1, 3]\,\text{cm}$.
    \item Degrader thickness: $t_{deg} \in [3, 5]\text{cm}$.
    \item Transverse aperture positions: The center of the aperture defined by the X and Y collimator pairs was shifted in the range $[-3.3, +3.3]\,\text{cm}$, while maintaining a fixed aperture opening of $3.6\,\text{cm}$.
\end{itemize}

The simulation output consists of the volumetric dose distribution recorded in a cylindrical mesh aligned with the beam axis. The scoring volume has a radius of $3\,\text{cm}$ and a height of $8\,\text{cm}$ starting from the phantom surface, discretized into $40 \times 30 \times 15$ voxels along the $z$ (depth), $r$ (radius), and $\theta$ (azimuth) coordinates, respectively. Additionally, the total secondary neutron yield is scored within both the Beryllium degrader and the Nickel collimators. It is worth noting that a complete and accurate simulation of the dose deposited by secondary neutrons within the phantom is computationally demanding and lies beyond the scope of this work. However, since the patient neutron dose is effectively proportional to the production rate, we adopted the total neutron yield in the hardware components as a proxy metric for minimization.

\subsubsection{Differentiable Surrogate Model}
The surrogate model consists of two distinct neural networks sharing the same input vector, defined by the four simulation parameters ($t_{coll}, t_{deg}, \Delta x, \Delta y$).

The first component is a dose engine designed to predict the volumetric dose distribution within the water phantom. This network is based on the \textit{Recursive Nearest Neighbours Graph U-Net} (ReNN GU-Net) architecture, previously validated for proton~\cite{Arsini2024-wt} and electron beam therapy~\cite{Arsini2024-tw}. The model employs a U-Net structure based on Graph Neural Networks (GNNs) to process data structured on a cylindrical mesh. It is composed by an Encoder, which processes density related information, and an Embedder, dedicated to beam setup, whose output are merged and passed to a Decoder which compute the dose distribution. The architecture was adapted for this application: the Embedder module, originally designed to process beam energy, now encodes the hardware configuration and collimator shifts. The latent features from the Encoder and the Embedder are concatenated and processed by a block of three convolutional layers before being passed to the decoder. The network treats the cylindrical voxel grid as a graph, where nodes represent voxels and edges connect nearest neighbors, learning edge weights to capture spatial dependencies in energy deposition.

The second component is a lightweight Multi-Layer Perceptron (MLP) tasked with predicting the secondary neutron yield in both the collimators and the degrader. The MLP consists of a single hidden layer with 128 nodes.

Both networks were trained simultaneously using a combined objective function defined as the sum of the dose Binary Cross-Entropy (BCE) loss and the neutron Mean Squared Error (MSE). Within the neutron loss component, a weighting factor of 10 was applied to the collimator yield relative to the degrader yield to prioritize accuracy for the components closest to the patient.

A limited hyperparameter optimization was performed, focusing on the learning rate, the number of convolutional layers following the Encoder-Embedder merging, and the number of hidden nodes in the neutron MLP.  
This process yielded the selected configuration described above.

Training was performed on an NVIDIA Tesla V100 GPU and lasted approximately 3.5 hours, using a learning rate of $5 \times 10^{-3}$ with a \textit{ReduceLROnPlateau} scheduler (patience of 20 epochs). The resulting dose engine achieves a relative error below 3\% across the region of interest. Regarding secondary particle production, the MLP predicts the Beryllium neutron yield with a mean relative error of 0.5\%, while the yield from the Nickel collimators is estimated with an error of 3\%. These metrics confirm the surrogate's suitability for the subsequent optimization tasks.

\subsubsection{Optimization Strategy}

The optimization methodology adheres to the framework established in Arsini et al.~\cite{Arsini2025-dm}, to which the reader is referred for comprehensive architectural details. In this implementation, the differentiable dose engine is queried in batches of $N=9$ pencil beams. A regridding algorithm maps the individual cylindrical outputs onto a shared Cartesian grid of $51 \times 51 \times 51$ voxels ($10.2 \times 10.2 \times 10.2\ \text{cm}$) representing the water phantom.

The optimization problem targets a total of 17 free parameters, divided into two categories:
\begin{enumerate}
    \item \textbf{Global Field Parameters (8):} Beryllium and Nickel thicknesses, grid pitch, four independent collimator jaw positions, and global fluence.
    \item \textbf{Pencil Beam Parameters (9):} Individual $\Delta$-fluences for each spot.
\end{enumerate}

Regarding collimation, we relax the fixed-gap constraint used in MC training. Each peripheral pencil beam is assumed to interact only with its proximal collimator jaw among the two on the same axis, while the central spot is treated as uncollimated. Variables are normalized for stability: spatial shifts are bounded to $[-1, 1]$, thicknesses and grid pitch to $[0, 1]$, and fluences are parameterized in log-space.

We employ a hybrid recursive strategy alternating between global exploration and local refinement:
\begin{itemize}
    \item \textbf{CMA-ES~\cite{Hansen2016-bx} (1800 iterations):} Optimization is restricted to the 8 global field parameters to navigate the non-convex landscape.
    \item \textbf{Adam~\cite{Kingma2014-uq} (1000 iterations):} Optimization is performed on all 17 parameters (field and pencil beams) to fine-tune the solution using gradient information.
\end{itemize}

The process is iterative. If a combined CMA-Adam cycle yields a lower loss than the previous best solution, the updated parameters are retained for the next loop. However, if the cycle fails to improve the result, the optimization reverts to the previous best configuration. Crucially, upon such a failure, the CMA exploration stage is disabled for all subsequent loops, and the optimization proceeds exclusively with the Adam optimizer to ensure convergence within the identified basin of attraction.

The optimization targets a $3 \times 3$ pencil beam field designed to treat a Planning Target Volume (PTV) defined as rectangular region of $1 \times 1 \times 11$ voxels ($2.6 \times 2.6 \times 0.2\ \text{cm}$). For this proof of concept, the prescription dose is normalized to unity ($D_{pre} = 1$). All voxels external to the PTV are collectively defined as a single Organ At Risk (OAR).

The process is guided by a composite loss function $\mathcal{L} = \mathcal{L}_{dose} + \lambda \mathcal{L}_{neutron}$, where $\lambda$ acts as a regularization parameter.

The dosimetric component $\mathcal{L}_{dose}$ comprises five terms:
\begin{enumerate}
    \item A voxel-wise Mean Squared Error (MSE) on the PTV to enforce the unitary prescription $D_{pre}$.
    \item A normalized MSE on the OAR voxels to minimize radiation spill-out.
    \item A coverage penalty $\mathrm{ReLU}(D_{pre} - D_{95\%})^2$, ensuring that 95\% of the target receives at least the prescribed dose.
    \item A hotspot penalty $\mathrm{ReLU}(D_{max}^{PTV} - 1.05 D_{pre})^2$, constraining the maximum target dose within 105\%.
    \item An OAR safety constraint $\mathrm{ReLU}(D_{max}^{OAR} - 0.8 D_{pre})^2$, limiting the maximum dose to the surrounding tissue.
\end{enumerate}

The hardware-specific penalty is defined as $\mathcal{L}_{neutron} = N_{Be}^2 + 100 \cdot N_{Coll}^2$. Here, $N_{Be}$ and $N_{Coll}$ represent the predicted neutron yields from the Beryllium degrader and Nickel collimators, respectively. To ensure numerical stability, both values are normalized to the range $[0, 1]$ using the minimum and maximum neutron yields observed for the Beryllium degrader. This common scaling basis is adopted because the neutron production in the degrader is approximately one order of magnitude higher than in the collimators. The weighting factor of 100 is then explicitly applied to the collimator term to prioritize the reduction of neutrons generated at the nozzle's distal end. Unlike upstream degrader interactions, which can be shielded, collimator-produced neutrons are generated in close proximity to the patient and contribute directly to the integral dose.

\subsubsection{Results}

To demonstrate the capabilities and flexibility of the proposed co-design framework, we present three optimization scenarios. These examples illustrate how the optimizer adapts both hardware geometry and treatment parameters to satisfy varying clinical priorities, governed by the regularization parameter $\lambda$:

\begin{itemize}
    \item \textbf{Plan A (Baseline, $\lambda = 0$):} A purely dosimetric optimization where neutron production is ignored. This serves as the reference for best achievable target coverage.
    \item \textbf{Plan B (Neutron-Averse, $\lambda = 0.1$):} A scenario with a severe penalty on secondary particle production, testing the system's behavior under extreme constraints.
    \item \textbf{Plan C (Balanced, $\lambda = 10^{-4}$):} An optimized trade-off incorporating the specific collimator weighting described in \autoref{ss:pp}, aiming to minimize local activation without compromising target coverage.
\end{itemize}

\begin{figure}[t!]
    \centering
    \begin{subfigure}[b]{0.48\linewidth}
        \includegraphics[width=\linewidth]{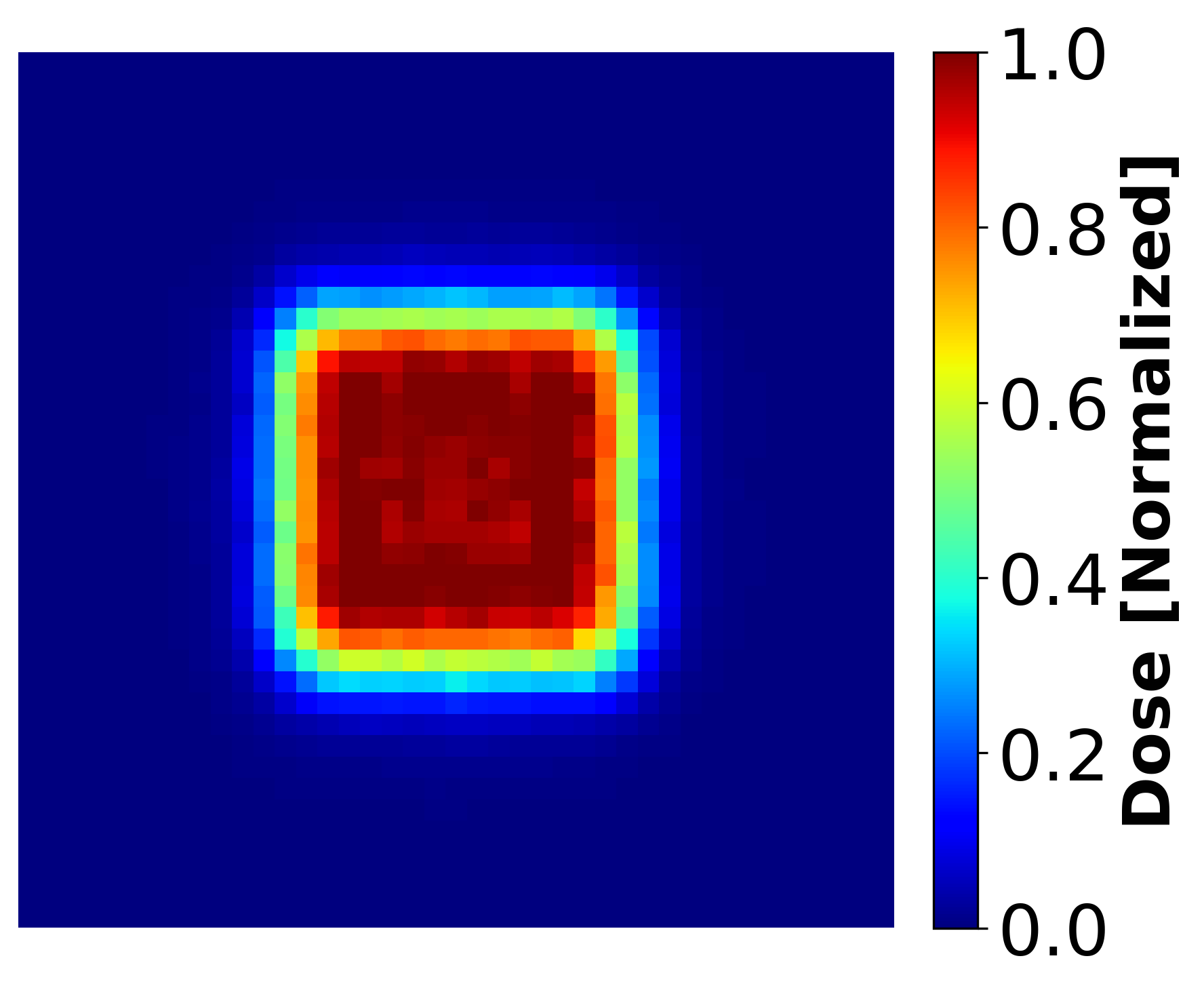}
        \caption{Plan A Dose}
        \label{fig:dose2d_a}
    \end{subfigure}
    \begin{subfigure}[b]{0.48\linewidth}
        \includegraphics[width=\linewidth]{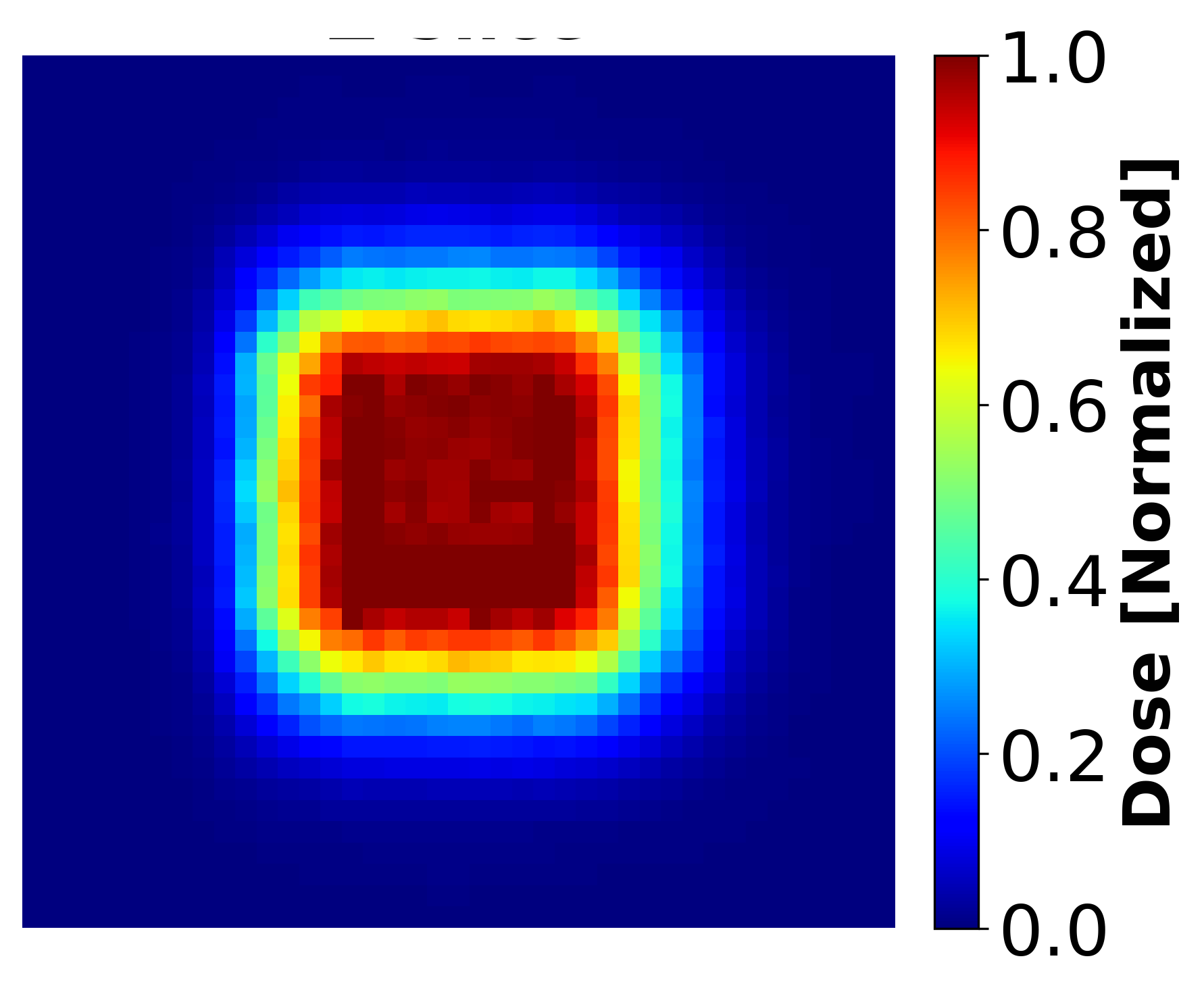}
        \caption{Plan C Dose}
    \end{subfigure}
    \caption{2D Transverse dose distributions for Plan A (dose-only) and Plan C (balanced).}
    \label{fig:dose2d}
\end{figure}

Table~\ref{tab:results} summarizes the quantitative metrics for the three scenarios.
Plan A achieves a highly conformal dose distribution ($D_{95} \approx 94\%$) and sharp field edges (\autoref{fig:dose2d_a}), but results in the highest neutron yield from the collimators ($0.017$).

Plan B demonstrates the physical consistency of the surrogate model through a failure mode. Driven by the high penalty on material interaction, the optimizer maximally thins the Beryllium degrader and fully retracts the collimators. As a physical consequence, the beam energy is insufficiently degraded, causing the Bragg peak to overshoot the PTV and land in the distal OAR (\autoref{fig:profileB}). While this plan is clinically invalid ($D_{95} = 37.5\%$), it confirms that the gradient-based optimization correctly identifies the causal link between degrader thickness and beam range.

Plan C represents the successful application of the co-design strategy. By utilizing a balanced penalty, the optimizer adopts a dual strategy: it retracts the collimator jaws to minimize direct beam interaction—thereby reducing secondary radiation—and compensates for the resulting loss in penumbra sharpness by decreasing the pencil beam spacing (grid pitch) compared to Plan A. Quantitatively, this trade-off is highly effective: the collimator neutron yield drops by approximately one order of magnitude ($0.017 \to 0.0020$) relative to the baseline. Although this configuration results in slightly inferior dosimetric coverage compared to the unconstrained scenario (Target $D_{95}$ decreases from $94.0\%$ to $93.6\%$), the plan remains robust and clinically acceptable.

\begin{table}[h]
\centering
\caption{Comparison of optimization metrics for Plans A, B, and C. Target Dose normalized to prescription (1.0).}
\label{tab:results}
\begin{tabular}{lccc}
\toprule
\textbf{Metric} & \textbf{Plan A} & \textbf{Plan B} & \textbf{Plan C} \\
\midrule
Neutron Yield (Be) & 0.099 & 0.073 & 0.099 \\
Neutron Yield (Coll) & 0.017 & 0.0009 & 0.0020 \\
\midrule
Target $D_{95}$ & 0.940 & 0.375 & 0.936 \\
Target Mean & 0.988 & 0.425 & 0.985 \\
Target Max & 1.058 & 0.502 & 1.058 \\
\midrule
OAR 1 $D_{1}$ & 0.436 & 0.506 & 0.447 \\
OAR 1 Max & 0.824 & 1.188 & 0.852 \\
\bottomrule
\end{tabular}%
\end{table}

\begin{figure}[t!]
    \centering
    \begin{subfigure}[b]{0.48\linewidth}
        \includegraphics[width=\linewidth]{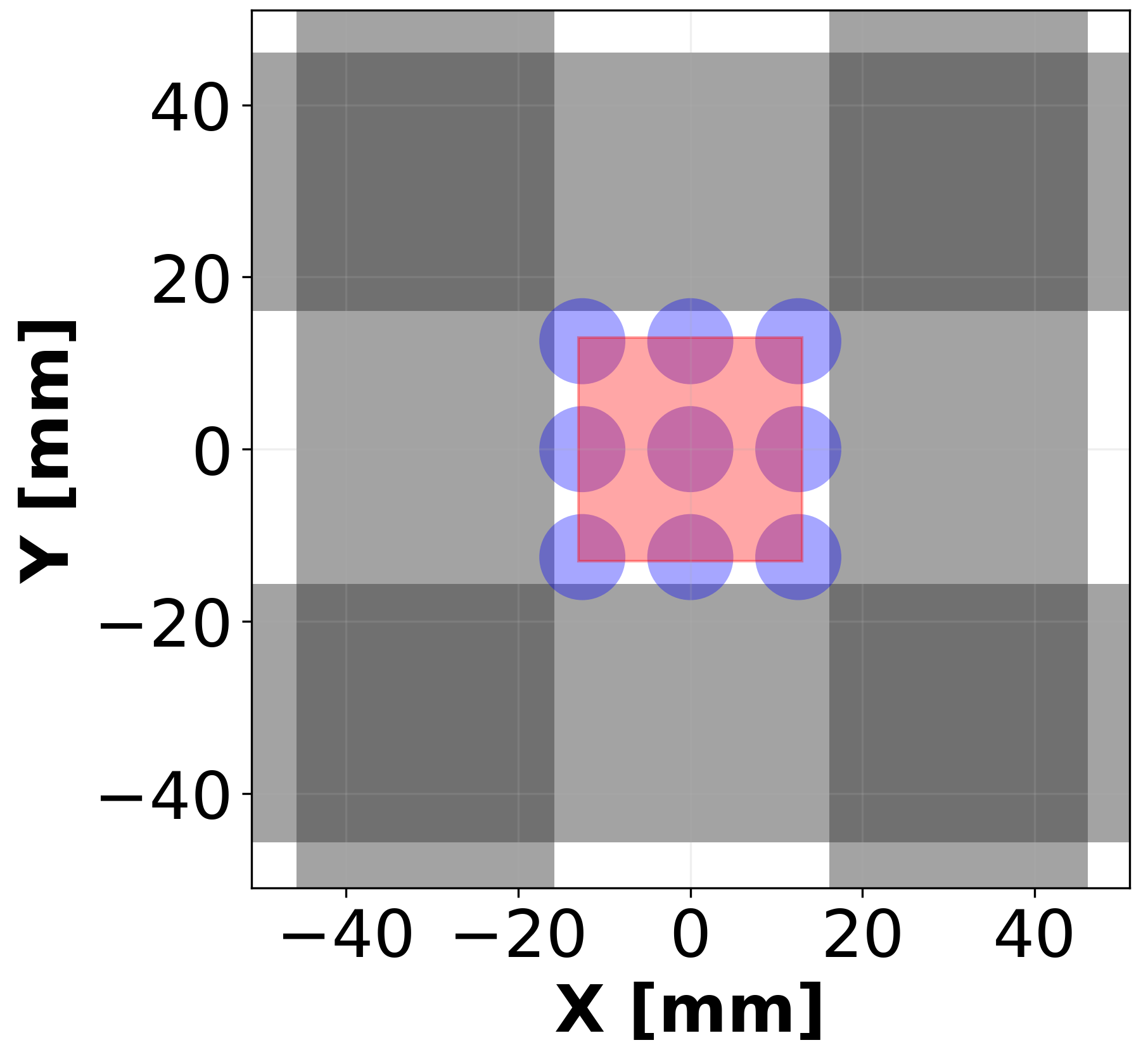}
        \caption{Plan A Setup}
    \end{subfigure}
    \begin{subfigure}[b]{0.48\linewidth}
        \includegraphics[width=\linewidth]{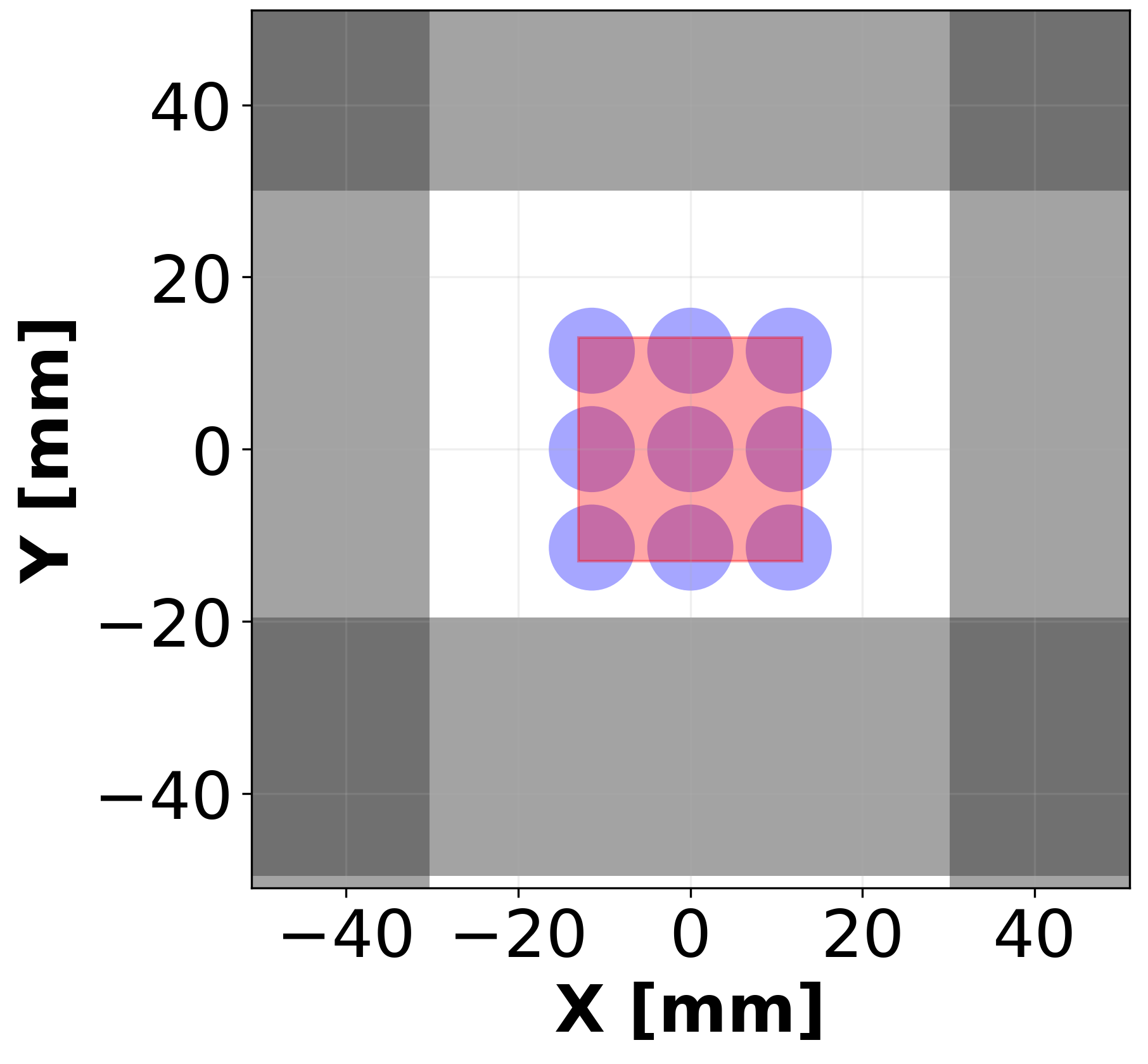}
        \caption{Plan C Setup}
    \end{subfigure}
    \caption{Visualization of collimator apertures (grey) and pencil beam positions (colored dots) relative to the tumor (red square).}
    \label{fig:collimators}
\end{figure}

\begin{figure}[t!]
    \centering
    \includegraphics[width=0.8\linewidth]{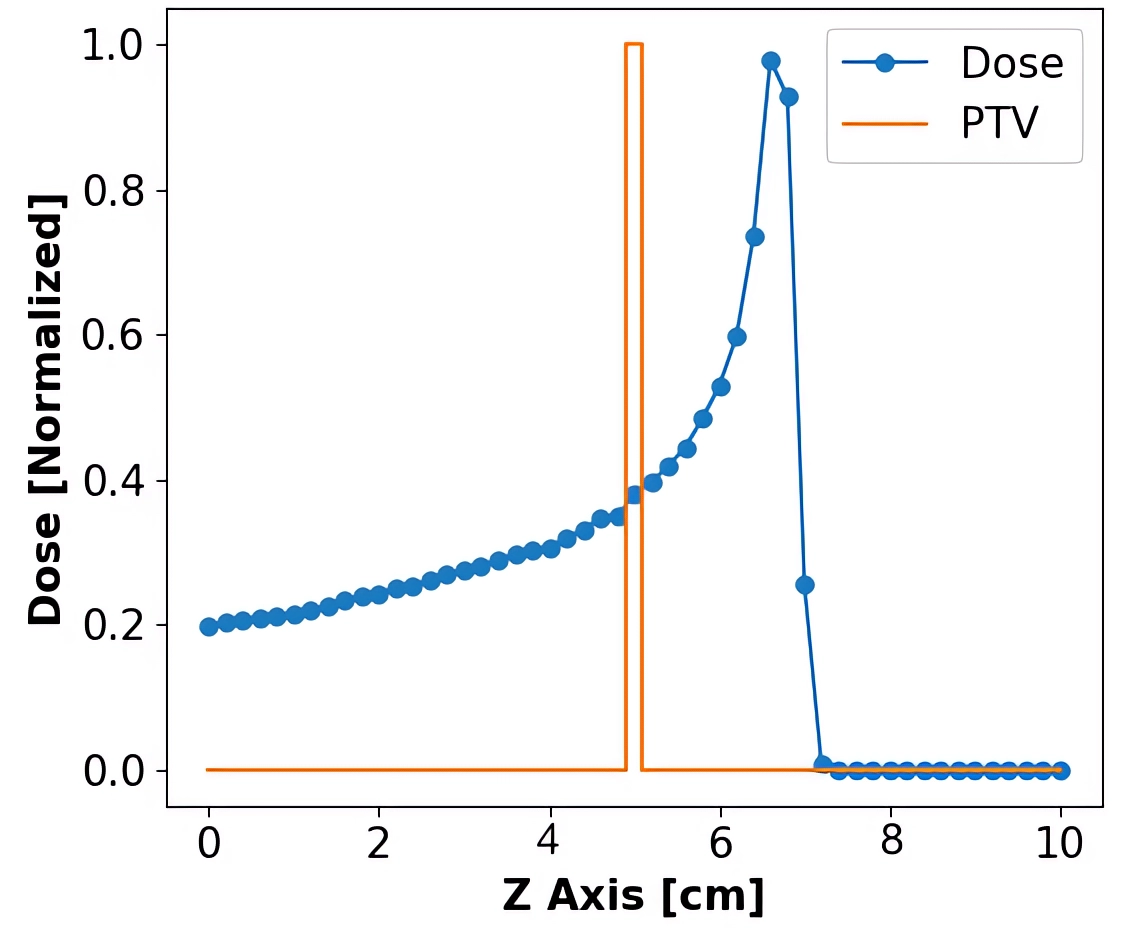}
    \caption{1D Dose profile along Z for Plan B. The Bragg peak (blue line) falls outside the PTV (shaded area).}
    \label{fig:profileB}
\end{figure}

\begin{figure}[t!]
    \centering
    \includegraphics[width=0.8\linewidth]{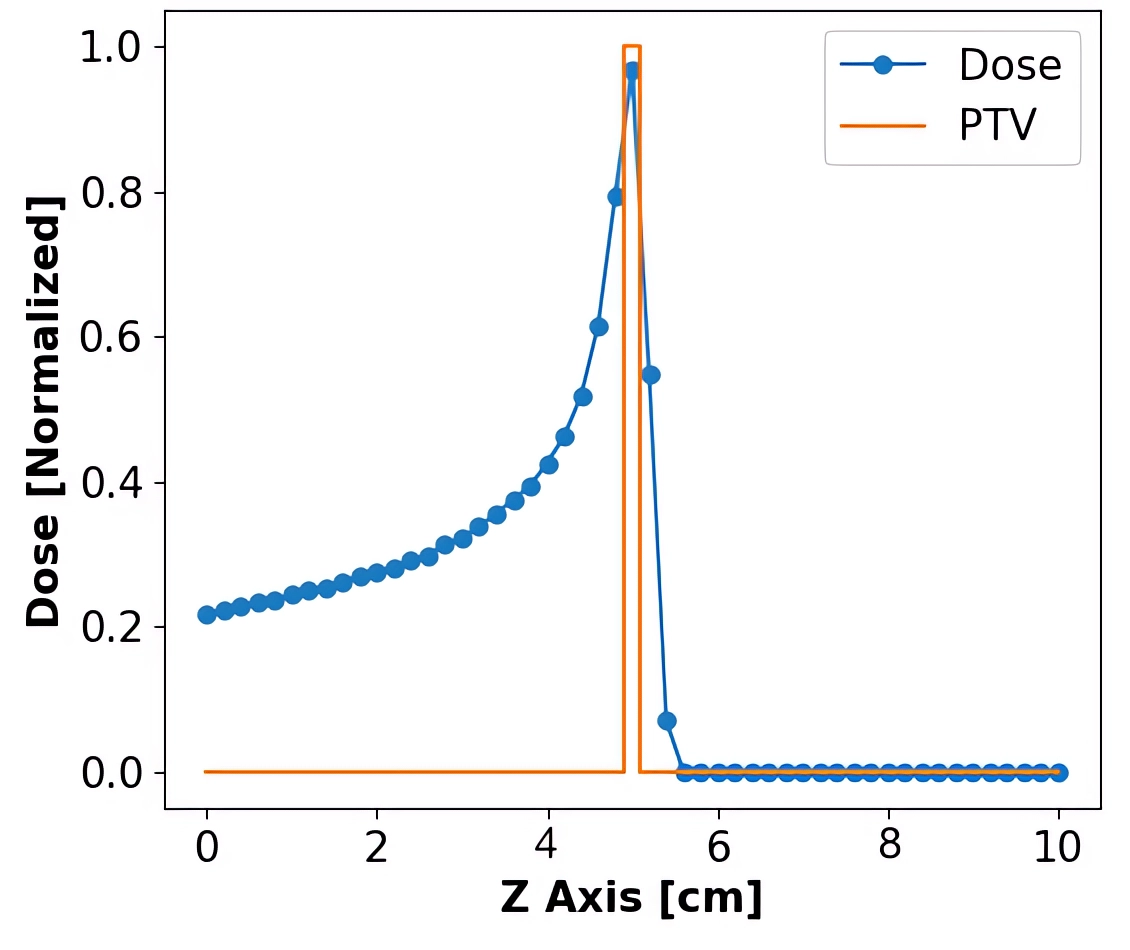}
    \caption{1D Dose profile along Z for Plan C. The Bragg peak is correctly positioned within the PTV.}
    \label{fig:profiles}
\end{figure}
\subsubsection{Discussion and Conclusion}

This study presents a proof of concept for a hardware-software co-design framework in proton therapy. While standard Treatment Planning Optimization (TPO) typically considers the nozzle geometry as a fixed constraint, our approach leverages differentiable programming to concurrently optimize both the treatment planning variables and the hardware configuration.

The results demonstrate the system's ability to adapt to varying clinical priorities. Specifically, the introduction of a penalty term for secondary neutron production allowed the optimizer to explore innovative trade-offs. This capability is particularly relevant for specific patient demographics, such as pregnant women, where minimizing the integral dose contribution from secondary radiation is a critical safety requirement. As observed in the balanced optimization scenario, the system autonomously compensated for the retraction of collimator jaws—necessary to limit neutron generation—by tightening the pencil beam spacing, thus preserving target coverage.

Current limitations of this work include the validation in a homogeneous water phantom and the specific adoption of the dynamic collimation geometry proposed by Hyer et al.~\cite{Hyer2014-gw}. However, the underlying differentiable modeling approach is generalizable and can be extended to more complex delivery systems or patient-specific geometries.

From a computational perspective, the optimization process currently requires approximately 10 minutes on a single GPU. It should be noted that the code is not yet optimized for performance, and the stopping criteria for the hybrid strategy require further refinement. These aspects, while outside the scope of this proof of concept, will be the focus of future research aimed at scaling the framework to realistic clinical cases based on patient CT data.

\subsection{Gravitational waves} 
\label{sec:GW}
Gravitational-wave (GW) detectors aim to measure extremely small distortions of spacetime caused by passing gravitational waves. Astrophysical waves are emitted by systems such as merging black holes and neutron stars~\cite{Abbott2016_GW150914, Abbott2017_GW170817}, while cosmological waves are expected to form a stochastic background originating from the early Universe~\cite{Maggiore2018_Stochastic}. When a GW reaches an interferometric detector, it produces a characteristic change in the interference pattern of the laser beams by stretching one arm and compressing the other. The size of this effect is described by the strain $h = \delta L / L$, typically of order $10^{-21}$, far smaller than the dimensions of an atomic nucleus. Given this extraordinary scale, GW detectors must achieve sensitivity sufficient to distinguish actual GW signals from a vast array of instrumental and environmental noise sources. 

Ground-based instruments like the LIGO-Virgo detector use kilometre-scale laser interferometers to measure variations in the distance between suspended mirrors~\cite{TheLIGOScientific2015_AdvLIGO, Acernese2015_AdvVirgo}, while the future space mission LISA will measure similar variations across millions of kilometres between freely flying spacecraft~\cite{Amaro-Seoane2017_LISA}. Extracting reliable signals requires both advanced optical and mechanical systems and equally sophisticated data analysis methods that model the noise, reconstruct the signals, and estimate the astrophysical properties of each event~\cite{Veitch2015_ParameterEstimation}.

\subsubsection{Multi-band gravitational-wave detectors}
Because the GW signals are faint and the noise environment is complex, detectors exhibit a stronger form of hardware--software co-design than many other scientific instruments. Their scientific performance depends on tight coupling between the physical design of the detector and the analysis pipelines that process the data. The measurement quality depends directly on the analysis methods, while the analysis performance is constrained by how the hardware is engineered. Examples from both space-based and ground-based detectors illustrate this interdependence clearly.

GW detectors are primarily distinguished by the range of frequencies to which they are most sensitive. Ground-based interferometers such as LIGO and Virgo operate in the $\sim 10$--$10^3\,\mathrm{Hz}$ band, allowing them to detect the mergers of stellar-mass black holes and neutron stars~\cite{Abbott2019_GWTC1}. Space-based detectors like LISA will probe much lower frequencies, from $\sim 10^{-4}$ to $10^{-1}\,\mathrm{Hz}$, making them suitable for observing massive black-hole binaries and compact galactic binaries with long orbital periods~\cite{Amaro-Seoane2017_LISA, Robson2019_LISASensitivity}. Even lower frequencies, down to $\sim 10^{-9}\,\mathrm{Hz}$, are monitored by pulsar-timing arrays, which are sensitive to supermassive black-hole binaries and a possible cosmological background~\cite{NANOGrav2023_Evidence, EPTA2023_Collaboration}.

\subsubsection{Gravitational-wave signals and detection methods}

GW signals are so weak that they are typically buried beneath noise that is many orders of magnitude larger than the signal itself. The signals have characteristic time-frequency patterns determined by the dynamics of the source, such as the chirping waveform produced by inspiraling compact binaries~\cite{Buonanno2009_Waveforms}. Detecting these signals requires accurate theoretical models (templates), which are generated using both analytical  and numerical  techniques~\cite{Blanchet2014_PostNewtonian, Husa2016_NRWaveforms}. 

In the traditional approach of matched filtering, the observed data are cross-correlated against these templates to identify events whose patterns align with the expected signal shape~\cite{Allen2012_MatchedFiltering}. Noise sources, such as seismic motion, thermal vibrations, or quantum fluctuations, are reduced using a combination of hardware design and dedicated subtraction methods~\cite{Driggers2019_NoiseSubtraction}. After a candidate signal is identified, further analysis estimates the source properties by comparing the data with theoretical templates using Bayesian inference techniques~\cite{Veitch2015_ParameterEstimation}.

\subsubsection{Co-design in gravitational-wave detection}

A first case of co-design in GW detection comes from LISA. The mission relies on a specialised method of data processing known as Time-Delay Interferometry (TDI) that removes large fluctuations in the laser frequency by combining delayed measurements along different optical paths~\cite{Armstrong1999_TDI, Tinto2021_TDIReview}. This technique sets strict requirements on how the spacecraft move, how clocks are synchronised, and how the laser links are arranged. Because the cancellation only works when these conditions are met, the analysis method and the hardware configuration must be designed and validated together. In practice, end-to-end simulations of the spacecraft motion, timing errors, optical performance, and reconstruction accuracy are run to find combinations that keep the residual noise below the required level~\cite{Bayle2019_LISASimulations}.

A second example is the use of squeezed light in LIGO and Virgo~\cite{Aasi2013_SqueezedLight, Tse2019_QuantumEnhanced}. This technique reduces quantum noise at high frequencies by redistributing the quantum fluctuations in the phase and amplitude of the light. However, implementing squeezed light changes the statistical properties of the noise, introduces new couplings between optical subsystems, and modifies the calibration behaviour. To benefit from the hardware upgrade, the analysis pipelines had to be adapted to the new noise characteristics and to the increased sensitivity to alignment and phase-control errors. Simulations of quantum noise, calibration stability, and signal recovery were used to tune the squeezing level, control loops, and optical layout~\cite{Yancey2022_FrequencyDependent}.

A related co-design problem appears in future underground detectors such as the Einstein Telescope~\cite{Punturo2010_ET, Maggiore2020_ETScience}. Fluctuations in ground and air density cannot be shielded and must be cancelled by combining environmental measurements with subtraction algorithms~\cite{Harms2013_Newtonian, Badaracco2019_ETNoise}. The performance of these algorithms depends on how well the sensor network samples the seismic and atmospheric fields, which in turn depends on the cavern layout and the allowed construction geometry. Co-design therefore requires joint simulations of underground seismic fields, sensor placement strategies, and subtraction algorithms to identify configurations that achieve sufficient noise reduction.

In next-generation detectors, optimal sensitivity requires tuning both the detector components and the analysis pipelines together. Changes in mirror coatings, laser power, or optical geometry alter the noise spectrum, which affects which filtering, calibration, or inference methods perform best~\cite{Adhikari2020_CosmicExplorer}. Constraints also flow in the opposite direction: analysis requirements such as low-latency detection or stable calibration over long timescales impose limits on sampling rates, actuator performance, and control-loop bandwidth~\cite{Nitz2020_RapidDetection}. This creates a feedback loop between hardware choices and pipeline capabilities.

Finally, environmental sensors used to identify and classify non-astrophysical disturbances must be developed together with the analysis methods that rely on them~\cite{Davis2021_GlitchMitigation}. The performance of glitch-identification pipelines depends on what sensors exist, but installing sensors that the analysis cannot exploit accomplishes little. Data-driven studies of noise transients, sensor correlations, and classification accuracy help determine both which sensors are needed and how the pipelines should use them~\cite{Zevin2017_GravitySpy, Cuoco2020_MLGlitches}.

Together, these examples show that GW detectors are not simply built and then analysed: they are co-designed systems in which hardware and software must be optimised together using simulations, noise models, and data-driven validation. This co-design approach is essential for achieving the sensitivity required for future discoveries.

\subsubsection{Machine learning methods for co-design}

Machine learning enters co-design of GW detectors mainly through surrogate models that emulate expensive simulations related to both detector and waveform modelling~\cite{Chua2019_Surrogates, Williams2021_MLWaveforms}. Such surrogates learn the mapping from design parameters ({\it e.g.}\ mirror coatings or seismic environment) to performance metrics ({\it e.g.}\ strain sensitivity or noise power spectrum) and then provide predictions in milliseconds instead of minutes or hours. This makes it feasible to explore large coupled hardware--software design spaces and to compare many candidate configurations within a realistic computational budget~\cite{Doctor2020_BayesianOptimization}.

However, standard optimisation methods from machine learning, such as stochastic gradient descent, are generally not suitable for this case. This is because the design space is high-dimensional and irregular, and often not differentiable. The latter means that small changes in configuration can cause abrupt changes in lock stability, control behaviour, or noise coupling, so reliable gradients are either unavailable or meaningless. The objective functions are also multi-objective (sensitivity, robustness, cost, operability) and constrained, with many hard engineering limits that do not fit easily into a smooth loss function~\cite{PhysRevD.101.042003_Vajente2020}. In addition, important noise sources are non-stationary and only partly modelled, so the mapping learned by a surrogate or assumed by an optimiser may drift over time.

For these reasons, co-design studies usually combine surrogate models with global, gradient-free strategies such as Bayesian optimisation, evolutionary algorithms, and structured parameter scans, rather than relying on gradient-based training alone~\cite{Brochu2010_BayesianOpt, Deb2002_NSGA}. These methods are better suited to handle the complex, multi-modal landscapes characteristic of detector design problems and can systematically explore trade-offs between competing objectives while respecting physical and engineering constraints.

\subsubsection{Practical case study}
\begin{figure*}[h!t]
    \centering
    \includegraphics[width=1.0\linewidth]{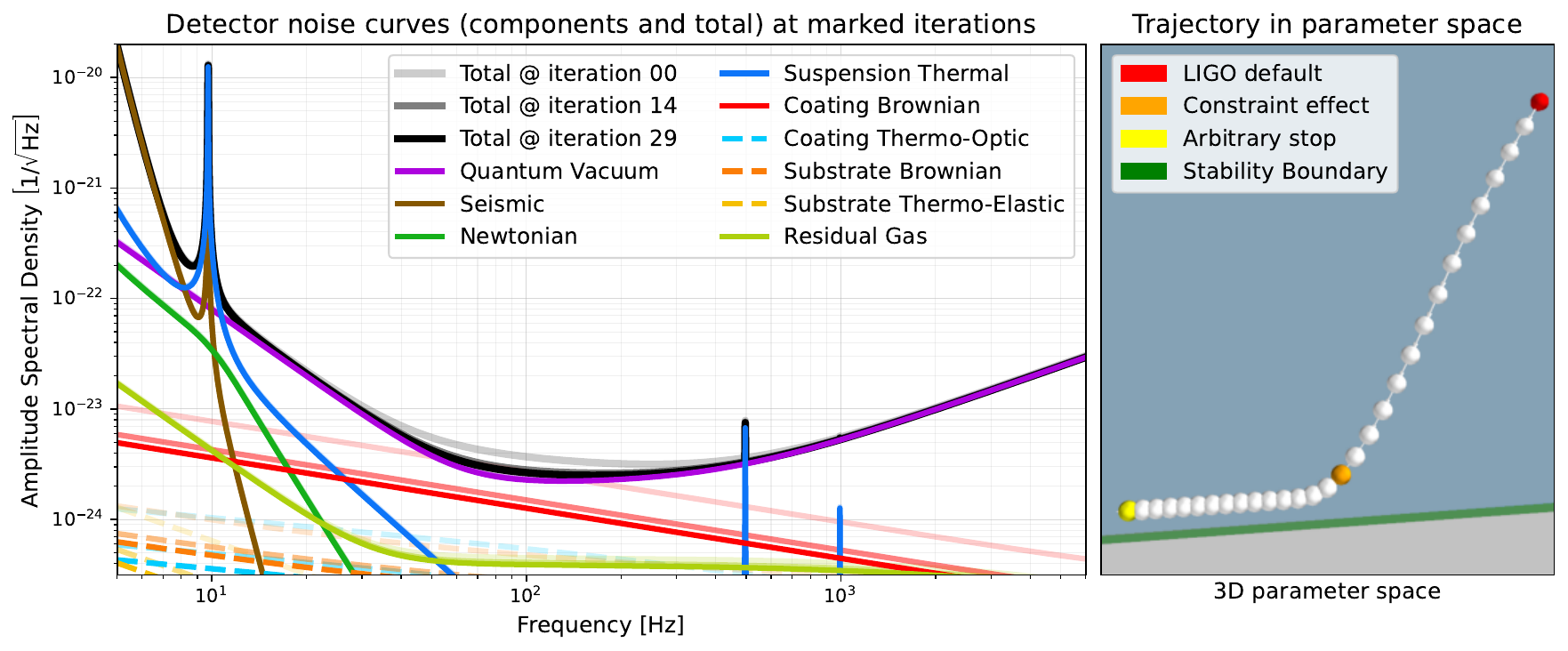}
    \caption{Preliminary demonstration of gravitational wave detector optimization.  The left plot is showing noise curves of our simulated LIGO-like detector and its individual components at different stages of the optimization. Note that increasing opacity on the noise curves implies further optimization. The right plot is showing an image of the optimizer’s trajectory through a reduced three-dimensional projection of the detector's parameter space, illustrating convergence from the initial configuration (current LIGO setup) towards a plane parallel to the cavity stability boundary (green line) thanks to gradient projection, whose effect starts being noticeable at iteration 14 (orange). }
    \label{fig:LIGO_optim}
\end{figure*}
In order to illustrate some of the concepts discussed so far in the realm of gravitational wave astronomy,  a naive optimization of a generic LIGO-like interferometer is presented, adjusting only the length of its arms and the radius of curvature of its main mirrors. These three parameters are arguably some of the most important when designing a GW detector, and exhibit a physical constraint that would make this approach to optimization seem infeasible. 

Fabry-Perot cavities such as those in a ground-based interferometer need to be kept stable\footnote{In a physical configuration that ensures the laser beam cross-section will not grow indefinitely and therefore escape the mirror.} for long periods of time to operate correctly as a detector. To remain stable, such a cavity needs the stability factors, $g_1$, for its Input Test Masses (ITMs) and $g_2$ for its End Test Masses (ETMs), to satisfy the condition:
\begin{equation}
     g^{2} := g_1g_2(1-g_1g_2) \geq 0\,,
     \begin{cases}
     g_1 := 1-L/R_{ITM}\\ g_2 := 1-L/R_{ETM}\,, 
\end{cases}
\end{equation}
which implies $g_1, g_2 \in [0, 1]$. As the stability factors are themselves functions of our three parameters, the border of our allowed region can be quite complex. Thankfully,  the constraint reduces to:  $\max(R_{ITM}, R_{ETM}) < L < R_{ITM} + R_{ETM}$, so the border can be constructed just by intersecting three planes. Because the sensitivity of an interferometer is proportional to its length, the optimization will only risk trespassing the upper length constraint. If left unchecked, Stochastic Gradient Descent (SGD) will accelerate towards the limit and cross it, forcing the simulator to terminate the program.

Many methods have been employed over the years to perform constrained optimizations~\cite{1965_complex_constrained_optimization, 1994_genetic_constrained_optimization, kotary2021_end2end_constrainedoptimization_learning_survey}. Penalty methods are the most ubiquitous~\cite{fletcher1983penalty, 1993_penalty_methods_NNs}, as they are comparatively straightforward to implement. However, our need to remain close to the edge (ideally, at a constant distance of our choosing) pushed us to experiment with more geometry-inspired solutions.

Since we aim to stop the optimization trajectory from ever crossing a known surface, and we have the gradient at our disposal, we can project out the component of the gradient that is perpendicular to our surface once a predefined proximity threshold if met. If we introduce a smoothly increasing parameter to control the amount of projection that is rejected, we get a smooth transition between our free and constrained regimes (right image of \autoref{fig:LIGO_optim}).

\begin{equation}
    \vec{\nabla}f_{\parallel} = \vec{\nabla}f - \alpha(g)(\vec{\nabla}f\cdot\vec{n})\vec{n} \hspace{0.3cm} \begin{cases}
        \alpha(g_{ini}) = 0\\
\alpha(g_{lim}) = 1
    \end{cases}\,.
\end{equation}
In this study, $\alpha(g) = \text{Lerp}^{-1}(g_{ini}, g_{lim}, t_{intrp}=g)$ is chosen. No other costs were considered, so as to provide a clear example of gradient projection in action. That been said, a naive economical cost-per-meter could be easily implemented for all three parameters.

The optimization has been achieved by partially adapting the gwinc python library~\cite{ligogwinc} to enable differentiable calculation of detector noise curves. From them, we can construct a loss function to optimize. Since it is a simple demonstration, we chose the base 10 logarithm of the area under the combined noise (black line on the right plot of \autoref{fig:LIGO_optim}). Minimizing the area leads to better sensitivities, though constructing proper objective functions will be an endeavour in its own right.

\begin{table}[]
    \centering
    \begin{tabular}{|c|ccc|}
    \hline
         &   ITM [m]& ETM [m]&L [m]\\
         \hline
         Iteration 0& 
     2245.00& 1934.00&3995.00\\
 Iteration 14& 2216.21& 1902.35&4099.45\\
 Iteration 20& 2234.92& 1919.93&4144.46\\
 \hline
 \end{tabular}
    \caption{Optimized parameter values at marked iterations (red, orange and yellow on \autoref{fig:LIGO_optim}).}
    \label{tab:LIGO_optim}
\end{table}

With 30 iterations (initial + 29), detector range\footnote{Detector range is a more promising optimization target than the one used here, tough the algorithms used to calculate it are complex and have not yet been implemented in differentiable programming paradigms.  Still, we can obtain it from our optimized noise curves.} was improved from the $185\text{ Mpc}$ of the Adv-LIGO design sensitivity to $236\text{ Mpc}$ ($28\%$ increase) by increasing the length by $200\text{ m}$ (only $5\%$) and tweaking the mirror radii slightly (\autoref{tab:LIGO_optim}). This was mainly achieved by working much closer to the stability edge, but since this distance is a hyperparameter of the optimization, it can be agreed upon with the teams responsible for detector operation early in the design process.

\subsection{Bridging the Hardware-Software Gap: A Case Study in Physical Generalization}
\label{subsec:hep_codesign}

\noindent
\paragraph{The Simulation Bottleneck}
In the lifecycle of modern experiments, the simulation pipeline acts as the ``digital twin'' of the physical detector, consuming a significant fraction of the computing budget. Traditionally, modeling charge transport in silicon sensors relies on computationally intensive Monte Carlo methods ({\it e.g.}, \textsc{Geant4}). However, the transition to the High-Luminosity LHC (HL-LHC) introduces a crisis of scale: next-generation 4D sensors (such as 3D trench-type pixels) possess complex geometries that are prohibitively expensive to simulate at the required volume. This creates a classic co-design tension: the hardware optimization (for precision) is decoupled from the software optimization (for throughput), creating a risk where advanced hardware designs become operationally unsustainable due to software latency.

\noindent
\paragraph{Bridging the Gap with ML Surrogates}
To resolve this deadlock, we deployed a co-design workflow to integrate Machine Learning (ML) surrogates directly into the experiment's C++ digitization framework. A key enabler was \textbf{SOFIE} (System for Optimized Fast Inference code Emit)~\cite{Brun:1997pa}, which transpiles models trained in flexible Python environments into dependency-free C++ header files. This approach effectively treats the neural network as a standard mathematical function, bridging the gap between scientific exploration and industrial-scale production constraints (latency and thread-safety).

\begin{figure}[h!t]
    \centering
    \begin{subfigure}[b]{0.48\textwidth}
        \centering
        \includegraphics[width=\linewidth]{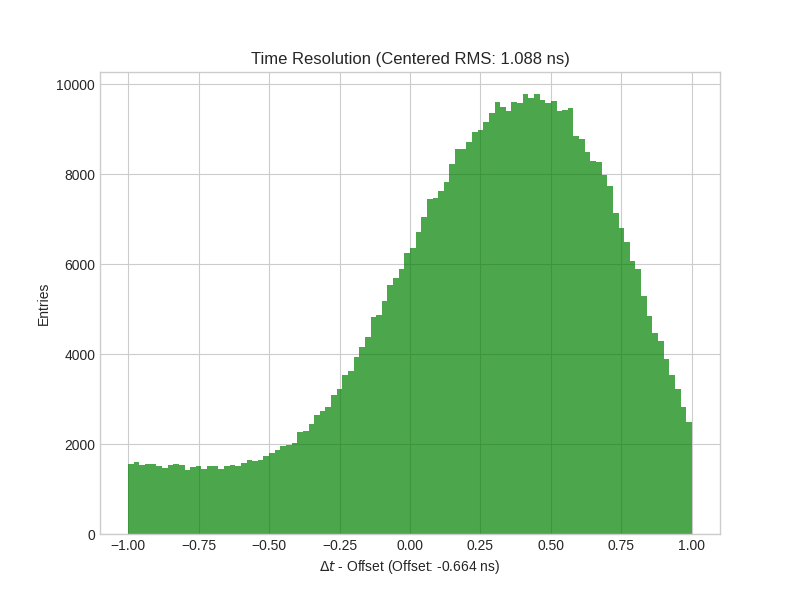}
        \caption{Time Resolution Shift}
    \end{subfigure}
    \hfill
    \begin{subfigure}[b]{0.48\textwidth}
        \centering
        \includegraphics[width=\linewidth]{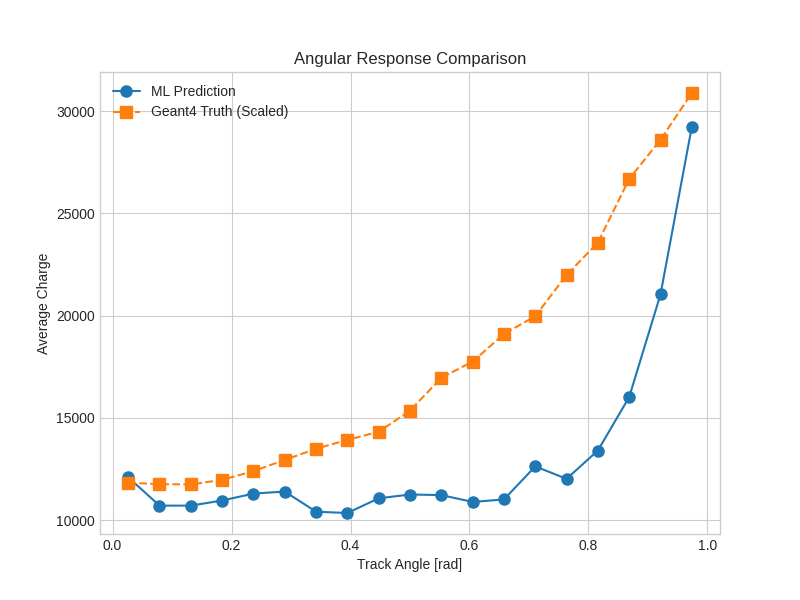}
        \caption{Angular Response Comparison}
    \end{subfigure}
    \caption{Manifestation of the ``Geometric Feature Gap'' acting as a probe for physical generalization. The ML surrogate, trained on $150\,\mu\text{m}$ sensors, was validated against a $200\,\mu\text{m}$ reference. (a) A systematic time offset of $\Delta t \approx 0.66\,\text{ns}$ reflects the increased drift distance. (b) The divergence in angular response confirms the model is sensitive to path length changes ($L \propto d/\cos\theta$). These residuals indicate the model successfully extrapolated the underlying charge transport physics.}
    \label{fig:hep_validation}
\end{figure}

\noindent
\paragraph{Validation as a Stress Test}
A critical insight into the hardware-software coupling emerged during validation. We encountered a ``Geometric Feature Gap'' where a model, trained on $150\,\mu\text{m}$ thick sensors, was inadvertently validated against a $200\,\mu\text{m}$ hardware design. 
Initial residuals suggested a failure: a systematic time offset of $\Delta t \approx 0.66\,\text{ns}$ (\autoref{fig:hep_validation}) and a broadened time distribution. However, a physics-driven analysis revealed that the surrogate had correctly generalized the underlying charge transport mechanism. By inverting the kinematic relationship for the thickness difference $\Delta d = 50\,\mu\text{m}$, we derived the effective drift velocity implied by the model's prediction:
\begin{equation}
    v_{drift} = \frac{\Delta d}{\Delta t} = \frac{50\,\mu\text{m}}{0.664\,\text{ns}} \approx 0.75 \times 10^7\,\text{cm/s}\,.
\end{equation}
This inferred value aligns consistently with the physical saturation velocity of electrons in silicon at room temperature ($v_{sat} \sim 1.0 \times 10^7\,\text{cm/s}$) under non-uniform fields. Furthermore, the increased RMS width was found to correctly reflect the broader uniform distribution of drift times inherent to the thicker sensor volume, demonstrating that the model learned the implicit physics rather than memorizing the training geometry.

\noindent
\paragraph{Implications for Co-Design}
This case study demonstrates that robustness requires parameterization. The observed mis-alignment acted as an unintentional stress test, proving that ML surrogates can bridge the gap between hardware parameters and software predictions if properly conditioned. To fully converge to the global maximum of the system's utility function, future co-design workflows should explicitly parameterize hardware geometry as a conditioning feature for neural networks. This ensures that the software pipeline evolves synchronously with hardware design iterations, effectively closing the loop between sensor engineering and experimental simulation.

\subsection{Automatic optimization of an Optical Parallel-Plate Avalanche Counter for neutron tomography}
\label{sec:neutron_tomography}

Neutron tomography is a powerful non-destructive imaging technique that exploits the high penetration power of neutrons to visualize the internal structure of dense materials such as metals and alloys~\cite{VONTOBEL2006475}. This capability, which makes it complementary to other tomography techniques such as X-rays that are attenuated easily by high-$Z$ materials, is particularly valuable in fields including additive manufacturing, metallurgy, and border security.

However, the large number of free parameters and their interdependent, non-trivial correlations make the global optimization of a full tomography setup a challenging task. As a first step toward this goal, this work presents a proof-of-concept study focused on a core component: the Optical Parallel-Plate Avalanche Counter (O-PPAC).

The O-PPAC is a gaseous detector designed for heavy-ion tracking and imaging. Its design consists of two parallel squared electrodes separated by a 3 mm gap filled with low-pressure scintillating gas, with arrays of collimated photo-sensors positioned along the edges. The position of the impinging particle is reconstructed from the photon distributions detected along the gas gap.

The characteristics of the photon distributions are determined by several design parameters such as the gas pressure and the collimator geometry. These variables have an impact on the scintillation light yield and the angular distribution of the photons reaching the sensors, directly influencing the precision of the position reconstruction. 

To solve this optimization problem, a framework was implemented~\cite{particles8010026}, based on a deep learning surrogate model trained to emulate both the detector response and the reconstruction process, enabling the minimization of an objective function through gradient-based optimization. The framework successfully identifies a stable and optimal parameter configuration, yielding results compatible with traditional optimization techniques and validating the effectiveness of the proposed methodology.

\paragraph{Detector concept and design parameters}

The detector design, originally introduced by Cortesi et al.~\cite{Cortesi_2018}, consists of two parallel metallized electrodes separated by a 3 mm gap filled with a scintillating gas, such as CF$_4$. When an ionizing particle traverses the active volume, it deposits energy by creating primary ionization electrons, which are subsequently multiplied in the gas through a Townsend avalanche driven by a uniform electric field established between the two plates.

\begin{figure}[ht]
    \centering
    \includegraphics[width=\linewidth]{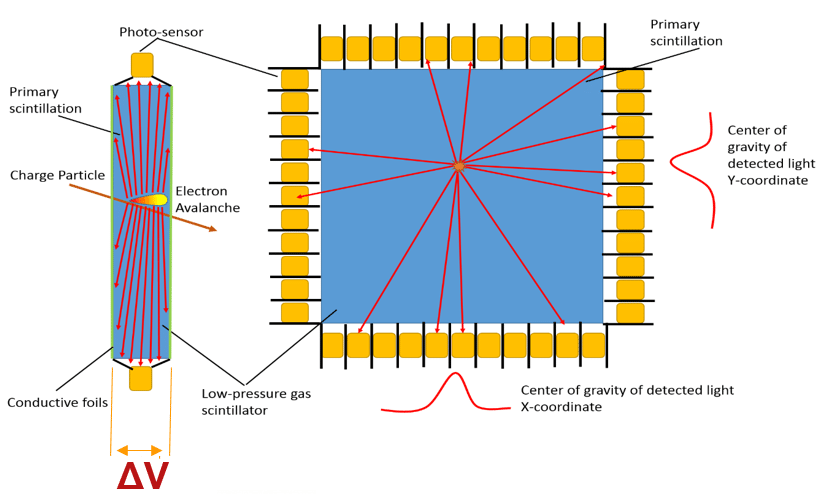}
    \caption{Schematic representation of the operational principle of the Optical Parallel-Plate Avalanche Counter (O-PPAC). Image reproduced from Ref.~\cite{Cortesi_2018}.}
    \label{fig:oppac_operational_principle}
\end{figure}

The scintillation light emitted during the avalanche process is reflected by the metallized electrode foils and guided toward arrays of collimated photo-sensors. The collimation is critical for the precise localization of the impinging particles, as it constrains the detected photon distributions so that their peaks are more heavily weighted near the avalanche position (see \autoref{fig:oppac_operational_principle}).

In this study, the optimization was focused on two operational parameters: the scintillating gas pressure ($p$) and the collimator length ($L$). While higher gas pressure increases the electroluminescence yield and improves signal statistics, the collimator length regulates the spatial spread of the light in each wall. A longer collimator improves localization but significantly reduces photon detection efficiency.

\paragraph{Position reconstruction}

As suggested in Ref.~\cite{Cortesi_2018}, the reconstruction of avalanche coordinates ($\hat{x}$, $\hat{y}$) is performed by computing the arithmetic mean between the light peaks recorded  by each pair of opposing arrays weighted by the total number of detected photons and the dispersion of the distribution:

\begin{equation}
\hat{x}= \left(\frac{\mathrm{P}_{\mathrm{x} 1} \cdot \mathrm{N}_{\mathrm{x} 1}}{\sigma_{\mathrm{x} 1}}+\frac{\mathrm{P}_{\mathrm{x} 2} \cdot \mathrm{N}_{\mathrm{x} 2}}{\sigma_{\mathrm{x} 2}}\right) /\left(\frac{\mathrm{N}_{\mathrm{x} 1}}{\sigma_{\mathrm{x} 1}}+\frac{\mathrm{N}_{\mathrm{x} 2}}{\sigma_{\mathrm{x} 2}}\right)\,,
\end{equation}

\begin{equation}
\hat{y}=\left(\frac{\mathrm{P}_{\mathrm{y} 1} \cdot \mathrm{N}_{\mathrm{y} 1}}{\sigma_{\mathrm{y} 1}}+\frac{\mathrm{P}_{\mathrm{y} 2} \cdot \mathrm{N}_{\mathrm{y} 2}}{\sigma_{\mathrm{y} 2}}\right) /\left(\frac{\mathrm{N}_{\mathrm{y} 1}}{\sigma_{\mathrm{y} 1}}+\frac{\mathrm{N}_{\mathrm{y} 2}}{\sigma_{\mathrm{y} 2}}\right)\,,
\end{equation}
where $P$ is the mean of the distribution on each wall, $N$ is the number of photons detected by the SiPM arrays, and $\sigma$ represents the standard deviation of each distribution. 

\begin{figure}[ht]
    \centering
    \includegraphics[width=0.98\linewidth]{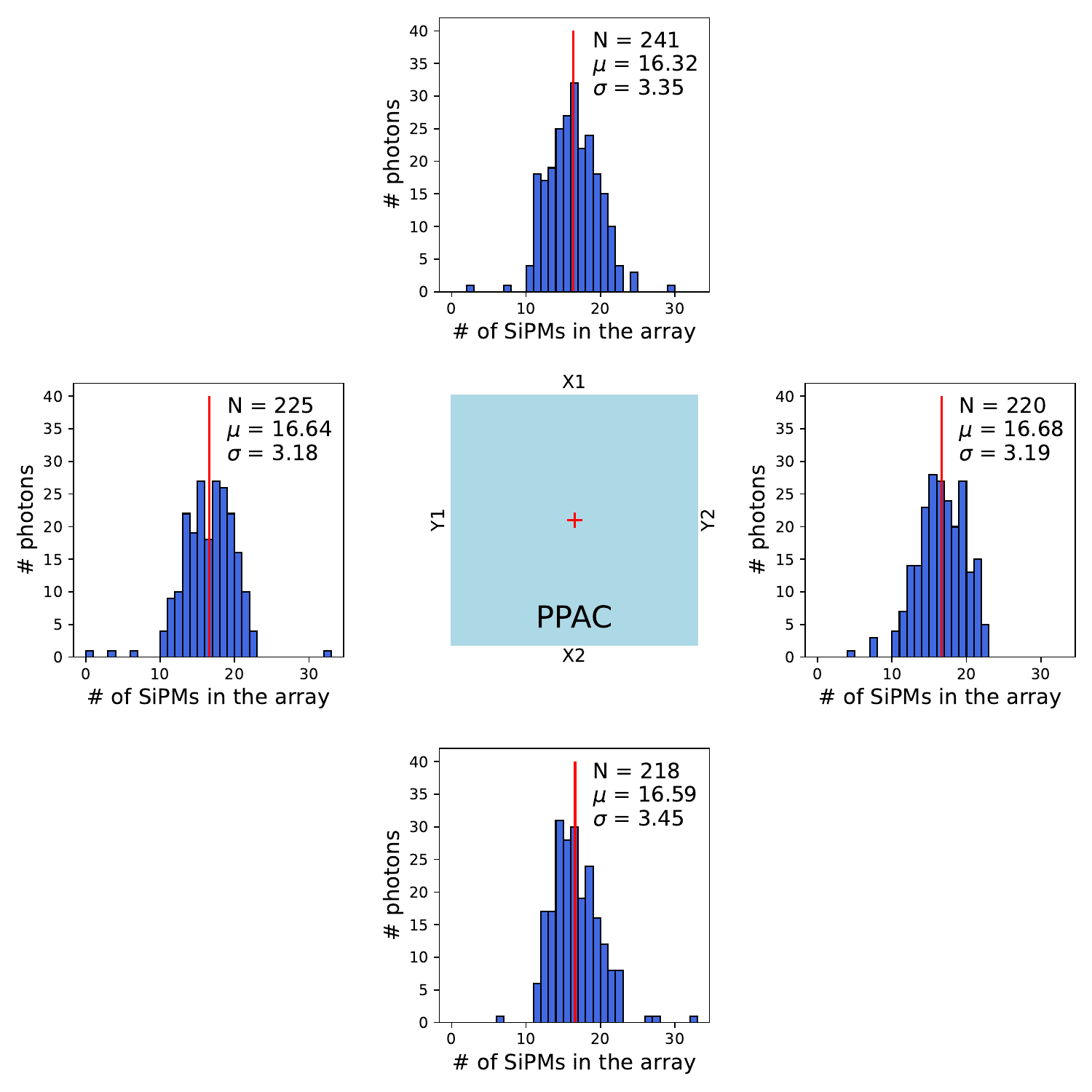}
    \caption{Illustration of the position reconstruction for a simulation event induced by a 5.5 MeV alpha particle. Image reproduced from Ref.~\cite{particles8010026}.}
    \label{fig:oppac_reconstruction_example}
\end{figure}

\paragraph{Surrogate Model}

The surrogate model was implemented as fully connected neural network trained to predict the reconstructed position ($\hat{x}, \hat{y}$) from 4 input variables: detector configuration ($p$, $L$) and particle initial position ($x$, $y$).

The training dataset was generated using \textsc{Geant4} simulations of a $10 \times 10$ cm$^2$ O-PPAC. The simulation setup utilized 5.5 MeV alpha particles as the impinging species, with events distributed over a central active area of $x, y \in [-4, 4]$ cm. To characterize the detector response, a grid of configurations was simulated for gas pressures $p \in [10, 50]$ Torr and collimator lengths $L \in [5, 50]$ mm.

To ensure best performance, the network architecture and hyperparameters such as learning rate, number of layers and hidden units were optimized using Optuna~\cite{optuna}. A detailed description of the neural network architecture, the dataset and the hyperparameter optimization strategy can be found in Ref.~\cite{particles8010026}.

\paragraph{Optimization}

The optimization stage aims to minimize the position reconstruction error by employing the surrogate model as a differentiable emulator of the O-PPAC response. With this aim, the objective function is defined as the Mean Squared Error (MSE) between the true particle coordinates $\mathbf{x}$ and the positions $\mathbf{\hat{x}}_{pred}$ predicted by the surrogate:

\begin{equation}
    \mathcal{L}(p, L) = MSE( \mathbf{x} - \hat{\mathbf{x}}_{pred}(p, L))\,.
\end{equation}

\begin{figure}[ht]
    \centering
    \includegraphics[width=0.98\linewidth]{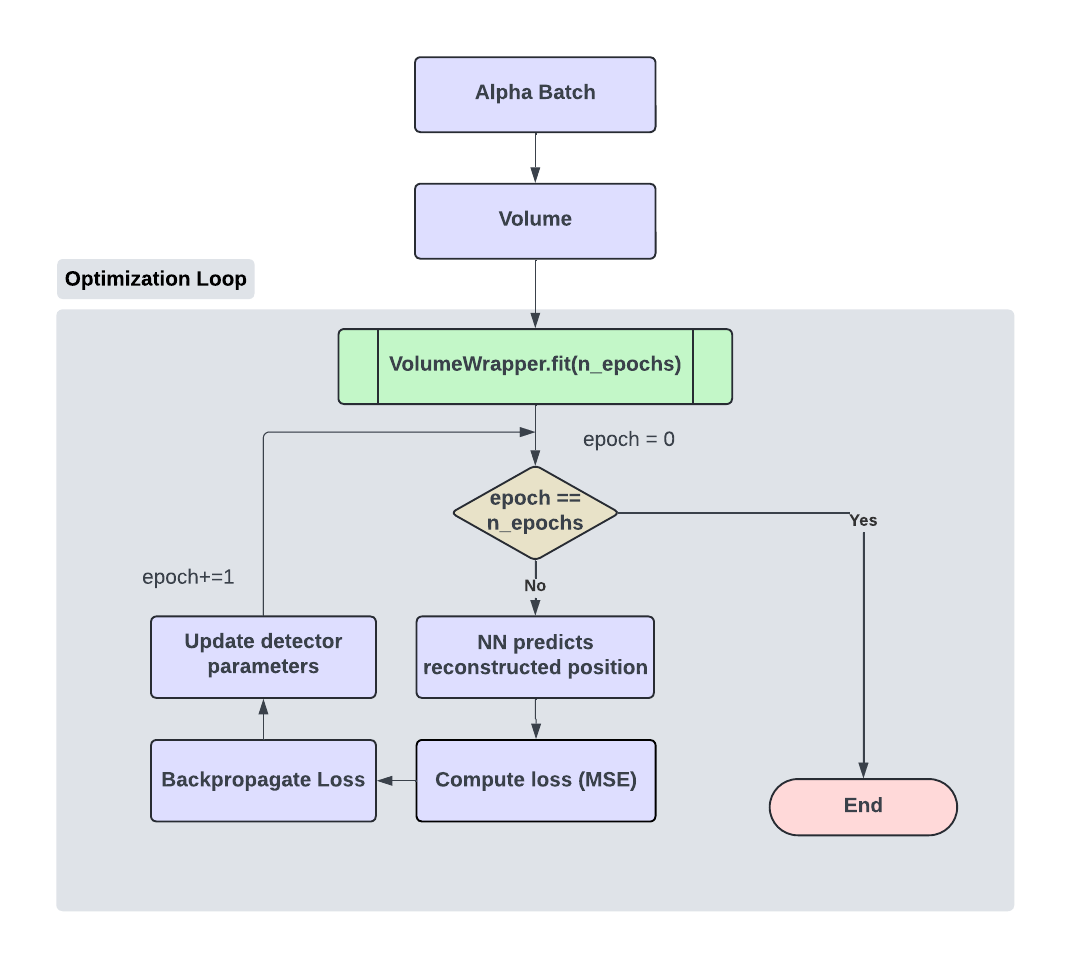}
    \caption{Breakdown of the detector optimization loop, outlining the process of initializing and updating the parameters (p and L) based on the gradient of the loss function. Image reproduced from Ref.~\cite{particles8010026}.}
    \label{figs_oppac:optimization_loop}
\end{figure}

As illustrated in \autoref{figs_oppac:optimization_loop}, the process begins with an initial configuration of design parameters and a batch of $10^5$ random particle positions. These values are processed by the surrogate model to compute the loss, after which we employ the Adam optimiser~\cite{adam} with a learning rate of $10^{-3}$. This cycle is repeated iteratively until the parameters and the loss stabilize.

\paragraph{Results and discussion}

\begin{figure}[htbp]
     \centering
     \begin{subfigure}[b]{0.32\textwidth}
         \centering
         \includegraphics[width=\textwidth]{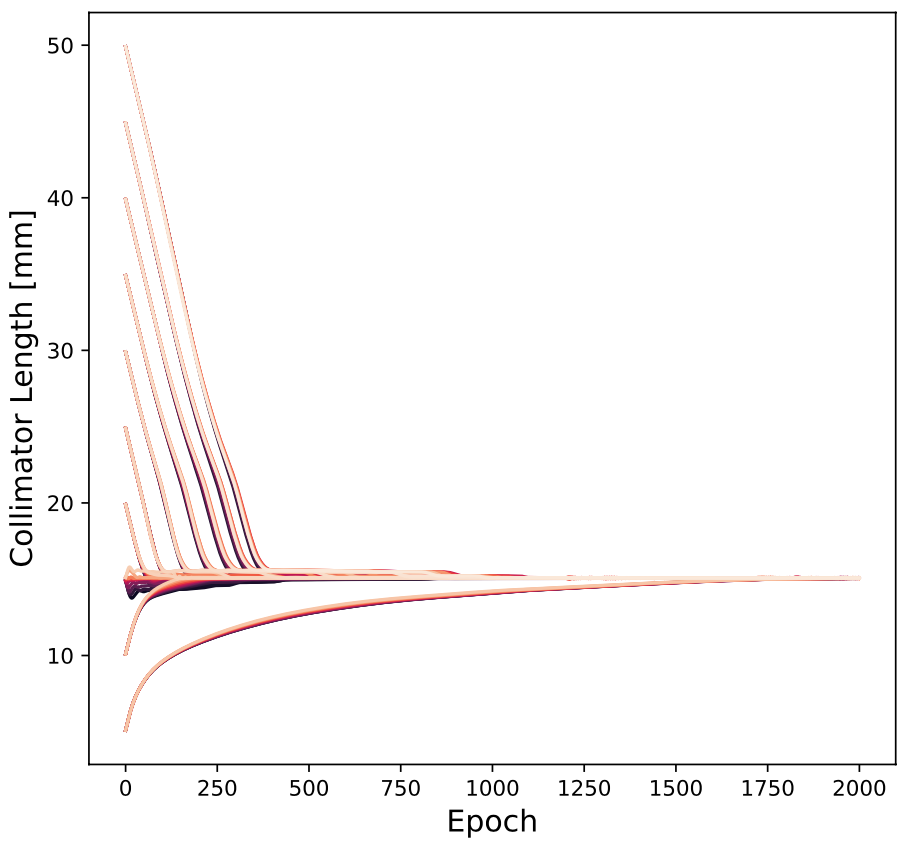}
         \caption{}
     \end{subfigure}
     \hfill
     \begin{subfigure}[b]{0.32\textwidth}
         \centering
         \includegraphics[width=\textwidth]{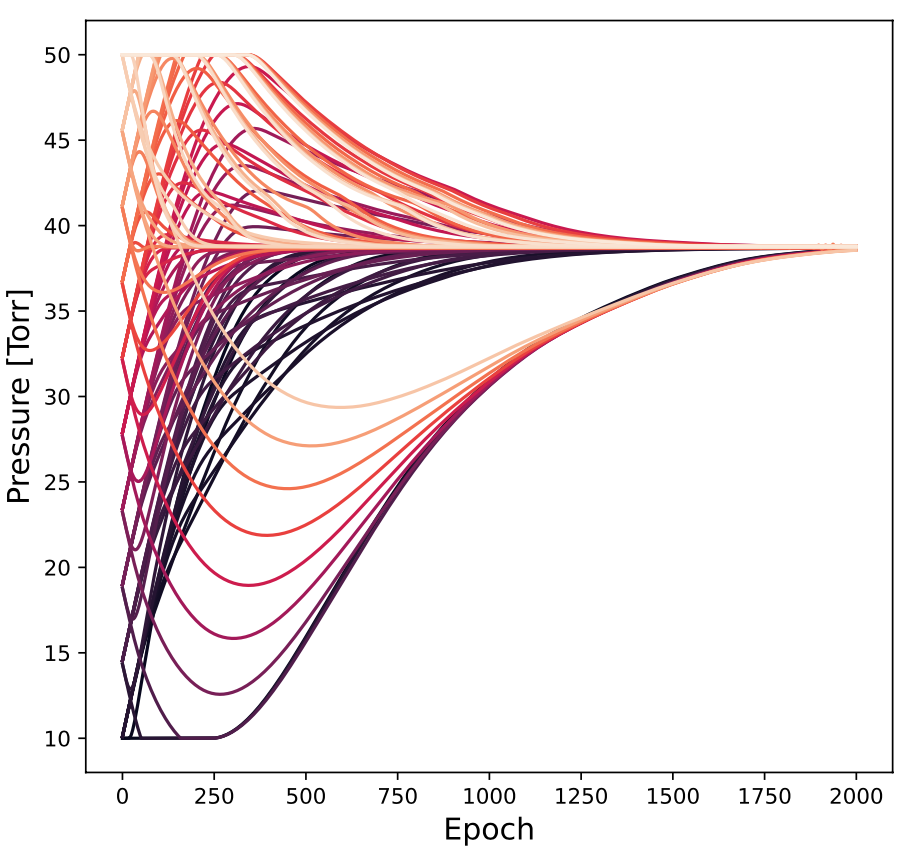}
         \caption{}
     \end{subfigure}
     \hfill
     \begin{subfigure}[b]{0.32\textwidth}
         \centering
         \includegraphics[width=\textwidth]{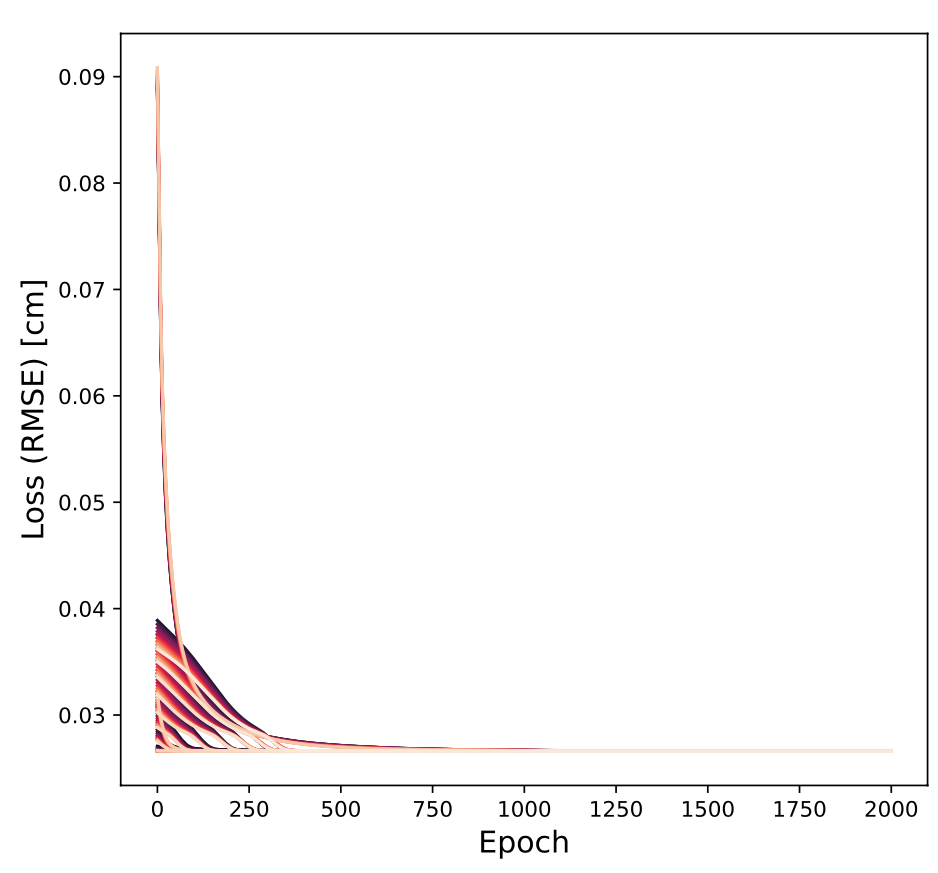}
         \caption{}
     \end{subfigure}
        \caption{Evolution of the (a) collimator length, (b) pressure, and (c) loss function
throughout the optimization loop for a grid of initial configurations of pressure and collimator length. Image reproduced from Ref.~\cite{particles8010026}}
        \label{figs_oppac:optimization_trajectories}
\end{figure}

The robustness of the proposed optimization approach was evaluated by initializing the optimization loop from 400 different starting points in the $(p, L)$ parameter space. Regardless of the initial condition, all trajectories successfully converged to a consistent optimal point at $p \approx 39.03$~Torr and $L \approx 15.11$~mm, as shown in \autoref{figs_oppac:optimization_trajectories}. This uniform convergence demonstrates that the identified solution corresponds to a global minimum.

Finally, the results obtained via differentiable programming were compared with traditional methods to validate the approach. The optimal collimator length identified by the framework is consistent with the traditional optimization studies~\cite{Cortesi_2018}, further supporting the robustness of the method.

Regarding the gas pressure, the optimal value was found to be relatively high within the specified range. While this can be partly attributed to the higher statistics observed at increased pressures, this result cannot be explained purely by photon yield, as that would suggest the optimal pressure should always be the maximum value allowed. We hypothesize that other factors, potentially related to the physical behavior of the detector or intrinsic limitations of the simulation, may be influencing this specific result. 

In conclusion, this proof-of-concept study shows that differentiable programming combined with surrogate models can effectively identify a stable global minimum in the case of the O-PPAC. 

Future studies will apply this methodology to optimize the full geometry of the neutron tomography system, with the O-PPAC as a central component. The goal is to jointly maximize detection efficiency, spatial resolution, and cost-effectiveness, finding the optimal balance between performance and economic viability for industrial applications.

\subsection{Muon scattering tomography scanners}
\label{sec:tomography}

Muon tomography (also known as muography) is a non-invasive imaging technique that uses naturally occurring cosmic-ray muons to probe the internal structure of large or otherwise inaccessible objects~\cite{tanaka2023muography,Bonechi2019muography}. 
Most of these muons, produced in the upper atmosphere, have energies on the order of \unit{\giga\electronvolt} or higher, enabling them to traverse hundreds of meters of rock or dense materials before stopping. By placing tracking detectors around a volume of interest (VoI) and measuring the scattering or flux attenuation that muons undergo as they pass through the volume, one can reconstruct a density map of the interior. When a scattering angle is recorded, its distribution further provides information on the atomic composition of the VoI. 
Applications of muon tomography range from volcano imaging and archeological surveys to monitoring industrial infrastructure and nuclear safeguards, offering a passive and cost-effective alternative to more conventional imaging methods~\cite{IAEA2022,Bonechi2019muography}.

Designing effective muon tomography systems requires striking non-trivial trade-offs between performance, cost, and physical constraints. To address this issue, the \textsc{TomOpt} project introduced a comprehensive framework for end-to-end optimization of muon scattering tomography detectors~\cite{TomOpt-Strong2023}. \textsc{TomOpt} models the entire measurement pipeline, from the generation of cosmic-ray muons and their scattering within the VoI to the detector response and final inference, as a fully differentiable chain. This approach allows for the application of gradient-based optimization techniques to simultaneously tune hardware parameters, such as the geometric configuration and spatial resolution of the tracking layers, alongside the reconstruction objectives. 
Recent studies have demonstrated \textsc{TomOpt}'s capacity to tailor detector designs for specific tasks, such as measuring the fill level in furnace ladles~\cite{TomOpt-Strong2023} and cargo scanning for border security~\cite{TomOpt-Zaher2025-MARESEC,TomOpt-Zaher2025-MODE,TomOpt-zaher2025-Muographers}, highlighting its ability to produce task-specialized detector designs rather than generic tracking systems.

In this Section we present three simplified case studies. 
First, we examine contrastive imaging using Bayesian optimization to jointly tune hardware geometry, inference hyperparameters, and exposure time.
Second, we study anomaly detection in cargo scanning, where a hypothesis-testing loss is optimized within the differentiable \textsc{TomOpt} framework to detect high-Z targets without full volumetric reconstruction. 
Finally, we explore a civil engineering application using Bayesian optimization to co-design a muography system for the imaging of a reinforced concrete pillar.

\subsubsection{Simultaneous optimisation of hardware, software, and exposure-time for contrastive imaging}
\label{subsec:muo1}
\paragraph{Overview}
The goal of this particular investigation is to explore, and demonstrate the potential of, simultaneous optimization of all aspects of muon-tomography-based imaging system: the detector-hardware, the inference software, and the way in which the system is used in application. In order to focus on the optimization potential, we take a simplified, general scenario that draws on the primary application of muon tomography systems: imaging inside a closed volume. The goal of the system is to produce an \textit{image} in which various sub-volume regions are clearly identifiable.

Two optimisation processes are considered:
\begin{itemize}
    \item Given a fixed cost, find the system that produces the best image -- a single-objective optimisation
    \item Given a minimal image quality, find the Pareto-optimal set of systems that minimize cost and imaging time -- a constrained multi-objective optimisation
\end{itemize}

\paragraph{Simulation}
All results in this Section are based on simulations of the physical processes. The \textsc{MUTE}~\cite{mute, mceq, corsika, sibyll_23c, HillasGaisser2012_H3a} package is used to sample the initial energy and polar angles of the muons; their subsequent scattering through the passive volume follows the ``PDG model"~\cite{pdg_model}.

\begin{figure}
    \begin{subfigure}[c]{\linewidth}
        \begin{center}
            \includegraphics[width=\linewidth]{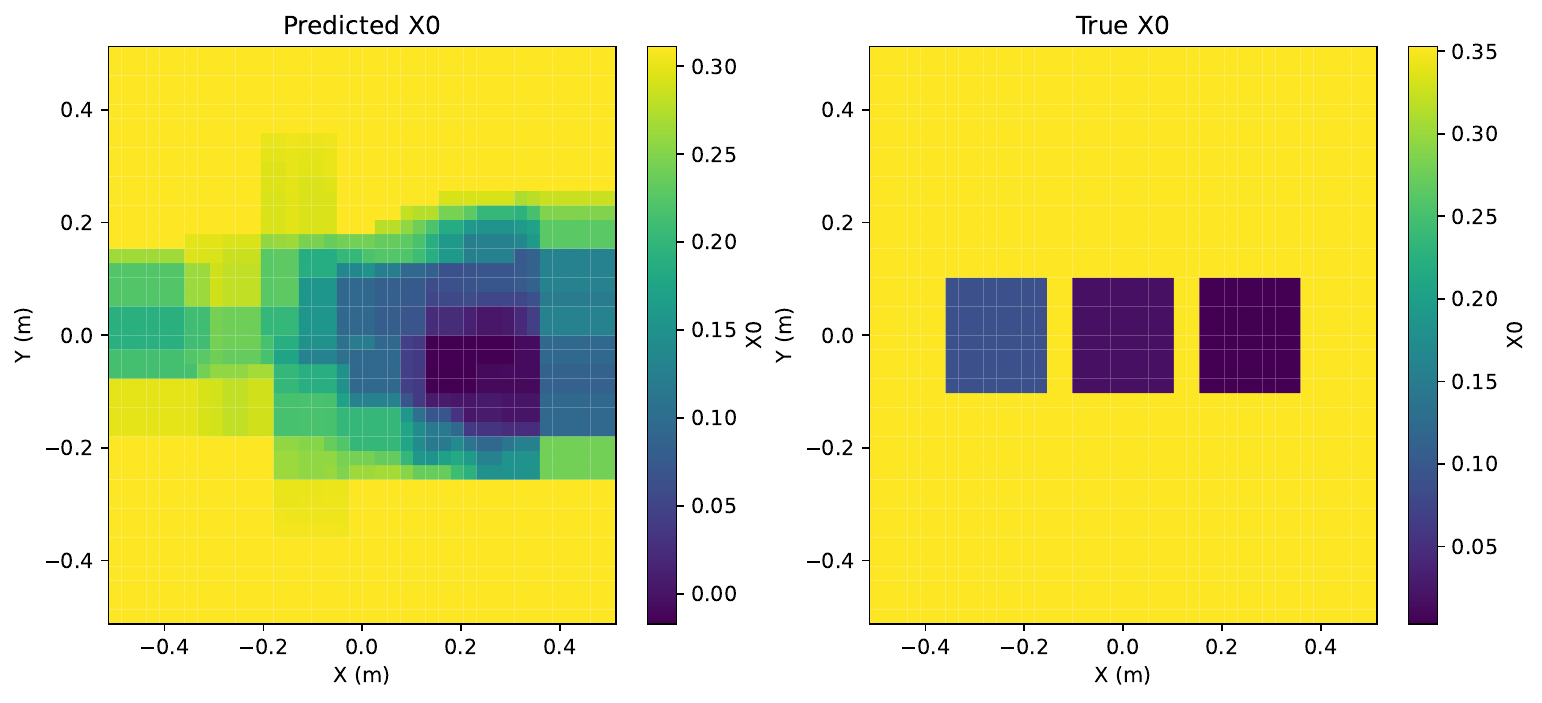}
                \caption{Single-objective optimisation (left), next to the true $x_0$ distribution of the passive volume (right)}
            \label{fig:kat:soo_x0_prediction}
        \end{center}
    \end{subfigure}
    \begin{subfigure}[c]{\linewidth}
        \begin{center}
            \includegraphics[width=\linewidth]{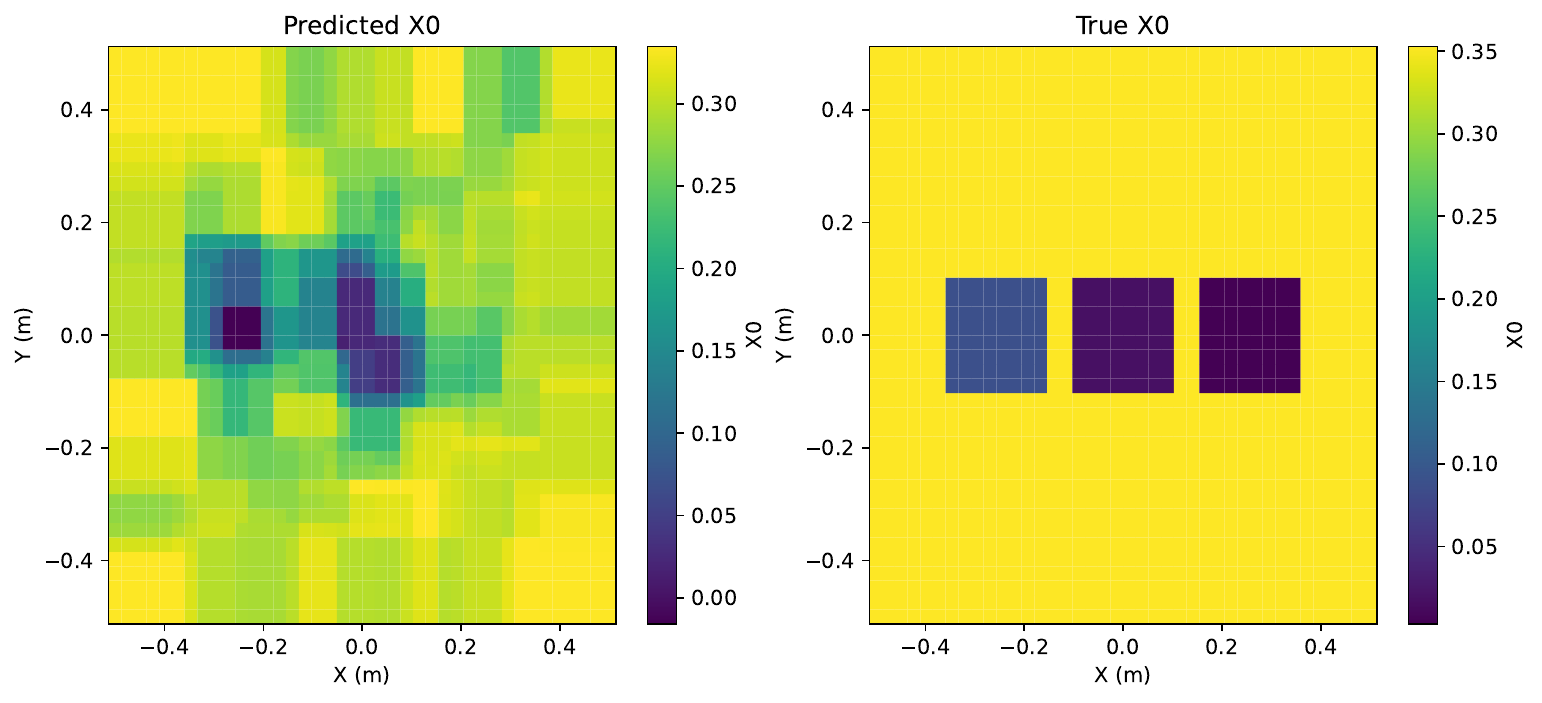}
                \caption{Medium exposure-time system resulting from the multi-objective optimisation (left), next to the true $x_0$ distribution of the passive volume (right). }
            \label{fig:kat:moo_x0_prediction}
        \end{center}
    \end{subfigure}
    \caption{Example images created by the optimised system for the two different optimisation runs performed. The images are constructed by averaging over $z$ in the region $-0.2<z\leq0.2$, in order to focus on the inner cubes. The left-hand-side images are produced by using the optimised systems to infer the X0 of the materials. The right-hand-side images are the true distributions of materials. The surrounding material is beryllium, and inside it are three inner cubes, which from left to right are made from aluminium, lead, and uranium.}
\end{figure}

\paragraph{The passive volume}
In order to factor out noise, that would otherwise interfere with the main goal of this investigation, a single passive volume is used. This consists of a  beryllium cube of $\SI{1}{\metre}$ side, centered at the origin. Inside the cube are three smaller cubes of $\SI{0.2}{\metre}$ sides, equally spaced along the $x$-axis at $y=z=0$. These smaller cubes are modelled as being composed of aluminium, iron, and uranium. The outer beryllium casing has a comparatively high $X_0$ compared to the inner metals, which allows for imaging of the inner structures by scattering angle measurement.  Figure~\ref{fig:kat:soo_x0_prediction} details a cross-section of the passive volume layout. This passive volume naturally provides high and low contrast parts. A decent scanning system should be able to distinguish between the inner cubes and the beryllium casing, but a very good system would be necessary to distinguish the inner cubes themselves.

\paragraph{The detector system}
The detector consists of several active panels, modelled as silicon pixel detector planes. These are kept centred at $x=y=0$, but otherwise free to translate along the $z$-axis. Whilst the total sensitive area of the panels is fixed, this area may be distributed to each panel individually, and the numbers of panels above and below the passive volume can also vary: between two and four, inclusive, each above and below.

Such a system already presents a non-trivial optimisation problem:
\begin{itemize}
    \item Increasing the distance between the detector panels allows for more precise muon-trajectory reconstruction, at the cost of a reduced number of muons whose trajectory passes through both panels;
    \item Increasing the number of detector panels further allows for a trade-off between trajectory reconstruction precision, and helping ensuring that muons whose trajectories miss later detectors, are detected at least twice; however this comes at the cost of reducing each panel area;
    \item Increasing size of the panels on the top increases the exposure to muons that pass through the detectors and the passive volume, at the cost of reduced detection of muons exiting the passive volume;
    \item Conversely, larger panels below the passive volume can help to detect muons that undergo large angular scatters (indicative of low $X_0$ material;
    \item There may also be interaction with the reconstruction software, in which the tuning of the hardware can act as a filter to deliberately miss detection of muons that due to deficiencies in the software would otherwise spoil the resulting image.
\end{itemize}

In total, there are 18 hardware parameters:
\begin{itemize}
    \item The $Z$ position and areas assignments of up to 8 panels (16);
    \item The number of panels above and below the passive volume (2).
\end{itemize}

Whilst all detectors are assumed to record muons that pass through them, each detector has a fixed measurement efficiency and spatial resolution. The former eventually feeds into a trajectory efficiency, which collapses to zero if only one muon trajectory point is measured. Spatial resolution is modeled through a random jitter to the reconstructed position of the muon hit, which makes trajectory fitting more difficult, unless multiple measurements (panels) are used to constrain the muon position.

\paragraph{The imaging system}
Producing an image follows a general pattern:  
\begin{enumerate}
    \item Detector hits are reconstructed to form incoming and outgoing trajectories;
    \item Trajectories are extrapolated inside the passive volume to identify the point of closest approach (PoCA);
    \item The PoCAs and other information of the muons are used to compute localised measurements of the passive volume density, the value of which is monotonic to the $X_0$ of the material;
    \item These measurements together constitute the resulting image; a calibration may be applied.
\end{enumerate}

Optimisation of the imaging system (software) includes both the choice of algorithm used to move from PoCAs to localised density measurements, and the hyper-parameters of the algorithms. We consider three different algorithms which are deliberately designed to have a shared set of hyper-parameters. Additionally, all algorithms localise the measurements within the passive volume by considering it to be discretised into a tetrahedral mesh.

\textbf{\textsc{PoCA-Single}} -- this algorithm only allows muons to contribute to the tetrahedron that contains its PoCA. After all PoCAs are localised, scattering angles, trajectory efficiencies, muon momenta, and their associated uncertainties\footnote{The system used does not allow for momentum measurements; in this study the true muon momentum is used for inference, with no uncertainty. Momentum measurement is aimed to be included in future developments, which will further expand the parameterisation space.} are used to compute $X_0$ predictions per tetrahedron via inversion of the PDG scattering model~\cite{pdg_model}. Whilst this algorithm is very quick to run, and memory-light, it results in sparse predictions, unless muon exposure is high.

\textbf{\textsc{PoCA-Spread}} -- this algorithm follows the same approach as the \textsc{PoCA-Single} algorithm, except that muons can now contribute to multiple tetrahedra. It does so by using the uncertainty on the PoCA position to form a 3D Gaussian PDF. The integral of this PDF across a given tetrahedron is used as a weight reflecting the probability that the muon scattered somewhere inside it. 

Next, per-muon scattering angles are computed for every tetrahedron's centroid, considering the detector-hit-centroids above and below the passive volume as start/end points.
As before, the properties of the muons, their scatterings, and their measurements are used to invert the PDG model per tetrahedron. However, now each muon's contribution to the aggregations is weighted by the probability of scattering in each tetrahedron.

This algorithm is slow to compute and very memory-heavy, however it allows for dense measurements of the volume, reducing the number of muons that may be necessary to form a high-quality image. Additionally, it produces a stronger software-hardware coupling: precise measurements of muons result in muons contributing significantly to fewer tetrahedra, which provides an interesting trade-off between precise $X_0$ measurements and smoothness of the surrounding measurements -- it may be beneficial to aim for a less constrained measurement in order to actually increase the quality of the final image.

\textbf{\textsc{PoCA-Density}} -- this algorithm attempts to strike a balance between the \textsc{PoCA-Single} and \textsc{PoCA-Spread} algorithms in terms of allowing a single muon to contribute to multiple tetrahedra, but without the large memory usage and computation time incurred by \textsc{PoCA-Spread}. Crucially, it does not attempt to directly estimate the $X_0$ of tetrahedra, but instead produces a measurement that is expected to be monotonic to the $X_0$. This measurement is based on the spatial density of PoCA points. Since muons are more likely to scatter in low-$X_0$ regions, the PoCA density in those regions should be higher.

Similar to \textsc{PoCA-Spread}, the uncertainty on the PoCA locations is used to construct an integral per tetrahedron. The muons' scattering angles, momenta, and measurement efficiencies are used to construct per-muon weights. The weighted average of PoCA integrals per tetrahedron is then computed by aggregating over the muons. This set of PoCA masses is then divided by the tetrahedra volumes to produce a PoCA density, the inverse of which should be monotonic to the tetrahedra $X_0$s. Similarly to \textsc{PoCA-Spread}, this algorithm can potentially exploit hardware-driven uncertainties to produce higher-quality images, and so may benefit from simultaneous optimisation.

The algorithms described above can function as they are, however not all muons may actually be beneficial to the image construction. To account for this possibility, software filters are added, controlled by optimisable hyper-parameters:
\begin{itemize}
    \item Minimum chord length -- this is the minimum distance that muons must travel inside the tetrahedron that contains their PoCA, in order to be included in the image. If a PoCA is very close to the surface of a tetrahedron, it most likely scattered in a different tetrahedron; and the uncertainty on the $X_0$ computation is likely to be high. This filter can be used to focus in on muons whose PoCA is well contained within a tetrahedron. This hyper-parameter used by all three algorithms, but only expected to be relevant when using \textsc{PoCA-Single} and \textsc{PoCA-Density}.
    \item Minimum and maximum momenta -- these control a simple window on the muon momentum, outside of which the muon is discarded. Low-momentum muons scatter more easily, and so may make it harder to distinguish differences in $X_0$. {\it Vice versa}, high-momentum muons do not scatter as much, again making it hard to distinguish $X_0$ differences. Tuning of these cuts allows the software to focus on the momentum range that is most beneficial. These parameters are used by all three algorithms.
    \item Minimum scattering angle -- this allows for muon selection based on the estimated amount of scattering they undergo. This parameter is expected to be closely tied with the momentum window cut, and the detector panel positions. It is used by all three algorithms.
    \item Minimum fractional tetrahedron integral -- this parameter is only used by the \textsc{PoCA-Spread} and \textsc{PoCA-Density} algorithms. It allows the software to constrain the effect that PoCAs have on far away tetrahedra by only allowing muons to contribute to a tetrahedron, if their PDF integral within the tetrahedron is above the minimum threshold. This is expected to be closely tied with the hardware parameters, which mainly influence the uncertainty on the PoCA positions.
\end{itemize}

In total, there are six software parameters:
\begin{itemize}
    \item The choice of inference method (1);
    \item The hyperparameters of the three algorithms (5).
\end{itemize}

\paragraph{Image quality quantification}\label{sec:kat:utility}
The goal of the system is to produce an image in which each sub-volume region is as clearly identifiable as possible from each other. {\it I.e.}, we do not care what values the measurements of each region are in terms of absolute scale, only that they are different between sub-volumes and similar within the same sub-volume. In order to quantify this we use a contrastive utility function, detailed below. 

While above it was said that the we do not care about the absolute values of the measurements, since the utility function will be based on absolute differences in the measurements, we want to ensure that different systems still produce measurements of the same scale. Therefore we first calibrate all measurements using a global linear rescaling that is fit via mean-squared error minimisation between the measurements and the true $X_0$ -- this may be done in practice for the final system by calibrating it on a set of known volumes. Next, we compute the differences in measurements for each combination of sub-volume pairs:
\begin{equation}
    \delta_{ij}=\frac{\left|\mu_i-\mu_j\right|}{\sqrt{\sigma^2_i+\sigma^2_j+\sigma^2_n}}\,,
\end{equation}
where $\mu_i$ and $\sigma^2_i$ are the mean and standard deviation of the measurements in sub-volume $i$. $\sigma^2_n$ is the base level of expected noise, which is estimated \textit{a priori}, using repeated measurements of the same system, and checking how much the measurements naturally vary. This distance increases if the measurements in different sub-volumes become more different, however it decreases if the measurements within the same-volume become very noisy; it quantifies how well we can distinguish sub-volumes $i$ and $j$.

To prevent the optimization from focusing on one specific pair at the expense of others, we compute a soft minimum of the set of distances:
\begin{equation}
    \mathcal{U}=-\frac{1}{\tau}\ln\left[\sum_{ij}e^{-\tau\delta_{ij}}\right]\,,
\end{equation}
where $\tau$ is tuned \textit{a priori} such that the second worst pair contributes at 5\% the value of the worst.

\paragraph{Single-objective optimisation}
The goal of this optimisation is to find the system that can produce the best quality image, given a fixed exposure time ($\SI{120}{\second}$) and a fixed total detector area ($\SI{100}{\metre\squared}$). Bayesian Optimisation is used to maximise the contrastive utility function described in \autoref{sec:kat:utility} by simultaneous optimisation of both the software and hardware parameters. 

The best resulting detector, detailed on the left in \autoref{fig:kat:systems}, achieves a metric score of just about 0.5. It uses two equally-sized and highly separated panels to measure incoming muons --a result unsurprisingly coincident with the optimal layout of the idealized strip tracker we found by analytic optimization in \autoref{sec:particle-tracking}. Outgoing muons are measured using four panels. Two are placed very close together, near to the passive volume. Their combined measurements allow the detector to measure the positions of outgoing muons very precisely, but not their angular trajectories. That task is accomplished by two further panels: the first is set at small spacing from the previous, presumably to catch muons with moderate angular scattering; the second is set at a very large vertical separation, presumably to offer very high angular precision on muons that only scatter slightly. The area of panels in the optimal configuration is approximately equal.

For the software side, the optimal system employs the \textsc{PoCA-Spread} inference algorithm, with a moderate cut on the fractional tetrahedron integral of 0.3, which allows muons to contribute to tetrahedra the are relatively far away, but not in tetrahedra where it is unlikely that the most significant scatter took place. The window cut on the muon momentum is $\SIrange{2.4}{892}{\giga\electronvolt}$. Given that the energy spectrum peaks at around $\SIrange{2}{3}{\giga\electronvolt}$, this is potentially cutting out a large number of low-momentum muons, which undergo large-angle scattering especially in high-$x_0$ materials. This suggests that the system realizes the need to filter out muons that carry high uncertainty on the PoCA and less information about the material they scattered in. The minimum chord length and minimum angular scattering were both set to their maximum upper values of $\num{1e-5}$, suggesting that they are useful in filtering out muons, but should have been allowed to vary to even higher values.

Figure~\ref{fig:kat:soo_x0_prediction} shows an example image created by the resulting system. In spite of the low muon statistics employed in this data-poor exercise, the uranium area is clearly visible, and the low-$x_0$ prediction region also extends back along the $x$-axis. The beryllium region is relatively smooth, although the inner cubes are blurred outside their true locations.

\paragraph{Multi-objective optimisation}
Having explored the maximum performance for a fixed budget detector (in terms of surface area and exposure time), and the typical ranges of our utility function, we can move to search for the set of detectors that achieve \textit{decent} results, whilst minimising cost and exposure time. To do so, we introduce the total area of the detector and the exposure time as parameters to the optimiser, and attempt to minimise them, whilst ensuring the utility function exceeds a minimum value. This constrained, multi-objective optimisation should result in a set of equally optimal systems: a Pareto-optimal front spanning exposure time and total area.
In this case, due to the potentially higher increased exposure time and detector area, memory limitations preclude the use of the \textsc{PoCA-Spread} algorithm. Therefore only \textsc{PoCA} and \textsc{PoCA-Density} algorithms are considered for optimisation.

\begin{figure}
    \centering
    \includegraphics[width=0.75\linewidth]%
    {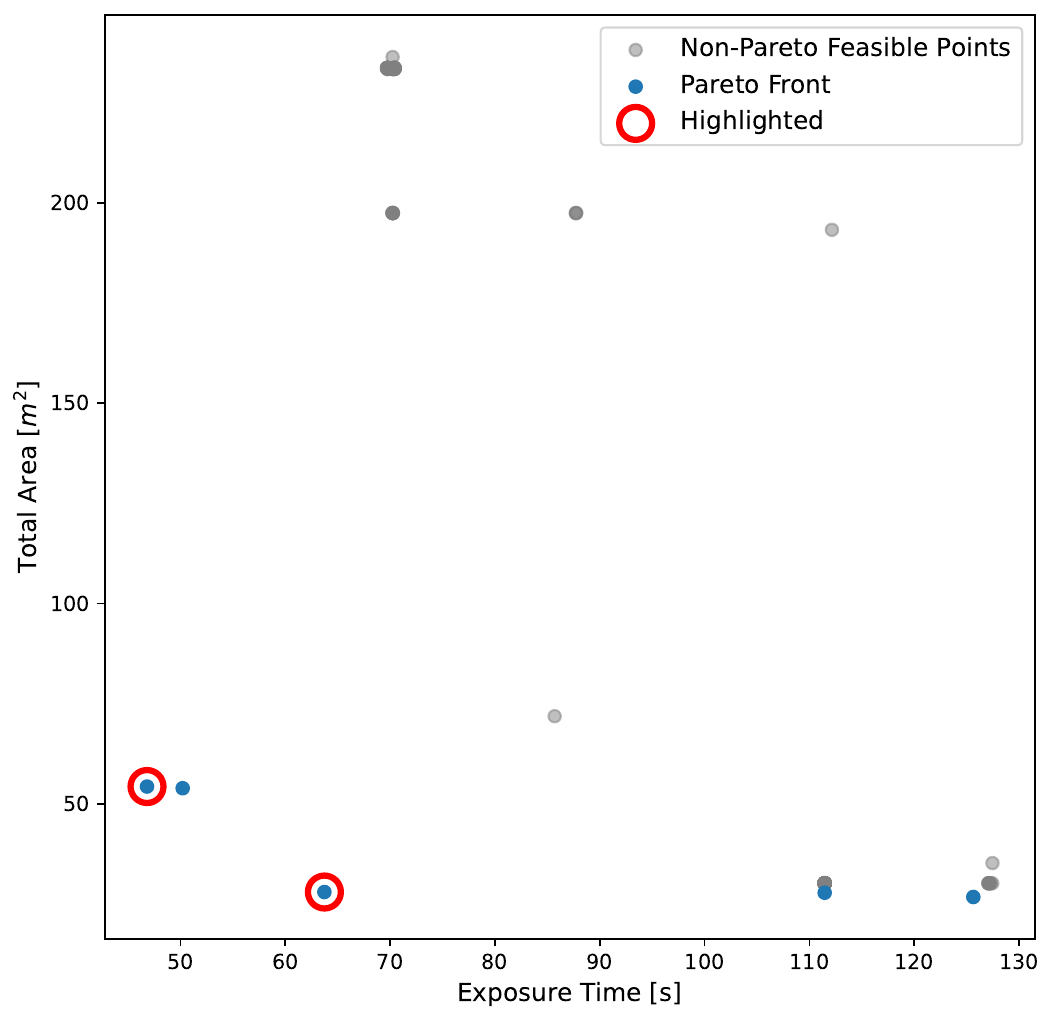}
    \caption{The Pareto-optimal front resulting from the multi-objective optimisation. 
    Total area is in units of $m^2$. The two circled systems are the ones detailed in \autoref{fig:kat:systems} (middle and right plots). Only points with image quality above 0.4 are shown.}
    \label{fig:kat:moo_Pareto_optimal_front}
\end{figure}

\begin{figure}[ht]
    \begin{center}
        \includegraphics[width=0.75\linewidth]{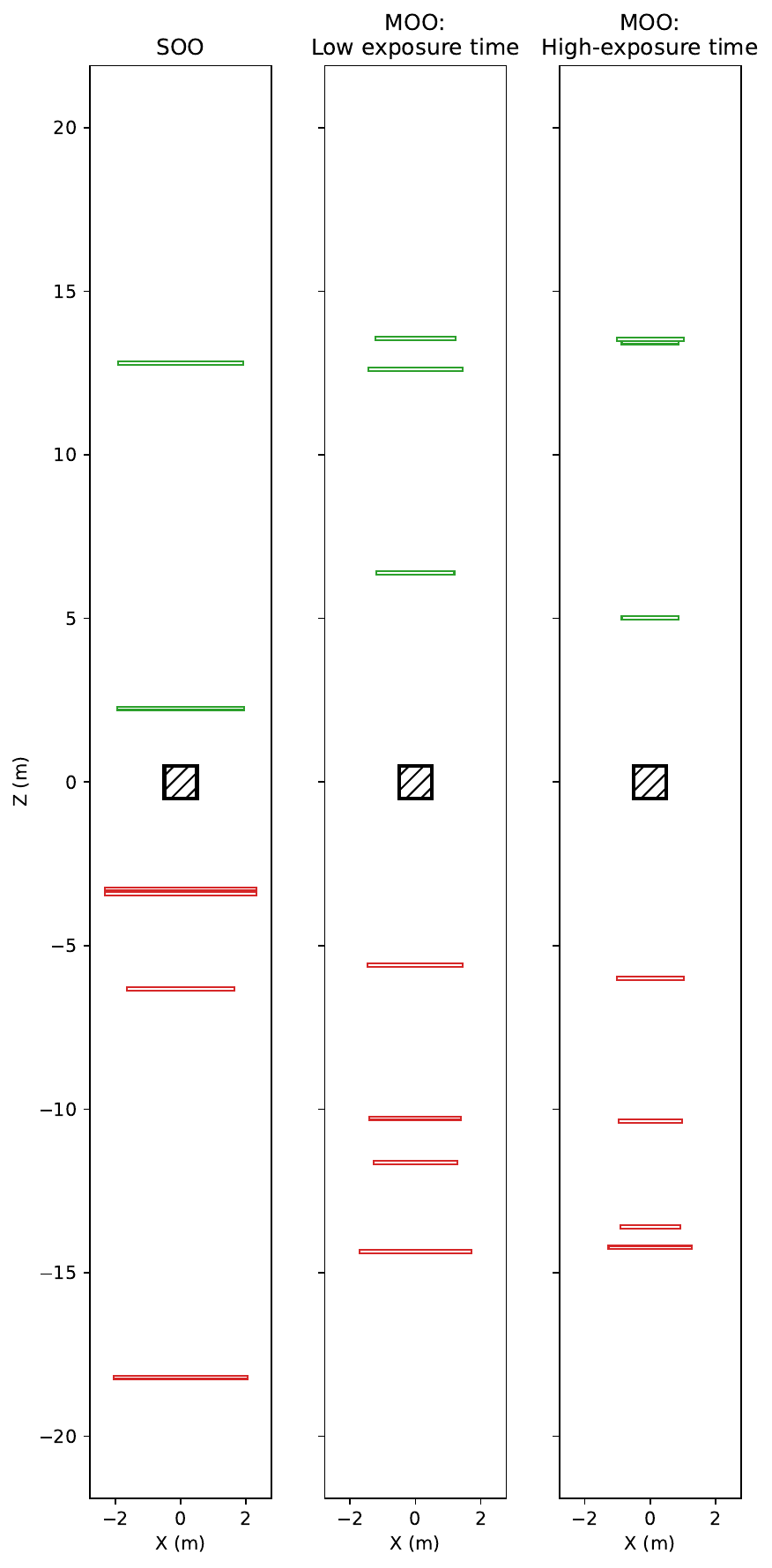}
        \caption{Side-on layouts of the systems resulting from the optimisation studies. The hashed box represents the passive volume, and the red/green boxes represent the detector panels. Left: single-objective optimisation: best resulting detector. Middle and right: selected detectors from the Pareto-optimal set of systems discovered during multi-objective optimisation. The middle panel shows the larger, low exposure-time system, and the right panel shows the smaller, higher exposure-time system.} 
        \label{fig:kat:systems}
    \end{center}
\end{figure}

A set of optimisers are used to explore the feasible space, and map out the rough location of the Pareto-optimal front: \textsc{NSGA2}~\cite{nsga2}, \textsc{NSGA3}~\cite{nsga3}, and \textsc{UNSGA}~\cite{unsga}. Their exploration is the used to initialise a Bayesian optimisation designed to push the Pareto-optimal front further. The resulting front is shown in \autoref{fig:kat:moo_Pareto_optimal_front}. 

In terms of software, all systems use \textsc{PoCA}, with a light cut on minimum muon momentum of around $\SI{0.5}{\giga\electronvolt}$. Within these software settings, the hardware can shift between a large-size, low-exposure-time detector, and a smaller, higher-exposure-time system. Since moving to a higher exposure-time only allows the detector to be reduced in size by very little, we focus on the two circled detector configurations in \autoref{fig:kat:moo_Pareto_optimal_front}. These are displayed in \autoref{fig:kat:systems}.

In terms of layout, the two optimal solutions are quite similar: three panels above, with a focus on measuring muon positions at the top half of the incoming trajectory, and four panels below, with a focus on measuring the bottom half of the outgoing trajectory. The only real difference is the size of the panels. Both detectors also show an intention to miss high-scattering muons, and instead focus on muons undergoing small scatterings, or passing straight through the passive volume. This becomes evident when we examine the images produced by the system (\autoref{fig:kat:moo_x0_prediction}): the lighter metals are well isolated, however the uranium section is difficult to distinguish -- those high-scattering muons that would be the most sensitive to lower $x_0$ are missed by the detector. This indicates that either the minimum constraint on the image quality is too lenient, or that the utility function needs to also encourage also some ordering to the mean prediction within volume regions.

\paragraph{Summary}
In this investigation we demonstrated that simultaneous optimisation of hardware, software, and application (exposure time) parameters was possible, and resulted in detection systems and software that make intuitive sense. This simultaneous optimisation is particularly useful, since it allows for constrained multi-objective optimisation of the system. This would have been very challenging if hardware, software, and application were to be optimised separately: if the outer constrained optimiser were to act on optimising hardware, then each sample's measurement of the constraint would imply a full constrained minimisation of the exposure time, in which each sample would imply a full unconstrained maximisation of the image quality via the optimisation of the software parameters. Whilst possible, this three-tiered optimisation approach requires significantly more runtime than simultaneous optimisation.

\subsubsection{Hypothesis-testing loss for anomaly detection}
\label{sec:muography_anomaly}
\paragraph{Overview}
While the previous application focuses on imaging the substructure of a volume, a common case study in scattering-based muography involves determining the presence of a high-Z target within a specified VoI, which is indicative of anomalous or potentially hazardous materials. This approach is particularly relevant in cargo inspection applications, and is pursued as an alternative to performing a full three-dimensional reconstruction of the underlying material density map. Such 3D reconstructions typically rely on the PoCA inference, which estimates the scattering location of the muon inside the VoI as a single point. Downstream voxelization of combined PoCA-derived spatial and angular deflection variables allows the inference of local radiation lengths. Although this PoCA-based reconstruction is simple and widely used, the resulting $X_0$ estimates become unstable in low-statistics, highly anisotropic geometries, and low-Z mediums, motivating task-driven alternatives that bypass full volumetric reconstruction. To avoid these issues, we instead pose the problem as a binary hypothesis test:

\begin{itemize}
   \item \(\mathbf{H_0:}\) 
    the VoI contains only background material;
   
   \item \(\mathbf{H_1:}\)
   the VoI contains an unknown high-Z target.
   
\end{itemize}

The detector design is then optimized to maximize the separation between these two hypotheses. We implement this as a Jensen–Shannon divergence ($\mathrm{D_{JS}}$)–based test, where TomOpt simulates two batches of events - one under \(\mathrm{H_0}\) and another under  \(\mathrm{H_1}\) - and computes the $\mathrm{D_{JS}}$ between the resulting PoCA scattering distributions. In this setup, $\mathrm{D}_{JS}$ between the background distribution and its mean reference defines the critical value that corresponds to a false-positive rate of 5\%. $\mathrm{D_{JS}}$ between the signal distribution and the same mean background is then used to quantify the statistical power of the test. Differentiability is maintained throughout every stage of the test, ensuring that gradients propagate cleanly and enabling gradient-based optimization of all relevant parameters.

\paragraph{Differentiable Hypothesis Test Loss}

Let $\mathcal{V}_\mathrm{sig}$ and $\mathcal{V}_\mathrm{bkg}$ denote the sets of signal and background volumes, respectively. Volumes $\mathcal{V}_\mathrm{bkg}$ are defined as an aluminium block, and  $\mathcal{V}_\mathrm{sig}$ volumes include in addition a small uranium block situated on the floor of the block.
Each volume generates a distribution of PoCA scattering events. 
We construct a kernel density estimate (KDE) of the scattering distribution for each volume, with a learnable software smoothing parameter $\sigma_\mathrm{sw}$ controlling the KDE bandwidth. Small values of $\sigma_{sw}$ produce sharp histograms that are highly sensitive to statistical fluctuations and noise, while large values yield smoother distributions that can blur fine details and reduce sensitivity to the presence of a target. 

The average background scattering distribution is denoted as $\bar{S}_\mathrm{bkg}$. The similarity of each volume to the background is quantified using $\mathrm{D_{JS}}$, which can be expressed in terms of Kullback-Leibler divergence ($\mathrm{D_{KL}}$):
\begin{equation}
\begin{split}
\mathrm{D_{JS}}\big(S_V, \bar{S}_\mathrm{bkg}\big) 
&= \frac{1}{2} \mathrm{D_{KL}}\big(S_V \,||\, M\big) 
  + \frac{1}{2} \mathrm{D_{KL}}\big(\bar{S}_\mathrm{bkg} \,||\, M\big)\,,\\
M &= \frac{1}{2}\big(S_V + \bar{S}_\mathrm{bkg}\big)\,,
\end{split}
\end{equation}
where $S_V$ is the KDE of the scattering distribution corresponding to volume $V$.

A differentiable threshold $T_c$ is computed from the background $\mathrm{D_{JS}}$ distribution at significance level $\alpha$, and the detection of a signal volume is defined with a sigmoid threshold:
\begin{equation}
\hat{y}_i = \frac{1}{1 + \exp\Big[-\,\tau_\mathrm{detect} \big(\mathrm{D_{JS}}(S_{V_i}^\mathrm{sig}, \bar{S}_\mathrm{bkg}) - T_c\big)\Big]}\,,
\end{equation}
where $\tau_\mathrm{detect}$ controls the sharpness of the transition.
The overall detection power,  denoted $\Pi$, is
\begin{equation}
\Pi = \frac{1}{N_\mathrm{sig}} \sum_{i=1}^{N_\mathrm{sig}} \hat{y}_i\,.
\end{equation}

The differentiable loss function optimized in our framework is
\begin{equation}
\mathcal{L}(\sigma_\mathrm{hw}, \sigma_\mathrm{sw}) = 1 - \Pi +\lambda\, \sigma_\mathrm{hw}\,,
\end{equation}
where the hardware angular resolution $\sigma_\mathrm{hw}$ is included to account for the finite detector resolution in the optimization. This angular resolution is computed via error propagation from the panel longitudinal positions $z_i$ (hardware free parameters) and their spatial resolutions $\sigma_x$ using a weighted variance: 
\begin{equation}
\sigma_\mathrm{hw} = \sqrt{ \frac{1}{\sum_i w_i (z_i - \bar z)^2} }\,, 
\quad w_i = \frac{1}{\sigma_x^2 + \epsilon}\,, 
\quad \bar z = \frac{\sum_i w_i z_i}{\sum_i w_i}\,,
\end{equation}
where $\epsilon>0$ is a small constant introduced to ensure numerical stability.

\paragraph {Optimization loop} At each optimization epoch, three batches, each consisting of 50 background and 50 signal passive volumes, are scanned. PoCA reconstruction is then performed to collect scattering samples, and the Jensen-Shannon-based utility loss is evaluated. $\sigma_{sw}$ is optimized in a block-iterative manner: at each epoch, an inner loop first adjusts $\sigma_{sw}$ to best fit the current scattering distributions, and then the hardware parameters are updated in the outer loop using gradients that reflect the optimal software setting. The hardware parameters consist of 6 longitudinal positions $z$, corresponding to a pair of detector modules, each of three panels, placed above and below the passive volume. Additionally, the area of each panel is optimized by rescaling with a learned fractional budget weight per panel, keeping the total detector budget fixed.

The same utility loss is used in both stages: it is first minimized with respect to $\sigma_{sw}$ in the inner loop, and subsequently backpropagated through the full differentiable pipeline to update the hardware parameters in the outer loop. This bi-level procedure accounts for the stochasticity of the reconstructed scattering samples, ensuring that hardware updates are guided, at each step, by a software configuration that is locally optimal for the current statistical realization of the data.

\paragraph{Loss landscape} Prior to optimization, we inspect the loss landscape by performing a two-dimensional grid scan of the utility loss over representative hardware and software parameters. On the hardware side, we vary the effective detector thickness, defined as the total longitudinal span of each detector module. This parameter directly controls the spacing between the constituent panels and therefore sets the angular lever arm and the hardware angular resolution $\sigma_\mathrm{hw}$. On the software side, we scan the smoothing parameter $\sigma_{sw}$ used to construct the differentiable PoCA scattering-angle distributions. The spatial resolution of each detector panel is held fixed at 1~mm, isolating the effect of detector thickness on the angular resolution. The panel areas are also fixed for a detector budget of 10 a.u., corresponding to around 1.29 $m^2$ for each panel with a cost of 1 $m^{-2}$. For each volume scan, a fixed sample of $3000$ muons is generated, which is the resampled number of muons reaching the VoI. 

\begin{figure}
    \centering
    \includegraphics[width=\linewidth]{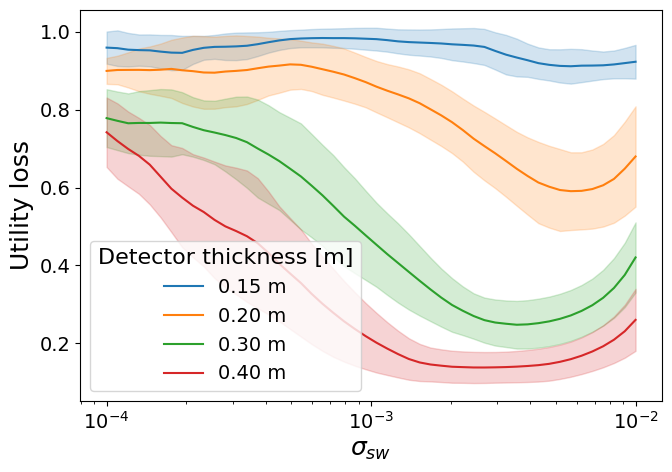}
    \caption{Differentiable hypothesis-test utility loss for high-Z anomaly detection as a function of software smoothing parameter $\sigma_{sw}$. Curves correspond to different detector thicknesses (longitudinal separation between first and last panels in each detector module). Intra-separations between panels within a detector are the same in this scan. Smaller $\sigma_{sw}$ preserves fine scattering features, enhancing separation between background and high-Z signal volumes, while larger $\sigma_{sw}$ blurs statistical fluctuations. The plot illustrates the interplay between hardware angular resolution (controlled by detector thickness) and software smoothing in shaping the detection utility. }
    \label{fig:loss-landscape}
\end{figure}

The resulting utility curves shown in \autoref{fig:loss-landscape} exhibit the expected qualitative behavior. Increasing the detector thickness initially leads to a monotonic improvement in utility, driven by the enhanced angular lever arm and corresponding improvement in hardware angular resolution. The dependence on the software smoothing parameter follows a complementary pattern. For configurations with smaller hardware angular resolution, lower values of $\sigma_{sw}$ are favored, as minimal smoothing preserves fine scattering features that enhance hypothesis separation. However, beyond a certain limit, statistical limitations and noise set a practical lower bound on $\sigma_\mathrm{sw}$.
Treating these degrees of freedom independently would lead to suboptimal or unstable operating points. By optimizing hardware and software parameters jointly within a unified, differentiable framework, the detector naturally converges to balanced configurations that maximize task-level performance, which is the central objective of this work.

\begin{figure}[ht]
    \centering
    \includegraphics[width=0.5\linewidth]{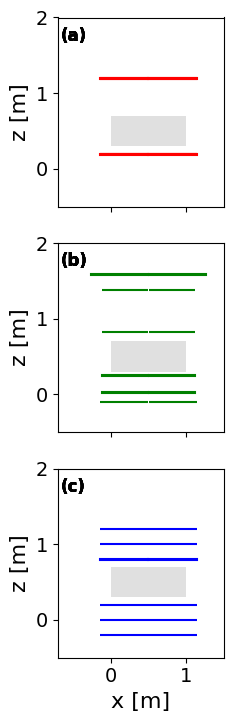}
    \caption{Comparison of detector geometries in the x–z plane.
(a) Initial detector configuration prior to optimization.
(b) Detector geometry obtained through gradient-descent optimization under the imposed budget constraint.
(c) Reference detector configuration with the same total detector budget as (b), designed based on human intuition.}
    \label{fig:detectors}
\end{figure}

Based on the observed utility landscape, we conceptualize an optimal reference detector (\autoref{fig:detectors} (c)) with a total thickness of 40~cm (corresponding to $\sigma_\mathrm{hw} = 0.005$) and a software smoothing parameter $\sigma_\mathrm{sw} = 0.002$. This detector serves as a reference to assess the performance achieved by the optimized detector.

\paragraph{Results} To demonstrate an optimization procedure, we initialize the detector with a fixed total budget of 10 a.u. and with suboptimal panel positions, tightly stacked along the $z$ axis as shown in \autoref{fig:detectors} (a), and a suboptimal software smoothing parameter of 0.001. All panels are assigned the same spatial resolution of 1~mm. After gradient-based optimization, the detector converges to a configuration with improved panel spacing (\autoref{fig:detectors} (b)) that produces an angular resolution of $\sigma_\mathrm{hw} = 0.0044$, and a slightly increased area of the upper panel improving acceptance. The achieved software smoothing parameter is $\sigma_\mathrm{sw} = 0.0032$.

To assess the performance of the optimized detector, we adopt a reference-based evaluation protocol inspired by practical cargo inspection workflows. For each set of VoI volumes, one background volume is designated as a reference, representing a first-line X-ray scan that identifies regions likely free of high-density anomalies. All other volumes are then compared against this reference using the $\mathrm{D_{JS}}$ of their scattering distributions. Signal and background labels are used only to compute the ROC curve and the corresponding AUC. By repeating this procedure across multiple reference volumes and averaging, we obtain a robust measure of the detector’s discriminative power in a realistic scenario where a scanned region is evaluated relative to a reference slice (see \autoref{tab:auc}). 

\begin{table}[h!]
\centering
\footnotesize
\caption{Reference-based ROC AUC for the reference and optimized detectors across different signal/background pairs. Values are reported as mean $\pm$ standard deviation reflecting the variability of AUC across repeated realizations of the scattering samples.}
\label{tab:auc}
\begin{tabular}{l l c c}
Signal & Background & Reference AUC ($\pm\sigma$) & Optimized AUC ($\pm\sigma$) \\
\hline
Uranium & Aluminium & 0.954 $\pm$ 0.026 & 0.954 $\pm$ 0.016 \\
Lead    & Aluminium & 0.790 $\pm$ 0.015 & 0.808 $\pm$ 0.017 \\
Iron    & Aluminium & 0.542 $\pm$ 0.034 & 0.518 $\pm$ 0.030 \\
Uranium & Lead      & 0.521 $\pm$ 0.016 & 0.501 $\pm$ 0.012 \\
Uranium & Iron      & 0.580 $\pm$ 0.041 & 0.579 $\pm$ 0.027 \\
\hline

\end{tabular}
\end{table}

\paragraph{Summary}
We reformulate scattering-based muography as a differentiable binary hypothesis test, replacing unstable voxel-level $X_0$ reconstruction with a task-driven objective that directly measures separability between background-only and high-Z-containing volumes. The results show that gradient-based co-optimization of hardware and software parameters converges to performance comparable to that of an optimal reference detector based on a partial grid scan.

\subsubsection{\textsc{Geant4}-based hypothesis-testing co-design for the PoCA reconstruction of a concrete pillar}
\label{subsec:muo3}

\paragraph{Overview} This application aims to optimise a scattering muography system for reconstructing the density distribution of a reinforced concrete pillar. We “focus” the system on the pillar through selection of significant muons for a PoCA reconstruction~\cite{Priedhorsky2003, Chatzidakis2016} of the volume. Using a single \textsc{Geant4} simulation, we generate multiple system configurations with different co-design parameters and a fixed spatial resolution. The utility function evaluates both the total number of PoCA points, rewarding higher acceptance, and the quality of the reconstructed image. Detector positions and sizes are optimized considering the full angular muon flux. Hardware choices influence reconstruction (SW) mainly through angular resolution, which is affected by the separation of the detector layers. 

\paragraph{Simulation} We simulate the reinforced pillar scenario using \textsc{Geant4} and generate muons using EcoMug~\cite{Pagano2021, Pagano2023}, fully covering the pillar and the detectors with a semi-spherical generation surface.

Note that the detection muon flux is bidirectional. In the simulation we include six detection panels in total, but only four (two on either side of the pillar) are considered in the co-designed system. The purpose of this simulation is to capture all the meaningful muon ﬂux with large detection layers positioned very close to each other and to the pillar, saving the actual trajectories ("perfect spatial resolution") of the detected muons. The simulation software can be found in Ref.~\cite{Lagrange2025CoDesign_Concrete_Geant4}.

\paragraph{Generation of system configurations} After saving the \textsc{Geant4}-simulated muon trajectories, we use a new and customized layer of code to set a spatial resolution for the detection panels and to generate different panel separations and sizes, as well as muon event filters. It is worth mentioning that when generating detection panel separations, we assume straight propagation in the air between. This new layer of code~\cite{Orio2025PostSimCoDesignTools} exploits reconstruction tools from "Muograph"~\cite{Lagrange2025Muograph}, among others, and is also used for the optimisation part of this study.

\paragraph{The passive volume} The passive volume corresponds approximately to the space between the detection panels closest to the pillar, which are fixed at $x = \SI{\pm 110}{\milli\meter}$ in this study. Its dimensions in the coordinate reference system are $\SI{0.2} \times \SI{0.6} \times \SI{0.6}{\meter^3}$ and it is subdivided into voxels, which are the maximum size that allows the necessary calculations for optimization to be performed, {\it i.e.} $\SI{0.2} \times \SI{0.2} \times \SI{0.2}{\meter^3}$. It is made up of the reinforced pillar and the air on either side of it. The dimensions of the pillar are $\SI{0.2} \times \SI{0.2} \times \SI{0.6}{\meter^3}$. The metal bars of the reinforcement are located inside it. Within the passive volume, in turn, we define a target volume ($A$) and a background volume ($B$). The target volume corresponds exactly to that of the pillar, while the background volume is composed of the six air voxels located on either side of it.

\paragraph{The active volume} The active volume, or, in other words, the volume that contains the detection system, could be unlimited since, as previously mentioned, to generate new system configurations we simply propagate muon tracks in a straight line. In any case, during optimization, only up to a certain limit is inspected, since the system starts to undergone a significant acceptance drop when separating the panels. Four squared detection panels, two at each sides of the pillar, are used to reconstruct the muon trajectories.

\paragraph{Imaging system and utility quantification} We reconstruct the pillar as a PoCA point density image, which should highlight high density areas. To quantitatively assess whether the scattering activity inside the target region differs from that of the surrounding background, we model the reconstructed PoCA events using a Poisson process with region-dependent exposure. 

A and B denote the two disjoint regions of interest (target and background volumes), and we define:

\begin{itemize}
  \item $N_A, N_B$: number of PoCA points reconstructed in regions $A$ and $B$, respectively.
  \item $E_A, E_B$: effective exposures of regions $A$ and $B$.
\end{itemize}

While $E$ can be estimated using the geometric surface or volume of a region, for a through-going muon beam the relevant exposure is more accurately represented by the total muon path length within the region, as the probability of scattering scales with the traversed material length. In all cases, $E$ represents the effective measure over which scattering events may occur. Both strategies have been tested, but in this document we present results in which the effective exposure of each region is estimated from the total muon path length traversing the region. 

For each reconstructed muon track, the path length inside a given region is computed assuming a Straight Line Path (SLP), defined as the straight segment connecting the measured points of the track in the detectors. The exposure is then defined as the sum of these path lengths over all reconstructed muons. Accordingly, the exposure for regions $A$ and $B$ is given by

\begin{equation}
E_A = \sum_{i=1}^{N_\mu} L_{i,A}\,, \qquad
E_B = \sum_{i=1}^{N_\mu} L_{i,B}\,,
\end{equation}
where $L_{i,A}$ and $L_{i,B}$ denote the SLP path length of the $i$-th muon track within regions $A$ and $B$, respectively, and $N_\mu$ is the total number of reconstructed muons.

The PoCA counts in each region are assumed to follow independent Poisson distributions:
\begin{equation}
N_A \sim \mathrm{Poisson}(\lambda_A E_A)\,, \qquad
N_B \sim \mathrm{Poisson}(\lambda_B E_B)\,,
\end{equation}
where $\lambda_A$ and $\lambda_B$ are the scattering rates per unit exposure in regions $A$ and $B$.

The null hypothesis corresponds to equal scattering density in both regions:
\begin{equation}
H_0:\ \lambda_A = \lambda_B = \lambda\,,
\end{equation}
while the alternative hypothesis allows for different scattering rates:
\begin{equation}
H_1:\ \lambda_A \neq \lambda_B\,.
\end{equation}
Under the null hypothesis $H_0$, a single common rate is fitted:
\begin{equation}
\hat{\lambda}_0 = \frac{N_A + N_B}{E_A + E_B}\,.
\end{equation}
Under the alternative hypothesis $H_1$, the maximum-likelihood estimators of the rates are\begin{equation}
\hat{\lambda}_A = \frac{N_A}{E_A}\,, \qquad
\hat{\lambda}_B = \frac{N_B}{E_B}\,.
\end{equation}
The corresponding log-likelihoods are given by
\begin{align}
\ell_1 &=
N_A \log\left(\hat{\lambda}_A E_A\right)
- \hat{\lambda}_A E_A
+
N_B \log\left(\hat{\lambda}_B E_B\right)
- \hat{\lambda}_B E_B\,,
\\
\ell_0 &=
N_A \log\left(\hat{\lambda}_0 E_A\right)
- \hat{\lambda}_0 E_A
+
N_B \log\left(\hat{\lambda}_0 E_B\right)
- \hat{\lambda}_0 E_B\,,
\end{align}
where $\ell_1$ and $\ell_0$ denote the maximised log-likelihoods under the alternative and null hypotheses, respectively.

The estimated scattering densities per unit exposure are defined as
\begin{equation}
\hat{\rho}_A = \frac{N_A}{E_A}\,, \qquad
\hat{\rho}_B = \frac{N_B}{E_B}\,, \qquad
\hat{\rho}_0 = \frac{N_A + N_B}{E_A + E_B}\,,
\end{equation}
where $\hat{\rho}_0$ corresponds to the common density under the null hypothesis of equal scattering rates.

Therefore, the Poisson Likelihood-Ratio Test (PLRT) statistic, our utility, is defined as

\begin{equation}
U =
2 \left[
N_A \log\left(\frac{\hat{\rho}_A}{\hat{\rho}_0}\right)
+
N_B \log\left(\frac{\hat{\rho}_B}{\hat{\rho}_0}\right)
\right]\,.
\end{equation}

By construction, $U \ge 0$, with $U = 0$ when the estimated scattering rates in the two regions are equal. $U$ can be interpreted in terms of statistical significance, or converted into an equivalent Gaussian significance via $Z = \sqrt{U}$, where $Z$ is the standard score, or z-score.

The exposure correction ensures that the differences in PoCA counts arise solely from the density distribution of the matter inside the passive volume. When exposure is defined as the total muon path length through each region, the test naturally incorporates detector geometry and directional effects. The resulting statistic provides both a formal hypothesis test and a scalar measure of contrast suitable for optimization procedures.

\paragraph{Configuration and parameters of the optimization problem} The configuration and parameters are common for the sequential HW-SW scan and the co-design of the full system, both presented below. The study is performed with data from a single simulation scenario, corresponding to a muon flux acquisition time of $\approx \SI{82}{\hour}$. It is also worth mentioning that we emulate the spatial resolution of the panels adding a tridimensional Gaussian spread to the actual muon hits provided by \textsc{Geant4}, which in this case is $\sigma_x = \SI{1}{\milli \meter}$, $\sigma_y = \SI{1}{\milli \meter}$, $\sigma_z = \SI{1}{\milli \meter}$.

The optimized parameters are detection panel separation ($x_1$) and area difference ($x_2$) on the HW side, and a deviation threshold to accept muon events ($x_3$) on the reconstruction SW side. The first parameter, $x_1$, refers to the separation between panels closer to the pillar (interior panels) and those further away (exterior panels). The second parameter, denoted as $x_2$, represents the relative area difference between the exterior panels and the interior panels for a single panel pair (the panels always keep a square shape). To calculate $x_2$, the difference in surface area between one exterior panel and one interior panel, normalized by the reference panel area used in the \textsc{Geant4} simulations, $A_{\mathrm{ref}}$, which corresponds to $\SI{0.36}{\meter^2}$. Mathematically, \( x_2 \) is expressed as:
\begin{equation}
x_2 = \frac{A_{\mathrm{ext}} - A_{\mathrm{int}}}{A_{\mathrm{ref}}} \times 100 \,,
\end{equation}
where \( A_{\mathrm{ext}} \) and \( A_{\mathrm{int}} \) denote the surface areas of the exterior and interior panels of a single pair, respectively. In all the configurations, the total detection area employed for the four panels, $A_{\mathrm{tot}}$, is kept constant, such that:
\begin{equation}
A_{\mathrm{tot}} = 2 \left(A_{\mathrm{ext}} + A_{\mathrm{int}}\right) = 4 A_{\mathrm{ref}}\,.
\end{equation}
ensuring that any increase in the exterior panel area is exactly compensated by a corresponding decrease in the interior panel area, and {\it vice versa}. Note that the system is symmetric; {\it i.e.}, the panel separation ($x_1$) and area difference ($x_2$) of the detectors on both sides of the pillar are the same and, therefore, are both modelled with a unique parameter. Finally, the third parameter, $x_3$, is the quantile value of the muon deviation distribution used to establish the high-pass threshold. 

\begin{figure}
    \centering
    \includegraphics[width=\linewidth]{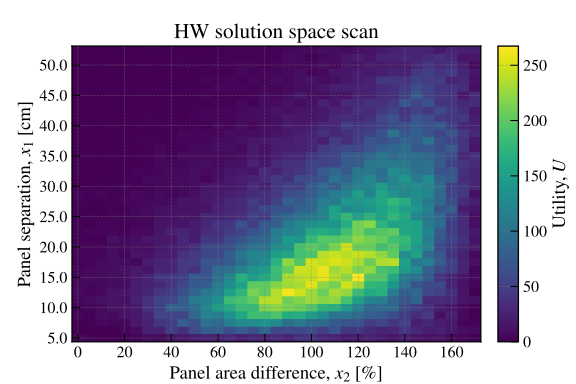}
    \caption{First step of the sequential scan of the solution space used to optimize the muography system. The figure shows the PLRT-based utility, $U$, for different values of the HW parameters ($x_1$ and $x_2$), which are related to the separation and area of the detection panels.}
    \label{fig:Mu_Concrete:SequentialOptimisation_HW}
\end{figure}

\paragraph{Sequential HW-SW solution space scan} To demonstrate the co-design paradigm in this application, we compare the optimal utility obtained by means of a sequential HW-SW solution scan with that yielded by a simultaneous co-design optimisation. To start with the demonstration, we scan the solution space of the HW parameters (see \autoref{fig:Mu_Concrete:SequentialOptimisation_HW}).

\begin{figure}
    \centering
    \includegraphics[width=\linewidth]{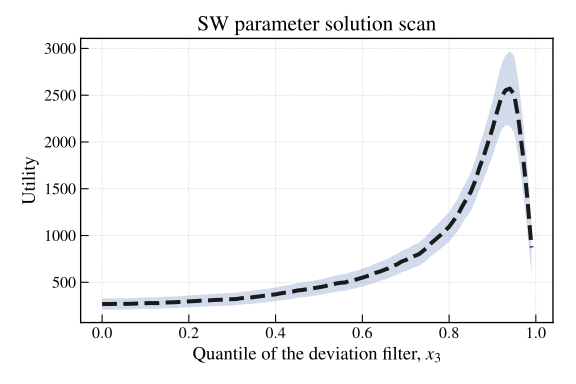}
    \caption{Second step of the sequential scan of the solution space used to optimize the muography system. The figure shows the PLRT-based utility, $U$, for different values of the SW parameter ($x_3$), which consists in a deviation threshold for accepting the muon events.}
    \label{fig:Mu_Concrete:SequentialOptimisation_SW}
\end{figure}

We set the HW parameters to the optimal values, and then run an exploration of the solutions for the SW parameter, obtaining the results shown in \autoref{fig:Mu_Concrete:SequentialOptimisation_SW}. They show a significant improvement in utility when increasing the values of the SW parameter.

The optimal parameters yielded by the sequential optimisation are $x_1 = \SI{15.0}{\centi \metre}$, $x_2 = \SI{120.0}{\%}$, and $x_3 = 0.940$, which correspond to an utility of $2571 \pm 414$.

\paragraph{Co-design} To confirm that certain crosstalk between HW and SW parameters exist, we encapsulate the full HW-SW system into a Bayesian optimization algorithm, which performs the co-design. It internally runs our system configuration generation functions and maximizes the utility function. Optimization is carried out using a Gaussian Process surrogate model via the \texttt{gp\_minimize} routine. The parameter search space is defined by the set of optimization dimensions, and a total of 100 function evaluations are performed (far fewer than we have used to perform sequential scanning). The optimisation is initialized with 20 randomly sampled points to provide an initial exploration of the parameter space, which corresponds with that inspected in the sequential scan. Subsequently, new evaluation points are selected by maximising the Expected Improvement acquisition function, which balances exploration and exploitation. 

\begin{figure}
    \centering
    \includegraphics[width=\linewidth]{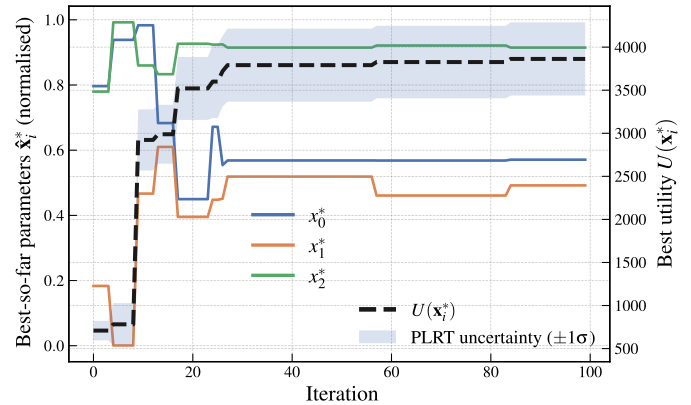}
    \caption{Simultaneous HW-SW Bayesian Optimisation. Evolution of the normalised best-so-far parameters, $\mathbf{x}^\ast_i$, and the corresponding PLRT utility, $U(\mathbf{x}^\ast_i)$, as a function of the optimisation iteration.}
    \label{fig:Mu_Concrete:CoDesign_BestSoFar}
\end{figure}

Several complementary diagnostics are carried out, which provide insight into the optimization trajectory, parameter stability, exploration properties, convergence behaviour, and structure of the objective function landscape. Specifically, \autoref{fig:Mu_Concrete:CoDesign_BestSoFar} shows the evolution of the \emph{best-so-far} parameter values together with the corresponding utility during the optimisation process. At each iteration $i$, the best-so-far parameter vector $\mathbf{x}^\ast_i$ is defined as
\[
\mathbf{x}^\ast_i = \arg\min_{j \leq i} U(\mathbf{x}_j),
\]
where $U(\mathbf{x})$ denotes the objective function evaluated by the optimiser. The associated best-so-far utility is given by $U(\mathbf{x}^\ast_i)$. In \autoref{fig:Mu_Concrete:CoDesign_BestSoFar} each parameter is min--max normalised over the full optimisation history to allow direct comparison on a common scale. The almost constant value of parameters from approximately iteration $30$ onwards indicate that the algorithm has converged towards an optimal solution, which in this case is $x_1 = \SI{12.1}{\centi \metre}$, $x_2 = \SI{83.6}{\%}$, and $x_3 = 0.957$, and provides a utility of $3863 \pm 425$. The performance of co-design optimisation is compared to that of the sequential scan by evaluating the separation between their utility values in units of the combined statistical uncertainty, which is of approximately $2.2 \sigma$. This indicates a meaningful improvement in the utility values (sub-optimality ratio $SR$ of the sequential scan relative to the co-design of $1.50$). Concerning the evolution of parameters, the co-designed system places the panels closer together to maximize the detected muon flux, reduces the size difference between the inner and outer panels to capture greater vertical length of the pillar, and slightly increases the event admission deviation threshold.

\begin{figure}[h!]
     \centering
     
     \begin{subfigure}[b]{0.48\textwidth}
         \centering
            \includegraphics[width=\linewidth]{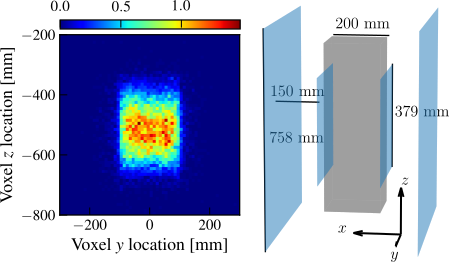}  
         \label{}
     \end{subfigure}
     
     \begin{subfigure}[b]{0.48\textwidth}
         \centering
            \includegraphics[width=\linewidth]{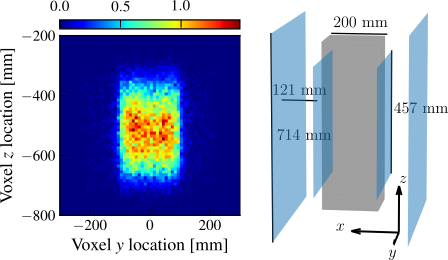}
     \end{subfigure}
    \caption{PoCA reconstructions (left) and detector panel dimensions and locations (right) for the two optimisation strategies: sequential scan [$U = 2571 \pm 414$] (top) and co-design [$U = 3863 \pm 425$] (bottom). The color scale image depicts the joint two-dimensional probability distributions of the POCA scattering centres. It is worth remembering that the panels are square in shape.}
    \label{fig:Mu_Concrete:CoDesign_POCA}
\end{figure}

\paragraph{Summary and next steps} In this study, we present the disadvantages of designing a muography system by sequentially optimising HW and SW. In this simple application example, we show how the HW parameters, which are related to the separation of the detection panels and their size, interact with the SW parameter, {\it i.e.}, muon filtering by deflection. Angular resolution, which depends on spatial resolution and panel separation, plays a key role in this regard. The effects of system acceptance and target geometry and density are also worth mentioning.

To avoid the aforementioned disadvantages of optimizing the system by means of sequential scans of the HW and SW solution spaces, we co-designed the full system using a Bayesian optimisation algorithm. Our results show that we improved the utility of the system. In addition, this improvement agrees with the enhancement of PoCA reconstruction, which indicates a good fit to the problem and a satisfactory design of the methods and the system's utility employed. 

An additional observation is that in all cases the system manages to differentiate the part of the passive volume occupied by the pillar (volume $A$) from that occupied by air (volume $B$), with z-scores of $Z = 51$ in the sequential scanning and $Z = 62$ in the co-design. 

The next steps we envisage in this \textsc{Geant4}-based co-design investigation include the development of more advanced models of the muography system to be optimized, adding new parameters --including those describing the performance of the electronic system (synchronisation, detection thresholds, etc.). Regarding the SW part, performing a complete $X_0$-distribution inference is an important objective, which could also improve the quality of the reconstruction and incorporate new co-design parameters. We also contemplate applying this co-design technique in the attenuation modality of muography. Finally, and in relation to this specific application in reinforced concrete pillars, it should be remembered that the ultimate goal is to co-design the system with the specific purpose of detecting the reinforcement bars. To this end, we plan to significantly increase the production of \textsc{Geant4} simulation data in order to perform meaningful co-design optimisations considering as targets the smaller volumes of the bars. In addition, the availability of additional data could also allow to quantify the Monte Carlo statistical uncertainties linked to \textsc{Geant4} simulations. 

\subsubsection{Conclusion}
In this subsection we demonstrated the capabilities of principled co-design of muon scattering tomography systems, spanning detector hardware, inference software, and operational parameters. Through contrastive imaging case studies, we illustrated how simultaneous optimization can yield detector layouts and reconstruction strategies that are intuitive yet non-obvious, effectively balancing angular resolution, acceptance, statistical efficiency, and exposure time under realistic constraints. Reframing scattering-based muography as a hypothesis-testing problem rather than a full volumetric reconstruction task, we then demonstrated end-to-end gradient-based optimization of detector performance with respect to statistical detection power of anomalous high-Z material. A Bayesian approach to co-design muography systems for the purpose of scanning reinforced concrete was also demonstrated, leveraging Monte Carlo simulations.

\subsection {Other applications}

To complete this Section, below we provide a short discussion of three additional applications for LHC physics where co-design is relevant.

\subsubsection{FPGA-Based Co-Design for Real-Time Triggers and Data Reduction}
\label{sec:FPGA}

Next-generation particle physics and astrophysics experiments are facing unprecedented challenges driven by rapidly increasing data rates, finer detector granularity, and stringent real-time decision requirements. Facilities such as the High-Luminosity LHC, next-generation neutrino experiments, radio astronomy arrays, and gravitational-wave observatories will operate in regimes where only a small fraction of the produced data can be retained for offline analysis. Meeting these demands requires computing architectures capable of fast, deterministic processing close to the data source. Field-Programmable Gate Arrays have therefore become a key enabling technology for future experiments. Their ability to deliver ultra-low-latency, high-throughput, and energy-efficient processing makes them uniquely suited for real-time triggering, data reduction, and near-detector inference. Unlike CPU- or GPU-based systems, FPGAs provide fine-grained control over parallelism, dataflow, and numerical precision, allowing algorithms to be tailored precisely to hardware constraints and timing budgets.

Realizing the full potential of FPGAs increasingly relies on a co-design paradigm in which algorithms, ML models, firmware, and hardware architectures are developed together. Hardware–software and algorithm–hardware co-design allows physics-driven optimization of performance while minimizing data movement and latency. This integrated approach ensures that detector capabilities, computing resources, and scientific objectives evolve coherently, thereby maximizing the physics reach of future experiments.

The importance of this approach is clearly illustrated by the trigger systems of LHC experiments. The LHC produces proton–proton collisions at a rate of 40 MHz, far exceeding the capacity for permanent data storage. Experiments such as ATLAS and CMS reduce this rate using a two-stage trigger system. The first stage, the Level-1 trigger, is implemented in custom FPGA-based hardware and uses information from calorimeters and muon detectors to select events at a rate of approximately 100 kHz within a fixed latency of about 4 microseconds. Events passing this stage are processed by the High-Level Trigger, a software-based system that performs more detailed reconstruction and reduces the event rate to roughly 1 kHz for storage.

Because the Level-1 trigger rejects the vast majority of events, its design has a direct impact on physics sensitivity. While the current trigger architecture has enabled landmark results such as the discovery of the Higgs boson, it is inherently limited by predefined selection criteria. The High-Luminosity LHC will increase collision rates by more than an order of magnitude and significantly raise event complexity, pushing existing trigger strategies to their limits and motivating more flexible and powerful real-time algorithms.

In response, collider experiments including ATLAS, CMS, LHCb, and Belle II are actively exploring FPGA-based ML for tasks such as event filtering, graph neural network based tracking, jet tagging, anomaly detection, radiation-tolerant low-latency compression, and real-time reconstruction. Similar efforts are underway in other fields, with the DUNE neutrino experiment investigating online ML for data acquisition and gravitational-wave observatories developing FPGA-based real-time inference and low-latency alert systems. 

Another example is a neural-network-based trigger system for ultra-high-energy neutrino detectors, currently being developed for arrays of autonomous radio detector stations deployed within the ice sheets at the South Pole (IceCube-Gen2~\cite{Gen2-TDR}) and in Greenland (RNO-G~\cite{RNO-G:2020rmc}). The main challenges are the high sampling rates of  approx. a Gsample/s per antenna, a limited power budget of only around 30~W per detector station, and limited data rate which requires data reduction of several orders of magnitude. Preliminary studies are promising~\cite{Arianna:2021vcx, RNO-G:2023oxb} and as part of the NuRadioOpt project, this approach has been shown to potentially double the neutrino detection rate~\cite{Glaser:2023udy}, which would be equivalent to doubling the array size. 

Although ML inference on FPGAs offers clear latency advantages, developing and deploying such models remains challenging. Neural networks must satisfy strict constraints on latency, resource usage, and power consumption, while maintaining sufficient physics performance. Tools such as \texttt{hls4ml}~\cite{Schulte_2025} facilitate the translation of trained networks into FPGA firmware using high-level synthesis. However, current workflows provide limited feedback before synthesis, making it difficult to determine whether a given model will fit on a target device. Since synthesis can take hours and failed attempts are common for large networks, this significantly slows the development cycle.

Recent studies have demonstrated the feasibility of deploying advanced ML models at the Level-1 trigger. Examples include FPGA-based anomaly detection using variational autoencoders in CMS~\cite{Gandrakota:2025X5} and the deployment of continuous normalizing flows for unsupervised anomaly detection~\cite{Vaselli_2025}. These results highlight the potential of ML to enhance real-time event selection, but they also underscore the need for systematic optimization across both algorithmic and hardware dimensions. 

Additionally, several studies have systematically explored ML architectures under explicit constraints on latency, throughput, and FPGA resource usage~\cite{Jiang2019FPGAAwareNAS}. Hardware-aware neural architecture search embeds device-level constraints—such as limits on DSP slices, LUTs, BRAM blocks, and pipeline depth—directly into the optimization objective, enabling a principled balance between predictive accuracy and hardware efficiency. Complementary advances in gradient-based mixed-precision quantization further integrate numerical precision into the training loop, allowing bit-widths to be co-optimized with model parameters to reduce resource utilization while preserving physics performance~\cite{Sun2024HGQ}. These studies demonstrate that architecture design, precision optimization, and structural simplification can be unified within a coherent co-design framework for resource-constrained, real-time deployment. A similar study is presented in \autoref{sec:codesign_ad_fpga}, demonstrating how ML parameters and FPGA resources can be jointly co-designed to satisfy stringent latency and resource constraints.

Co-design of ML model architectures, hyperparameters, and FPGA hardware resources provides a path forward. By jointly optimizing accuracy, latency, energy consumption, and resource utilization, co-design enables ML deployments that satisfy the stringent constraints of real-time systems, including latency budgets as low as a few hundred nanoseconds. Such approaches will be essential for fully exploiting FPGA-based ML in the trigger and data acquisition systems of next-generation experiments.

\subsubsection{Co-Design and Optimization of Unsupervised Anomaly Detection for Real-Time FPGA Deployment}
\label{sec:codesign_ad_fpga}

At the LHC, unsupervised anomaly detection (AD) is an emerging paradigm in the search of unexpected physics processes, which avoids reliance on predefined signal models. Instead of being trained on specific new-physics signatures, a model may learn the distribution of Standard Model data and thus the capability to flag events that significantly deviate from it —-an approach that complements traditional physics-driven searches in HEP. This model-agnostic paradigm, pioneered at the Tevatron at the turn of the century~\cite{knuteson}, has been explored in multiple recent studies highlighting the need for model-independent strategies in new physics searches~\cite{Belis_2024, DAvanzo_2025}\footnote{ For a list of over 100 works focusing on AD for HEP searches see the living review here: \url{https://iml-wg.github.io/HEPML-LivingReview/}.}. Model independence is particularly critical at trigger level, where supervised algorithms may fail to capture unforeseen phenomena. The low latency of Level-1 Triggers in ATLAS and CMS and their resource constraints demand highly optimized architectures of useful machine learning models for real-time deployment~\cite{Vaselli_2025}.

Here we discuss a brief case study demonstrating the co-design of ML parameters for an unsupervised anomaly detection algorithm alongside FPGA resource constraints, illustrating a hardware-aware optimization strategy suitable for real-time trigger deployment. A Continuous Normalizing Flow model is employed for unsupervised anomaly detection and is trained via Flow Matching on Standard Model data. 
The model maps input events to a Gaussian latent space and assigns anomaly scores based on their likelihood under the learned Gaussian distribution. The training dataset consists of 100k SM events, with 20k SM events reserved for testing. Performance is evaluated using a benchmark new-physics signal, $A \rightarrow 4\ell$ (50~GeV), in which a neutral scalar boson $A$ with mass 50~GeV decays into two off-shell $Z$ bosons. Each event is represented using low-level kinematic variables and flattened into a fixed 57-dimensional input vector. The same dataset sizes are used consistently throughout the full optimization pipeline. For more information about the dataset, features and  architecture, refer to Ref.~\cite{Vaselli_2025}.

We adopt a global hyperparameter search strategy to systematically explore the design space of flow-matching–based anomaly detection models. Rather than relying on manual tuning, the problem is formulated as a bi-objective optimization task:
\[
\text{AUC} \uparrow \;\; , \;\; \text{BOPs} \downarrow
\]
where AUC measures predictive performance and Bit Operations (BOPs) serve as a hardware-aware proxy for FPGA computational cost.

The search is performed using a Non-Dominated Sorting Genetic Algorithm (NSGA-II) evolutionary algorithm, which approximates the Pareto front and identifies optimal trade-offs between accuracy and computational cost. The search space spans architectural, training, and hardware-related parameters. Hidden dimensions for the multi-layer perceptron backbone are sampled as powers of two, depth is varied, and dropouts and activations are explored. Training hyperparameters include learning rate ranges, batch sizes, and flow matching coefficients.

After completing 400 global trials, Pareto dominance filtering is applied to remove configurations that are strictly inferior in both objectives. Eight Pareto-optimal architectures are retained and then refined locally using the Tree-Structured Parzen Estimator sampler. Quantization-aware training (QAT) and sparsification techniques are then applied for hardware-efficient deployment. QAT allows networks to adapt to low-precision arithmetic during training to facilitate FPGA inference~\cite{Schulte_2025}. Unstructured pruning and structured pruning strategies are used for weight sparsity and reduced architectural size. 
Analytical hardware estimations replace BOPs with latency (cycles and nanoseconds) as the hardware objective. Pareto filtering yields five FPGA-feasible models.

Exported models are converted to high-level synthesis C++ using \texttt{hls4ml}, which bridges ML frameworks and FPGA synthesis—enabling real-time neural inference. All five models successfully pass FPGA synthesis evaluation, as seen in \autoref{tab:stage4_results}.

\begin{table*}[!t]
\centering
\caption{FPGA resource utilization and latency for the final Pareto-selected models.}
\label{tab:stage4_results}
\footnotesize
\setlength{\tabcolsep}{3pt}
\begin{tabular}{l c c c c c c c c c c c}
\hline
Model & Hidden & Layers & Act. & Bits & AUC & Spars. & Lat.\,(ns) & DSP & LUT & FF & BRAM \\
 & Dimensions &  & &  &  &  & &  &  &  &  \\
\hline
M1 & 256 & 4 & ReLU & 32 & 0.9169 & 0.050 & 40 & 5928 & 17784 & 1650 & 280 \\
M2 & 256 & 3 & ReLU & 32 & 0.9155 & 0.053 & 30 & 4904 & 14712 & 1138 & 166 \\
M3 (pruned) & 256 & 3 & ReLU & 32 & 0.8892 & 0.986 & 30 & 3312 & 9936 & 626 & 55 \\
M4 & 16 & 3 & ReLU & 16 & 0.8865 & 0.683 & 45 & 555 & 1665 & 142 & 0 \\
M5 & 16 & 2 & ReLU & 4 & 0.8852 & 0.780 & 30 & 506 & 1518 & 128 & 0 \\
\hline
\end{tabular}
\end{table*}

Despite significant variation in model size and numerical precision, all final designs achieve sub-50 ns latency. This demonstrates that structured pruning and reuse-factor scheduling effectively constrain inference delay. 

\subsubsection{Agentic coding for experiments}
\label{sec:agentic}
The longevity and evolutionary scope of the codebases of the Large Hadron Collider (LHC), consisting of tens of millions of lines of code developed and maintained over decades by thousands of physicists, pose significant challenges in terms of maintenance and long-term sustainability of large-scale scientific software. The need to modernise and sustain software systems of such complexity and durability requires the adoption of radically new computational paradigms. 

Beyond source code, similar concerns apply to the extensive corpus of technical documentation produced within these collaborations, which increasingly relies on lightweight markup languages such as Markdown and LaTeX as de facto programming languages for writing. Treating documentation artefacts as first-class, executable objects in the software ecosystem foregrounds issues of long‑term preservation, format obsolescence, and reproducibility that parallel those faced by the underlying code, and suggests the need for explicit documentation architectures, curation workflows, and sustainability policies. Studies on the preservation of TeX/LaTeX documents and on LaTeX as an archival format show that markup‑based formats can support long‑term accessibility and migration to future representation standards, but only when embedded in well‑defined repository infrastructures and conversion pipelines. At the scale of LHC‑like collaborations, documentation must therefore be designed, versioned and validated with the same engineering rigour as the software it describes, leveraging institutional document‑management systems and community guidelines to ensure continuity over the multi‑decadal lifetimes of the experiments~\cite{CERN2014documentationHL,CERN2024DMSguidelines,Wuttke2022collaborativeSoftware}.

In this context, Agentic AI Systems represent an emerging and rapidly expanding area of research that, catalysed by LLMs, offers a potential solution.
An AI agent is a computational entity that perceives its environment and takes actions to achieve goals. Although the concept of autonomous software agents dates back to the 1990s, their definition and capabilities underwent a profound transformation with the advent of Large Language Models~\cite{wang2025aiagenticprogramming,raza2025trismagenticai}.

Traditional AI agents were generally task-specific and deterministic, focused on automating limited tasks such as retrieving information, summarizing data or responding to dialogues. They often operated through single-step logic or pre-defined rules, lacking the deep reasoning, adaptability and persistence required to solve complex, multi-step problems. Their cognitive foundations were often based on symbolic logic, finite state machines (FSM) or cognitive models. Classic examples include MYCIN, ELIZA, SOAR and ACT-R.

In contrast, modern AI Agents, which have gained significant traction since 2023 thanks to LLMs, are autonomous or semi-autonomous software entities capable of interpreting high-level goals, decomposing tasks into sub-goals, planning execution strategies and dynamically adapting their behaviour based on feedback and results, requiring minimal human intervention.

This capacity for continuous reasoning, planning and adaptation positions AI agents as promising tools for the evolutionary management and sustainable maintenance of extremely large scientific codebases.

However, these promises are not without risk~\cite{pane2025impatto}. The excitement of faster code development could lead developers to neglect fundamental practices and oversight of what the AI does, resulting in the accumulation of technical debt and erosion of code quality in the long term.

Following the energy‑landscape perspective on machine learning~\cite{Ballard_2017}, where the loss function defines an effective energy over model parameters, we can heuristically extend the analogy to LLM‑based software development workflows. 

In this view, the joint system comprising the model, the evolving codebase, and the task specification explores a high‑dimensional landscape with basins corresponding to partially‑satisfactory code states. Due to architectural biases and training‑induced inductive priors, a single LLM may become trapped in suboptimal basins, analogous to local minima~\cite{2025arXiv250520340Z}. 

Collaborative or ensemble schemes, in which distinct LLMs with different inductive biases sequentially or jointly modify the code, can be interpreted as perturbations of the effective landscape, facilitating escape from such basins in a way reminiscent of multi‑funnel energy landscapes and landscape ensembles~\cite{shen2024learningdecodecollaborativelymultiple}. 

Experiments could in principle develop their own LLMs and science‑driven agentic workflows, tightly adapted to their software stacks, data formats and documentation practices, with the goal of enabling collaborators to devote more of their limited time to physics, while preserving safety and code quality. Rather than repeatedly fine‑tuning separate models on evolving, experiment‑specific corpora—an approach that quickly becomes costly, brittle and prone to obsolescence as the underlying code and documentation change—retrieval‑augmented, agentic architectures offer a more sustainable path, in which a strong general‑purpose model is grounded at inference time on versioned repositories, technical notes and operational procedures through domain‑aware retrieval pipelines. This design externalises most experiment‑specific knowledge into maintainable knowledge bases and configuration artefacts, allowing updates to propagate via standard software and documentation workflows, while reserving light‑weight fine‑tuning for stable, high‑value adaptations such as domain‑specific terminology, style or safety constraints. Such agentic systems can be staged: starting from documentation‑centric assistants that help keep manuals, READMEs and configuration files consistent and up to date, and progressively evolving towards multi‑agent workflows capable of refactoring code, orchestrating CI pipelines and supporting complex analysis tasks, always under explicit human oversight, in line with emerging empirical evidence on agentic coding tools and configuration practices.

Building on these considerations, it is instructive to distinguish between domain knowledge that is structurally stable over long timescales and information that evolves rapidly with the software and documentation ecosystem. 

Empirical studies comparing RAG and fine-tuning pipelines indicate that retrieval-augmented generation systematically improves answer quality and that domain-specific fine-tuning, when applied selectively, can further consolidate the gains provided by contextual grounding—the two approaches thus proving complementary rather than mutually exclusive~\cite{Microsoft2026augmentLLMsRAG,Pingua2025medicalLLMs,IBM2024ragvsfinetuning,balaguer2024ragvsfinetuningpipelines}.

For LHC‑scale collaborations, a pragmatic strategy is therefore to reserve fine‑tuning for slowly varying or effectively immutable components of the knowledge base—such as terminology, core physics concepts, safety norms, or enduring architectural patterns—while relying on RAG over curated, version‑controlled repositories to expose frequently updated code, configuration and documentation to agentic workflows. In this hybrid regime, the model embodies a stable prior over the experiment’s conceptual landscape, whereas retrieval pipelines continuously adapt the accessible context as the software and its documentation evolve, reducing the need for repeated re‑training cycles and aligning AI‑assisted development with existing maintenance and review processes.

Despite recent extensions of context windows into the million‑token regime, which enable models to ingest entire monographs or sizeable code modules in a single pass, this working memory remains fundamentally finite and subject to performance degradation at extreme lengths. For agentic workflows operating on modular software ecosystems and documentation spanning decades, this limitation further motivates architectures where long‑term knowledge is stored and indexed externally, and agents iteratively retrieve and summarise only the most relevant fragments, instead of relying on a single, ever‑growing prompt~\cite{DeMiliani2025understandingLLMperformance,DataAnnotation2026contextwindow,balaguer2024ragvsfinetuningpipelines}.

From a systems perspective, the co‑evolution of large scientific codebases, LLM‑based assistants and the hardware platforms on which they run naturally suggests a joint optimisation problem over software, model and infrastructure parameters. In a simplified view, one may consider an effective objective function $L(s,h,a)$ that encodes problem‑dependent trade‑offs between developer productivity, code quality, latency, throughput, energy consumption and resource costs, where $s$ denotes software‑level parameters ({\it e.g.} build configurations, dependency graphs, testing policies), $h$ hardware‑level parameters ({\it e.g.} accelerator type, precision, memory hierarchy, scheduling policies) and $a$ agentic workflow parameters ({\it e.g.} task decomposition strategies, tool‑use policies, review depth). Research on hardware–software co‑design and multi‑objective optimisation in machine‑learning systems shows that treating these layers jointly, rather than in isolation, can yield Pareto‑optimal configurations that significantly improve performance‑per‑watt and system‑level efficiency. For LHC‑scale collaborations, parametrising codebases, models and agentic workflows with explicit graph‑like structures and exposing them to such multi‑objective search would support more controlled, measurable and quality‑assured deployment of AI assistants, while making explicit the trade‑offs between energy efficiency, responsiveness and the rigorous software‑engineering standards required in high‑stakes scientific environments~\cite{horikawa2025agenticrefactoringempiricalstudy,Wali2024hardwaresoftwarecoedgeai,EtemadiIdgahi2014qualitydriven}.

\vskip 1cm

\section{Co-Design in Industrial Applications \label{s:cd_industry}}

In profit-driven industrial research the exploration of solutions for the simultaneous optimization of systems and software design has reached a more advanced stage than in fundamental science research, mainly because of larger investments as well as a higher predisposition to disruptive innovation by the industrial sector. It is precisely because of this emerging gap that we wish to examine in this section a few industrial use cases, to try and foster a synergy in the area of design optimization between experimental science and technology  --where at this point of time we believe it is the latter than may help the former move forward. 

\subsection{Digital pen}

The development of a digital pen provides a concrete example of the challenges and opportunities of hardware–software co-design. Unlike abstract benchmarks or controlled prototypes, this case is defined by real industrial constraints: form power consumption, latency requirements, and memory limitation. These constraints are not optional design choices but inherent to the product’s function and user expectations.

In what follows, we present how the project team addressed these challenges. We describe the solutions deployed, covering both hardware-level design decisions and software strategies, and highlight how co-design shaped the overall system architecture.

\subsubsection{Project presentation} 
\begin{figure*}[!h]
    \begin{center}
    \includegraphics[width=0.9\linewidth]{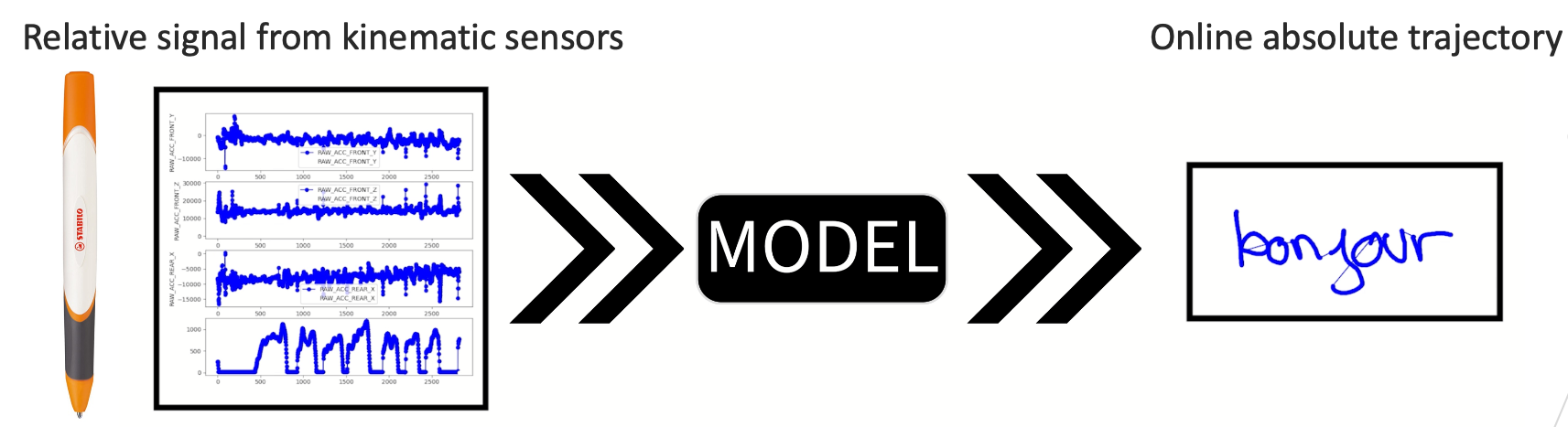}
    \caption{The goal of the digital pen project is to embed a model into the pen to enable the reconstruction of handwriting trajectories from the pen's kinematic sensors.}
    \label{fig:Digital_pen_goal}
    \end{center}
\end{figure*} 

The objective of the digital pen project~\cite{KIHT} was to design a device that integrates the familiar advantages of handwriting with the capabilities of digital systems (see \autoref{fig:Digital_pen_goal}). Handwriting remains a widely accessible and efficient mode of input, providing flexibility and immediacy compared to keyboards or touch interfaces. Reproducing these qualities while enabling reliable digital capture introduced several technical challenges.

\begin{figure*}[!h]
    \begin{center}
    \includegraphics[width=0.8\linewidth]{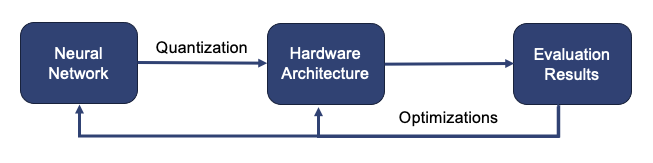}
    \caption{Hardware-software co-design workflow integrating neural-network quantization, system-on-chip requirement derivation, and system-level simulation to guide iterative HW-SW optimizations.}
    \label{fig:HW_SW_codesign}
    \end{center}
\end{figure*}

From the early stages, the project had to account for constraints related to ergonomics, battery capacity, and the need to achieve accurate, low-latency handwriting recognition. Addressing these requirements necessitated a hardware–software co-design strategy (see \autoref{fig:HW_SW_codesign}): sensing, data acquisition, and energy-use constraints were managed at the hardware level, while recognition algorithms, adaptability, and interaction mechanisms were implemented in software.

This coordinated methodology enabled performance and efficiency optimizations across system layers, without compromising the expected writing experience. The following sections describe the solutions employed at both the hardware and software levels and examine how co-design influenced the resulting system architecture

\subsubsection{Constraints on hardware, software, and data}
\label{subsec:pen-constraints}

\begin{figure*}[!h]
    \begin{center}
    \includegraphics[width=0.8\linewidth]{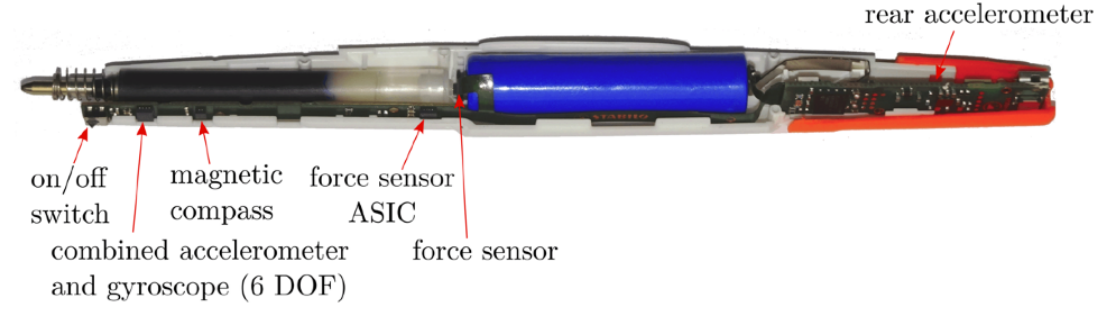}
    \caption{ \textcopyright STABILO International Digipen’s sensor location.}
    \label{fig:Digipen}
    \end{center}
\end{figure*}

The design of the digital pen is governed by strict product-level constraints that directly couple hardware, embedded software and learning algorithms. On the hardware side, the pen must remain close to the form factor of a conventional writing instrument: low weight, balanced mass distribution and comfortable grip for children and adults. These ergonomics limit battery size and therefore the available energy budget for sensing, processing and wireless communication. The current Digipen generation (see \autoref{fig:Digipen}) integrates multiple kinematic sensors (two 3-axis accelerometers, a 3-axis gyroscope, a 3-axis magnetometer and a force sensor) sampled up to 400\,Hz, and streams data over Bluetooth Low Energy (BLE), with transmission delays in the order of 10–40\,ms~\cite{ijdar_icdar_2023,igs2025_digipen} At the same time, the on-board microcontroller is constrained to sub-megabyte non-volatile memory and a few hundred kilobytes of RAM, as typical for low-power IoT-class SoCs, which sets an upper bound on model size and intermediate buffers~\cite{harbaum2024dl4iot,hal-04568785}.

On the software and algorithmic side, the main requirement is to reconstruct and interpret handwriting in (near) real time in order to provide immediate pedagogical feedback in applications such as KIHT/Kaligo~\cite{KIHT}. This excludes very deep models or algorithms with high latency or irregular control flow, and favours architectures that are both efficient and streamable ({\it e.g.}, Temporal Convolutional Networks)~\cite{ijdar_icdar_2023,harbaum2024dl4iot}. The BLE link and tablet impose additional constraints on bandwidth and latency: only a limited subset of raw or preprocessed signals can be transmitted, and any computation offloaded to the tablet must not introduce a noticeable delay from the user’s perspective, especially in learning scenarios with continuous guidance.

Finally, the project is constrained by data availability and domain variability. Ground-truth trajectories can only be collected when writing on a tablet with an EMR layer, not directly on paper, which motivated the construction of dedicated dual-acquisition datasets such as IRISA–KIHT~\cite{ijdar_icdar_2023,igs2025_digipen}. These datasets exhibit several sources of variability: different writers (children {\em vs.}\ adults), surfaces (tablet vs.\ paper) and acquisition conditions, with misaligned sampling rates and asynchronous timestamps between sensors and tablet~\cite{ijdar_icdar_2023}. Coping with this variability requires sophisticated preprocessing ({\it e.g.}, Dynamic Time Warping alignment), domain adaptation techniques and careful evaluation metrics.

\subsubsection{A priori ideas and initial design assumptions}
\label{subsec:pen-apriori}

At the beginning of the project, the team approached the problem with several “reasonable” but ultimately limiting assumptions. From an application point of view, it was tempting to treat the Digipen as a drop-in replacement for a tablet stylus: the idea was to treat Inertial Measurement Unit (IMU) signals as another form of online handwriting and reuse existing Kaligo/KIHT recognition pipelines originally designed for tablet trajectories.  Early prototypes~\cite{KIHT} relied on classical sequence models adapted from speech recognition ({\it e.g.}, Kaldi-style HMM/NN hybrids) to map sensor streams directly to characters or words, in line with previous work on IMU-based handwriting recognition.  In this view, trajectory reconstruction~\cite{S0031320324009828} was seen mainly as an intermediate by-product, and recognition accuracy was the primary metric.

On the hardware side, the initial expectation was that all “intelligent” processing could run on a companion device (tablet or PC), leaving the pen firmware responsible only for sampling, basic filtering and BLE streaming. This is consistent with many existing pen-based systems, where the pen is essentially a sensor head and all computation is offloaded. Under this assumption, model size and computational cost~\cite{hal-04358219,hal-04568785} were not considered first-order constraints; the main concern was ensuring sufficient sampling rate and robust wireless transmission. At the data level, it was also assumed that models trained on tablet-based acquisition would generalise with minimal additional effort to handwriting on paper, with the surface playing a secondary role.

The KIHT and DL4IoT studies, along with the first dual-acquisition experiments, quickly showed the limits of these assumptions: IMU-only signals are significantly noisier and structurally different from tablet trajectories; paper and tablet domains differ in friction and dynamics; and some degree of on-device processing is required to respect latency and bandwidth constraints while keeping the system scalable to large deployments in schools. These insights~\cite{ijdar_icdar_2023,igs2025_digipen,harbaum2024dl4iot}  motivated a shift towards a genuine hardware–software co-design strategy.

\subsubsection{Software choices}
\label{subsec:pen-software}

The software stack has been organised around a dedicated trajectory reconstruction pipeline, rather than treating reconstruction as a by-product of recognition. The first pillar of this stack is a Temporal Convolutional Network (TCN) architecture that maps multi-channel IMU signals and force measurements to 2D displacement vectors, trained with DTW-based alignment and evaluated using the Fréchet distance~\cite{ijdar_icdar_2023}. This TCN model demonstrated a clear advantage over earlier CNN baselines on multiple datasets (FAU–EINNS, IRISA–KIHT), particularly in its ability to exploit a temporal receptive field~\cite{ijdar_icdar_2023} adapted to the 400\,Hz sensor rate while remaining compact enough for eventual deployment.

Building on this, a Mixture-of-Experts (MoE) architecture was introduced to explicitly disentangle touching and hovering parts of the trajectory. One expert module focuses on the fine-grained reconstruction of pen-down strokes, incorporating temporal context and physical constraints, while a second expert module leverages an additional height dimension to better model pen-up repositioning.  The two expert modules~\cite{hal-04605593,igs2025_digipen} are trained jointly to produce a coherent trajectory, significantly improving the placement of new strokes, which is crucial for reconstructing complex words and multi-stroke characters. 

Domain adaptation is another key software choice. Signals acquired on tablets and on paper differ markedly, especially in noise characteristics and dynamics. A dedicated domain adaptation framework based on representation learning and adversarial / discrepancy-based losses was therefore adopted to reduce domain shift between acquisition conditions, leveraging earlier work on tablet–paper adaptation in online handwriting recognition.  This approach~\cite{hal-04605593,igs2025_digipen} proved important when moving from controlled tablet experiments to realistic paper-writing scenarios, as reported in both the ADAPDA workshop paper and in the IGS 2025 contribution dedicated to “writing on paper and getting the online trace”.

Finally, given the tight on-device constraints, the team invested in model compression and hardware-aware architecture search for the MoE experts. Using zero-cost proxies (NASWOT) and multi-objective Bayesian optimisation, TCN variants with as few as $\sim 11$\,k parameters were identified, fitting within $\approx 66$\,kB of ROM and 74\,kB of RAM while preserving acceptable reconstruction quality.  This compression step is essential to make on-pen inference feasible and strongly influenced ~\cite{igs2025_digipen,harbaum2024dl4iot,S0031320324009828} the final structure of the software models described in the thesis~\cite{tel-05037646}. 

\subsubsection{Hardware choices}
\label{subsec:pen-hardware}

The hardware architecture of the Digipen has been iteratively refined to support the needs of trajectory reconstruction while respecting industrial constraints. The sensing front-end is built around a set of low-cost, low-power inertial sensors: two 3-axis accelerometers placed at different positions in the barrel, a 3-axis gyroscope, a 3-axis magnetometer, and a unidimensional force sensor on the tip. This configuration~\cite{hal-04568785,hal-04358219,ijdar_icdar_2023,igs2025_digipen} provides sufficient redundancy to infer fine movements and pen orientation, while remaining compatible with cost and assembly constraints for mass production. Mechanical features such as moulded grips were introduced to encourage a stable hand posture and reduce orientation variability, which simplifies downstream learning.

From a processing perspective, the pen uses an ultra-low-power microcontroller with limited flash and RAM, complemented by a BLE radio to communicate with tablets and smartphones. This choice reflects the need to run at least a subset of the reconstruction and recognition pipeline locally in future revisions, without compromising battery life. Early system-level studies on hardware-aware workload distribution between pen and host device demonstrated that carefully splitting feature extraction and classification across both ends can reduce energy consumption and latency while maintaining accuracy.  These results~\cite{harbaum2024dl4iot} informed the decision to reserve a fraction of the MCU budget for on-device TCN inference, and to standardise data formats for efficient transfer of intermediate representations.

The team investigated specialised hardware accelerators for temporal models as part of a broader DL4IoT effort. Co-designed accelerators such as ATLAS (for LSTM-like models) and LOTTA (for TCNs) illustrate how approximate computing and FPGA-based implementations can dramatically reduce energy per inference in time-series workloads, including handwriting-related tasks.  While these accelerators~\cite{harbaum2024dl4iot,S0031320324009828} are not yet integrated into the current Digipen, they provide a roadmap for future versions or for docking stations where more powerful, yet energy-efficient, processing could be offloaded. The general co-design guidelines discussed in the DL4IoT paper – notably the emphasis on early hardware/software co-exploration and explicit performance–energy trade-offs – directly influenced these hardware choices.

\subsubsection{Joint hardware–software optimisation}
\label{subsec:pen-hwsw-optim}

The final architecture of the digital pen emerged from several cycles of joint hardware–software optimisation, rather than from independent tuning of sensors, firmware and neural networks.
At the sensing and pre-processing level, the decision to adopt DTW-based alignment between IMU streams and tablet trajectories was not purely algorithmic: it was enabled by precise time-stamping in the firmware and by constraining BLE packetisation to preserve temporal structure, thereby ensuring that alignment remains robust in the presence of 10–40\,ms transmission jitter. Similarly, the choice of sampling rate (400\,Hz for Digipen v6) and quantisation formats~\cite{ijdar_icdar_2023,igs2025_digipen}  was made jointly with the TCN receptive field design, so that models could exploit enough context without exploding memory requirements.

At the inference level, hardware-aware Neural Architecture Search and compression were driven by explicit hardware constraints (flash, RAM, MAC throughput) while using Fréchet distance as a task-specific performance objective~\cite{igs2025_digipen,ijdar_icdar_2023}. This resulted in a family of Pareto-optimal models that span from large, high-accuracy experts for server-side training and evaluation to compact, on-device variants capable of running on the pen’s MCU. The KIHT and “Towards the on-device reconstruction” studies~\cite{ijdar_icdar_2023, igs2025_digipen} further refined this partitioning by exploring quantisation schemes and log-based multipliers, showing that mixed-precision implementations can preserve reconstruction quality while fitting into realistic power budgets. 

Finally, the integration step for “writing on paper” crystallises the co-design loop. The IGS 2025 work showed that simply reusing tablet-trained models on paper data leads to a marked degradation, and that both the data pipeline (dual acquisition, noise filtering) and the model (domain adaptation, MoE specialisation) must be adapted to the new operating conditions.  This adaptation was carried out~\cite{hal-04605593,igs2025_digipen} while respecting the same hardware envelope, demonstrating how domain-shift handling and model compression cannot be treated separately in such a constrained system. More broadly, the Digipen use case exemplifies the pattern highlighted in recent co-design studies: only by jointly optimising sensing, embedded compute, communication and learning algorithms~\cite{harbaum2024dl4iot,S0031320324009828} can one reach the global optimum of the system’s utility function – here, accurate, low-latency handwriting reconstruction on everyday paper, at scale. 

\subsubsection{Results}
\label{subsec:pen-hwsw-results}

The digital pen project led to the development of a functional hardware–software demonstrator, validating the feasibility of real-time trajectory reconstruction and on-device inference under stringent embedded constraints. This proof of concept integrates the complete sensing front-end, the firmware pipeline ensuring temporally consistent acquisition, and a compressed TCN-based reconstruction model running within the limited memory budget of the microcontroller. The demonstrator confirms that accurate handwriting reconstruction is achievable even on everyday paper, closing the loop between sensing, embedded computation and learning algorithms.

As summarised in \autoref{tab:demo-results}, the optimized model meet strict energy, ROM and RAM constraints. The comparison with the baseline model highlights the progress achieved: compact architectures in the 11k parameter range fit within tens of kilobytes of memory while still delivering usable inference throughput, in contrast with the baseline model nearly two orders of magnitude larger which exceeds the memory envelope of the target MCU.

\begin{figure*}[!h]
    \begin{center}
    \includegraphics[width=0.99\linewidth]{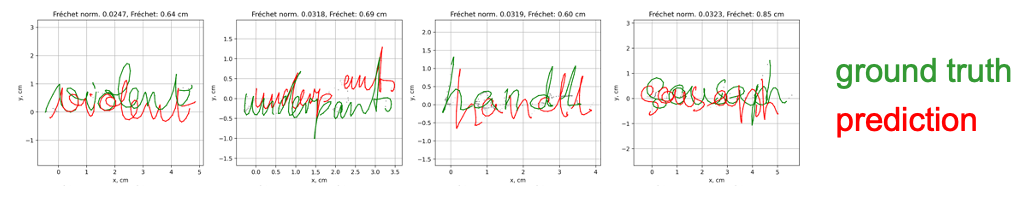}
    \caption{Some handwriting trajectory reconstruction using the optimised network. Illustration reproduced from Ref.~\cite{10773805}}
    \label{fig:Reconstruction}
    \end{center}
\end{figure*}

Reconstruction quality was evaluated using the Fréchet distance, a trajectory-level metric that captures both spatial alignment and temporal consistency. As shown in \autoref{tab:demo-results}, the optimized model achieves a Fréchet distance of 0.2456, only marginally higher than the 0.2269 score of the much larger baseline. This small degradation illustrates that the proposed architecture preserves the essential of the pen trajectories while operating under tight energy and memory constrain. Qualitative examples in \autoref{fig:Reconstruction} further show that stroke ordering, curvature, and overall letter shapes are reliably reproduced, demonstrating the practical viability of the reconstruction pipeline on real writing data.

\begin{table}[!h]
\centering
\caption{Comparison of Baseline and optimized models.}
\label{tab:demo-results}
\small
\begin{tabular}{lcc}
\hline
\textbf{Param.} & \textbf{Baseline} & \textbf{optimized TCN} \\
\hline
Kernel size & 3 & 7 \\
Layer config. & 100 / layer & [15,5,10,20] \\
Layers/stack & 2 & 4 \\
Stacks & 4 & 1 \\
\hline
\multicolumn{3}{c}{\textit{Model size and compute}} \\
\hline
Receptive field & 49 & 181 \\
Params & 467k & 10.9k \\
MFLOPs & 45.39 & 3.78 \\
\hline
\multicolumn{3}{c}{\textit{Runtime and memory}} \\
\hline
Throughput & -- & 0.61 s$^{-1}$ \\
ROM & 1850 kB* & 66 kB \\
RAM & 144 kB* & 74 kB \\
\hline
\multicolumn{3}{c}{\textit{Reconstruction quality}} \\
\hline
Fréchet distance & \textbf{0.2269} & 0.2456 \\
\end{tabular}
\end{table}

Overall, the demonstrator validates the core thesis of the project: a deliberate hardware–software co-design process can unlock the necessary trade offs to embed machine learning-based handwriting reconstruction into a low-power, pen-sized device. The integration of compact temporal models, tailored sampling strategies, BLE-aware firmware and domain-adapted learning resulted in a system that is both deployable at scale and capable of delivering high-fidelity trajectories in real-world scenarios. This proof of concept establishes a foundation for future iterations, including mixed-precision inference, lightweight accelerators, and extended on-device intelligence.

\subsection{Industrial design of hardware w.r.t. model efficiency}
\label{subsec:industrial-model-efficiency}

Artificial neural networks (ANNs) in industry have been largely optimized around \emph{parallel matrix multiplication} with floating- or low-precision activations. This has driven accelerator designs toward wide arrays and Tensor Cores, prioritizing throughput of dense (or structured-sparse) matrices at FP16/BF16/INT8 precision~\cite{jouppi2017tpu,nvidiaA100, choquette2021nvidia}. In parallel, dataflow research shows that energy is dominated by \emph{moving} activations/weights/partial sums rather than computing them, favouring architectures and loop nests that maximize on-chip reuse~\cite{eyeriss2016,scnn2017,eie2016}. Thus, "simpler" operations from a human perspective, such as integer addition, are dominated by read/write operation latency, and are comparable in speed to same byte size float multiplication~\cite{choquette2021nvidia}. In parallel, TinyML applications under hardware constraints, such as edge computing, require networks optimised for memory and processing constraints~\cite{tinyML_review}. To integrate findings from big data processing, knowledge distillation is used~\cite{distill_review} with the goal to maintain performance while scaling down model size. However, these approaches operate primarily under the assumption of frozen hardware, being limited in optimisation space to software-dependent parameters inspired by models and architectures developed for TPU-accelerated computing.

\paragraph{Event-driven alternatives}
Spiking neural networks (SNNs) and event-based perception shift computation to \emph{asynchrony} and \emph{binary events}, with neuromorphic chips such as TrueNorth, Loihi/Loihi~2, SpiNNaker, and BrainScaleS illustrating the hardware advantages of sparse, time-coded processing and co-located memory/compute~\cite{merolla2014truenorth,davies2018loihi,orchard2021loihi2,furber2014spinnaker,brainscales2022,gallego2020survey}. These platforms highlight a distinct point in the design space where model efficiency stems from event sparsity and temporal coding rather than purely from arithmetic precision.

\paragraph{Is there a “missing middle” of operations?}
Between dense FP multiply-accumulate (MAC)-heavy ANNs and event-driven SNNs lies a promising but under-explored region of \emph{operator and representation choices} that mainstream chips underserve today. There are three main axes of optimisation that can be integrated in a co-design loop:

\begin{enumerate}
    \item Precision: Recent ANN trends point towards reducing precision as a design choice in parts of the network, {\it e.g.} int8 quantatizated inference~\cite{moon2025yolov6}, which can be further explored in a hardware-software optimisation loop.
    \item Operators: MAC and matrix multiplications in general are the dominating operations in ANNs, unlike the time-encoded operations of brain-inspired SNNs.
    \item Temporal aspects: Virtually all ANNs are feedforward only at inference and time-agnostic, resulting in scaling issues in for instance continual learning and always-on sensing. Our brains and brain-inspired SNN hardware have time dependent activations and continuos analogue inputs, facilitating real-time strengthening of neurons in an always-on mechanism.
\end{enumerate}

There are existing works investigating one axis, though from a software-only perspective:
\begin{enumerate}
    \item \emph{Adder/shift networks} replace multiplications by absolute-difference plus additions or bit-shifts, maintaining accuracy while cutting energy substantially~\cite{chen2020addernet,you2020shiftaddnet};
    \item \emph{bitwise} networks approximate convolutions with XNOR+popcount~\cite{rastegari2016xnornet}; 
    \item \emph{structured sparsity} ({\it e.g.}, 2:4) enables double-throughput tensor operations but requires training-time constraints and layout-aware compilers~\cite{nvidiaA100}; 
    \item \emph{time-coded or continuous-time} nets amortize precision over time steps, bringing them closer to event-driven execution and opening room for bit-serial datapaths~\cite{chen2018neuralode}.
\end{enumerate}

We posit that these families become truly competitive when \emph{co-designed} with hardware whose primitive datapath is \emph{not} a MAC: {\it e.g.}, L1-distance engines (abs+add), shift-add PEs, XNOR+POPCOUNT tiles, bit-serial accumulators, and event-aware NoCs. 

\paragraph{Co-design loop}
Rather than developing models in isolation, we advocate a loop where model hyperparameters and hardware features are \emph{jointly} optimized. Such a loop would optimise over data precision, 
:
(1) \emph{Operator set} $\mathcal{O}$: \{MAC, ABS+ADD, SHIFT+ADD, XNOR+POPCOUNT\}; 
(2) \emph{Numeric set} $\mathcal{P}$: \{FP32/TF32/BF16, INT8/INT4, 1-bit\};
(3) \emph{Temporal regime} $\mathcal{T}$: \{frame-synchronous, event-driven, continuous-time\}.
Given a task $\mathcal{D}$ and constraints (latency/energy/area), hardware-in-the-loop tools select $(o,p,t)\in\mathcal{O}\!\times\!\mathcal{P}\!\times\!\mathcal{T}$ and train the network while measuring \emph{actual} latency/energy on a simulator or dev board~\cite{haq2019,proxylessnas2019,fbnet2019,ofa2020}. This closes the gap between FLOPs proxies and real hardware cost models. 

\subsection{Co-design in microbial metabolic modeling}
\label{subsec:co_design_microbial_modeling}

In industrial biotechnology, microbes and microbial communities can be leveraged to convert feedstocks such as waste streams into value-added products. This bioconversion is enabled by their inherent metabolic networks: the collection of intracellular enzyme-catalyzed chemical reactions that can vary substantially between different species. However, these bioprocesses often operate far from their theoretical optimal performance. Important tools that can inform on how to optimize the productive performance of microbes in bioprocesses are constraint-based models of metabolism and their integration with experimental measurements.
A clear example is butyrate production by \textit{Clostridium butyricum}, whose natural metabolism~\cite{LouisFlint2017} and pathway engineering~\cite{Fu2011} can be aided through computational flux models~\cite{Weinberger2013}.
Genome-scale metabolic models (GEMs) have been widely applied to industrial organisms such as \textit{Saccharomyces cerevisiae} and \textit{Escherichia coli} for the production of fuels, chemicals, and biopharmaceuticals~\cite{behravan2022genome, bro2006silico, chhetri2015efficient, topalouglu2023saccharomyces}.
In this context, our objective is to improve production of target compounds such as butyrate or polyhydroxybutyrate (PHB), while maintaining robust growth and process stability. Experimental approaches such as fluxomics and metabolomics provide quantitative data to constrain and validate models~\cite{orth2010flux, bordbar2014constraint}, whereas computational tools help identify bottlenecks and missing metabolic pathways. Machine learning and large language models further accelerate this process by supporting annotation, gap-filling, and literature integration. Together, these elements form an iterative co-design loop in which experiments refine models and models guide subsequent experiments~\cite{nielsen2017systems, carbonell2018automated}.

\subsubsection{Conceptual framework for bioprocess development}

Bioprocess optimization has traditionally relied on extensive experimental work. Because biological systems involve many interacting variables, approaches that vary one factor at a time are generally insufficient. Design of Experiments (DoE) provides a more systematic way to explore parameter interactions, although it still depends on prior assumptions and can require substantial experimental effort~\cite{niedz2016design, mandenius2008bioprocess}.
Advances in genome sequencing have enabled the development of predictive models such as GEMs~\cite{edwards1999systems}, which represent cellular metabolism as a network of biochemical reactions subject to physicochemical and environmental constraints~\cite{o2015using}. These models support hypothesis generation and help reduce experimental burden.
Despite this, GEMs remain incomplete and require iterative refinement, largely due to gaps or uncertainties in biological knowledge~\cite{thiele2010protocol}. Model reconstruction is therefore an ongoing process involving cycles of automated generation, curation, and validation, often informed by both literature and experimental data~\cite{mandenius2008bioprocess}. This iterative refinement process highlights the limitations of purely manual approaches and makes GEM development well suited to co-design strategies supported by ML.

\subsubsection{Reconstruction Tools}

To address these limitations and reduce manual effort, a range of tools has been developed for semi-automated and automated GEM reconstruction. For example, CarveMe~\cite{machado2018fast} and the DEMETER framework~\cite{heinken2021demeter} generate draft models from genomic data, while other tools such as ModelSEED, RAVEN, and AuReMe rely on curated biochemical databases including KEGG and MetaCyc~\cite{henry2010high, wang2018raven, aite2018traceability, kanehisa2000kegg, caspi2018metacyc}.
In practice, these approaches produce draft models that require further refinement through curation and simulation. Early methods relied on optimization and flux-based analyses to detect inconsistencies such as dead-end metabolites~\cite{bautista2013semi}, whereas more recent workflows incorporate structured curation procedures and large-scale reconstruction strategies~\cite{marin2023protocol}.
More recently, ML-based approaches have been introduced to prioritize refinement steps and identify high-impact corrections. For example, AMMEDEUS highlights reactions that strongly influence model behavior~\cite{medlock2020guiding}, while Architect improves annotation quality~\cite{nursimulu2022architect}. Emerging methods such as AlphaGEM further integrate deep learning and structural information to improve model completeness~\cite{han2025alphagem}.
Even with these advances, draft models must still be evaluated using constraint-based simulations such as flux balance analysis (FBA) within the COBRA (Constraint-Based Reconstruction and Analysis) framework~\cite{orth2010flux}, highlighting the need for continued integration of complementary approaches, including LLMs.

\subsubsection{Automated Gap-Filling with Language Models}

One key challenge in this context is gap-filling, which identifies missing reactions that prevent models from reproducing essential metabolic functions. Traditional approaches such as GapFind and GapFill~\cite{satish2007optimization} rely on constraint-based optimization and curated biochemical databases to add reactions required for growth or metabolite production, which limits their ability to propose novel or context-specific reactions.
LLMs offer a complementary strategy by leveraging unstructured biochemical knowledge from literature, annotations, and pathway descriptions. They can suggest plausible reactions, transport processes, or pathway variants that are not explicitly represented in existing databases. This capability is consistent with earlier text-mining systems such as Argo~\cite{rak2014text} and more recent LLM-based annotation tools such as GeneWhisperer~\cite{li2025genewhisperer}, as well as agent-based systems including GeneGPT and GeneAgent~\cite{jin2023genegpt, toro2024dynamic, wang2024geneagent, lobentanzer2025platform}.
Combining constraint-based diagnostics with LLM reasoning provides an effective way to identify and prioritize curation targets. 
Preliminary results in \textit{Clostridium butyricum} suggest that such hybrid approaches can significantly reduce manual effort by focusing expert attention on validation rather than discovery.
Although not yet implemented in the current study, encoding reactions using SMILES representations offers a promising extension, allowing LLMs to process biochemical transformations as structured sequences. This opens the door to sequence-based inference of missing substrates, products, or cofactors using transformer models developed for reaction prediction~\cite{andronov2023reagent, wang2022theory}.
Overall, LLM-guided gap-filling complements existing reconstruction and refinement frameworks ({\it e.g.}, DEMETER, Architect, AMMEDEUS), reducing manual bottlenecks and strengthening the integration between computational modeling and literature-derived biochemical knowledge~\cite{heinken2021demeter, nursimulu2022architect, medlock2020guiding, faria2018methods}.

\subsubsection{Hybrid ML--LLM Pipeline for GEM Curation}
\label{Hybrid ML--LLM Pipeline for GEM Curation}

\begin{figure*}[t]
\centering
\includegraphics[width=\textwidth]{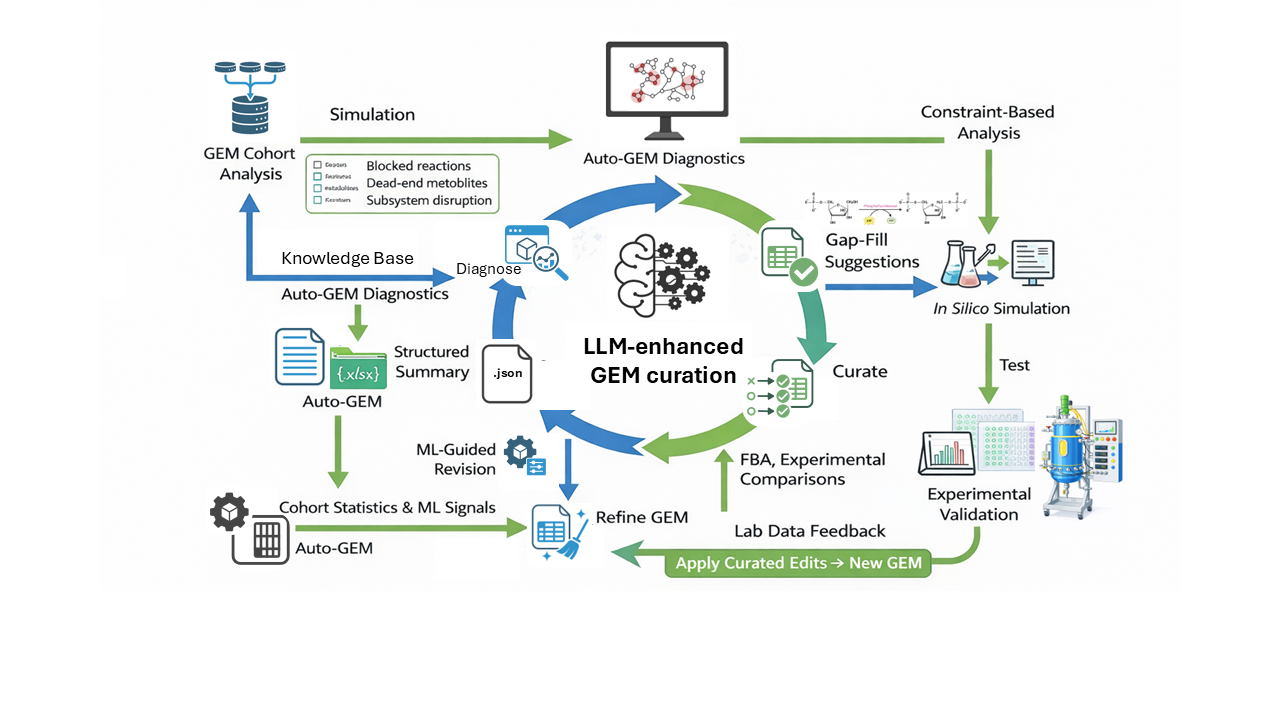}
\caption{AI-Assisted Genome-Scale Metabolic Model (GEM) Curation Pipeline. This diagram illustrates the integrated hybrid ML--LLM co-design loop. Constraint-based diagnostics (blue) identify structural inconsistencies such as blocked reactions and dead-end metabolites. These signals, combined with Auto-GEM cohort statistics, are synthesized by an LLM-enhanced curation engine to provide biologically grounded gap-filling suggestions. The process is iteratively validated through \textit{in silico} simulations (FBA/COBRA) and experimental bioreactor flux data, shifting the curation workflow from manual discovery toward automated, data-driven refinement (see \autoref{subsec:co_design_microbial_modeling}, in particular \autoref{Hybrid ML--LLM Pipeline for GEM Curation}).}
\label{fig:curation_pipeline}
\end{figure*}

To support scalable model refinement across large GEM cohorts, we developed a hybrid pipeline that combines ML-based diagnostics with LLM-driven reasoning (see ~\autoref{fig:curation_pipeline}). Constraint-based analyses identify key inconsistencies, including blocked reactions, dead-end metabolites, and abnormal exchange behavior, while ML-derived features such as subsystem coverage, gene–rule completeness, and reaction prevalence quantify deviations from expected metabolic organization across model cohorts.
Rather than operating directly on full SBML files, the pipeline extracts a concise representation of high-value information, including the most affected subsystems, representative blocked reactions, dead-end metabolites, cohort-level reaction gaps, and ML-predicted candidates for reaction addition or removal. This structured summary is passed to the LLM via an automated Python interface using a strict JSON-based schema.
The LLM interprets these diagnostics by integrating curated biochemical databases ({\it e.g.}, AGORA, KEGG, MetaCyc) with broader literature-derived knowledge~\cite{magnusdottir2017generation, kanehisa2000kegg, caspi2018metacyc}. The model then proposes targeted corrections, including transport processes, cofactor balancing steps, or missing pathway components.
Proposed modifications are iteratively evaluated through simulation and comparison with experimental data, improving agreement with observed phenotypes and flux distributions. LLM-derived suggestions can also be fed back into ML models, creating a feedback loop in which simulations detect inconsistencies, ML quantifies them, and LLMs provide biochemical interpretation. This integrated workflow supports continuous model refinement and improves predictive performance.

\subsubsection{LLMs in the Experimental–Computational Co-Design Loop}

Within this broader co-design framework, beyond the specific pipeline described above, experimental data and computational models continuously inform one another. Fluxomics and metabolomics provide quantitative constraints, while GEMs and ML tools such as Medusa and AMMEDEUS identify discrepancies between predicted and observed behavior~\cite{medlock2020medusa, medlock2020guiding}.
LLMs contribute by proposing biologically plausible model adjustments grounded in curated databases and literature. In the case of \textit{Clostridium butyricum}, discrepancies between experimentally measured fluxes and FBA predictions were addressed through ML diagnostics followed by LLM-guided refinement.
Incorporating LLMs into this workflow transforms GEMs from static reconstructions into adaptive metabolic representations that evolve as new data accumulate. As phenotypes are measured or environmental conditions change, LLMs can propose updated transporters, alternative pathways, and literature-supported revisions. Existing large-scale frameworks such as HumanGEM/THG already demonstrate the value of systematic curation~\cite{marin2023protocol}, and LLMs extend this capability by reasoning over unstructured biochemical knowledge. When combined with text-mining systems such as Argo~\cite{rak2014text}, they can support multiple stages of the reconstruction process, from annotation to validation, thereby improving alignment between in vivo metabolism and in silico models and enabling more reliable prediction for metabolic engineering applications.
 
\subsubsection{Summary}
GEM curation is increasingly moving from manual, expert-driven workflows toward hybrid approaches that integrate automated reconstruction, ML-based diagnostics, and LLM-driven reasoning. Within this co-design paradigm, experiments inform models, models guide experiments, and LLMs help translate biochemical knowledge from both structured and unstructured sources. This iterative loop supports the development of more predictive and adaptable metabolic models for biotechnology applications. 

\subsection{Benchmarking and Optimization of Deep Feature
Matchers for Real-Time Applications}\label{subsec:deepfeature}

We explore below the deployment of deep learning-based feature matchers in real-time computer vision pipelines, particularly in environments constrained by edge hardware. Feature matching is a fundamental task in computer vision, essential for applications such as visual localization~\cite{couturier2021review}, SLAM (Simultaneous Localization and Mapping), 3D reconstruction~\cite{rad2018feature}, scene understanding~\cite{mokayed2021anomaly, mokayed2014car}, and augmented reality~\cite{sarlin2022lamar}.

\begin{figure}[h!t]
    \centering
    \includegraphics[width=0.9\linewidth]{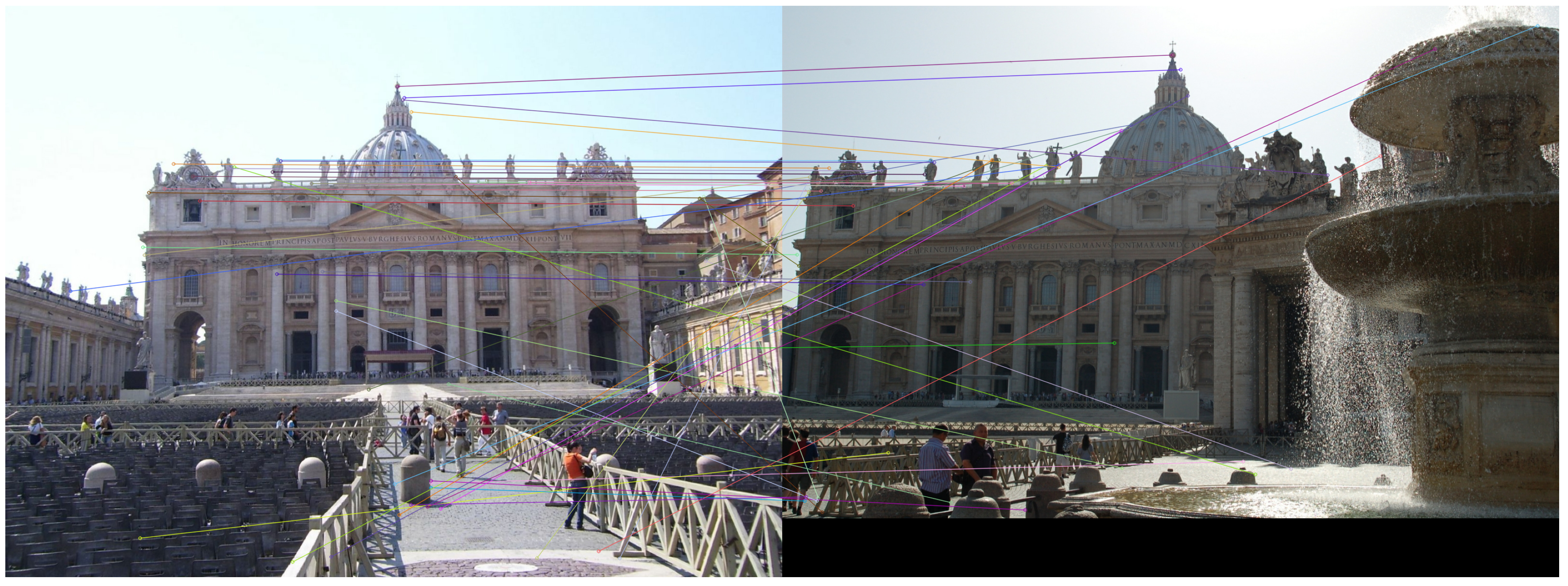}
    \caption{Visual explanation of classical feature descriptor used for feature mapping}
    \label{fig:Feat_Map}
\end{figure}

In early years, feature matching primarily relied on handcrafted descriptors. Methods such as Scale-Invariant Feature Transform (SIFT)~\cite{Lowe2004SIFT, ng2015detection}, Speeded-Up Robust Features (SURF)~\cite{bay2006surf}, and Oriented FAST and Rotated BRIEF (ORB)~\cite{Rublee2011ORB} dominated the field for over a decade due to their robustness and efficiency. More recently, state-of-the-art deep learning-based matchers such as SuperGlue~\cite{sarlin2020superglue} and LightGlue~\cite{lindenberger2023lightglue} have significantly improved matching robustness and accuracy. However, their deployment in real-time and embedded systems is limited by computational complexity, latency, and memory requirements-constraints that are especially critical in edge scenarios.

The objective of the study reported here is to benchmark and analyze the performance of SuperGlue and LightGlue across different hardware configurations. Although testing on the NVIDIA Jetson Nano was originally planned, outdated library versions and compatibility issues prevented successful execution on this device. As part of future work, the benchmarking pipeline can be extended to newer Jetson platforms ({\it e.g.}, Jetson Xavier or Orin), which offer updated software stacks and more capable hardware for deep feature matching.

Key research questions include:
\begin{itemize}
\item How do SuperGlue and LightGlue perform in terms of matching accuracy and runtime across various datasets?
\item How does the underlying hardware (CPU/GPU architecture, memory capacity, compute throughput) influence matcher performance?
\item Can lighter architectures such as LightGlue provide a viable accuracy-efficiency trade-off on edge devices?
\item What do these findings imply for hardware-software co-design in practical deployment scenarios?
\end{itemize}

This use case provides a concrete example of a non-decomposable optimization problem, where matcher selection (software) and hardware characteristics must be evaluated together to achieve optimal system behavior.

A modular benchmarking framework was developed to evaluate the performance of deep feature matchers across different datasets, hardware platforms, and runtime conditions. The framework is structured to ensure reproducibility, extensibility, and compatibility with both desktop and edge environments. The system consists of the following components:

\begin{itemize}
\item \textbf{Dataset loaders}: Implementations for HPatches~\cite{balntas2017hpatches}, MegaDepth~\cite{li2018megadepth}, and ScanNet~\cite{dai2017scannet}, providing standardized preprocessing and pair selection. These datasets cover planar scenes, outdoor 3D environments, and indoor RGB-D scans respectively, enabling a diverse evaluation of matcher behavior.
\item \textbf{Unified matcher interface}: SuperGlue and LightGlue are wrapped under a common API that handles keypoint extraction, descriptor processing, and matching. This abstraction ensures that both matchers can be evaluated under identical conditions.

\item \textbf{Evaluation metrics:} The framework computes a homography estimation error, a reprojection error, PCK-based accuracy metrics, and runtime (per pair latency). Runtime-specific information such as CPU utilization, GPU availability, compute throughput, and memory footprint are recorded when supported by the target device.
\end{itemize}

Although the framework was initially intended to run on the Jetson Nano, outdated library versions and compatibility issues prevented successful deployment. The design, however, remains fully compatible with newer Jetson devices, making the benchmarking pipeline adaptable for future evaluations on modern edge hardware.

This modular architecture allows for a transparent comparison of matcher performance across platforms and facilitates extensions such as adding new datasets, integrating additional matchers, or profiling advanced hardware configurations.

\begin{figure*}[h!t]
    \centering
    \includegraphics[width=0.9\linewidth]{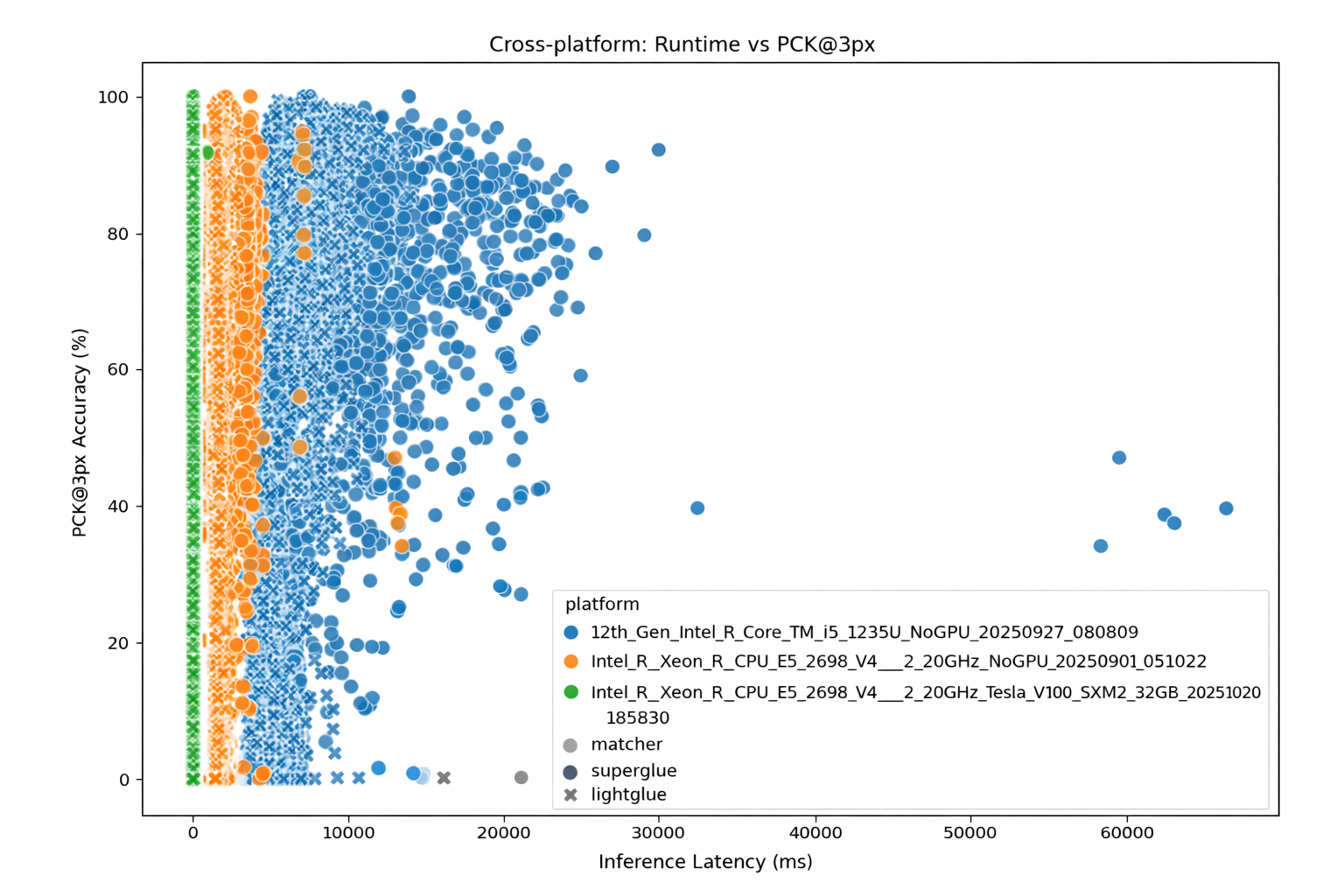}
    \caption{Cross-platform runtime vs. PCK@3px comparison.}
    \label{fig:Cross_Platform}
\end{figure*}

The benchmarking results highlight the trade-offs between accuracy, efficiency, and hardware constraints when deploying deep feature matchers in real-time settings.
SuperGlue consistently achieved the highest matching accuracy across all evaluated datasets, particularly in 3D environments such as MegaDepth and ScanNet. LightGlue, while slightly less accurate in some scenarios, demonstrated substantially lower latency. This efficiency makes it more suitable for applications where real-time performance is critical; see  ~\autoref{fig:Cross_Platform}.

We note that runtime performance varied significantly across hardware platforms. On desktop GPUs, both matchers operated well within real-time bounds. However, SuperGlue’s computational demands made it impractical for constrained devices. LightGlue, with its lighter architecture and reduced attention overhead, provided a more favorable accuracy–latency balance under limited compute resources. Due to outdated library versions and incompatibilities, neither matcher could be benchmarked on the Jetson Nano as originally planned. This limitation emphasizes the importance of software stack support in embedded deployment. Future evaluations using newer Jetson devices ({\it e.g.}, Xavier or Orin) are expected to yield more meaningful insights into edge performance.

The experiments showed that matcher performance depends not only on architectural design but also on factors such as input resolution, available GPU memory, and CPU–GPU concurrency. These dependencies demonstrate that the hardware–software interaction plays a larger role in overall performance than model design alone.

Overall, the results indicate that LightGlue provides a viable path for real-time operation on constrained devices, while SuperGlue remains preferable when accuracy is prioritized and sufficient compute resources are available (\autoref{tab:summary_results}).

\begin{table*}[ht]
    \centering
    \begin{tabular}{lccc}
        \toprule
        \textbf{Metric} & \textbf{i5 CPU} & \textbf{Xeon CPU} & \textbf{Xeon + V100 GPU} \\
        \midrule
        Runtime (ms) & LightGlue $<$ SuperGlue & LightGlue $<$ SuperGlue & LightGlue $<$ SuperGlue \\
        PCK@3px & Comparable & Comparable & LightGlue $\gtrsim$ SuperGlue \\
        Median Error (px) & LightGlue $<$ SuperGlue & LightGlue $<$ SuperGlue & LightGlue $<$ SuperGlue \\
        Efficiency (PCK@3px / runtime) & LightGlue $>$ SuperGlue & LightGlue $>$ SuperGlue & LightGlue $>$ SuperGlue \\
        \bottomrule
    \end{tabular}
    \caption{Comparison of runtime and matching accuracy across hardware platforms highlights a key deployment trade-off: SuperGlue maximizes accuracy, while LightGlue significantly reduces latency, offering a more practical solution for real-time systems.}
    \label{tab:summary_results}
\end{table*}

In conclusion, this study underscores the importance of evaluating deep feature matchers within the context of their target hardware. The benchmarking results show that matcher selection should not be made in isolation: accuracy, latency, and resource usage are tightly coupled with hardware characteristics, making the problem inherently non-decomposable.

SuperGlue delivers strong accuracy across diverse datasets but is difficult to deploy on constrained platforms due to its computational demands. LightGlue offers a more favorable balance between efficiency and performance, making it a viable candidate for real-time applications, particularly on edge devices. Although the Jetson Nano could not be evaluated due to outdated software support, the benchmarking framework remains compatible with newer Jetson architectures, providing a clear path for future experimentation.

Overall, the presented benchmarking pipeline and analysis contribute to a practical understanding of hardware–software co-design for real-time vision systems. By jointly considering model behavior and deployment constraints, practitioners can make more informed decisions about matcher–hardware combinations tailored to specific application requirements.

\subsection{LLM training}
\label{llm_training}

\paragraph{Introduction}
Here we wish to discuss the onset of strong hardware-software coupling in the training of LLMs in a manner that is grounded in the recent literature,
addressing 
the inter-relation of model parameters, hardware parameters and data.
Ignoring this inter-relation certainly leads to sub-optimal performance, as exemplified in Ref.~\cite{hoffmann2022training}.
LLMs are usually deep neural networks with tens or hundreds of billions of parameters and training them is known to be computationally expensive~\cite{nakamura-etal-2025-aurora}.

The scaling laws have shown that we need to optimize the hardware (especially for FLOPs (Floating-Point Operations per second), the training data size (or expressed as longer training duration) and the algorithm (with respect to model parameters).
These factors form the scope of our use case discussion, including energy and CO2e (carbon dioxide emissions).
Essentially, this use case shows that co-design leads to better model performance if there is (1) increased model size and hardware power or (2) increased model size and training data or (3) increased model size, hardware power and training data~\cite{wu2024towards,chen2025sub,pearce2024scaling}.
The co-design ablation studies (based on scaling) with regards to the 3 main resources are confirmed in several papers, as established in this work.
The main contributions of our use case discussion are two-fold:

\begin{enumerate}
    \item We factor in the energy costs and CO2e for the projections from the scaling laws, which has not been done in previous studies (to the best of our knowledge).
    
    \item We confirm that better model performance results from co-design through the relatively extensive literature survey.

\end{enumerate}

\paragraph{Method}
\label{llms_method}
To get insight from the literature about how co-design is treated in the LLM community, we followed the Preferred Reporting Items for Systematic Reviews and Meta-Analyses (PRISMA) standard in surveying recent literature~\cite{page2021prisma}.
The method consists of the following key steps.

\begin{enumerate}
    \item We identified papers in two databases: Google Scholar and Scopus, using the search criteria (without case sensitivity): \textit{Optimal} AND \textit{LLMs} AND \textit{scaling} AND \textit{law}.\footnote{Conducted on 2025-09-26.}
    Since so many papers were returned in the search, we limited ourselves to the initial 5 pages of results, due to time constraint and since the most relevant papers appear in the initial pages~\cite{adewumi2024fairness}.
    
    \item We removed papers based on the exclusion criterion of repetition or non-relevance by using the paper titles to assess.
    
    \item Thereafter, we screened the remaining papers and again applied the exclusion criterion of repetition or non-relevance by using the abstracts to assess.
    
    \item The final papers remaining in the review were then studied. We provide the CSV files of the papers for validation.\footnote{\url{https://drive.google.com/drive/folders/ 1lzlSf20rsEty0\_KRMUswZka\_mUpu Qa5S?usp=sharing}}
    
\end{enumerate}

\paragraph{Findings}
\label{llms_findings}

Table~\ref{llms_searchtable} summarizes the results of the search.
Most of the Scopus articles were review papers and many did not match our inclusion criteria.
After careful review of the final 63 remaining papers, we observed that there are 3 main categories of the scaling laws.
These are pretraining scaling, post-training scaling and test-time scaling (or long thinking) laws.
We describe each category in the sub-sections below.

\begin{table}[!ht]
\centering
\caption{PRISMA search outcome}
\label{llms_searchtable}
\begin{tabular}{l|r|r}
\hline
      \textbf{PRISMA Step}   & \textbf{Scholar} & \textbf{Scopus} \\
      \hline
      Initial search results & 26,100 & 914 \\
      \hline
      Limitation to initial 5 pages & 50 & 50 \\
      \hline
      After removal - based on exclusion & & \\
      criteria or repetition in titles & 47 & 36 \\
      \hline
      Reviewed - After screening, based on & & \\
      exclusion criteria or non-relevance in & & \\
      abstracts & 47 & 16 \\
 \hline
\end{tabular}
\end{table}

\paragraph{Pretraining scaling}
Papers that addressed the pretraining scaling generally identify that model performance improves but the scaling trend is becoming unsustainable~\cite{verma2025uncovering,goel2025position,hoffmann2022training}.
Some researchers have even introduced new scaling laws within this category~\cite{bi2024deepseek,wu2025inference,xiao2024densing}. Ref.~\cite{hossain2025generalizing} proposed a unified framework for both dense and sparse models. The study in Ref.~\cite{NEURIPS2024_cf5a019a} went further and showed that jointly scaling the vocabulary and model sizes leads to efficient performance.
Finally, Ref.~\cite{ge2025bytescale} introduced \textit{ByteScale} for scalable LLM training for large-scale mixed training of long and short sequences. Further, Ref.~\cite{NEURIPS2024_1cded4f9} proposed an observational approach, which bypasses training, and instead creates scaling laws from about 100 publicly available models, thereby showing that several
emergent phenomena follow a smooth, sigmoidal behavior and they are predictable
from small models.

\paragraph{Post-training scaling}
Ref.~\cite{lin2024selecting} explored the scaling laws for supervised finetuning (SFT) and Ref.~\cite{roberts2025compute} showed that scaling laws are task-dependent.
The laws have also been investigated in other downstream tasks~\cite{wyatte2025scaling}.
Ref.~\cite{zhang2024scaling} explored full model finetuning and parameter-efficient finetuning (PEFT).
Ref.~\cite{chen2024scaling} introduced a two-stage solution for predicting downstream performance.
Ref.~\cite{liu2025paretoq} presented \textit{ParetoQ}, comparing quantization settings of 1-bit, 1.58-bits, 2-bits, 3-bits, and 4-bits.

\paragraph{Test-time scaling}
The application of more compute at inference in test-time scaling has been observed to improve accuracy for reasoning models~\cite{wu2025inference,liu2025can,singhi2025solve}.
Also, combining smaller models with advanced inference algorithms offers Pareto-optimal trade-offs in performance and cost.
The study in Ref.~\cite{snell2025scaling} identified the strategy to accomplish such compute-optimal scaling as the one that chooses hyperparameters for maximal performance on a given prompt.
There have been other works on post-training scaling~\cite{PAN2026104394}.

\begin{table*}[!ht]
\centering
\caption{Estimated optimal training FLOPs with energy costs and CO2e, extended from Ref.\cite{hoffmann2022training}.}
\label{llms_matrix}
\begin{tabular}{s | s | s | p{0.15\linewidth} | p{0.1\linewidth}}
\hline
      \textbf{Parameters}   & \textbf{FLOPs} &  \textbf{Tokens} &  \textbf{Energy (GWh)}  &  \textbf{CO2e} (t) \\
      \hline
       400 M & 1.92e+19 & 8.0 B & 0.27 MWh & 0.11 \\
      1 B & 1.21e+20 & 20.2 B & 1.68 MWh & 0.67 \\
      10 B & 1.23e+22 & 205.1 B & 0.17 & 0.07 K\\
      67 B & 5.76e+23 &  1.5 T & 8.00 & 3.20 K\\
      175 B &  3.85e+24 & 3.7 T & 53.47 & 21.39 K \\
      280 B & 9.90e+24 & 5.9 T & 137.5 & 55.00 K\\
    520 B & 3.43e+25 & 11 T & 476.39 & 190.56 K\\
      1 T & 1.27e+26 & 21.2 T & 1,763.89 & 705.56 K\\
      10 T &  1.30e+28 & 216.2 T & 180,555.56 & 72.22 M\\
      \rowcolor{white} 100 T & 1.30e+30 & 2162 T & 18,055,555.56 & 7,222 M\\

 \hline
\end{tabular}
\end{table*}

\paragraph{Discussion}
Co-design is useful in many areas and it has been demonstrated that the same types of scaling laws for LLMs arise in world modeling and imitation learning though the coefficients are heavily influenced by the tokenizer, task and architecture~\cite{pearce2024scaling}.
To achieve compute-optimal training, Ref.~\cite{hoffmann2022training} further noted that equal scaling is required.
This equal proportion is far more than what was earlier recommended by Ref.~\cite{kaplan2020scaling}, who were of the view that very large model parameters should go with modest data when coupling the design elements.
It has also been observed that performance reaches saturation, especially if any of the elements is fixed while still increasing the other~\cite{kaplan2020scaling,wu2025inference}.
The general scaling law is given by Ref.~\cite{hoffmann2022training} and expressed in the equation below, where \textit{N} is the number of model parameters, \textit{D} is the number of training tokens, \textit{L} is the final pretraining (cross-entropy) loss, and \textit{C} is the hardware computation budget or FLOPs.
Minimizing the loss consequently maximizes the utility function, which, for LLMs, may be evident in increased scores of different metrics, depending on the task.

\begin{equation}
N_{opt} (C), D_{opt} (C) = argmin\: L (N,D)\,.
\end{equation}

\begin{figure}[!ht]
\centering
\includegraphics[width=0.45\textwidth]{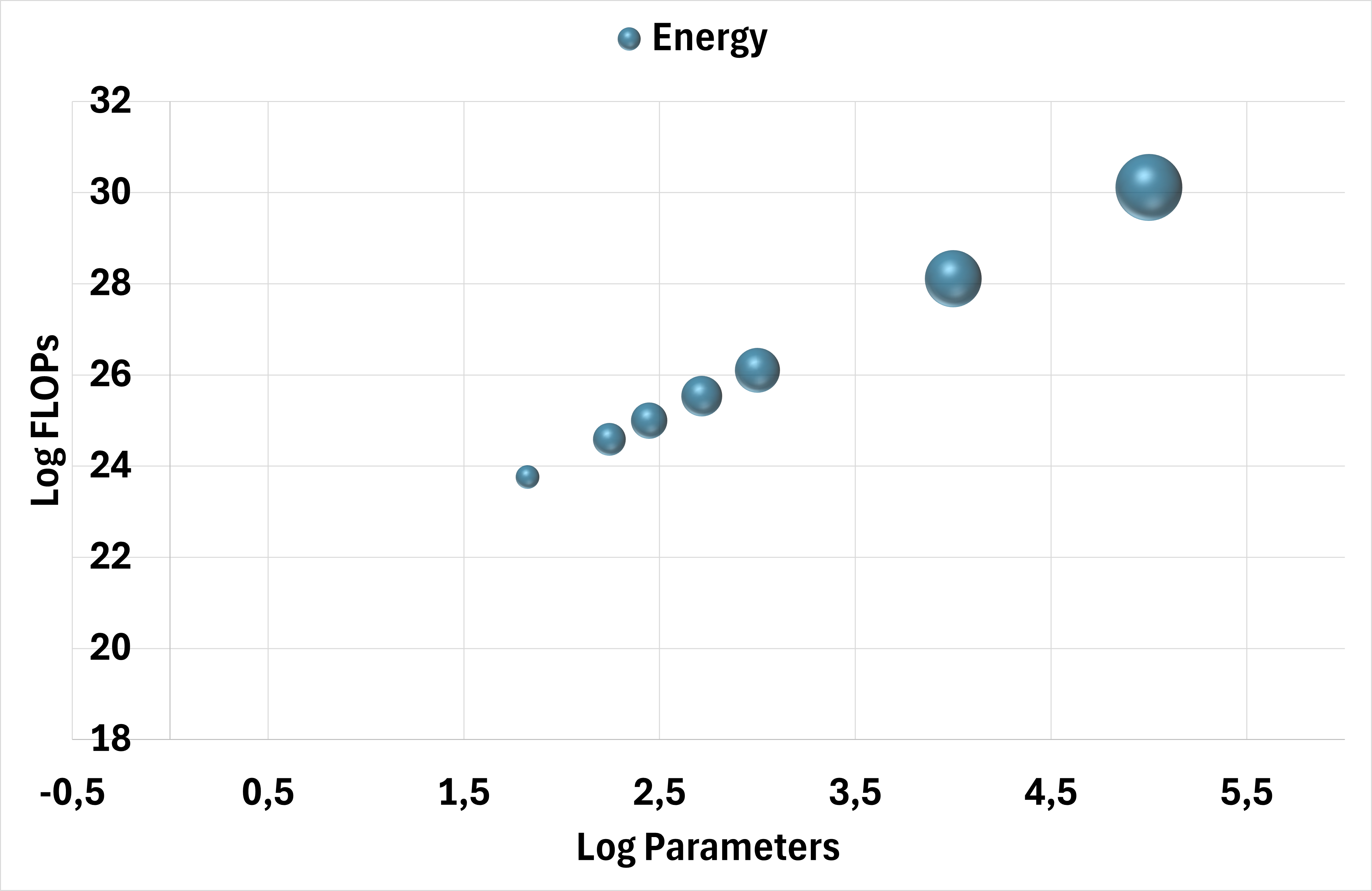}
\caption{Bubble plot of energy in log10. (Bubble shapes merely indicative.)} 
\label{llms_bubbles}
\end{figure}

\begin{figure}[!ht]
\centering
\includegraphics[width=0.45\textwidth]{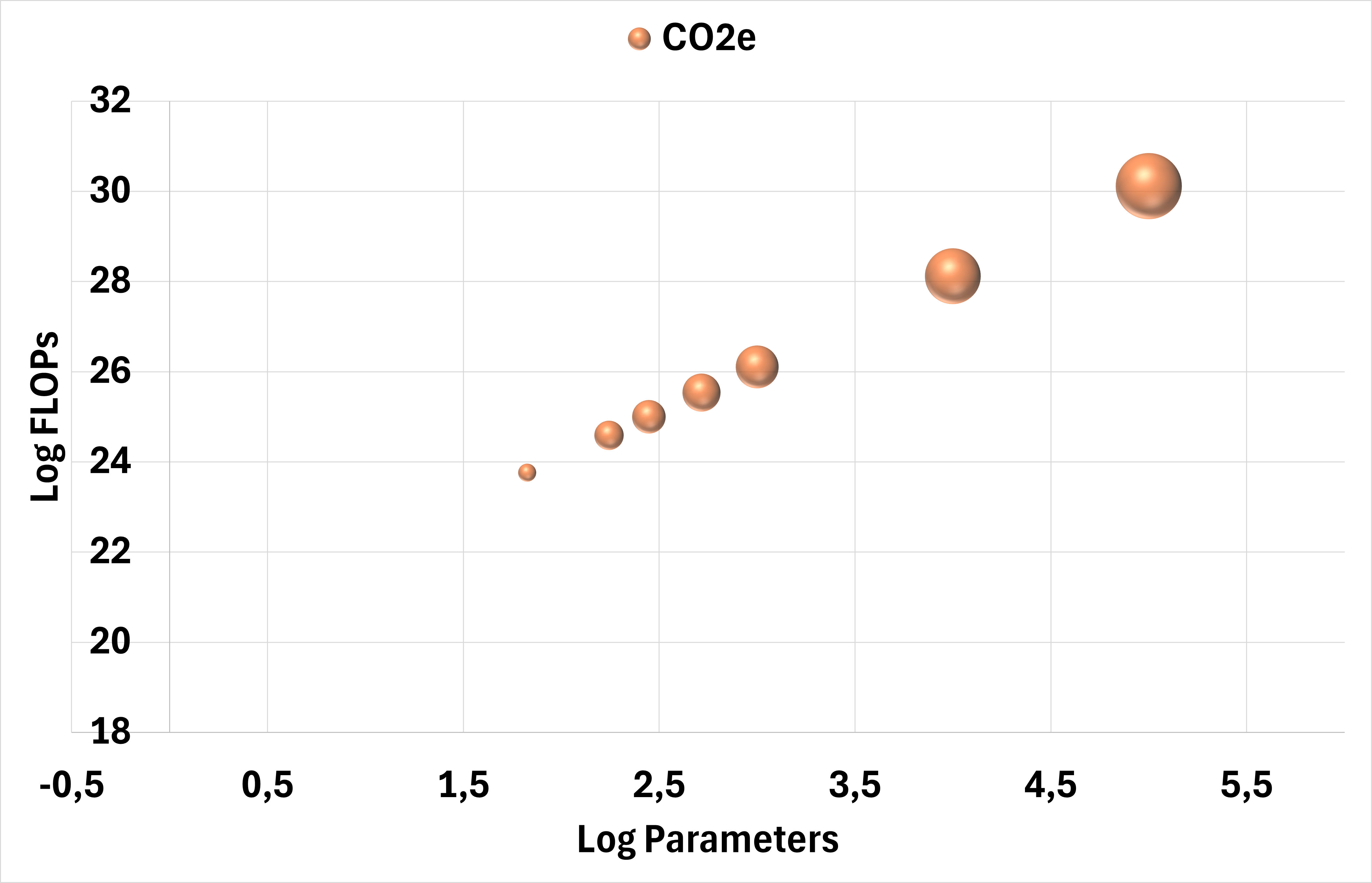}
\caption{Bubble plot of CO2e in log10. (Bubble shapes merely indicative.)} 
\label{llms_bubbles_co2}
\end{figure}

In \autoref{llms_matrix}, we projected the extra row of 100 trillion parameters, the energy column and the CO2e column from the original table in Ref.~\cite{hoffmann2022training}.
We also provide the Bubble plots of the energy and CO2e in (base 10) logarithmic scale in \autoref{llms_bubbles} and \autoref{llms_bubbles_co2}, respectively.
The projected FLOPs estimate, tokens estimate, energy estimates, and CO2e estimates were calculated using the equations below.\footnote{CO2e based on the International Energy Agency projection.}
The energy values were estimated with the assumption that training consumes 0.05 nJ per FLOP.
The extensive demand for electricity by
LLMs are linked to multiple factors like software, hardware, data center, power grid, and external societal factors~\cite{ji2026systematic}.
Given the energy consumption per capita (11.9 GWh) and CO2e (37.77 Mt) in 2024 for Sweden\footnote{\url{https://edgar.jrc.ec.europa.eu/country_profile/SWE \& worlddata}}, we can observe that, for one year, one 175B model ({\it e.g.} GPT3) consumed the energy of more than 4 Swedes and  a single 5T model's CO2e is roughly that of the whole country.
Clearly, this represents a huge environmental challenge.

\begin{align}
FLOPs(100T) & = 1.30 \times 10^{28} \times (100T/10T)^{2} \nonumber \\
 & =  1.30 \times 10^{30} FLOPs\,,
\end{align}

\begin{align}
Tokens(100T) &= 216.2T \times (100T/10T) \nonumber \\
 &= 2,162T\,,
\end{align}

\begin{align}
E(kWh) &= FLOPs \times 5 \times 10^{-11} J/FLOP \ldots \nonumber \\
& \times (1/ 3.6  \times 10^{6}) kWh/J\,,
\end{align}

\begin{equation}
CO2e (kg) = E(kWh) \times 0.4 kg/kWh\,.
\end{equation}

\paragraph{Benefits of Co-Design in LLM training}
In training LLMs, several key factors play crucial roles, including model architecture, hardware, and data.
Co-design enables us to ensure that these factors work synergistically to enhance various aspects, including environmental impact, efficiency, scalability, and cost.
We briefly discuss some benefits below.

\begin{enumerate}
    \item Environmental Impact: Making LLMs more environmentally sustainable is very important as the models continue to grow in size and complexity. Hence, reducing the power needed to train a model or choosing a well-trained model (with fewer parameters) can lead to lowering carbon emissions. 
    
    \item Efficiency: Matching hardware strengths with software needs or model size with training data size, as was done by Ref.~\cite{hoffmann2022training} in reducing model size and matching it with the right amount of tokens for a given compute budget, leads to efficiency.
    This potentially reduces slowdowns.
    
    \item Scalability: Adequate scalability can result from training differently-sized models by ensuring both hardware and algorithms meet requirements. It helps support scaling across different settings and maximizes the use of available resources.
    
    \item Cost: Cost may be reduced by optimizing how the models utilize hardware, reducing training time, conserving energy, and minimizing the use of excessive hardware.
    
\end{enumerate}

\paragraph{Use Case Limitations}
We note that the projected 100 trillion estimate comes with uncertainty, like in previous work, and at an immense hardware computation budget, energy cost and CO2e.
Also, it is uncertain if saturation will be reached before getting to this point.

\vskip 1cm

\section{Concluding Remarks \label{s:conclusions}}

In this work we have considered the optimization of complex systems in scientific and industrial applications, focusing on the interplay between their hardware and software elements. In many of the scientific applications we discuss in \autoref{s:cd_science} the problem can be cast under the guise of optimal measurement, which allows us to quantitatively assess to what degree the simultaneous consideration of hardware and software in the optimization procedures are necessary. In other systems such as those discussed in \autoref{s:cd_industry}, where economic cost is the final utility, the quantitative advantage of co-design procedures in non-decomposable problems can also be assessed clearly. However, due to the demanding nature of complete optimization solutions, we can only produce quantitative estimates in selected use cases. In others, we discuss the conditions when co-design becomes important.

In analytically tractable measurement models where the statistical likelihood is explicitly known and efficient estimators achieving the Rao-Cramér–Fréchet bound are available, system performance is determined primarily by the Fisher information of the hardware configuration. In such regimes, hardware design and inference algorithms can often be optimized sequentially, since the optimal estimator is known and independent of the particular algorithm used to implement it.

When stochastic complexity destroys this analytical tractability, however, the likelihood becomes implicit and inference must rely on approximate algorithms. In that case the achievable measurement precision depends jointly on hardware fidelity and software approximation quality. This coupling defines the regime where hardware–software co-design becomes necessary. In modern sensing systems the inference algorithm itself becomes part of the measurement apparatus, making hardware and software parameters inseparable components of the design space.

The discussion of a breadth of scientific use cases encompassing particle detectors, telescope arrays, tomography imagers, and other apparatus at the forefront of experimental investigation of fundamental science, as well as a few selected industrial use cases in computer vision, natural language processing, chemical reaction modeling, and others, provides a cross-sectional view of the effect of complexity of the systems on the optimization procedures that are required to improve their overall performance. A few commonalities between these systems indicate what synergies can be leveraged for sharing of solutions, quicker development time, and better performance. 

Although co-design has been a topic of research for a few decades, it is only since the advent of deep learning and AI technologies that a deeper understanding of the most complex applications and a real possibility to achieve holistic optimization has become possible, with the concrete chance of obtaining full alignment of the different domains of design in systems at the highest levels in the complexity scale. As research in this area involves specific competences in computer science and significant investments --both in terms of development time and in terms of GPU resources, {\it e.g.}--, the scientific community needs to thwart the danger that only market-driven applications may benefit from the required AI solutions. Organized efforts by the scientific community are therefore advisable. One such effort in Europe is the EUCAIF initiative, wherein this work was originally conceived.

\section*{Acknowledgments}
 Nicolas Gauger, Alexander Schilling, and Tobias Kortus gratefully acknowledge the funding of the German National High Performance Computing (NHR) association for the Center NHR South-West.
Luigi Favaro, Andrea Giammanco and Zahraa Zaher are supported by the Fonds de la Recherche Scientifique - FNRS under Grant No. 4.4503.16.
The work of Xabier Cid Vidal and María Pereira Martínez is supported by the Spanish Research State Agency under projects PID2022-139514NA-C33, PCI2023-145984-2 and EUR2025-164824; by the “María de Maeztu” grant CEX2023-001318-M, funded by MICIU/AEI/10.13039/501100011033; and by the Xunta de Galicia (CIGUS Network of Research Centres and project 2025-CP119). The work of María Pereira Martínez is supported by Xunta de Galicia under Programa de axudas á etapa predoutoral Grant ED481A-2025-057.
András Bóta has been supported by the Swedish Research Council through grant no. 2025-06617.
Pietro Vischia’s work was supported by the “Ramón y Cajal” program under Project No. RYC2021-033305-I funded by the MCIN MCIN/AEI/10.13039/501100011033 and by the European Union NextGenerationEU/PRTR. Pietro Vischia and Daniel Lanchares Álvarez gratefully acknowledge the use of the resources at the Artemisa computing infrastructure, funded by the European Union ERDF and the Comunitat Valenciana as well as the technical support provided by the Instituto de Fisica Corpuscular, IFIC (CSICUV).
Ippocratis D. Saltas is funded by the Czech Grant Agency (GACR) under grant number 21-16583M. The views and opinions expressed are solely those of the authors and do not necessarily reflect those of the European Union or the European Commission. Neither the European Union nor the European Commission can be held responsible for them.
Zlatan Dimitrov is supported by the GATE project under grant agreement No. 857155 and PRIDST 2021-2027 under grant agreement No. BG16RFPR002-1.014-0010-C01.
\clearpage

\bibliographystyle{elsarticle-num-5}
\bibliography{biblio}

\appendix

\end{document}